\documentclass[12pt]{article}
\pdfoutput=1 
\usepackage{amsmath}
\usepackage{geometry}
\usepackage{cmap}

\usepackage[pdfpagelabels=false,linktocpage=true,bookmarksnumbered=true,breaklinks=true,pagebackref=false,colorlinks=true,urlcolor=Mahogany,citecolor=Mahogany,linkcolor=Mahogany]{hyperref}
\usepackage{url}

\usepackage[T2A,T1]{fontenc}
\usepackage[utf8]{inputenc}  % newer: inputenx
\usepackage[ukrainian,russian,english]{babel}  % last = primary language
\addto\captionsenglish{\renewcommand\refname{\normalsize Further references:}}

\usepackage{hyphenat}
\usepackage{fancyhdr}
\makeatletter
\renewcommand{\@pnumwidth}{2.5em}% default is 1.55em
\makeatother
\usepackage{enumitem}
\makeatletter
\AddEnumerateCounter{\asbuk}{\russian@alph}{щ}
\makeatother

\newcounter{ppage}
\renewcommand{\theppage}{\mbox{P-{\arabic{ppage}}}}
\renewcommand{\thepage}{\theppage}
\usepackage[dvipsnames]{xcolor}
\usepackage{pdfpages,textcomp,multicol,tikz}

\usepackage{isoent,amssymb,revsymb}

\newcommand{\stexttt}{\fontfamily{lmtt}\selectfont\hyphenchar\font=45\relax\texttt}

\usepackage{graphicx}
\DeclareMathSizes{14.4}{13}{9}{7}  % !

\newcommand{\handcolor}{\color{blue}}
\newcommand{\ith}{\it\handcolor}
\newcommand{\udensdash}[1]{%
     \tikz[baseline=(todotted.base)]{
        \node[inner sep=1pt,outer sep=0pt] (todotted) {#1};
        \draw[thick,dash pattern=on 5.78pt off 1.63pt,dash phase=-1.43pt] 
             (todotted.south west) -- (todotted.south east);
   }%
}%

\newcommand{\uloosdot}[1]{%
    \tikz[baseline=(todotted.base)]{
        \node[inner sep=1pt,outer sep=0pt] (todotted) {#1};
        \draw[loosely dotted] (todotted.south west) -- (todotted.south east);
    }%
}%

\begin{document}
\refstepcounter{ppage}
\parindent0.cm
\thispagestyle{empty}

\ \vspace{1.2cm}

\begin{center}
\LARGE
The application of the electrodynamics of Born to the theory
of the propagation of light in electromagnetic fields [Engl.\ 
transl.\ of kand.\ diss.\ (Ph.D.\ thesis), 1936]\\[1cm]

\large
Adrian Anatol'evich Smirnov [16.11.(O.S.\ 3.11.)1908-6.12.1992]

Ural Physico-Technical Institute, Sverdlovsk

\end{center}

\vspace{5cm}

Full source details:

\vspace{3mm}

\begin{otherlanguage}{russian}
Адриан Анатольевич Смирнов
\end{otherlanguage}
%{\cyrrm Adrian Anatol\cprime evich Smirnov} 
[Adrian Anatol'evich Smirnov]:
\begin{otherlanguage}{russian}
Применение электродинамики 
Борна к теории распространения света 
в электромагнитных полях
\end{otherlanguage}
%{\cyrrm Primenenie \protect{\`{e}}lektrodinamiki 
%Borna k teorii rasprostraneniya sveta v
%\protect{\`{e}}lektromagnitnykh polyakh} 
[Primenenie \`{e}lektrodinamiki 
Borna k teorii rasprostraneniya sveta v
\`{e}lektromagnit\-nykh polyakh]/[The application of the 
electrodynamics of Born to the theory
of the propagation of light in electromagnetic fields].
\begin{otherlanguage}{russian}
Кандидатская диссертация
\end{otherlanguage}
%{\cyrrm Kandidat{s}kaya dissertatsiya}
[Kandidat$\cdot$skaya dissertatsiya]/[Ph.D.\ thesis],
\begin{otherlanguage}{russian}
Московский государственный университет
\end{otherlanguage}
%{\cyrrm Moskovski\u i gosudarstvenny\u i universi\-tet}
[Moskovski\u\i\ gosudarstvenny\u\i\ universitet]/[Moscow
State University], Moscow, 1936, 67 pp.. [in Russian]

\vspace{0.7cm}

Translator:

\vspace{3mm}

K.\ Scharnhorst (E-mail: {\tt k.scharnhorst@vu.nl},\hfill\ \linebreak
ORCID: \url{http://orcid.org/0000-0003-3355-9663}),\hfill\ \linebreak
Vrije Universiteit Amsterdam,
Faculty of Sciences, Department of Physics and Astronomy,
De Boelelaan 1081, 1081 HV Amsterdam, The Netherlands

\newpage
\refstepcounter{ppage}
\ \vspace{4.3cm}

\tableofcontents

\newpage
\refstepcounter{ppage}

\phantomsection
\addcontentsline{toc}{section}{English translation of the thesis}

\renewcommand{\thesection}{\Roman{section}}
\renewcommand{\thesubsection}{\arabic{subsection}}
\renewcommand{\thesubsubsection}{\alph{subsubsection}}
\renewcommand{\caption}{\refstepcounter{figure}}

\ \vspace{3.2cm}

\begin{center}
{\Large Adrian Anatol'evich Smirnov}\\[0.2cm]

[16.11.(O.S.\ 3.11.)1908-6.12.1992] \cite{1988smirnov,1996smirnov}\\[1cm]

{\Large The application of the 
electrodynamics of Born to the\\[0.3cm]

theory of the propagation of light in electromagnetic fields}\\[1.cm]

\end{center}

English translation of:

\begin{otherlanguage}{russian}
Применение электродинамики 
Борна к теории распространения света 
в электромагнитных полях
\end{otherlanguage}
%{\cyrrm Primenenie \protect{\`{e}}lektrodinamiki 
%Borna k teorii rasprostraneniya sveta v
%\protect{\`{e}}lektromagnitnykh polyakh}
[Primenenie \`{e}lektrodinamiki 
Borna k teorii rasprostraneniya sveta v
\`{e}lektromagnitnykh polyakh].\vspace{2.8cm}

\begin{otherlanguage}{russian}
Кандидатская диссертация
\end{otherlanguage}
%{\cyrrm Kandidat{s}kaya dissertatsiya}

[Kandidat$\cdot$skaya dissertatsiya]/[Ph.D.\ thesis],

\begin{otherlanguage}{russian}
Московский государственный университет
\end{otherlanguage}
%{\cyrrm Moskovski\u i gosudarstvenny\u i universi\-tet}

[Moskovski\u\i\ gosudarstvenny\u\i\ universitet]/[Moscow
State University],

Moscow, 1936, 67 pp.. [in Russian]

\vspace{2.3cm}

Thesis advisor: 
\begin{otherlanguage}{russian}
С.\ П.\ Шубин
\end{otherlanguage}
%{\cyrrm S.\ P.\ Shubin} 
[S.\ P.\ Shubin] 
[31.7.(O.S.\ 18.7.)1908-20.11.1938] \cite{shub1,vons1}.

\vspace{0.4cm}

Thesis defense: 28.6.1936 (approved: 17.2.1937), Moscow State University

\hspace{2.9cm}(source: \cite{zaio1}, p.\ 120, item 656).

\newpage
\refstepcounter{ppage}

\phantomsection
\addcontentsline{toc}{subsection}{Contents of the English 
translation of the thesis}
\ \\[1cm]

{\bf\large Contents of the English translation of the thesis}\\

\begin{center} % just for vertical spacing and killing indent
\begin{tabular*}{\textwidth}{@{}l@{\extracolsep{\fill}}r@{}}
{First page}&\pageref{tpage1}\\[1cm]
{\bf Chapter I}&\pageref{tglavaI}\\[3mm]
\ \ \S\ 1\ \uloosdot{\ \hspace{11.5cm}\ }&\pageref{tglavaIpara1}\\[3mm]
\ \ \S\ 2\ \uloosdot{\ \hspace{11.5cm}\ }&\pageref{tglavaIpara2}\\[3mm]
\ \ \S\ 3\ \uloosdot{\ \hspace{11.5cm}\ }&\pageref{tglavaIpara3}\\[3mm]
\ \ \S\ 4\ \uloosdot{\ \hspace{11.5cm}\ }&\pageref{tglavaIpara4}\\[3mm]
\ \ \S\ 5\ \uloosdot{\ \hspace{11.5cm}\ }&\pageref{tglavaIpara5}\\[3mm]
\ \ \S\ 6\ \uloosdot{\ \hspace{11.5cm}\ }&\pageref{tglavaIpara6}\\[3mm]
\ \ \S\ 7\ \uloosdot{\ \hspace{11.5cm}\ }&\pageref{tglavaIpara7}\\[3mm]
\ \ \S\ 8\ \uloosdot{\ \hspace{11.5cm}\ }&\pageref{tglavaIpara8}\\[1cm]
{\bf Chapter II}&\pageref{tglavaII}\\[3mm]
\ \ \S\ 1. Introduction&\pageref{tglavaIIpara1}\\[3mm]
\ \ \S\ 2. A plane wave of light in a homogeneous
electrostatic field&\pageref{tglavaIIpara2}\\[3mm]
\ \ \S\ 3. {"}Scattering"\ of light off the constant field of a
plane capacitor&\pageref{tglavaIIpara3}\\[3mm]
\ \ \S\ 4. A plane wave of light in a homogeneous magnetic field 
&\pageref{tglavaIIpara4}\\[3mm]
\ \ \S\ 5. Two plane waves in vacuo&\pageref{tglavaIIpara5}\\[1cm]
{\bf Note}&\pageref{tpage67}
\end{tabular*}
\end{center}

\newpage
\refstepcounter{ppage}

\phantomsection
\addcontentsline{toc}{subsection}{Text [pp.\ T-1 - T-67 (= P-5 - P-71)]}

\renewcommand{\thepage}{\mbox{T-{\arabic{page}}}}
\chead{{\rm\color{gray}- \theppage\ -}\hspace{3.3cm}\ }

\setcounter{page}{1}
\label{tpage1}

% italic equation numbering
\makeatletter
\let\mytagform@=\tagform@
\def\tagform@#1{\maketag@@@{$(#1)$}}
\makeatother

{\large\stexttt\

\udensdash{THE$\ \atop\ $\hspace{-0.1cm}APPLICATION OF THE 
ELECTRODYNAMICS OF BORN TO THE 
THEORY\hspace{-0.05cm}\ }\vspace{0.2cm}

\udensdash{OF THE$\ \atop\ $\hspace{-0.1cm}PROPAGATION OF LIGHT IN 
ELECTROMAGNETIC FIELDS.\hspace{-0.05cm}\ }\vspace{0.5cm}

\hspace{7.4cm}\udensdash{A.A.$\ \atop\ $\hspace{-0.1cm}Smirnov.
\hspace{-0.05cm}\ }\vspace{1.6cm}

The subject of the present dissertation consists in the consideration of some
problems of the theory of the electromagnetic field proposed recently by
\udensdash{B$\ \atop\ $\hspace{-0.33cm}orn\hspace{-0.05cm}\ } and 
\udensdash{I$\ \atop\ $\hspace{-0.33cm}nfeld\hspace{-0.05cm}\ 
}$^{\mbox{\large\stexttt{1/}}}$. We will pursue these considerations 
exclusively within the classical /not the quantum/ variant of the theory;
the question how the set-up of the problems studied changes in course
of the quantization of the field is not discussed by us. The dissertation
is split into two parts. In the first part, we give a general review 
of the current state of the Born theory hereby not dwelling on
certain particularities of the calculations, but trying to reveal
the basic course of thought, only. In particular, we are trying to
find out which general questions should be clarified yet in order to 
obtain a better understanding of the value of the theory and its 
further perspectives - hereby, of course, by far not aiming 
at giving an answer onto these questions within the present dissertation.
The second part contains details of the solution of three examples for
the application of the electrodynamics of Born

-----------------------------

1/
{\it M.\ Born. Proc.\ Roy.\ Soc.\ }\hspace{2.cm} 
\underline{143}.\ 1934.\ 410\hfill\ 

\hspace{0.5cm} 
{\it M.\ Born and L.\ Infeld. Proc.\ Roy.\ Soc.\ }\hspace{1.2cm}
\underline{144}.\ 1934.\ 425

\hspace{1.5cm} {\large\it''\ \hspace{2.6cm}''\ \hspace{2.55cm}''\ }
\hspace{2.77cm}\underline{147}.\ 1934.\ 522\hfill\

\hspace{1.5cm} {\large\it''\ \hspace{2.6cm}''\ \hspace{2.55cm}''\ }
\hspace{2.77cm}\underline{150}.\ 1935.\ 141\hfill\

\hspace{0.5cm} {\large\it E.\ Schr\"odinger  Proc.\ Roy.\ Soc.\ }
\hspace{2.5cm} \underline{150}.\ 1935.\ 465\hfill\

\pagebreak
\label{tpage2}
\refstepcounter{ppage}
\chead{{\rm\color{gray}- \theppage\ -}\hspace{3.3cm}\ \linebreak
\large\tt - \thepage\ -\hspace{3cm}\ }

onto the theory of the 
propagation of light in electromagnetic fields. In these examples,
the specific deviations \linebreak from the laws of Maxwell electrodynamics, 
predicted by the Born theory, show up in a particularly 
clear way, deviations of a completely different nature than those 
that are related to the well-known quantum effects.\\

\refstepcounter{section}
\label{tglavaI}
\hspace{3.5cm}
\udensdash{\ C$\ \atop\ $\hspace{-0.1cm}h\ a\ p\ t\ e\ r\ \ {\tt I}.\ 
\hspace{-0.1cm}\ }\\[-0.2cm]

\refstepcounter{subsection}
\label{tglavaIpara1}
\hspace{1cm}\S\ {\tt 1}. For the first time, the works of 
\udensdash{B$\ \atop\ $\hspace{-0.33cm}orn\hspace{-0.05cm}\ } and 
\udensdash{I$\ \atop\ $\hspace{-0.33cm}nfeld\hspace{-0.05cm}\ } \linebreak
gave a solution of a problem having occupied much
the minds of theoreticians in the first twenty years of the XX century -
the problem, within the framework of classical electrodynamics, 
of constructing a so-called 
{"}\udensdash{c$\ \atop\ $\hspace{-0.33cm}lassical /or unitary/\hspace{-0.05cm}\ } \linebreak
\udensdash{t$\ \atop\ $\hspace{-0.33cm}heory of field and matter"\hspace{-0.05cm}\ }
\ $^{\mbox{\large\stexttt{1/}}}$.

\hspace{1cm}The basic idea of an unitary theory - in its most
radical /partially even deliberately overstretched/ understanding -
consists therein that the existence of any electrically charged 
\udensdash{p$\ \atop\ $\hspace{-0.33cm}article\hspace{-0.05cm}\ } can
completely be described as a special \linebreak state of the 
\udensdash{e$\ \atop\ $\hspace{-0.33cm}lectromagnetic field\hspace{-0.05cm}\ }.
The standard scheme, more or less uniquely dictated, 
for the construction of such a theory looks, in the most general terms, 
- the following way.

----------------------------------

1/ We are applying here the terminology established among \linebreak physicists 
according to which  {"}field"\ and {"}matter"\ are contrasted. From out a 
general philosophical point of view, such a terminology, of course, does
not withstand a critique because the electromagnetic field itself is
also \udensdash{m$\ \atop\ $\hspace{-0.33cm}atter\hspace{-0.05cm}\ },
as any reality, existing objectively outside of consciousness.

\pagebreak
\label{tpage3}
\refstepcounter{ppage}

The laws of the electromagnetic field, best expressed in the form
of a variational principle, are taken at the foundation. As usual, the field
is described by two vectors - an electric and a magnetic one;
its Lagrangian is a function of these quantities. From the 
variational principle, as 
\udensdash{E$\ \atop\ $\hspace{-0.33cm}uler-\hspace{-0.05cm}\ }\linebreak
\udensdash{L$\ \atop\ $\hspace{-0.33cm}agrange\hspace{-0.05cm}\ }
equations, the so-called field equations derive \linebreak
/whereby, in performing the
variation, usually the additional assumption of the existence 
of a four-dimensional vector potential is introduced whose components
also serve as variable functions/. These equations immediately allow
to obtain the 
\udensdash{c$\ \atop\ $\hspace{-0.33cm}onservation laws\hspace{-0.05cm}\ } 
for the electromagnetic field by introducing the concept of the 
electromagnetic energy and the electromagnetic force of motion /momentum/.

The decisive step - the inclusion of matter into the theory 
is done the following way. A certain class of solutions of the field
equations is being considered, and in particular such solutions for which
in some chosen coordinate system the magnetic vector is equal to zero
and the electric vector - relatively symmetric. Due to the very 
structure of the field equations these two conditions are sufficient 
to determine the type of the solution completely$^{\displaystyle\ast)}$; 
just one constant factor remains arbitrary. The physical state described 
by some solution of this class is interpreted as a state corresponding to the
presence of an electrically charged
\udensdash{m$\ \atop\ $\hspace{-0.33cm}aterial\hspace{-0.05cm}\ }\linebreak 
\udensdash{p$\ \atop\ $\hspace{-0.33cm}article\hspace{-0.05cm}\ }
at rest at the origin of the chosen reference system. The remaining
arbitrary factor determines, as is being said, the total charge of
our particle; this quantity is absolutely not fixed by the theory.
\linebreak\nopagebreak[5]
\ \hspace{-1.3cm}
\underline{\ \hspace{12.5cm}\ }\linebreak\nopagebreak[5]
\ \hspace{-1.5cm}
$\ast)$ 
Recently, a new variant of an unitary theory has 
been proposed by Infeld where this uniqueness does not occur (cf.\
the example at the end).

\pagebreak
\refstepcounter{ppage}

The concrete meaning of the phrase: {"}the given state is 
being interpreted as a state corresponding to the presence of a
particle"\ mainly consists therein that the coordinates 
of the particle /or, its {"}center{"}/ are defined in terms of
the coordinates of the {"}special point"\ of the solution and the 
\udensdash{m$\ \atop\ $\hspace{-0.33cm}echanical energy\hspace{-0.05cm}\ }
and the
\udensdash{m$\ \atop\ $\hspace{-0.33cm}echanical momentum\hspace{-0.05cm}\ }
of the particle are defined in terms of the total energy and 
the total momentum corresponding to the considered state 
of the field, respectively. The meaning of these
quantities of course depends on the choice of the reference system;
in particular, in the initial system, i.e.\ in the system relative
to which the particle is at rest, the total momentum is zero /without
this fact the mechanical interpretation would of course be impossible/,
the total energy just represents the relativistic rest energy of 
the considered particle.

After the basic 
\udensdash{c$\ \atop\ $\hspace{-0.33cm}oncepts\hspace{-0.05cm}\ }
of mechanics have been introduced this way, it is necessary to establish 
its \udensdash{l$\ \atop\ $\hspace{-0.33cm}aws\hspace{-0.05cm}\ }
/of mechanics/ in order to have a complete theory. Within the given 
framework, one can obviously demand an answer
from these laws on one question only: How do the energy and the 
momentum of the particle, or, more generally the state of the 
field characterising the presence of a particle change under the 
influence of external electromagnetic fields?

In such a limited set-up of the problem, the laws of mechanics 
can be obtained, at least in principle, as a special case of the variational 
principle of electromagnetism established at the very beginning as
starting point, and this way the theory can be considered complete.

Such a kind of framework, if used in a sensible manner, 
is {"}ideal"\ in that sense that 
\udensdash{o$\ \atop\ $\hspace{-0.33cm}nly\hspace{-0.05cm}\ } 
quantities characterizing the 
field are playing the role of the basic quantities introduced into the theory
from the very beginning. 

\pagebreak
\refstepcounter{ppage}
As within the framework of Maxwell electrodynamics, the realization 
of this scheme has been impossible for well-known reasons,
besides it also other schemes of constructing an unitary theory 
have been considered which one could call more {"}compromising{"}, in a 
well-known relative sense. In particular, the 
theories of \udensdash{A$\ \atop\ $\hspace{-0.33cm}braham\hspace{-0.05cm}\ }
and \udensdash{L$\ \atop\ $\hspace{-0.33cm}orentz\hspace{-0.05cm}\ }
laid out in many textbooks are constructed on the basis of 
such {"}compromising"\ schemes. 
These theories deviate from the sketched programme in that
sense that, from the very beginning, besides the concepts of  
electric and magnetic vectors concepts characterizing 
\udensdash{c$\ \atop\ $\hspace{-0.33cm}harges\hspace{-0.05cm}\ } -
of course only covering their electric properties -
and, foremost, the concept of the charge density have been introduced. 
Instead of more abstract concepts of particles as special states
of a field, Abraham and Lorentz held ideas that were more descriptive 
at first sight: A particle was considered as a {"}charged ball{"}, 
i.e.\ as being characterized by a certain electric density distribution in 
a relatively small volume /whereby the size of 
its {"}radius"\ has been determined by means of the well-known 
relation $r_0\sim\frac{e^2}{mc^2}$/. Hereby, in difference to
an {"}ideal"\ scheme a well-known ambiguity entered into the theory
because even under the assumption of a radial symmetry of the 
particle the charge distribution {"}inside"\ of it could exhibit 
a most diverse character. Of course, this additional freedom has
been watered down by the fact that different hypotheses on the internal
{"}structure"\ of particles did not have any influence on the principle
side of the conclusions drawn from the theory.

\pagebreak
\refstepcounter{ppage}
As is well-known, the development of the ideas of 
\udensdash{A$\ \atop\ $\hspace{-0.33cm}braham\hspace{-0.05cm}\ }
and \udensdash{L$\ \atop\ $\hspace{-0.33cm}orentz\hspace{-0.05cm}\ }
has led to a number of difficulties later on /s.\ \S\ 
{\tt\ref{tglavaIpara3}} of this chapter/ and the only 
successful unitary theory existent so far - Born-Infeld theory -
is just built on the basis of an {"}ideal scheme{"}. We have recalled here
the {"}compromising"\ theories mainly for the purpose of emphasizing 
that also in these theories the introduction of the basic
\udensdash{c$\ \atop\ $\hspace{-0.33cm}oncepts\hspace{-0.05cm}\ } \linebreak
\udensdash{o$\ \atop\ $\hspace{-0.33cm}f mechanics\hspace{-0.05cm}\ }
has been carried out based on exactly the same foundation just mentioned,
namely, by means of the identification of the energy and momentum
of the electromagnetic field of a particle with its mechanical energy
and mechanical momentum. This way, just this identification should
be considered as the central point of any unitary theory. For this
reason, a unified theory of matter and field is often called a 
{"}theory of electromagnetic mass{"}.

\refstepcounter{subsection}
\label{tglavaIpara2}
\hspace{1cm}\S\ 2. To which extent can the guiding idea of an unitary theory
be considered to be physically justified?

Within the framework of classical physics, the answer onto this
question seems to be more or less clear. The fact that any 
electric charge has an electromagnetic energy and an 
electromagnetic momentum, which necessarily have to be taken
into account in writing down the conservation laws for any
system this charge belongs to, is for a classical theory
completely indisputable.

\pagebreak
\refstepcounter{ppage}
Indeed, it derives from the most fundamental foundations of Maxwell
electrodynamics.

From this, however, it does by far not follow that the 
\udensdash{w$\ \atop\ $\hspace{-0.33cm}hole\hspace{-0.05cm}\ } 
mass of any charged particle is of a purely electromagnetic origin as
an unitary theory maintains. Indeed, experience
shows that no material particle exists all whose properties
are exhausted by the fact that these particles are electrical charges.
Even putting aside gravitation /in general, in this dissertation, problems of 
gravitational actions are absolutely not considered/, we know for sure
that besides electromagnetism and gravity another completely different
type of force exists which binds the atomic nucleus together. Incidentally,
the classical theory asserts that the {"}source"\ of the mass of a 
material  particle is not only the electromagnetic field
but also any force field created by it (gravity, as said further above,
is excluded).

This way, in the radical form spelled out by us in 
\S\ {\tt\ref{tglavaIpara1}} an unitary
theory does certainly not correspond to physical reality. Irrespective 
of this, its study nevertheless is useful taking into 
account the following considerations. First, it exists such a
type of material particles - electrons /and positrons/ which, as much as
presently is known to us, are a source of only two types of force fields:
electromagnetic and gravitational. Hereby, there are all reasons to assume that
the gravitational field created by the electron exerts a comparatively
small influence on the laws governing the behaviour of it. 
Therefore, an unitary theory basically has to be considered to be 
\udensdash{a$\ \atop\ $\hspace{-0.33cm}pplicable\hspace{-0.05cm}\ }
to \udensdash{e$\ \atop\ $\hspace{-0.33cm}lectrons\hspace{-0.05cm}\ }; 
in any case, it is completely natural to try 
to construct such a theory for electrons. It is not strange that
the, in essence, completely obvious fact that any 

\pagebreak
\refstepcounter{ppage}
unitary, purely electromagnetic theory, in particular also 
\udensdash{B$\ \atop\ $\hspace{-0.33cm}orn-Infeld\hspace{-0.05cm}\ } 
theory, is the {"}largest"\ theory of the electron is being disputed in the 
literature up to now$^{\mbox{\large\stexttt{1/}}}$.
Second, an unitary theory has besides a direct physical also a
deep \linebreak
\udensdash{m$\ \atop\ $\hspace{-0.33cm}ethodological\hspace{-0.05cm}\ } 
significance. In fact, the whole development of theoretical physics 
allows to believe that the construction of a theory for different
force fields can 
\udensdash{b$\ \atop\ $\hspace{-0.33cm}e performed\hspace{-0.05cm}\ } \linebreak 
\udensdash{b$\ \atop\ $\hspace{-0.33cm}y means of methods 
basically analogous\hspace{-0.05cm}\ } to each other. Therefore,
the theory of {"}pure electricity{"}, i.e.\ imagined particles which are
\udensdash{o$\ \atop\ $\hspace{-0.33cm}nly\hspace{-0.05cm}\ } 
electrical charges, can be considered as a prototype of a theory 
of real particles serving as \linebreak sources of highly diverse fields.
Presently, the conviction rules among theoreticians that this 
assertion is applicable in general also to those recently discovered
forces acting between heavy particles - protons and neutrons and
which are responsible for the stability of the atomic 
nucleus.$^{\mbox{\large\stexttt{2/}}}$ With respect to the latter generalization
one has to mention, of course, that the proton-neutron interaction 
can apparently, in general, be expressed in terms of a field theory
using the quantum language only. We will not dwell on the
analysis of this question here because it does not have any direct
relation to our subject; we only mention that the comment made just serves 
as a reminder of the basic fact that the 
true laws of the microworld are not 

-----------------------------------

1/ See the comment by
\udensdash{B$\ \atop\ $\hspace{-0.33cm}orn\hspace{-0.05cm}\ } and
\udensdash{S$\ \atop\ $\hspace{-0.33cm}chr\"odinger\hspace{-0.05cm}\ } 
in {\it Nature 1935}.

2/ See, e.g.\ the article of W.\ Heisenberg in the {\it Zeeman}

\hspace{0.7cm}{\it Festschrift}, 1935.

\pagebreak
\refstepcounter{ppage}
classical but 
\udensdash{q$\ \atop\ $\hspace{-0.33cm}uantum\hspace{-0.05cm}\ } 
laws, a fact which plays a decisive role in all these questions 
discussed here also without any mention of it. Therefore, any theory
dealing with the behaviour of the elementary particles of matter and not taking
into account these laws can at best be considered as something
temporary, i.e.\ as a transitory stage to another, more correct theory.

Today, a relativistic quantum electrodynamics does not exist yet. 
It is well-known that the basic difficulty standing in the way
of its development - the so-called {"}infinite self-energy problem"\ - 
is the analogue of that difficulty that up to the works of Born
classical physics could not, at least not in any satisfactory manner, 
overcome in attempting the construction of an unitary theory of field
and matter. A priori, two points of view are possible on the approach
by means of which this difficulty is being resolved in the following.
One can hold to the point of view that first a correct theory of the
electromagnetic mass should be built within the framework of classical
ideas whose translation to the quantum language performed by means of
recipes analogous to the well-known recipes of {"}quantization"\ also
will yield the true relativistic micro-electrodynamics. From this
point of view, the study of the construction of Born is of fundamental
interest for all the further development of science.  However, one 
also may apply that point of view according to which the path
of the classical analogue, however fruitful it has been during
all the preceding period of the quantum theory, has fundamentally 
exhausted itself and the removal of the difficulties 
standing in the way

\pagebreak
\refstepcounter{ppage}

of the further development of this theory should and will be 
achieved along certain new, on principle, paths. From out this point 
of view, the study of the theory of Born 
can be considered more of less decidedly as a waste of time.

Presently, we do not have any clear criterion which would allow us
to decide which of these two mentioned points of view is closer
to the truth. In any case, there are good reasons to first explore
the known path of the analogy with classical physics - either this 
leads to success, or in the lesser case, will help to clarify 
at which point this path ceases to be reliable.

\refstepcounter{subsection}
\label{tglavaIpara3}
\hspace{1cm}\S\ 3. After this small excursion we return directly
to our subject again.

In \S\ {\tt\ref{tglavaIpara1}}, we have sketched the {"}ideal"\ scheme for
constructing an unitary theory of matter and field. According to this
scheme, the only freedom which remains in the hands of a theoretician
trying to construct an unitary theory consists in the choice of the 
Lagrangian of the field. Hereafter, once this choice has been made, 
everything else already follows more or less automatically /if one 
assumes the existence of a four-dimensional vector potential, in order
to not depart too much from the Maxwellian scheme/.
In this context, it is naturally to ask oneself: Which requirements
should the Lagrangian of the theory fulfill in order to in fact realize 
the programme sketched further above? Essentially, this question has 

\pagebreak
\refstepcounter{ppage}
to be answered first in critically studying the theory. It 
would mean pedantry attempting to list all these requirements; it is 
more useful to mainly concentrate onto those \linebreak 
which really significantly
constrain the class of admissible Lagrangians. One can list at least
\udensdash{s$\ \atop\ $\hspace{-0.33cm}ix\hspace{-0.05cm}\ } of them. 
We will number them by means of Latin letters. 

\setcounter{subsubsection}{0}
\refstepcounter{subsubsection}
\label{tsuba}
\ \udensdash{\ \ /a/. 
The theory should be relativistically and 
gauge in-\hspace{-0.05cm}\ }

\udensdash{\ \ v$\ \atop\ $\hspace{-0.33cm}ariant.\hspace{-0.05cm}\ }\\[-0.2cm]

\hspace{1cm}
The first half of the requirement does not need any comments. The second
half of it means that the four-dimensional vector-potential should play
a purely auxiliary role in the theory. The existence of this potential
should be considered as an abbreviated notation of certain properties
inherent to the electromagnetic 
\udensdash{f$\ \atop\ $\hspace{-0.33cm}ield\hspace{-0.05cm}\ } 
/concretely - to the first group
of the Maxwell equations/.

\refstepcounter{subsubsection}
\label{tsubb}

\ \udensdash{\ \ /b/ The equations should allow such solutions
as men-\hspace{-0.05cm}\ } \linebreak
\udensdash{t$\ \atop\ $\hspace{-0.33cm}ioned 
in \S\ {\tt\ref{tglavaIpara1}},\hspace{-0.05cm}\ }
namely,  if the magnetic field is equal to zero the electric field is
radially symmetric and the solution contains one arbitrary constant
which can be identified with the charge. Without this it would be
extremely difficult to introduce charged particles into the theory
in a natural manner.

\refstepcounter{subsubsection}
\label{tsubc}
/c/ In the coordinate system in which the condition /\ref{tsubb}/
is fulfilled 
\udensdash{t$\ \atop\ $\hspace{-0.33cm}he total angular momentum should be 
equal to zero\hspace{-0.05cm}\ } \linebreak
\udensdash{a$\ \atop\ $\hspace{-0.33cm}nd the total energy should be finite.\hspace{-0.05cm}\ }
As is well-known, the Maxwell theory did not fulfill the second
condition. The \linebreak Abraham-Lorentz theories

\pagebreak
\refstepcounter{ppage}
being based on an artificially introduced concept of the {"}radius of the 
electron"\ fulfilled it. But, on the other hand they were bad in that
they did not fulfill the following \linebreak fourth requirement without whose
applicability, again, the theory cannot be constructed. This
fourth requirement reads:

\setcounter{subsubsection}{3}
\refstepcounter{subsubsection}
\label{tsubd}
\udensdash{/d/. T$\ \atop\ $\hspace{-0.33cm}he identification of the electromagnetic 
quantities\hspace{-0.05cm}\ } \linebreak
\udensdash{w$\ \atop\ $\hspace{-0.33cm}ith the mechanical angular momentum 
and energy should have\hspace{-0.05cm}\ }
\udensdash{a r$\ \atop\ $\hspace{-0.33cm}elativistically invariant character.\hspace{-0.05cm}\ }\\[-0.2cm]

As we will see further below, this condition whose necessity is
obvious is by far not fulfilled automatically. Its violation leads
to a violation of the relativistic invariance of the theory in its
mechanical part what occurred, as is well-known, within the theory 
of Lorentz. If these four conditions are fulfilled an unitary theory
can, in principle, be constructed. 
However, in order to give it a physical meaning 
it is necessary that those basic equations of electromagnetism and 
mechanics which it leads to, in any case, approach the Maxwell-Lorentz 
and Einstein equations, respectively, in the known limit. Hereby,
the limit, of course, is not determined by quantum arguments
/which are absolutely not discussed here/ but by certain other
considerations. We will, therefore, demand the following yet:

\refstepcounter{subsubsection}
\label{tsube}
/e/ If, in constructing the theory, one has to modify the 

\pagebreak
\refstepcounter{ppage}
Maxwell equations these equations should, despite this, remain 
correct in the known limit, i.e.\ 
\udensdash{c$\ \atop\ $\hspace{-0.33cm}lassical electrody-\hspace{-0.05cm}\ } \linebreak
\udensdash{n$\ \atop\ $\hspace{-0.33cm}amics should this or that way be contained in the new 
theory.\hspace{-0.05cm}\ }\\[-0.3cm]

\refstepcounter{subsubsection}
\label{tsubf}
/f/. An analogous condition should apply also with respect to 
the new equations of 
\udensdash{m$\ \atop\ $\hspace{-0.33cm}echanics:\hspace{-0.05cm}\ }
In a certain limit they should go over into the equations of
\udensdash{r$\ \atop\ $\hspace{-0.33cm}elativistic 
mechanics\hspace{-0.05cm}\ } \linebreak
\udensdash{o$\ \atop\ $\hspace{-0.33cm}f Einstein.\hspace{-0.05cm}\ }

\refstepcounter{subsection}
\label{tglavaIpara4}
\hspace{1cm}\S\ 4. 
\udensdash{B$\ \atop\ $\hspace{-0.33cm}orn\hspace{-0.05cm}\ }
and \udensdash{I$\ \atop\ $\hspace{-0.33cm}nfeld\hspace{-0.05cm}\ }
succeeded in constructing a Lagrange function fulfilling all six
requirements listed above.

We will try to briefly recall their line of thought. The variational
principle serving, as always, as starting point of the theory has
the form
\begin{eqnarray}
\label{teq1}
\ \hspace{-4cm}\delta\int L\ d\tau &=&0\ ,\\[0.3cm]
\ \hspace{-6cm}{\it where}\hspace{2cm}d\tau &=&dx^1 dx^2 dx^3 dx^4\ . 
\nonumber
\end{eqnarray}
In order to fulfill the first half of the requirement /\ref{tsuba}/
the quantity $L\ d\tau$ should be an invariant in the sense of the 
general theory of relativity. Consequently, we should have
\begin{eqnarray}
\label{teq2}
\ \hspace{-5cm}L\ d\tau &=&Inv\ . 
\end{eqnarray}

\hspace{1cm}How should $L$ transform to achieve this? The well-known
answer to this

\pagebreak
\refstepcounter{ppage}
reads: $L$ should transform as an expression of the form
\begin{eqnarray}
\label{teq3}
\sqrt{\vert a_{ik}\vert}&&  
\end{eqnarray}
where $\vert a_{ik}\vert$ is the determinant constructed from the
covariant components of an arbitrary 2nd rank tensor.

\hspace{1cm}According to the basic idea of the theory 
/see \S\ {\tt\ref{tglavaIpara1}}/,  the Lagrange function should
depend on the components of the \linebreak field tensor 
$f_{ik}$\ \ $^{\mbox{\large\stexttt{1/}}}$ and, obviously, on the components
of the metric tensor $g_{ik}$ because the theory is constructed within
the framework of the general principle of relativity. Taking this 
into account it is natural to search $L$ among the expressions of 
the type 
\begin{eqnarray}
\label{teq4}
\sum_\alpha\ A_\alpha\ 
\sqrt{\sum_\beta B_{\alpha\beta}\ \vert a_{ik}^{\ (\alpha\beta)}\vert}\ \ ,
\end{eqnarray}
where $A_\alpha$ and $B_{\alpha\beta}$ are invariants, 
and $a_{ik}^{\ (\alpha\beta)}$ are tensor components somehow depending
on $f_{ik}$ and $g_{ik}$.

\udensdash{B$\ \atop\ $\hspace{-0.33cm}orn\hspace{-0.05cm}\ }
and \udensdash{I$\ \atop\ $\hspace{-0.33cm}nfeld\hspace{-0.05cm}\ }
give two expressions of the type /\ref{teq4}/ which, standing at
the foundation of the theory, fulfill all six requirements 
laid out above, namely:

------------------------------

1/ Other quantities characterizing the electromagnetic
field, e.g.\ the components of the vector-potential, cannot
enter for the requirement of gauge invariance which, this way,
is fulfilled in this theory automatically.

\pagebreak
\refstepcounter{ppage}
\begin{eqnarray}
\label{teq5}
L_1&=&\sqrt{-\vert g_{ik} + f_{ik} \vert}\ -\ 
\sqrt{-\vert g_{ik} \vert}\\[0.3cm]
\label{teq6}
L_2&=&\sqrt{-\vert g_{ik} + f_{ik} \vert + \vert f_{ik} \vert }\ -\ 
\sqrt{-\vert g_{ik} \vert}
\end{eqnarray}
or
\begin{eqnarray}
\label{teq7}
L_1&=&\sqrt{-\vert g_{ik} \vert}\ \left(\sqrt{1 + F^2 - G^2} - 1\right)\\[0.3cm]
\label{teq8}
L_2&=&\sqrt{-\vert g_{ik} \vert}\ \left(\sqrt{1 + F^2} - 1\right)\ \ ,
\end{eqnarray}
where
\begin{eqnarray}
\label{teq9}
F&=&\frac{1}{2}\ f_{ik} f^{ik}\\[0.3cm]
\label{teq10}
G&=&\left(f_{23}\, f_{14} + f_{31}\, f_{24} + f_{12}\, f_{34}\right)
\frac{1}{\sqrt{-\vert g_{ik} \vert}}
\end{eqnarray}
also are spatial invariants.

\hspace{1cm}$\bigl[$ In these formulas the fields are, of course, 
expressed in relative units. For the transition to the usual 
units, we should introduce into the formulas a factor {"}$b$"\ having
the dimension of a field strength. Then, for example, the formula
/\ref{teq8}/ assumes the form
\begin{eqnarray}
\label{teq11}
L_2&=&\sqrt{-\vert g_{ik} \vert}\ 
\left(\sqrt{1 + \frac{f_{ik} f^{ik}}{2 b^2}} - 1\right)\ .
\end{eqnarray}
Also the formula /\ref{teq7}/ can be rewritten in an analogous manner.
In the following, it turns out that the constant {"}$b$"\ plays the  
role of a {"}critical"\ field, in a sense in an analogous manner 
as the velocity of light plays the role of a critical velocity 
in relativity theory. The order of magnitude of {"}$b$"\  

\pagebreak
\refstepcounter{ppage}
turns out to be extremely large - of the order of magnitude of
I0$^{\mbox{\large\stexttt{I6}}}$ abs.\ units. $\bigr]$

One easily sees that for fields small in comparison with the 
{"}critical{\it "\ } field $b$ both Lagrange functions of the theory of Born
go over into the Lagrange function of Maxwell, i.e.\ the 
requirement /\ref{tsube}/ is fulfilled. We recall that in the 
Maxwell theory the Lagrangian assumes the form:
\begin{eqnarray}
\label{teq12}
&&\frac{1}{4}\ \sqrt{-\vert g_{ik} \vert}\ f_{ik} f^{ik}
\end{eqnarray}
which is also in accord with the general formula /\ref{teq4}/.
The principle difference between this formula /\ref{teq12}/ 
and /\ref{teq5}/ and /\ref{teq6}/
consists in the fact that for Maxwell in /\ref{teq12}/ under the root
only the metric tensor remains while the components of the field
tensor enter one of the invariants $A_\alpha$ only.
In the theory of Born, as is visible from the formulas 
/\ref{teq5}/ and /\ref{teq6}/, quantities characterising the field
and the metric as well are standing under the root and the 
invariants  $A_\alpha$ and $B_{\alpha\beta}$ \linebreak
turn out to be simple
numbers. As will become clear in the following, this fact 
essentially impacts the shape of the new field equations making
them nonlinear in the fields in difference to the Maxwell equations.

\hspace{1cm}To which extent is the choice of the functions 
/\ref{teq5}/ and /\ref{teq6}/ unique and how, at all, can 
the question concerning the uniqueness be posed in the present case? 

It is completely clear that just the requirement of the relativistic 
invariance of the action integral is by far not sufficient to
choose uniquely even the type of the Lagrangian.$^{\mbox{\large\stexttt{1/}}}$

-----------------------

\hspace{-0.8cm}$1)$\ The comment by Born, that the usual choice of 
the Lagrangian in the form $Inv \cdot \sqrt{-\vert g_{ik} \vert}$
where $Inv$ only depends on $f_{ik}$ is {"}formal{"}, hardly is important
because even the criterion of {"}formality"\ is fairly unclear. 
We just mention that one of the expressions recommended by Born
himself - namely the expression /\ref{teq6}/ - also belongs to
the type criticised by him as is particularly clearly visible 
from /\ref{teq8}/. 

\pagebreak
\refstepcounter{ppage}
If we supplement it by the requirement of the limiting transition
to the Maxwell theory yet this already considerably narrows the class
of admissible Lagrangians.

However, also this does not yield, of course, a complete \linebreak 
uniqueness of the choice: It is sufficient just to mention the 
existence of the three functions  /\ref{teq7}/, /\ref{teq8}/, and  
/\ref{teq12}/.

It would be interesting to find out whether it would be sufficient
to add to these two requirements the requirement of the finiteness 
of the energy of a particle yet in order to fix the 
general type of the Lagrange function at least. More narrowly, this question
can be put as follows: Do such Lagrangians exist which obey all the six 
requirements given in \S\ {\tt\ref{tglavaIpara3}} and which, at the same time, 
differ essentially from the expressions /\ref{teq5}/ and 
/\ref{teq6}/.$^{\displaystyle \ast)}$ We note that the expressions /\ref{teq5}/ and 
/\ref{teq6}/ are essentially of the same type and lead, in a number
of cases, even to completely identical results.

Indeed, this is the case in the elementary static problem.
Also, in all the cases of weak fields which we will be interested in
in the second chapter the difference between these results is
apparently inessential. Only for strong fields /near charged particles/
the results which both Lagrangians lead us to are essentially 
different and the question, which of these functions one has to prefer,
can only be solved by means of a comparison of the consequences of the
theory with the experimental facts. In the following, we will 
exclusively use the function /\ref{teq8}/. 

\vspace{1.5cm}

\underline{\ \hspace{10cm}\ }

$*)$\ See the comment at the end.

\pagebreak
\refstepcounter{ppage}
\refstepcounter{subsection}
\label{tglavaIpara5}
\hspace{1cm}\S\ 5.\ And now, the Lagrange function is chosen.

In the Descartes coordinate system of special relativity it
has the form:
\begin{eqnarray}
\label{teq13}
L&=&\sqrt{1\, +\, F}\ -\ 1\ .
\end{eqnarray}
The further development of the theory proceeds according to the 
{"}ideal"\ scheme explained in \S\ {\tt\ref{tglavaIpara1}}.

The existence of a four-vector-potential $\varphi_i$ is postulated
obeying the condition:
\begin{eqnarray}
\label{teq14}
f_{ik}&=&\frac{\partial\varphi_k}{\partial x^i}\ -\ 
 \frac{\partial\varphi_i}{\partial x^k}\ \ \ .
\end{eqnarray}
This is equivalent to establishing the first system of field 
equations:
\begin{eqnarray}
\label{teq15}
\frac{\partial f_{ik}}{\partial x^l}\ +\
\frac{\partial f_{kl}}{\partial x^i}\ +\
\frac{\partial f_{li}}{\partial x^k}&=&0\ .
\end{eqnarray}
The Euler equations for the variational principle 
$\delta \int L\, d\tau = 0$ yield the second group of field
equations:
\begin{eqnarray}
\label{teq16}
\frac{\partial\ }{\partial x^k}\, \frac{\partial L}{\partial f_{ik}}&=&0\ \ \ ,
\end{eqnarray}
which are, as mentioned, nonlinear in the  
fields in view of the fact that $L$ 
\udensdash{i$\ \atop\ $\hspace{-0.33cm}s not\hspace{-0.05cm}\ }
a quadratic function of the fields as this is the case in the 
Maxwell theory. Introducing the induction tensor
\begin{eqnarray}
\label{teq17}
P_{ik}&=&\frac{\partial L}{\partial f_{ik}}\ \ \ ,
\end{eqnarray}
we can rewrite the second group of field equations in the form
\begin{eqnarray}
\label{teq18}
\frac{\partial P^{ik}}{\partial x^k}&=&0\ \ \ .
\end{eqnarray}

\pagebreak
\refstepcounter{ppage}

One can construct the energy-momentum tensor $T_{ik}$
starting from the field equations /\ref{teq15}/ and /\ref{teq16}/
by means of exactly the same method as in the classical
electrodynamics of Maxwell-Lenard and, by means of it, obtain
the conservations laws:
\begin{eqnarray}
\label{teq19}
\frac{\partial T_i^{\, l}}{\partial x^l}&=&0\ \ \ ,
\end{eqnarray}
where
\begin{eqnarray}
\label{teq20}
T_i^{\, l}&=&\delta_i^{\, l} L\ -\ f_{ik} P^{lk}\ .
\end{eqnarray}
We find now the field equations and the conservation laws in 
vector form. We denote the spatial vectors characterizing the
electromagnetic field in the usual Heaviside units by B, E, $D$ and 
H$^{\displaystyle\ \ast)}$. We set:
\begin{eqnarray}
\label{teq21}
&\left.\begin{array}{lllcc}
f_{23}\ ,&f_{31}\ ,&f_{12}&\ \ \longrightarrow\ \ &B\\[0.2cm]
f_{14}\ ,&f_{24}\ ,&f_{34}&\ \ \longrightarrow\ \ &E\\[0.2cm]
P_{23}\ ,&P_{31}\ ,&P_{12}&\ \ \longrightarrow\ \ &H\\[0.2cm]
P_{14}\ ,&P_{24}\ ,&P_{34}&\ \ \longrightarrow\ \ &D
\end{array}\right\}&\ .
\end{eqnarray}
Then,
\begin{eqnarray*}
L&=&\sqrt{1 + F}\ -\ 1\hspace{0.5cm},
\end{eqnarray*}
\hspace{1cm}where
\begin{eqnarray}
\label{teq22}
F&=&\frac{1}{b^2}\left(B^2 - E^2\right).
\end{eqnarray}
Furthermore,
\begin{eqnarray}
\label{teq23}
&\left.\begin{array}{rcccl}
H&=&b^2\ \frac{\displaystyle\partial L}{\displaystyle\partial B}
&=&\frac{\displaystyle B}{\displaystyle\sqrt{1+F}}\\[0.8cm]
D&=&-\ b^2\ \frac{\displaystyle\partial L}{\displaystyle\partial E}
&=&\frac{\displaystyle E}{\displaystyle\sqrt{1+F}}
\end{array}\right\}&
\end{eqnarray}
\hspace{-1cm}{\underline{\ \hspace{10cm}\ }}

\hspace{-1cm}$*)$\ For the vector quantities, here and
in the following no special notation is introduced.

\pagebreak
\refstepcounter{ppage}
and the field equations assume the form of the Maxwell
equations in a medium, however, without charges and currents:
\begin{eqnarray}
\label{teq24}
&\left.\begin{array}{rcccccl}
rot\, E + \frac{\displaystyle 1}{\displaystyle c}\;\dot{B}&=&0&
\hspace{3cm}&div\, B&=&0\\[0.3cm]
rot\, H - \frac{\displaystyle 1}{\displaystyle c}\;\dot{D}&=&0&
\hspace{3cm}&div\, D&=&0
\end{array}\right\}&
\end{eqnarray}
or, if in the two latter equations, instead of H and $D$, their
expressions  /\ref{teq23}/ in terms of B and E are inserted,
finally the system of field equations assumes the form:
\begin{eqnarray}
\label{teq25}
&\left.\begin{array}{l}
rot\, E + \frac{\displaystyle 1}{\displaystyle c}\;\dot{B}\ =\ 0\\[0.3cm]
div\, B\ =\ 0\\[0.3cm]
rot\, B - \frac{\displaystyle 1}{\displaystyle c}\;\dot{E}\ =\ 
\frac{\displaystyle 1}{\displaystyle 2}\
\frac{\displaystyle 1}{\displaystyle 1 + F}\ \left\{\left[ grad\, F, B\right]\ 
-\, \frac{\displaystyle 1}{\displaystyle c}\;\dot{F} E\right\}\\[0.6cm]
div\, E\ =\ \frac{\displaystyle 1}{\displaystyle 2}\ 
\frac{\displaystyle 1}{\displaystyle 1 + F}\ 
\left( grad\, F, E\right)
\end{array}\right\}&\ .
\end{eqnarray}
For the energy-momentum tensor, we obtain the table of 
components
\begin{eqnarray}
\label{teq26}
T^{il}&=&\left[\begin{array}{cccc}
X_x&X_y&X_z&c\, G_x\\[0.3cm]
Y_x&Y_y&Y_z&c\, G_y\\[0.3cm]
Z_x&Z_y&Z_z&c\, G_z\\[0.3cm]
\frac{\displaystyle 1}{\displaystyle c}\, 
S_x&\frac{\displaystyle 1}{\displaystyle c}\, 
S_y&\frac{\displaystyle 1}{\displaystyle c}\, S_z&U
\end{array}\right]\ \ \ ,
\end{eqnarray}
where
\begin{eqnarray}
\label{teq27}
&\left.\begin{array}{rcl}
X_x&=&H_y\, B_y\ +\ H_z\, B_z\ -\ D_x\, E_x\ -\ b^2\, L\\[0.3cm]
Y_x&=&X_y\ =\ -\  H_y\, B_x\ -\ D_x\, E_y\\[0.3cm]
\frac{\displaystyle 1}{\displaystyle c}\, S_x&=&c\, 
G_z\ = D_y\, B_z\ -\ D_z\, B_y\\[0.3cm]
U&=&D\, E\ +\ b^2\, L
\end{array}\right\}&\ .
\end{eqnarray}

\pagebreak
\refstepcounter{ppage}
Hereby, the conservation laws assume the form:
\begin{eqnarray}
\label{teq28}
&\left.\begin{array}{rcl}
div\, X&=&-\ \frac{\displaystyle 1}{\displaystyle c^2}\ 
\frac{\displaystyle\partial S_x}{\displaystyle\partial t}\\[0.3cm]
.\ \ .\ &.&\ .\ \ .\ \ .\ \ .\\[0.3cm]
.\ \ .\ &.&\ .\ \ .\ \ .\ \ .\\[0.3cm]
div\, S&=&-\ \frac{\displaystyle\partial U}{\displaystyle\partial t}
\end{array}\right\}&\ .
\end{eqnarray}
As Born emphasizes, any pair of the four field vectors E,B,H and $D$
can of course be chosen as independent variables. In dependence on
this choice, we can obtain apparently different formulations of the 
theory with different Lagrangians \linebreak which, however, of course
lead in essence to one and the same field equations. Clearly,
4 of such formulations are possible. In the following, the most 
convenient formulation of the theory for us is that which we
have also considered so far /i.e.\ where E and B are selected
as independent vectors/, although also for any other choice 
the calculations would only become insignificantly more involved.

We note that for the construction of the quantum theory the choice
of $D$ and B as independent vectors turns out to be more convenient.
Hereby, the role of the Lagrange function is \linebreak played by the energy 
density $U$ which according to the last formula /\ref{teq27}/ is equal
to:
\begin{eqnarray}
\label{teq29}
U&=&b^2\ \left(\sqrt{\left(1 + \frac{D^2}{b^2}\right)
\left(1 + \frac{B^2}{b^2}\right)}\ -\ 1\right)\ .
\end{eqnarray}

\refstepcounter{subsection}
\label{tglavaIpara6}
\hspace{1cm}\S\ 6. By means of the recipe laid out
in \S\ {\tt\ref{tglavaIpara1}}, we will find the solution of the field 
equations corresponding to the existence of a charged particle at rest
at the origin of the coordinates.

According to what has been said in \S\ {\tt\ref{tglavaIpara1}},
here we shall primarily set 
\begin{eqnarray*}
B&=\ H\ =&0\ .
\end{eqnarray*}

\pagebreak
\label{tpage22}
\refstepcounter{ppage}

By virtue of /\ref{teq24}/, this yields
\begin{eqnarray}
\label{teq30}
rot\, E&=&0\hspace{2cm};\ \ \ \ div\, D\ =\ 0 
\end{eqnarray}
whereby $D$ /and, consequently, also E/ do not depend on time.
We are looking for a radially symmetric solution of these
equations, i.e.\ for such a solution for which both vectors E and $D$ 
are directed radially and for which their value depends on $r$ only.

The second of the equations /\ref{teq30}/ reads in spherical coordinates
\begin{eqnarray}
\label{teq31}
\frac{1}{r^2}\ \frac{d\ }{d r}\left(r^2\, D_r\right)&=&0\ .
\end{eqnarray}
From it, completely uniquely, we obtain
\begin{eqnarray}
\label{teq32}
D_r&=&\frac{e}{r^2}\ ,
\end{eqnarray}
where e is the only integration constant identified with the charge.
This solution has a singularity at $r$=0.
We also interprete the state of the field described by it as 
corresponding to the existence of a particle at the origin of coordinates.

\hspace{1cm}The whole course of reasoning shows us that the condition
/\ref{tsubb}/ is fulfilled in the Born theory as it is in the 
Maxwell theory. Calculating the field E we find
\begin{eqnarray}
\label{teq33}
E_r&=&\frac{e}{\sqrt{r^4 + r_0^4}}\ ,
\end{eqnarray}
where
\begin{eqnarray}
\label{teq34}
r_0&=&\sqrt{\frac{e}{b}}
\end{eqnarray}
also represents a constant with the dimension of a length which,
as becomes clear further below, will be, in a well-known
sense, analogous to the electron radius in the Lorentz theory.

\pagebreak
\refstepcounter{ppage}

From equation (\ref{teq33}), we see that the field E does not 
have any singularities. However, just the fact of the finiteness of E
by far does not mean that there are no divergencies in the theory. 
In order to check whether for our solution the condition  
/\ref{tsubc}/ is fulfilled, one has to turn to direct calculations
on the basis of the formulas /\ref{teq27}/.

It is easy to see, and here we have a strong success in principle 
of the theory, that in the present case this condition is 
\udensdash{f$\ \atop\ $\hspace{-0.33cm}ulfilled\hspace{-0.05cm}\ }.
Indeed, in our case the momentum vanishes by virtue of B=0;
the finiteness of the total energy $W = \int U dV$ is confirmed 
by direct calculations yielding for $W$ the value:
\begin{eqnarray}
\label{teq35}
W&=&{\tt I,2361}\ \frac{e^2}{r_0}\ .
\end{eqnarray}
As said, this point is fairly essential. It shows that Born,
staying within the frame of classical ideas, succeeded in overcoming the 
difficulty of the infinite self-energy of the electron hereby    
not relying on any artificial ideas about its \linebreak {"}structure{"}.
Knowing the charge and the mass of the electron we can calculate the
numerical value of the constant $r_0$:
\begin{eqnarray}
\label{teq36}
r_0&=&1,2361\ \frac{e^2}{m_0 c^2}\ =\ 2,28 \cdot 10^{-13}\ {\it cm.}
\end{eqnarray}
and, consequently, the numerical value of the {"}critical"\ field
{"}$b${"}:
\begin{eqnarray}
\label{teq37}
b&=&\frac{e}{r_0^2}\ =\ 9,18 \cdot 10^{15}\ {\it abs.\ units}.
\end{eqnarray}
From these formulas, it is clear that in the terminology of Lorentz
we can interprete the quantity $r_0$ as the {"}radius"\

\pagebreak
\refstepcounter{ppage}
of the electron, and the {"}critical"\ field {"}$b${"}\ as the electrical
field strength at its {"}surface{"}. The value of this {"}critical field{"}\
is clearly an universal constant of the theory not depending on the 
charge e which, as already mentioned, is not fixed by any theory.
Consequently, for any fixed e we obtain a
\udensdash{c$\ \atop\ $\hspace{-0.33cm}ompletely determined\hspace{-0.05cm}\ }
$r_0$, and by virtue of /\ref{teq36}/, a 
\udensdash{c$\ \atop\ $\hspace{-0.33cm}ompletely 
determined mass $m_0$\hspace{-0.05cm}\ }.
This remark convinces us that for a 
\udensdash{g$\ \atop\ $\hspace{-0.33cm}iven\hspace{-0.05cm}\ }
charge e the theory of Born leads to
\udensdash{o$\ \atop\ $\hspace{-0.33cm}ne\hspace{-0.05cm}\ } \linebreak
value of the mass $m_0$ only, i.e.\ it is a theory of only one sort
of electron-particles. As is visible from this, to obtain a proton 
theory turns out to be impossible within the framework of the above 
considerations.

\refstepcounter{subsection}
\label{tglavaIpara7}
\hspace{1cm}\S\ 7. We will now consider somewhat more closely 
the question dealt with in the formulation of the condition
/{\it\ref{tsubd}}/, namely the question as how the electromagnetic 
quantities can be related to mechanical ones in a relativistically
invariant manner.

As we have clarified already, for the construction of the mechanics 
in an unitary theory we need to identify, for the given type of the 
states of the field, the volume integral
of the electromagnetic angular momentum, i.e.\ 
$c\, G_x,$ $c\, G_y,$\ and $c\, G_z,$ 
with the mechanical angular momentum of particles,
and the energy density $U$ with the total energy of the particles.
In order to guarantee that such an identification was relativistically
invariant it is necessary that these integrals would transform 
as components of a four-dimensional vector under Lorentz transformations.
One can easily convince oneself that a necessary condition 

\pagebreak
\label{tpage25}
\refstepcounter{ppage}

for this consists in the vanishing of the volume integrals of the 
diagonal components $X_x,$ $Y_y$ and $Z_z,$ of the Maxwell \linebreak
stress tensor
in a coordinate system at rest relative to the particles, i.e.\
one has the condition
\begin{eqnarray}
\label{teq38}
\int\, X_x\, dV&=&\int\, Y_y\, dV\ =\ \int\, Z_z\, dV\ =\ 0\ .
\end{eqnarray}
In the 
\udensdash{A$\ \atop\ $\hspace{-0.33cm}braham\hspace{-0.05cm}\ } -
\udensdash{L$\ \atop\ $\hspace{-0.33cm}orentz\hspace{-0.05cm}\ }
theory, this conditions was 
\udensdash{n$\ \atop\ $\hspace{-0.33cm}ot\hspace{-0.05cm}\ }
fulfilled and the difficulty of constructing a relativistically
invariant mechanics consisted, again, herein.
The necessity of introducing forces of non-electric origin holding
together the electron have been connected with just this problem
because the resulting Maxwell stresses were different from zero 
$^{\mbox{\large\stexttt{1/}}}$.

\hspace{1cm}The Born-Infeld theory is free of such difficulties.
Directly inserting the solution /\ref{teq32}/-/\ref{teq38}/
into /\ref{teq27}/ shows that the condition /\ref{teq38}/ is
\udensdash{f$\ \atop\ $\hspace{-0.33cm}ulfilled\hspace{-0.05cm}\ }
for this solution and, consequently, we can construct a 
relativistically invariant mechanics. In other words, there is 
no need of introducing \linebreak forces holding together the electron because
the resulting Maxwell stresses vanish. This way, in the electrodynamics 
of Born also the last condition imposed on an unitary theory
is fulfilled on which even the best of the previous theories -
the theory of Lorentz - showed its inconsistency.

\refstepcounter{subsection}
\label{tglavaIpara8}
\hspace{1cm}\S\ 8. We sketch now the way by means of which one
can obtain the
\udensdash{l$\ \atop\ $\hspace{-0.33cm}aws\hspace{-0.05cm}\ }
of mechanics within the Born-Infeld theory. For this purpose,
we clearly should consider the question how the solutions of

\hspace{4cm}\underline{\hspace{0.7cm}}
\hspace{0.7cm}\underline{\hspace{0.7cm}}
\hspace{0.7cm}\underline{\hspace{0.7cm}}
\hspace{0.7cm}\underline{\hspace{0.7cm}}
\hspace{0.7cm}\underline{\hspace{0.7cm}}
\hspace{0.7cm}\underline{\hspace{0.7cm}}
\hspace{0.7cm}\underline{\hspace{0.7cm}}

\vspace{0.5cm}

\underline{\ \hspace{4cm}\ }
{\rm\small written on the backside of the manuscript leaf}
\underline{\ \hspace{2cm}\ }

\vspace{0.5cm}

{\it 1) The integrals of the non-diagonal components of the tensor
of the Maxwell stresses vanish in the theory of Lorentz as well as in the
theory of Born.}

\pagebreak
\label{tpage26}
\refstepcounter{ppage}

the field equations interpreted by us in terms of the presence
of an electron behave under the influence of an external field.

We know that the equations of Born allow solutions corresponding
to an electron  
\udensdash{a$\ \atop\ $\hspace{-0.33cm}t rest\hspace{-0.05cm}\ }.

Performing a Lorentz transformation of the coordinate system, we
can obtain solutions corresponding to an electron moving 
rectilinearly and uniformly. Right, for the mechanics it would 
be desirable to know also such solutions which correspond to an
\udensdash{a$\ \atop\ $\hspace{-0.33cm}cceleratedly\hspace{-0.05cm}\ }
moving electron, because such a motion is namely performed 
by the electron under the influence of external electromagnetic fields.
In principle, one could construct a perturbation theory by taking
the solution corresponding to an uniform and rectilinear motion as 
zeroth approximation and consider the influence of the 
acceleration as a small perturbation. However, it is well-known
that taking into account the acceleration is important for the
consideration of the radiation reaction of the own radiation of the 
electron only. If one is not interested in this effect it is completely 
sufficient to consider the solution corresponding to a vanishing acceleration.
This is what we also will do. Clearly, such solutions of the equations 
of Born should serve as starting point of the theory which correspond
to the simultaneous presence of an electron and an external field.
Right, we have seen further above that in an unitary theory only
certain solutions, which do not entail the presence of an additional
external field, can be viewed as entailing the existence of a particle.
This somehow changes our solution /\ref{teq32}/ and  /\ref{teq33}/ 
(or those obtained from it by means of a Lorentz transformation)
and it is a priori unclear of whether one can continue to speak
of the presence of a particle at all. In other words, in the theory
of Born such a case is conceivable in principle where the electron 
would apparently loose its 

\pagebreak
\refstepcounter{ppage}

individuality under the influence of a field, where,
completely roughly speaking, it would apparently be {"}destroyed{"}.  
It is clear physically, however, that this sort of things can only
happen in very strong fields (comparable to $b$), or - both are
connected to each other, of course - for large accelerations of 
the electron. For small fields and accelerations, basically the influence
of the field can apparently be described in classical terms, i.e.\ simply as
{"}bringing the electron into movement{"}. The mentioned difficulty 
is expressed in mathematical terms as follows. Let
$D_0\ {\it and}\ B_0$\ $^{\displaystyle 1)}$ 
be solutions of the equations 
of Born corresponding to the presence of an electron moving uniformly
and rectilinearly along some world trajectory (The quantities characterizing
the form of this trajectory play the role of parameters in the 
following considerations.). Let further  
$D_{\it ext}$ and $B_{\it ext}$ 
be a solution of these equations 
corresponding to an external field. Then in general, the sum of the 
form 
$D_0\hspace{-0.005cm}+\hspace{-0.03cm}D_{\it ext}$ and
$B_0\hspace{-0.005cm}+\hspace{-0.03cm}B_{\it ext}$ does 
\udensdash{n$\ \atop\ $\hspace{-0.33cm}ot\hspace{-0.05cm}\ }
represent any solution of the equations of Born - in view of 
their nonlinearity - for any choice of the world trajectory. Here,
the effect of the distortion of the properties of the electron
by the external field, specific to the theory of Born and 
alien to the classic theory, makes its appearance.

If $D_{\it ext}\ {\it and}\ B_{\it ext}$ 
are comparable to $b$ - for example,
if at a distance from our electron comparable to $r_0$ another electron
is located -

-----------------------

\ \hspace{-1cm}$1)$ It turns out that for doing the calculations
it is here perhaps most convenient to use that formulation of the theory
where $D$ and B are chosen as independent variables and where the role
of the Lagrangian is played by the energy density $U$.

\pagebreak
\label{tpage28}
\refstepcounter{ppage}

this distortion effect plays a fundamental role in
all phenomena, and it is not clear, a priori, how to perform 
any calculations. If $D_{\it ext}$ and $B_{\it ext}$ are small in comparison
with $b$ one can obviously put
\begin{eqnarray}
\label{teq39}
D&=&D_0 + D_{\it ed} + D^\prime\hspace{0.8cm};
\hspace{0.8cm}B\ =\ B_0 + B_{\it ext} + B^\prime\hspace{0.8cm}
\end{eqnarray}
and search for $D^\prime\ and\ B^\prime$ in terms of small corrections
to the initial fields. The effect of distortion will be there, but
it will be small. This is the simplest case 
considered by I.E.\ Tamm $^{\mbox{\large\stexttt{1/}}}$ when the solution 
$D_0\ and\ B_0$ corresponds to an electron at rest, and the 
solution $D_{\it ext}\ and\ B_{\it ext}$ to a constant homogeneous
electrostatic field. In this case $B = 0$; for the correction $D^\prime$
fairly simple equations are obtained which show that this correction
is really small in a well-known sense: Namely, while at the center
of the electron it is becoming infinite it remains much smaller than 
$D_0$ all the time; for large distances from the center  
holds $D^\prime \ll D_{\it ext}$. Of course, for  
$b\rightarrow\infty\ ,\ D^\prime$ approaches zero.
It is important to have in mind that the solution of the type
/\ref{teq39}/ apparently exists for 
\udensdash{a$\ \atop\ $\hspace{-0.33cm}ny\hspace{-0.05cm}\ }
world trajectory and the appearance of the corrections $D^\prime\ and\ B^\prime$
does not have any direct connection to the fact that the electron
gets into motion. Here from follows that starting with the field 
equations only, in general one cannot obtain the laws of the mechanics
of the electron. Particularly clearly this is visible just from
the example elaborated by Tamm which shows that the equations of 
Born also admit such solutions for which the electron is at rest
irrespectively of the presence of the external field. 
This fact just in itself
does not represent anything strange 

---------------------------------

1/ As I learned from a kind communication by him.

\pagebreak
\label{tpage29}
\refstepcounter{ppage}

\ \\[-1.5cm]

- in fact, it is physically 
completely conceivable to imagine such a case where the electron is 
kept on its place by certain forces of non-electric origin.
Obviously, the existence of such forces cannot be taken into account
in the very field equations which, in general, determine all 
electromagnetically possible states of the field and cannot, of course,
say anything about under which conditions this or \linebreak that of these 
states is realizable mechanically.

This situation, however, by far does not mean that for obtaining 
the laws of mechanics it would be necessary to introduce into the theory
certain independent new principles. As has been pointed out repeatedly 
in the course of the whole exposition, the explanation
is that the basic
foundation of the theory of Born is not represented by the field equations
but by the variational principle. The content of this principle is 
by far not exhausted by the field equations; the latter determine
the whole ensemble of extrema of the variational integral, but this
integral does by far not have the same value for all extrema. Just thanks to
this circumstance the variational principle is sufficient to not only 
obtain the laws of electrodynamics but also the laws of mechanics.

\hspace{1cm}Concretely, one has obviously to proceed as follows: Insert
the solution /\ref{teq39}/ into the Lagrangian /\ref{teq7}/ and  
/\ref{teq8}/ and find out for which of these solutions the variational
integral has an extremal value. The role of the parameters on the 
basis of which the variation is performed is played by the quantities
characterizing the shape of the world line of the electron on which
the expressions /\ref{teq39}/ depend. It is clear that the equation
obtained as result of this variational method is indeed analogous in 
type to the usual variational principle of mechanics.
It is also clear that one can, in principle, also take into account 
the action of non-electric forces on the charge by adding 
corresponding terms to the Lagrangian. If one is only interested in
the {"}zeroth approximation{"}, i.e.\ in those cases when one can
expect the applicability of the usual laws of mechanics - one can
clearly omit the corrections $D^\prime\ and\ B^\prime$ and put

\pagebreak
\label{tpage30}
\refstepcounter{ppage}
the sum of the field of the electron and the external field directly
into the variational integral. Born and Infeld have shown that, by proceeding
along the path just detailed, one can indeed obtain a variational
principle analogous to the mechanical one which - under the assumption
that the external field changes slowly on distances of the order of $r_0$
- goes over into the usual variational principle of relativistic 
mechanics. Herewith, the task of constructing a mechanics within
the framework of an unitary theory can be considered as solved,
at least in general. 

Of most interest, however, would of course be the study of those
\udensdash{d$\ \atop\ $\hspace{-0.33cm}eviations\hspace{-0.05cm}\ }
from the usual form of the mechanical equations which shall
emerge from the theory of Born once more exact calculations are
performed. Unfortunately, the approximations one should get rid of
apparently depend on each other - if the gradient of the external
field on distances of the order of $r_0$ is not small the corrections
$D^\prime\ and\ B^\prime$ as well as the impact of the radiation
reaction on the electron are also hardly small. However, if
one takes into account all these effects in the first approximation
the calculations can still be performed in principle although they
become fairly involved then. 

\pagebreak
\refstepcounter{ppage}
\refstepcounter{section}
\label{tglavaII}
\hspace{3.5cm}
\udensdash{\ C$\ \atop\ $\hspace{-0.1cm}H\ A\ P\ T\ E\ R\ \ II.\ 
\hspace{-0.1cm}\ }\\[-0.2cm]

\refstepcounter{subsection}
\label{tglavaIIpara1}
\hspace{0.4cm}
\udensdash{\ \S\ 1. Introduction\ \ \hspace{-0.05cm}\ }\\[-0.2cm]

\hspace{1.cm}
Thus, shortly summarizing the content of chapter {\tt\ref{tglavaI}}
one may say that Born succeeded in constructing an unitary theory 
of field and matter by giving up the Maxwell equations and 
introducing new field equations in their place /formulas 
/\ref{teq24}/ or /\ref{teq25}// whose characteristic feature is their
\linebreak
\udensdash{n$\ \atop\ $\hspace{-0.33cm}onlinearity\hspace{-0.05cm}\ }.
In the transition from the 
\udensdash{M$\ \atop\ $\hspace{-0.33cm}axwell\hspace{-0.05cm}\ } 
equations to the equations of 
\udensdash{B$\ \atop\ $\hspace{-0.33cm}orn\hspace{-0.05cm}\ }, this way,
the {"}mechanical"\ results of the theory essentially change:
The energy of a point charge becomes finite etc.. However, one
easily recognizes that simultaneously with this also a number of other 
results of the usual electrodynamics cease to be correct
which, on first sight, are not connected with the difficulties of 
constructing an unitary theory at all. But just because the theory
becomes nonlinear, as we already mentioned in passing
in \S\ {\tt\ref{tglavaIpara8}} of the {\tt\ref{tglavaI}}-st chapter, the 
\udensdash{s$\ \atop\ $\hspace{-0.33cm}uperposition principle\hspace{-0.05cm}\ }
playing such an important role in all of the Maxwell electrodynamics
no longer applies.

In other words, according to the theory of Born the deviations from the 
usual laws of electrodynamics should show up not only in the 
{"}mechanical"\ phenomena but also in the effects of the {"}mutual
distortion"\ of two /or more/ electromagnetic fields.
The most interesting of these effects is perhaps the distortion effect
of the field of a  
\udensdash{l$\ \atop\ $\hspace{-0.33cm}ight wave\hspace{-0.05cm}\ }
by other external fields. As can directly be recognized from the equations
/\ref{teq24}/, in connection with /\ref{teq22}/ and /\ref{teq23}/,
in the theory of Born the
\udensdash{p$\ \atop\ $\hspace{-0.33cm}lane\hspace{-0.05cm}\ } and 
\udensdash{s$\ \atop\ $\hspace{-0.33cm}pherical\hspace{-0.05cm}\ }
electromagnetic waves are two
\udensdash{e$\ \atop\ $\hspace{-0.33cm}xact\hspace{-0.05cm}\ }
solutions of the field equations
/ This also applies to the second variant of the theory dealing
with a different Lagrange function. One can convince oneself of this

\pagebreak
\refstepcounter{ppage}

by recalling that for fields of the wave zone both spatial
invariants $F$ and $G$ vanish identically./. This way, as long
as this type of waves are concerned the electromagnetic theory of
light is formulated according to  
\udensdash{B$\ \atop\ $\hspace{-0.33cm}orn\hspace{-0.05cm}\ }
exactly the same way as according to Maxwell. However, matters 
essentially start to change as soon as we go over to somewhat 
more involved situations, in particular to those when besides
the field, say, of a plane wave of light, also some other
electromagnetic field exists, just the electrostatic field of a plane
capacitor, or the field of another wave of light.
According to 
\udensdash{M$\ \atop\ $\hspace{-0.33cm}axwell\hspace{-0.05cm}\ },
the presence of the second field does not exert any influence
whatsoever on the properties of the initial wave while at the
same time, according to 
\udensdash{B$\ \atop\ $\hspace{-0.33cm}orn\hspace{-0.05cm}\ },
such an influence 
\udensdash{o$\ \atop\ $\hspace{-0.33cm}f course\hspace{-0.05cm}\ }
exists in general because, for example, the sum of two 
electromagnetic fields of two plane waves is no longer a solution
of the field equations of Born. So far, as long as all fields
we are dealing with here are weak in comparison with the
{"}critical"\ field {"}$b$"\ the impact of the nonlinearity should
apparently be weak, and it makes sense to say that in the presence 
of a background field we still have just a plane wave, some 
somewhat distorted only.

This chapter is just concerned with the study of three examples of 
such a distortion.

It should be mentioned that a remarkable analogy 
between the electrodynamics of 
\udensdash{B$\ \atop\ $\hspace{-0.33cm}orn\hspace{-0.05cm}\ }
and a theory built on apparently completely different physical
foundations, namely - the theory of the positron of 
\udensdash{D$\ \atop\ $\hspace{-0.33cm}irac\hspace{-0.05cm}\ },
makes it appearance here. Indeed, as has been first pointed
out by 
\udensdash{H$\ \atop\ $\hspace{-0.33cm}alpern\hspace{-0.05cm}\ }

\pagebreak
\refstepcounter{ppage}

and Debye$^{\mbox{\large\stexttt{1/}}}$,
a mutual distortion effect between two waves of light and, 
as one can convince oneself, also a number of other distortion 
effects analogous to those of Born should occur according 
to the latter theory too. Moreover,
\udensdash{E$\ \atop\ $\hspace{-0.33cm}uler\hspace{-0.05cm}\ }
and \linebreak
\udensdash{K$\ \atop\ $\hspace{-0.33cm}ockel\hspace{-0.05cm}\ }, 
students of Heisenberg, have shown$^{\mbox{\large\stexttt{2/}}}$
that this analogy has also a quantitative character in a well-know sense: 
{"}The scattering
of light by light"\ calculated according to Dirac turns 
out to be the same in general terms as one would expect it to be according 
to the quantized electrodynamics of Born. In connection with the 
existence of this analogy a number of interesting questions arise
which, however, we will not consider here because they are all 
essentially connected with the transition to the field quantization
which lies beyond the framework of the present dissertation (and, 
even further, because a detailed analysis of the results by Euler
and Kockel is not possible so far because their calculations are not 
published yet). In any case, the very existence of the analogy between
the results of 
\udensdash{B$\ \atop\ $\hspace{-0.33cm}orn\hspace{-0.05cm}\ } and
\udensdash{D$\ \atop\ $\hspace{-0.33cm}irac\hspace{-0.05cm}\ }, 
also in the domain of nonlinear effects, makes the investigation of 
this side of the electrodynamics of Born particularly interesting
also from out the classical point of view.\\

\refstepcounter{subsection}
\label{tglavaIIpara2}
\hspace{1cm}\S\ 2.$^{\mbox{\large\stexttt{3/}}}$\ 
\udensdash{A$\ \atop\ $\hspace{-0.1cm}plane wave of light in a  
homogeneous\hspace{-0.05cm}\ }

\hspace{4.5cm}\udensdash{e$\ \atop\ $\hspace{-0.33cm}lectrostatic
field\hspace{-0.05cm}\ }\\

\hspace{1cm}As first example, we consider the problem of the 
propagation of a plane wave of light in a homogeneous 
electrostatic field - say in the field of a plane capacitor. 

----------------------------

1/\ {\it cf.\ Heisenberg, Z.\ f.\ Phys.\ 
\underline{90}. 1934. 209.}

2/\ {\it Naturwiss.\ 23. 1935. 246.}

3/\ The results of this paragraph are published in an

\hspace{0.8cm}article by S.\ Shubin and A.\ Smirnov. DAN 1936.

\pagebreak
\refstepcounter{ppage}

We consider this problem, as most simple and typical, in particular detail.

Let us consider, initially, a plane wave of light whose 
electromagnetic field is determined by the equations:
\begin{eqnarray}
\label{teq40}
&\left.\begin{array}{rcccl}
E&=&D&=&E_0\ \cos\omega 
\left(t - \frac{\displaystyle rn}{\displaystyle c}\right)\\[0.3cm]
B&=&H&=&B_0\ \cos\omega 
\left(t - \frac{\displaystyle rn}{\displaystyle c}\right)
\end{array}\right\}&\ \ \ ,
\end{eqnarray}
where, as usual
\begin{eqnarray}
\label{teq41}
E_0\ =\ \left[n,B_0\right]\hspace{0.8cm}&;&\hspace{0.8cm}B_0
\ =\ \left[E_0,n\right]\ .
\end{eqnarray}
As just mentioned, the expression /\ref{teq40}/ is an 
\udensdash{e$\ \atop\ $\hspace{-0.33cm}xact\hspace{-0.05cm}\ }
solution of the equations of Born.

According to this theory, the wave represented by this solution can,
in general, propagate in vacuum only, i.e.\ in the absence of 
any other fields. In particular, if in the parts
of the space considered by us we switch on the field of a plane 
capacitor $N$ then the expression of the type
\begin{eqnarray}
\label{teq42}
&\left.\begin{array}{rcl}
E&=&E_0\ \cos\omega 
\left(t - \frac{\displaystyle rn}{\displaystyle c}\right)\ +\ N\\[0.3cm]
B&=&B_0\ \cos\omega 
\left(t - \frac{\displaystyle rn}{\displaystyle c}\right)
\end{array}\right\}&
\end{eqnarray}
does no longer represent a possible state of the field. It is 
clear, however, that as long as $N\ll b$ applies solutions will
exist which in any case are close to /\ref{teq42}/. We are
putting to ourselves the task of finding these solutions, or,
at least a sufficiently large class 

\pagebreak
\refstepcounter{ppage}

of them$^{\mbox{\large\stexttt{1/}}}$.

Hereby, we initially do not touch the question under which 
conditions one or another of these solutions is in fact realized;
we defer the consideration of this more concrete task to  
\S\ {\tt\ref{tglavaIIpara3}}.

\hspace{1cm}Relying on the inequality $N\ll b$ which is, as 
already said, essential we will discard in course of the 
calculations all powers of the 
quantity $\frac{\displaystyle N}{\displaystyle b}$ except
the lowest one. As becomes clear in the following, this 
lowest power is the 
\udensdash{s$\ \atop\ $\hspace{-0.33cm}econd\hspace{-0.05cm}\ } one.
This way, all expressions having $b^4$, $b^6$ etc.\ in the denominator  
will have to be crossed out by us.$^{\mbox{\large\stexttt{2/}}}$

-------------------------

1/ Of course, one could start, instead of with the vectors E and B,
with some other pair of vectors, $D$ and H for example, and put
the task this way: Find the solution close to
\begin{eqnarray*}
&\left.\begin{array}{rcl}
D&=&D_0\ \cos\omega 
\left(t - \frac{\displaystyle rn}{\displaystyle c}\right)\ +\ N\\[0.3cm]
H&=&H_0\ \cos\omega \left(t - \frac{\displaystyle rn}{\displaystyle c}\right)
\end{array}\right\}&\ .
\end{eqnarray*}
It is clear that the final results must be
\udensdash{t$\ \atop\ $\hspace{-0.33cm}he same\hspace{-0.05cm}\ }
in both settings of the problem.

We will convince ourselves in the 
next paragraph that this is indeed the case.

2/ If we encounter expressions of the type
$\frac{\displaystyle E_0^4}{\displaystyle b^4}$,
or $\frac{\displaystyle E_0^2 N^2}{\displaystyle b^4}\ ,$ 
in the following, we obviously can also neglect them.

\pagebreak
\refstepcounter{ppage}

It is clear that in this approximation the difference between the first
and the second variant of the theory of Born disappears because 
the Lagrange function /\ref{teq7}/ goes over into /\ref{teq8}/.

On first sight, it might seem most natural so search the
solution of the equations of Born /\ref{teq25}/ in the form
\begin{eqnarray}
\label{teq43}
&\left.\begin{array}{rcl}
E&=&E_0\ \cos\omega 
\left(t - \frac{\displaystyle rn}{\displaystyle c}\right)\ 
+\ N\ +\ E^\prime\\[0.3cm]
B&=&B_0\ \cos\omega 
\left(t - \frac{\displaystyle rn}{\displaystyle c}\right)\ +\ B^\prime
\end{array}\right\}&\ ,
\end{eqnarray}
considering $E^\prime\ and\ B^\prime$ as small quantities 
approaching zero at 
$\frac{\displaystyle N}{\displaystyle b}\ \rightarrow\ 0$. 
However, inserting the expression /\ref{teq43}/ into
the equation /\ref{teq25}/ it turns out that for 
$E^\prime\ and\ B^\prime$ equations are obtained which do not 
have any solutions that are finite everywhere. Namely, the calculations
show that $E^\prime\ and\ B^\prime$ contain terms proportional 
to the quantity $t + \frac{\displaystyle rn}{\displaystyle c}$,
i.e.\ becoming $\infty$ for  $r = \infty$ and $t = \infty$.

The failure of this calculation necessarily leads to the 
conclusion that such solutions of the equations of Born which would 
be close to the corresponding Maxwell solutions
/\ref{teq42}/ for all $r\ and\ t$ do not exist. In other words,
for the given class of problems the usually applied perturbative
method (the unknown quantity is searched for as known quantity + small 
addition) is not applicable in this form.

It is worth noting that we have a similar situation also in 
some nonlinear problems of classical mechanics. Let us consider,
for example, the equation of motion of an anharmonic oscillator
under a force containing terms proportional to\hfill\ \linebreak
$x\ and\ x^3$:
\begin{eqnarray}
\label{teq44}
\ddot{x}\ +\ \omega^2 x&=&\kappa x^3\ , 
\end{eqnarray}
and we are trying to find its solution in the form
\begin{eqnarray}
\label{teq45}
x&=&A\ \cos\omega t\ +\ x^\prime\ .
\end{eqnarray}

\pagebreak
\refstepcounter{ppage}

Considering $\kappa\ and\ x^\prime$ as small quantities 
we obtain for $x^\prime$ the equation
\begin{eqnarray}
\label{teq46}
\ddot{x}^\prime\ +\ \omega^2 x^\prime&=&
\kappa\ A^3 \cos^3\omega t = \kappa\ A^3\ 
\frac{3\cos\omega t + \cos 3 \omega t}{4}\ .                  
\end{eqnarray}
This equation can formally be viewed as an equation of forced
oscillations under a force containing terms with exactly 
the same frequency that the proper oscillations have. As a result,
a {"}resonance"\ is found, i.e.\ $x^\prime$ turns out to be proportional
to $t$ and for $t \rightarrow \infty$ we obtain a diverging solution
not having any physical meaning and indicating that one cannot search
for $x$ in the form /\ref{teq45}/.

In the present case, it is well-known how one has to change
the method of solving the equation  /\ref{teq44}/. Namely,
one has to change the zeroth approximation itself and to search
the solution in the form
\begin{eqnarray}
\label{teq47}
x&=&A\ \cos\nu t\ +\ x^\prime                  
\end{eqnarray}
where $\nu$ differs (if little) from $\omega$ and it should be 
chosen such a way that in the equation for $x^\prime$ the term
yielding the {"}resonance"\ disappears. As is not difficult to see,
for this purpose one has to put in 1st approximation 
\begin{eqnarray}
\label{teq48}
\nu&=&\omega \left(1 - \frac{3}{8} \frac{\kappa A^2}{\omega^2}\right)\ .      
\end{eqnarray}
If $\nu$ is chosen this way the equation for $x^\prime$ will then
have a solution which remains in fact small all the time. This 
simple example provides us with the key for the solution of our 
problem. We will search the solution of the equations of Born not 
in the form /\ref{teq43}/, but we will try to somehow deform the
zeroth approximation itself, i.e.\ we will assume that the 
presence of the constant electric field somehow changes the 
properties of the initial plane wave itself. It is most natural
to assume that this field exhibits a certain anisotropy in space and
to try to satisfy the equations of Born by means of expressions
that are analogous to the solutions of the Maxwell equations for light
propagating in an uniaxial crystal.

\pagebreak
\refstepcounter{ppage}

We choose as direction of the wave vector the $ox$-axis and
as the plane spanned by this vector and the vector of the external 
field $N$ the plane $xy$.

We denote the angle between these vectors by $\alpha$. Concerning
the field components of the wave we assume so far only 
that they are periodic functions of $x\ and\ t$, i.e.\ we will search
the solution of the equations /\ref{teq25}/ in the form 
\begin{eqnarray}
\label{teq49}
&\left.\begin{array}{rcl}
\hspace{-0.9cm}E_x&=&
E^0_{\ x} \cos\omega_1 
\left(t - \frac{\displaystyle x}{\displaystyle v_1} + \varphi_1\right)
 + N \cos\alpha\ =\ E^0_{\ x} \cos\gamma_1 + N \cos\alpha\\[0.3cm]
\hspace{-0.9cm}E_y&=&
E^0_{\ y} \cos\omega_2 
\left(t - \frac{\displaystyle x}{\displaystyle v_2} + \varphi_2\right)
+ N \sin\alpha\ =\ E^0_{\ y} \cos\gamma_2 + N \sin\alpha\\[0.3cm]
\hspace{-0.9cm}E_z&=&
E^0_{\ z} \cos\omega_3 
\left(t - \frac{\displaystyle x}{\displaystyle v_3} + \varphi_3\right)
\ =\ E^0_{\ z} \cos\gamma_3\\[0.3cm]
\hspace{-0.9cm}B_x&=&
B^0_{\ x} \cos\nu_1 
\left(t - \frac{\displaystyle x}{\displaystyle w_1} + \theta_1\right)
\ =\ B^0_{\ x} \cos\gamma_4\\[0.3cm]
\hspace{-0.9cm}B_y&=&
B^0_{\ y} \cos\nu_2 
\left(t - \frac{\displaystyle x}{\displaystyle w_2} + \theta_2\right)
\ =\ B^0_{\ y} \cos\gamma_5\\[0.3cm]
\hspace{-0.9cm}B_z&=&
B^0_{\ z} \cos\nu_3 
\left(t - \frac{\displaystyle x}{\displaystyle w_3} + \theta_3\right)
\ =\ B^0_{\ z} \cos\gamma_6\hspace{5.1cm},\hspace{-5.1cm}
\end{array}\right\}&\ 
\end{eqnarray}
where $\varphi\ and\ \theta$ are constant phases.

This way, we assume in advance that different field components
of the wave may have different amplitudes, frequencies, velocities, and
initial phases.

We see that the equations of Born can be satisfied by solutions of 
the type /\ref{teq49}/ if one takes care of the parameters entering it 
in an appropriate way; as it turns out, here we cannot supplement 
them by additive corrections $ E^\prime_x\ ,\ E^\prime_y$ etc..

\pagebreak
\refstepcounter{ppage}

We insert /\ref{teq49}/ into /\ref{teq25}/ and performing the 
calculations with a precision up to the lowest power of the 
small quantities we determine the ratios between the amplitudes,
the ratios between the frequencies and velocities entering 
/\ref{teq49}/. Here, we note that we can consider all quantities
of the type $\frac{\displaystyle v_1 -c}{\displaystyle c}$,
$\frac{\displaystyle E^0_{\ x}}{\displaystyle\sqrt{E^{0\, 2}_{\ y} +
E^{0\, 2}_{\ z}}}$ etc.\ as small, as in general all constant quantities
characterizing the deviations of our solution from the Maxwellian
one.

\hspace{1cm}We rewrite, in our approximation, the equations of Born
/\ref{teq25}/:
\begin{eqnarray}
\label{teq50}
\hspace{-2cm}&&\
rot\, E\ +\ \frac{\displaystyle 1}{\displaystyle c}\, 
\dot{B}\ =\ 0\hspace{0.5cm};\hspace{0.5cm}
div\, B\ =\ 0\\[0.3cm]
\label{teq51}
\hspace{-2cm}&&\left.\begin{array}{l}
rot\, B\ -\ \frac{\displaystyle 1}{\displaystyle c}\, \dot{E} = 
\frac{\displaystyle 1}{\displaystyle 2}\, \left\{\left[grad\, F,B\right]
\ -\ \frac{\displaystyle 1}{\displaystyle 2}\, \dot{F}E\right\}\\[0.3cm]
div\, E = \frac{\displaystyle 1}{\displaystyle 2}\, \left(grad\, F\cdot E\right)
\hspace{5cm},\hspace{-5cm}
\end{array}\right\}\hspace{3cm}\ \\[0.3cm]
\label{teq52}
\ \hspace{-0.45cm}\textnormal{\tt where}
\hspace{0.45cm}&&\hspace{3cm}
F\ =\ \frac{\displaystyle B^2 - E^2}{\displaystyle b^2}
\end{eqnarray}
The equation $div\, B = 0$ gives us
\begin{eqnarray}
\label{teq53}
B^0_x&=&0\ .
\end{eqnarray}
We are inserting now /\ref{teq49}/ into the remaining equations /\ref{teq50}/.
The first of them:
\begin{eqnarray*}
\frac{\partial E_z}{\partial y}\, -\, \frac{\partial E_y}{\partial z}
\ +\ \frac{1}{c}\, \dot{B_x}&=&0
\end{eqnarray*}
is satisfied automatically.

The second:
\begin{eqnarray*}
\frac{\partial E_x}{\partial z}\, -\, \frac{\partial E_z}{\partial x}
\ +\ \frac{1}{c}\, \dot{B_y}&=&0
\end{eqnarray*}
yields
\begin{eqnarray*}
-\ \frac{\omega_3}{v_3}\, E^0_z\, \sin\gamma_3
\ -\ \frac{\nu_2}{c}\, B^0_y\, \sin\gamma_5&=&0\ .
\end{eqnarray*}
\pagebreak
\label{tpage40}
\refstepcounter{ppage}

It follows:
\begin{eqnarray}
\label{teq54}
&\left.\begin{array}{rcl}
\nu_2\ =\ \omega_3&;&w_2\ =\ v_3\ ;\ \theta_2\ =\ \varphi_3\\[0.3cm]
B^0_y&=&-\ \frac{\displaystyle c}{\displaystyle v_3}\ E^0_z
\end{array}\right\}&\ .
\end{eqnarray}
The third:
\begin{eqnarray*}
\frac{\partial E_y}{\partial x}\, -\, \frac{\partial E_x}{\partial y}
\ +\ \frac{1}{c}\, \dot{B_z}&=&0
\end{eqnarray*}
yields
\begin{eqnarray*}
\frac{\omega_2}{v_2}\, E^0_y\, \sin\gamma_2
\ -\ \frac{\nu_3}{c}\, B^0_z\, \sin\gamma_6&=&0\ .
\end{eqnarray*}
It follows
\begin{eqnarray}
\label{teq55}
&\left.\begin{array}{rcl}
\nu_3\ =\ \omega_2&;&w_3\ =\ v_2\ ;\ \theta_3\ =\ \varphi_2\\[0.3cm]
B^0_z&=&-\ \frac{\displaystyle c}{\displaystyle v_2}\ E^0_y
\end{array}\right\}&\ .
\end{eqnarray}
Taking into account  /\ref{teq53}/, /\ref{teq54}/, and /\ref{teq55}/, we can
rewrite /\ref{teq49}/ in the form
\begin{eqnarray}
\label{teq56}
&\left.\begin{array}{rcl}
\hspace{-0.9cm}E_x&=&
p \cos\nu \left(t - \frac{\displaystyle x}{\displaystyle w} + \varphi_1\right)
 + N \cos\alpha\ =\ p \cos\gamma_1 + N \cos\alpha\\[0.3cm]
\hspace{-0.9cm}E_y&=&
L \cos\omega \left(t - \frac{\displaystyle x}{\displaystyle v} 
+ \varphi_2\right) + N \sin\alpha\ =\ L \cos\gamma_2 + N \sin\alpha\\[0.3cm]
\hspace{-0.9cm}E_z&=&
L^\prime \cos\omega^\prime 
\left(t - \frac{\displaystyle x}{\displaystyle v^\prime} + \varphi_3\right)
\ =\ L^\prime \cos\gamma_3\\[0.3cm]
\hspace{-0.9cm}B_x&=&0\\[0.3cm]
\hspace{-0.9cm}B_y&=&M^\prime \cos\gamma_3\\[0.3cm]
\hspace{-0.9cm}B_z&=&M \cos\gamma_2
\end{array}\right\}&\ ,
\end{eqnarray}
where the notations:

-------------------------------------

1/ Some letters here have been used for other quantities in
chap.\ {\tt\ref{tglavaI}} already. However, this, of course, will 
not lead to any confusion.

\pagebreak
\refstepcounter{ppage}

\hspace{1.2cm}$E^0_x = p$\hspace{0.7cm}, 
$\omega_1 = \nu$\hspace{1.3cm}$etc.$

have been introduced. Here,
\begin{eqnarray}
\label{teq57}
M&=&\frac{c}{v}\, L\hspace{0.7cm};\hspace{0.7cm}
M^\prime\ =\ - \frac{c}{v^\prime}\, L^\prime\ .
\end{eqnarray}
We are inserting /\ref{teq56}/ into the second system of the equations
of Born /\ref{teq51}/. Within the framework of our approximation,
we first calculate the right-hand sides of the equations. 
We find 
\begin{eqnarray*}
F&=&\frac{\displaystyle B^2 - E^2}{\displaystyle b^2}\ .
\end{eqnarray*}
In calculating $F$, we can put
\begin{eqnarray*}
p&=&0\hspace{0.7cm},\ M\ = L\hspace{0.7cm},\hspace{0.7cm}
M^\prime\ =\ -L^\prime
\end{eqnarray*}
because $F$ contains in the denominator $b^2$, i.e.\ a large 
quantity. 

Then
\begin{eqnarray}
\label{teq58}
F&=&-\ \frac{1}{b^2}\ \left(2 L N\, \sin\alpha\, \cos\gamma_2 + N^2\right)\ .
\end{eqnarray}
We find the components of $grad\; F$:
\begin{eqnarray}
\label{teq59}
&\left.\begin{array}{rcl}
\frac{\displaystyle\partial F}{\displaystyle\partial x}&=&
- 2\ \frac{\displaystyle\omega}{\displaystyle v}\ 
\frac{\displaystyle L N}{\displaystyle b^2}\
\sin\alpha\, \sin\gamma_2 
\\[0.3cm]
\frac{\displaystyle\partial F}{\displaystyle\partial y}&=&
\frac{\displaystyle\partial F}{\displaystyle\partial z}\ =\ 0
\end{array}\right\}&
\end{eqnarray}
and, finally, we determine $\dot{F}$:
\begin{eqnarray}
\label{teq60}
\dot{F}&=&2 \omega\ \frac{L N}{b^2}\ \sin\alpha\, \sin\gamma_2\ .
\end{eqnarray}
We formally denote the right-hand sides of the equations 
/\ref{teq51}/ by $\frac{\displaystyle 1}{\displaystyle c}\, j\ 
and\ \rho$. Then
\begin{eqnarray}
\label{teq61}
&\left.\begin{array}{rcl}
\frac{\displaystyle 1}{\displaystyle c}\, j_x&=&
- \frac{\displaystyle\omega}{\displaystyle c}\ 
L\ \frac{\displaystyle N^2}{\displaystyle b^2}\ \sin\alpha\,
\cos\alpha\, \sin\gamma_2\\[0.3cm]
\frac{\displaystyle 1}{\displaystyle c}\, j_y&=&
- \frac{\displaystyle\omega}{\displaystyle c}\ 
L\ \frac{\displaystyle N^2}{\displaystyle b^2}\ \sin^2\alpha\, \sin\gamma_2
\\[0.3cm]
\frac{\displaystyle 1}{\displaystyle c}\, j_z&=&0
\end{array}\right\}&
\end{eqnarray}

\pagebreak
\refstepcounter{ppage}

\begin{eqnarray*}
\ \hspace{2cm}\rho&=&- \frac{\displaystyle\omega}{\displaystyle c}\ 
L\ \frac{\displaystyle N^2}{\displaystyle b^2}\ \sin\alpha\,
\cos\alpha\, \sin\gamma_2\ .\hspace{5cm}(61a)
\end{eqnarray*}
The first of the equations /\ref{teq51}/
\begin{eqnarray*}
\frac{\displaystyle\partial B_z}{\displaystyle\partial y}
\ -\ \frac{\displaystyle\partial B_y}{\displaystyle\partial z}
\ -\ \frac{1}{c}\, \dot{E}_x&=&\frac{1}{c}\, j_x
\end{eqnarray*}
yields\hspace{2cm} 
$\frac{\displaystyle \nu}{\displaystyle c}\ p\ \sin\gamma_1 = 
- \frac{\displaystyle \omega}{\displaystyle c}\ L\
\frac{\displaystyle N^2}{\displaystyle b^2}\ \sin\alpha\, 
\cos\alpha\,\sin\gamma_2$\ .

It follows
\begin{eqnarray}
\label{teq62}
&\left.\begin{array}{rcl}
\nu&=&\omega\hspace{0.5cm};\ \ w\ =\ v\hspace{0.5cm};\ 
\varphi_1\ =\ \varphi_2\\[0.3cm]
p&=&-\ L\ \frac{\displaystyle N^2}{\displaystyle b^2}\ \sin\alpha\, \cos\alpha
\end{array}\right\}&\ .
\end{eqnarray}
This way, the perpendicular component of the electric field of the wave
is found to be different from zero.

The second equation
\begin{eqnarray*}
\frac{\displaystyle\partial B_x}{\displaystyle\partial z}
\ -\ \frac{\displaystyle\partial B_z}{\displaystyle\partial x}
\ -\ \frac{1}{c}\, \dot{E}_y\ =\ \frac{1}{c}\, j_y
\end{eqnarray*}
yields
\begin{eqnarray*}
-\ \frac{\omega}{v}\ M\ \sin\gamma_2\ +\ \frac{\omega}{c}\ L\ \sin\gamma_2&=&
-\ \frac{\omega}{c}\ L\ \frac{N^2}{b^2}\ \sin^2\alpha\,\sin\gamma_2
\end{eqnarray*}
where from we determine $v$:
\begin{eqnarray}
\label{teq63}
\frac{v^2}{c^2}&=&1\ -\ \frac{N^2}{b^2}\ \sin^2\alpha\ .
\end{eqnarray}
The third equation
\begin{eqnarray*}
\frac{\displaystyle\partial B_y}{\displaystyle\partial x}
\ -\ \frac{\displaystyle\partial B_x}{\displaystyle\partial y}
\ -\ \frac{1}{c}\, \dot{E}_z&=&0
\end{eqnarray*}
yields
\begin{eqnarray*}
\frac{\omega^\prime}{v^\prime}\ M^\prime\ \sin\gamma_3\ +\ 
\frac{\omega^\prime}{c}\ L^\prime\ \sin\gamma_3&=&0
\end{eqnarray*}
where from, taking into account /\ref{eq57}/, we obtain: 
\begin{eqnarray}
\label{teq64}
v^\prime&=&c\ .
\end{eqnarray}

\pagebreak
\refstepcounter{ppage}

The last equation $div\; E = \rho$ yields:
\begin{eqnarray*}
\frac{\nu}{w}\ p\ \sin\gamma_1&=& 
-\ \frac{\omega}{c}\ L\ \frac{N^2}{b^2}\ \sin\alpha\,
\cos\alpha\, \sin\gamma_2\ \ \ \ ,
\end{eqnarray*}
which is automatically satisfied by virtue of /\ref{teq62}/.

Thus, taking into account /\ref{teq62}/, /\ref{teq63}/, and /\ref{teq64}/,
we finally obtain the following solution of the equations of Born:
\begin{eqnarray}
\label{teq65}
&\left.\begin{array}{rcl}
E_x&=&p\ \cos\omega 
\left(t - \frac{\displaystyle x}{\displaystyle v} + \varphi\right)
\ +\ N\ \cos\alpha\\[0.3cm]
E_y&=&L\ \cos\omega 
\left(t - \frac{\displaystyle x}{\displaystyle v} + \varphi\right)
\ +\ N\ \sin\alpha\\[0.3cm]
E_z&=&L^\prime\ \cos\omega^\prime 
\left(t - \frac{\displaystyle x}{\displaystyle c} + \varphi^\prime\right)
\\[0.3cm]
B_x&=&0\\[0.3cm]
B_y&=&- L^\prime\ \cos\omega^\prime 
\left(t - \frac{\displaystyle x}{\displaystyle c} + \varphi^\prime\right)
\\[0.3cm]
B_z&=&M \ \cos\omega 
\left(t - \frac{\displaystyle x}{\displaystyle v} + \varphi\right)
\hspace{4cm},\hspace{-4cm}
\end{array}\right\}&\
\end{eqnarray}
where
\begin{eqnarray}
\label{teq66}
&\left.\begin{array}{rcl}
M&=&\frac{\displaystyle c}{\displaystyle v}\, L\\[0.3cm]
\frac{\displaystyle v^2}{\displaystyle c^2}&=&
1\, -\, \frac{\displaystyle N^2}{\displaystyle b^2}\ \sin^2\alpha\\[0.3cm]
p&=&- L\ \frac{\displaystyle N^2}{\displaystyle b^2}\ \sin\alpha\, \cos\alpha
\hspace{1.5cm},\hspace{-1.5cm}
\end{array}\right\}&\
\end{eqnarray}
whereby $L,\ L^\prime,\ \omega,\ \omega^\prime,\ \varphi\ 
and\ \varphi^\prime$ remain arbitrary.

Comparing these formulas with the known formulas of electromagnetic
crystal optics allows us to express their content in the following
illustrative way: In the presence of a homogeneous electrostatic
field of strength $N$, the vectors\hfill\ \linebreak
 $E\ and\ B$ of a plane
wave of light behave the same way, in first approximation, as they
would behave in an uniaxial crystal the optical axis of which would
be directed along $N$ and the main dielectric constants would be 
equal to
\begin{eqnarray}
\label{teq67}
\epsilon&=&1\ +\ \frac{N^2}{b^2}
\end{eqnarray}
and 1, respectively. 

\pagebreak
\refstepcounter{ppage}

Indeed, one can understand the solution /\ref{teq65}/ as 
consisting of the {"}ordinary"\ ray:
\begin{eqnarray*}
E_x&=\ 0\hspace{4.85cm}&B_x\ =\ 0\\[0.3cm]
E_y&=\ 0\hspace{4.85cm}&B_x\ =\ - L^\prime\ 
\cos\omega^\prime 
\left(t - \frac{\displaystyle x}{\displaystyle c} 
+ \varphi^\prime\right)\\[0.3cm]
E_z&=\ L^\prime\ 
\cos\omega^\prime 
\left(t - \frac{\displaystyle x}{\displaystyle c} 
+ \varphi^\prime\right)\hspace{1cm}&B_z\ =\ 0
\end{eqnarray*}
and the {"}extraordinary"\ one:
\begin{eqnarray*}
E_x&=\ p\ \cos\omega \left(t - \frac{\displaystyle x}{\displaystyle v} 
+ \varphi\right)\hspace{1cm}&B_x\ =\ 0\\[0.3cm]
E_y&=\ L\ \cos\omega \left(t - \frac{\displaystyle x}{\displaystyle v} 
+ \varphi\right)\hspace{01cm}&B_y\ =\ 0\\[0.3cm]
E_z&=\ 0\hspace{4.6cm}&B_z\ =\ M\ \cos\omega 
\left(t - \frac{\displaystyle x}{\displaystyle v} + \varphi\right)\ ,
\end{eqnarray*}
propagating in the field of the plane capacitor.

Incidentally speaking, from the formulas /\ref{teq65}/ obtained by us it can 
be seen that between the $x$ and $y$ components of the electric field 
of the ray of light passing through the capacitor of length $l$
the well-known phase difference equal to \linebreak
(putting $\varphi = \varphi^\prime$, for simplicity)
\begin{eqnarray*}
\Delta\varphi&=&\omega\, l\ \left(\frac{1}{v} - \frac{1}{c}\right)
\ =\ \omega\, l\ \frac{c - v}{c v}
\end{eqnarray*}
is accumulating.

Noting that $\frac{\displaystyle v}{\displaystyle c} = 
1 - \frac{\displaystyle N^2}{\displaystyle 2 b^2}$,
we obtain (for the case, when the light is propagating perpendicularly 
to the field of the capacitor):
\begin{eqnarray*}
c - v&=&c\ \frac{N^2}{2 b^2}
\end{eqnarray*}
and
\setcounter{equation}{68}  % equationnumber 67-->68
\begin{eqnarray}
\label{teq69}
\Delta\varphi&=&\frac{\omega}{c}\ l\ \frac{N^2}{2 b^2}\ =\ 
\pi\ \left(\frac{N}{b}\right)^2\ \frac{l}{\lambda}\ .
\end{eqnarray}
In other words, if the ray entering the capacitor is 
linearly polarized it should exit from it elliptically 
polarized. A factical check of this insight, however, can of \linebreak course
not be done because 
$b \equiv$\ I0$^{\mbox{\large\stexttt{I6}}}$ CGSE
and the effect

\pagebreak
\refstepcounter{ppage}

is by far too small for all practically realizable fields.
Irrespective of this, from out a purely theoretical point of view
the effect expressed by formula /\ref{teq69}/ is, without any \linebreak 
doubt, interesting because it gives an example for a completely
peculiar deviation from the laws of Maxwell electrodynamics:
The deviation does not  appear on atomic scales and not for
strong fields but in the comparison of observations performed at
\udensdash{l$\ \atop\ $\hspace{-0.33cm}arge\hspace{-0.05cm}\ } 
distances from each other. In fact, for sufficiently large $l$
the quantity $\Delta\varphi$ can be arbitrarily large for 
arbitrarily small $N/b$. This feature of the theory is in essence
connected with its nonlinearity: The corrections introduced by
small nonlinear terms can have a very different character
than small linear corrections.

Departing from the formula /\ref{teq65}/, it is not difficult 
to calculate for our solution also the second pair of Born vectors
$D$ and H. For this purpose one has to use the formulas 
\begin{eqnarray*}
D&=&\frac{\displaystyle E}{\displaystyle\sqrt{1+F}}\hspace{1.2cm};
\hspace{1.2cm}
H\ =\ \frac{\displaystyle B}{\displaystyle\sqrt{1+F}}\ \ \ .\hspace{5cm}(23)
\end{eqnarray*}
For simplicity, we put $\omega = \omega^\prime$ and 
$\varphi = \varphi^\prime = 0$.

We have:
\begin{eqnarray}
\label{teq70}
\frac{\displaystyle 1}{\displaystyle\sqrt{1+F}}&=&
1 - \frac{1}{2}\, F\ =\ 
1\ +\ \frac{\displaystyle L N}{\displaystyle b^2}\ \sin\alpha\,
\cos\omega \left(t -\frac{\displaystyle x}{\displaystyle v}\right)
\ +\ \frac{N}{2 b^2}\ . \ \ \ 
\end{eqnarray}
Inserting /\ref{teq65}/ into /\ref{teq23}/ we obtain:
\begin{eqnarray}
\label{teq71}
\hspace{-0.5cm}D_x&=&\frac{\displaystyle E_x}{\displaystyle\sqrt{1+F}} =
\left(p + L 
\frac{\displaystyle N}{\displaystyle b^2}\sin\alpha\,\cos\alpha\right) 
\cos\omega \left(t -\frac{\displaystyle x}{\displaystyle v}\right)
+ N \cos\alpha\cdot\left(1 + \frac{N^2}{2 b^2}\right).\ \ \ 
\end{eqnarray}
Taking into account  /\ref{teq62}/, we find the following expression
for $D_x$
\begin{eqnarray}
\label{teq72}
D_x&=&N\, \left(1 + \frac{N^2}{2 b^2}\right)\, \cos\alpha\ .
\end{eqnarray}
Analogously,
\begin{eqnarray}
\label{teq73}
\hspace{-2cm}D_y&=&\frac{\displaystyle E_y}{\displaystyle\sqrt{1+F}} =
N\,\left(1 + \frac{N^2}{2 b^2}\right)\,\sin\alpha +
L\,\left(1 + \frac{N^2}{2 b^2}\right)\, \cos\omega 
\left(t -\frac{\displaystyle x}{\displaystyle v}\right) 
+\ \ \ \nonumber\\[0.3cm]
&&+ N\,\frac{L^2}{b^2}\,\sin\alpha\
\frac{1 + \cos 2\omega 
\left(t -\frac{\displaystyle x}{\displaystyle v}\right)}{2}
+ L\,\frac{N^2}{b^2}\,\sin^2\alpha\,\cos\omega
\left(t -\frac{\displaystyle x}{\displaystyle v}\right)\ .
\end{eqnarray}

\pagebreak
\label{tpage46}
\refstepcounter{ppage}

Furthermore,
\begin{eqnarray}
\label{teq74}
D_z&=&\frac{\displaystyle E_z}{\displaystyle\sqrt{1+F}} =
L^\prime\,\left(1 + \frac{N^2}{2 b^2}\right)\,\cos\omega 
\left(t -\frac{\displaystyle x}{\displaystyle c}\right) 
+\ \ \ \nonumber\\[0.3cm]
&&+ N\,\frac{L L^\prime}{b^2}\,\sin\alpha\
\frac{\cos 2\omega 
\left[t -\frac{\displaystyle x}{\displaystyle 2}
\left(\frac{\displaystyle 1}{\displaystyle c} + 
\frac{\displaystyle 1}{\displaystyle v}\right)\right] +
\cos\omega\, x \left(\frac{\displaystyle 1}{\displaystyle c} - 
\frac{\displaystyle 1}{\displaystyle v}\right)}{2}\ .\ \ 
\end{eqnarray}
Exactly the same way, one can show that
\begin{eqnarray}
\label{teq75}
H_x&=&0\\[0.3cm]
\label{teq76}
H_y&=&- L^\prime\,\left(1 + \frac{N^2}{2 b^2}\right)\,\cos\omega 
\left(t -\frac{\displaystyle x}{\displaystyle c}\right) 
-\ \ \ \nonumber\\[0.3cm]
&&- N\,\frac{L L^\prime}{b^2}\,\sin\alpha\
\frac{\cos 2\omega 
\left[t -\frac{\displaystyle x}{\displaystyle 2}
\left(\frac{\displaystyle 1}{\displaystyle c} + 
\frac{\displaystyle 1}{\displaystyle v}\right)\right] +
\cos\omega\, x \left(\frac{\displaystyle 1}{\displaystyle v} - 
\frac{\displaystyle 1}{\displaystyle c}\right)}{2}\ \ \ \\[0.3cm]
\label{teq77}
H_z&=&\frac{\displaystyle c}{\displaystyle v}\, L
\,\left(1 + \frac{N^2}{2 b^2}\right)\,
\cos\omega\left(t - \frac{\displaystyle x}{\displaystyle v}\right)
+ N\,\frac{L^2}{b^2}\,\sin\alpha\
\frac{1 + 
\cos 2\omega\left(t - \frac{\displaystyle x}{\displaystyle v}\right)}{2}
\ .\ \ \ \ 
\end{eqnarray}
From the obtained formulas /\ref{teq72}/-/\ref{teq77}/, we see
that the 
\udensdash{p$\ \atop\ $\hspace{-0.33cm}er-\hspace{-0.05cm}\ } \linebreak
\udensdash{pendicular\hspace{-0.05cm}\ }
components of both the vectors $D_x$ and $H_x$ of our \underline{wave} turn out
to be equal to zero$^{\mbox{\large\stexttt{1/}}}$. Indeed, one easily sees that
the formula /\ref{teq72}/ yields the quantity $D_x$ determined by
the field of the capacitor $E_x = N \cos\alpha$ alone because in this
case 
\begin{eqnarray*}
D_x&=&\frac{\displaystyle N \cos\alpha}{\displaystyle\sqrt{1+F}}
\ =\ \frac{\displaystyle N \cos\alpha}{\displaystyle\sqrt{1- N^2/b^2}}
\ =\ N\,\left(1 + \frac{N^2}{2 b^2}\right)\,\cos\alpha\ . 
\end{eqnarray*}
This fact shows that, in difference to crystal optics, in the present case
the direction of the Poynting vector of the wave 
- recalling that in the theory of Born the Poynting vector is determined
by the product $[D \times B]$ - coincides with the direction of the 
wave vector. 

Furthermore, the vectors $D$ and H differ from $E\ and\ B$

----------------------------------------

1/ Which, incidentally speaking, one could have said in advance
on the basis of the equation $div\; D = 0$.

\pagebreak
\refstepcounter{ppage}

by the presence of small additive contributions, which constant terms
as well as terms periodic in $x\ and\ t$ belong to, \linebreak
whereby also terms 
with doubled frequency exist among the latter. Here from it can be shown
that in the present case no solution of the equations of Born
exists at all for which, instead of the components $E\ and\ B$, 
the components $D$ and H\linebreak  could be represented in the form of 
plane waves of the form  /\ref{teq65}/. On first sight, this situation 
seems to be somewhat strange because (as has been said already in a
comment at the beginning of this \S) we could construct the whole
theory in general, instead of on the basis of the vectors $E\ and\ B$,
on the basis of the vectors $D$ and H whereby, in particular for
our problem, all calculations would differ very little from the
calculations presented above. However, in order to understand 
the situation one has to take into account the following.

Imagine that we found some solution of the equations of Born that is valid
within the well-known approximation, for example, similarly as
our solution /\ref{teq65}/, valid up to terms having in the 
denominator $b^2$. Let this solution be determined by the vectors 
$E^0\ and\ B^0$ or, in the induction vector representation, by
the vectors $D^0$, $H^0$. Furthermore, $E_1\ and\ B_1$ are any
electromagnetic field having the following properties: /1/ Both
vectors $E_1\ and\ B_1$ are small in size,
i.e.\ they have in the denominator $b^2$ and, consequently, in 
our approximation $D_1 = E_1$, and $H_1 = B_1$\ ; /2/ this field is
in our approximation a solution of the Maxwell equations,
i.e.\ the quantities

\hspace{1cm}$rot\; E_1\ 
+\ \frac{\displaystyle 1}{\displaystyle c}\,\dot{B_1}$
\hspace{2cm};\hspace{2cm}
$rot\; B_1\ -\ \frac{\displaystyle 1}{\displaystyle c}\,\dot{E_1}
\hspace{1cm}etc.$

are quantities of higher order (They have in the denominator
$b^4$ or $b^6$ etc..). Then, it is clear that the sum of both of these 
fields, i.e.\ the field represented by the vectors $E\ and\ B$ as
 
\hspace{3cm}$E_0 + E_1$\hspace{1cm}and\hspace{1cm}$B_0 + B_1$\hspace{1cm},

and by the vectors $D$ and H (in our approximation) as

\pagebreak
\label{tpage48}
\refstepcounter{ppage}

\hspace{3cm}$D^0 + E_1$\hspace{1cm}and\hspace{1cm}$H^0 + B_1$

will also be a solution of the equations of Born (irrespective
of their nonlinearity) up to terms of higher order. \linebreak
This assertion is so obvious that it does not make any sense 
to specifically be concerned with its proof.

This way, one can convince oneself easily that the mentioned
{"}additions"\ to our fields $D$ and H determined by the formulas
/\ref{teq72}/-/\ref{teq77}/ belong to the type of fields $E_1\ and\ B_1$
again we just have discussed. Indeed, let us consider the field
\begin{eqnarray}
\label{teq78}
&\left.\begin{array}{rcl}
E_{1z}&=&D_{1z}\ =\ \frac{\displaystyle L L^\prime}{\displaystyle2 b^2}
\ N\ \sin\alpha\,
\cos 2\omega\left[t - \frac{\displaystyle x}{\displaystyle 2}
\left(\frac{\displaystyle 1}{\displaystyle c} 
+ \frac{\displaystyle 1}{\displaystyle v}\right)\right]
\\[0.3cm]
B_{1y}&=&H_{1y}\ =\ - \frac{\displaystyle L L^\prime}{\displaystyle 2 b^2}
\ N\ \sin\alpha\,
\cos 2\omega\left[t - \frac{\displaystyle x}{\displaystyle 2}
\left(\frac{\displaystyle 1}{\displaystyle c} 
+ \frac{\displaystyle 1}{\displaystyle v}\right)\right]
\end{array}\right\}&\ .
\end{eqnarray}
The amplitude itself of this wave is a quantity of first order in size;
furthermore, it clearly satisfies the Maxwell equations with
an accuracy up to second order terms (i.e.\ terms having $b^4$
in the denominator). Consequently, in accordance with the remarks just
made, we can safely deduct from /\ref{teq65}/ the expression
/\ref{teq78}/; as result, we obtain an expression which will also be
a solution of the equations of Born with the same accuracy as 
/\ref{teq65}/. It is clear, that this way one can construct,
in particular, also a solution for which the variable components of
$D$ and $N$ would be represented in the form of simple harmonic
functions of $x\ and\ t$.\\

\refstepcounter{subsection}
\label{tglavaIIpara3}
\hspace{1cm}\S\ 3.\ 
{"}\udensdash{Scattering{"}$\ \atop\ $\hspace{-0.1cm}of light
off the constant field of a\hspace{-0.05cm}\ }

\hspace{2.2cm}
\udensdash{\ plane capacitor$\ \atop\ $\hspace{-0.33cm}.\ 
\ \hspace{-0.05cm}\ }

\hspace{1cm}Up to now, we have set ourselves the task of simply
finding a known class of solutions of the field equations hereby
not clarifying under which conditions this or that of these solutions
is in fact realized. We now set the task somewhat more concretely.

We imagine that in the space between the plates of the plane capacitor

\pagebreak
\refstepcounter{ppage}

a light ray is allowed into. What does then happen? The answer on this
question should obviously be given on the basis of the boundary 
conditions for the electromagnetic field.

First, we formulate what these boundary conditions should consist in.
As we saw, the equations of 
\udensdash{B$\ \atop\ $\hspace{-0.33cm}orn\hspace{-0.05cm}\ } have
a form analogous to the 
\udensdash{M$\ \atop\ $\hspace{-0.33cm}axwell\hspace{-0.05cm}\ }
equations in a medium with a dielectric constant and a magnetic
permeability different from unity, but in the absence of charges and
currents, i.e.\ 
\begin{eqnarray*}
rot\, E\ +\ \frac{1}{c}\,\dot{B}\ =\ 0\hspace{2cm}div\, B\ =\ 0
\\[0.3cm]
rot\, H\ -\ \frac{1}{c}\,\dot{D}\ =\ 0\hspace{2cm}div\, D\ =\ 0\ .
\end{eqnarray*}
Here from, in analogy with ordinary electrodynamics, one can
immediately conclude that the boundary conditions must have the form:
\begin{eqnarray}
\label{teq79}
&\left.\begin{array}{rclcrcl}
E_{t1}&=&E_{t2}&\hspace{2.5cm}&B_{n1}&=&B_{n2}\hspace{1cm}\\[0.3cm]
H_{t1}&=&H_{t2}&\hspace{2.5cm}&D_{n1}&=&D_{n2}\hspace{1cm}
\end{array}\right\}&\ ,
\end{eqnarray}
where the labels $t\ and\ n$ indicate the tangential and normal 
components of the vectors relative to the interface of our
two {"}media"\ - vacuum and the field of the capacitor.

We will apply the boundary conditions /\ref{teq79}/ in answering
the question asked. For simplicity, we assume that the light ray
propagates perpendicularly to the field lines of the capacitor.
Then, our coordinate trihedron is given in the form

\begin{figure}[h]
\caption{\label{tris1}}
\unitlength1.mm
\begin{picture}(150,50)
\put(30,30){\line(1,0){70}}
\put(30,30){\line(0,1){20}}
\put(30,30){\line(-1,-1){20}}
\put(30,43){\line(1,0){48}}
\put(30,15){\line(1,0){52.5}}
\put(65,22){\line(0,1){17}}
\put(100,30){\line(-2,-1){3.6}}
\put(100,30){\line(-2,1){3.6}}
\put(30,50){\line(-1,-2){1.8}}
\put(30,50){\line(1,-2){1.8}}
\put(65,39){\line(-1,-2){1.8}}
\put(65,39){\line(1,-2){1.8}}
\put(10,10){\line(1,3){1.2}}
\put(10,10){\line(3,1){3.6}}
\put(30,43){\line(1,1){3.6}}
\put(32.5,43){\line(1,1){3.6}}
\put(35,43){\line(1,1){3.6}}
\put(37.5,43){\line(1,1){3.6}}
\put(40,43){\line(1,1){3.6}}
\put(42.5,43){\line(1,1){3.6}}
\put(45,43){\line(1,1){3.6}}
\put(47.5,43){\line(1,1){3.6}}
\put(50,43){\line(1,1){3.6}}
\put(52.5,43){\line(1,1){3.6}}
\put(55,43){\line(1,1){3.6}}
\put(57.5,43){\line(1,1){3.6}}
\put(60,43){\line(1,1){3.6}}
\put(62.5,43){\line(1,1){3.6}}
\put(65,43){\line(1,1){3.6}}
\put(67.5,43){\line(1,1){3.6}}
\put(70,43){\line(1,1){3.6}}
\put(72.5,43){\line(1,1){3.6}}
\put(75,43){\line(1,1){3.6}}
\put(32.5,15){\line(-1,-1){3.6}}
\put(35,15){\line(-1,-1){3.6}}
\put(37.5,15){\line(-1,-1){3.6}}
\put(40,15){\line(-1,-1){3.6}}
\put(42.5,15){\line(-1,-1){3.6}}
\put(45,15){\line(-1,-1){3.6}}
\put(47.5,15){\line(-1,-1){3.6}}
\put(50,15){\line(-1,-1){3.6}}
\put(52.5,15){\line(-1,-1){3.6}}
\put(55,15){\line(-1,-1){3.6}}
\put(57.5,15){\line(-1,-1){3.6}}
\put(60,15){\line(-1,-1){3.6}}
\put(62.5,15){\line(-1,-1){3.6}}
\put(65,15){\line(-1,-1){3.6}}
\put(67.5,15){\line(-1,-1){3.6}}
\put(70,15){\line(-1,-1){3.6}}
\put(72.5,15){\line(-1,-1){3.6}}
\put(75,15){\line(-1,-1){3.6}}
\put(77.5,15){\line(-1,-1){3.6}}
\put(80,15){\line(-1,-1){3.6}}
\put(82.5,15){\line(-1,-1){3.6}}
\put(70,0){\it Fig.\ 1.}
\put(68,37){$N$}
\put(14,8){$z$}
\put(32,49){$y$}
\put(103,29){$x$}
\end{picture}
\end{figure}

As the normal components of the fields B and $D$ are absent here

\pagebreak
\label{tpage50}
\refstepcounter{ppage}

the boundary conditions /\ref{teq79}/ are transformed into
\begin{eqnarray}
\label{teq80}
&\left.\begin{array}{rclcrcl}
E_{y1}&=&E_{y2}&\hspace{2.5cm}&H_{y1}&=&H_{y2}\hspace{1cm}\\[0.3cm]
E_{z1}&=&E_{z2}&\hspace{2.5cm}&H_{z1}&=&H_{z2}\hspace{1cm}
\end{array}\right\}&\ ,
\end{eqnarray}
where the label 1 corresponds to the field in vacuum and the label
2 - in the capacitor.
 
The first of these conditions entails the requirement that the 
electrostatic field of the capacitor itself is not cut off 
immediately but {"}approaches zero"\ gradually as it would happen
also in the usual electrodynamics. As the calculation of this
fall-off of the field is not of any interest to us we will impose 
the boundary conditions on the field of the light wave 
\udensdash{o$\ \atop\ $\hspace{-0.33cm}nly\hspace{-0.05cm}\ }
not taking into account hereby the field of the capacitor.
Then, the calculations can be performed without any difficulties.

\hspace{1cm}We choose an arbitrary wave from the two waves representing
our solution, for example, that one corresponding to the 
{"}extraordinary"\ ray in crystal optics. The boundary conditions
/\ref{teq80}/ for it will have the form: 
\begin{eqnarray}
\label{teq81}
E_{y1}&=&E_{y2}\hspace{1.5cm};\hspace{1.5cm}H_{z1}\ =\ H_{z2}\ .
\end{eqnarray}
Let in vacuum be
\begin{eqnarray}
\label{teq82}
&\left.\begin{array}{rcl}
E_{y1}&=&A\ \cos\omega\left(t -
\frac{\displaystyle x}{\displaystyle c}\right)\hspace{1cm}\\[0.3cm]
H_{z1}&=&A\ \cos\omega\left(t -
\frac{\displaystyle x}{\displaystyle c}\right)\hspace{1cm}\hspace{1cm}
\end{array}\right\}&\ .
\end{eqnarray}
We know that in the capacitor the field of the wave is characterized
by the quantities
\begin{eqnarray}
\label{teq83}
&\left.\begin{array}{rcl}
\hspace{-1.5cm}E_{y2}&=&L\, \cos\omega\left(t -
\frac{\displaystyle x}{\displaystyle v}\right)\\[0.3cm]
\hspace{-1.5cm}H_{z2}&=&\frac{\displaystyle c}{\displaystyle v}\, L\,
\left(1 + \frac{\displaystyle N^2}{\displaystyle 2 b^2}\right)\,
\cos\omega\left(t -
\frac{\displaystyle x}{\displaystyle v}\right)
+ N\, \frac{\displaystyle L^2}{\displaystyle 2 b^2}\,
\cos 2\omega\left(t -
\frac{\displaystyle x}{\displaystyle v}\right)
+ N\, \frac{\displaystyle L^2}{\displaystyle 2 b^2}
\end{array}\right\}.&\ 
\end{eqnarray}

\pagebreak
\refstepcounter{ppage}

We have seen further above that it makes sense to write down
solutions of the equations of Born up to small fields only that are,
in our approximation, solutions of the Maxwell equations in vacuo.
In particular, we can, therefore, in /\ref{teq83}/ discard the last,
constant term because it is small and it is an exact solution of the
Maxwell equations. Obviously, we cannot simply discard also the second
term in the expression for $H_{2z}$ because the {"}purely magnetic"\ wave, 
varying in $x\ and\ t$, is not a solution of the Maxwell equations.
On a \linebreak first, superficial glance, this fact fairly complicates the 
construction of a solution satisfying the boundary conditions. In reality,
however, such a solution can be constructed in a fairly simple manner
by means of the following ansatz.

We add to the solution /\ref{teq83}/ the following additional field
obeying both conditions written down at the end of 
\S\ {\tt\ref{tglavaIIpara2}}:
\begin{eqnarray}
\label{teq84}
&\left.\begin{array}{rcl}
E_{y2}^\ast&=&-\ N\ \frac{\displaystyle L^2}{\displaystyle 4 b^2}\
\cos 2\omega\left(t -
\frac{\displaystyle x}{\displaystyle v}\right)\\[0.3cm]
H_{z2}^\ast&=&-\ N\ \frac{\displaystyle L^2}{\displaystyle 4 b^2}\
\cos 2\omega\left(t -
\frac{\displaystyle x}{\displaystyle v}\right)\hspace{0.8cm}
\end{array}\right\}&\ .
\end{eqnarray}
Then, we obtain the following solution.
\begin{eqnarray}
\label{teq85}
&\left.\begin{array}{rcl}
E_{y2}&=&L\ \cos\omega\left(t -
\frac{\displaystyle x}{\displaystyle v}\right)
- N\ \frac{\displaystyle L^2}{\displaystyle 4 b^2}\
\cos 2\omega\left(t -
\frac{\displaystyle x}{\displaystyle v}\right)\\[0.3cm]
H_{z2}&=&\frac{\displaystyle c}{\displaystyle v}\, L\,
\left(1 + \frac{\displaystyle N^2}{\displaystyle 2 b^2}\right)\,
\cos\omega\left(t -
\frac{\displaystyle x}{\displaystyle v}\right)
+ N\ \frac{\displaystyle L^2}{\displaystyle 4 b^2}\  
\cos 2\omega\left(t -
\frac{\displaystyle x}{\displaystyle v}\right)
\end{array}\right\}.&\ \ 
\end{eqnarray}
We now demand that for $x = 0$ the solution in vacuo should agree with
the solution /\ref{teq85}/ in the capacitor. For this, it 
is sufficient to introduce two reflected waves: one with frequency
$\omega$ and one with doubled frequency $2 \omega$. Instead of 
/\ref{teq82}/, the solution in vacuo will be

\pagebreak
\refstepcounter{ppage}

\begin{eqnarray}
\label{teq86}
&\left.\begin{array}{rcl}
E_{y1}&=&A\ \cos\omega\left(t - \frac{\displaystyle x}{\displaystyle c}\right)
- m\ \cos\omega\left(t + \frac{\displaystyle x}{\displaystyle c}\right)
- N\ \frac{\displaystyle L^2}{\displaystyle 4\; b^2}\ 
\cos 2\omega\left(t + \frac{\displaystyle x}{\displaystyle c}\right)\\[0.3cm]
H_{z1}&=&A\ \cos\omega\left(t - \frac{\displaystyle x}{\displaystyle c}\right)
+ m\ \cos\omega\left(t + \frac{\displaystyle x}{\displaystyle c}\right)
+ N\ \frac{\displaystyle L^2}{\displaystyle 4\; b^2}\ 
\cos 2\omega\left(t + \frac{\displaystyle x}{\displaystyle c}\right)
\end{array}\right\}.&\ \ \ \ 
\end{eqnarray}

Setting /\ref{teq85}/ and /\ref{teq86}/ equal to each other, 
we find furthermore
\begin{eqnarray*}
A - m&=&L\\[0.3cm]
A + m&=&\frac{\displaystyle c}{\displaystyle v}\ L\ 
\left(1 + \frac{\displaystyle N^2}{\displaystyle 2 b^2}\right)
\end{eqnarray*}
and noting that 
\begin{eqnarray*}
\frac{\displaystyle c}{\displaystyle v}&=&1
+ \frac{\displaystyle N^2}{\displaystyle 2 b^2}
\end{eqnarray*}
finally we find
\begin{eqnarray}
\label{teq87}
&\left.\begin{array}{rcl}
L&=&A\ \left(1 - \frac{\displaystyle N^2}{\displaystyle 2 b^2}\right)\\[0.5cm]
m&=&A\ \frac{\displaystyle N^2}{\displaystyle 2 b^2}
\end{array}\right\}&\ . \ 
\end{eqnarray}
This way, knowing the amplitude of the wave in vacuo and the
field strength of the capacitor we can determine the amplitudes
of the {"}incoming"\ wave and the wave {"}reflected"\ by this field.
The problem is solved.

The obtained expressions show that according to the theory of Born,
besides the distortion effect described in \S\ {\tt\ref{tglavaIIpara2}}
for the wave propagating through the capacitor, also a 
completely peculiar physical phenomenon occurs which can be
characterized as a scattering of light off the constant field
of the capacitor. In fact, we found that the
wave of light with frequency $\omega$ and amplitude $A$ 
incoming to this field sets
off two propagating waves: one with frequency $\omega$ and an
amplitude almost equal to $A$ and a second - with frequency
$2 \omega$ and a small amplitude (having $b^2$ in the denominator) 
and in addition two small 
\udensdash{r$\ \atop\ $\hspace{-0.33cm}eflected\hspace{-0.05cm}\ }
waves with frequencies $\omega\ and\ 2 \omega$. In principle, these
waves could be observed.

\pagebreak
\label{tpage53}
\refstepcounter{ppage}

The very course of the performed calculations shows that this
indeterminacy of the solution of the field equations met in the
preceding paragraph and consisting in a tolerance of
solutions of the equations of Born against adding small expressions of
a special type completely vanishes in course of the 
concrete setting of the problem we are dealing with.
In any case, on the basis of the boundary conditions we can
completely uniquely say which of the expressions of the mentioned type
have to be added to the initial solution$^{\mbox{\large\stexttt{1/}}}$.

The question of the uniqueness of the solutions obtained here can be
asked more broadly, however, - namely, one can ask how unique 
the choice of the zeroth approximation in the form /\ref{teq49}/ is,
i.e.\ does one have to assume that the components of the {"}large"\
propagating wave are harmonic functions of the coordinates and time.
We do not attempt here to give a mathematically strict solution of 
this problem and present few consideration of a purely physical
character only which, incidentally, 
\udensdash{i$\ \atop\ $\hspace{-0.33cm}n essence\hspace{-0.05cm}\ }
completely solve, as seems to us, the problem.

\hspace{1cm}Let the field components $E_y\ and\ B_z$ of the
incoming wave in vacuo have the form
\begin{eqnarray*}
A\ \cos\omega\left(t - \frac{\displaystyle x}{\displaystyle c}\right)&=&
A\ \cos\omega t\ \cos\frac{\displaystyle \omega x}{\displaystyle c}
+ A\ \sin\omega t\ \sin\frac{\displaystyle \omega x}{\displaystyle c}\ .
\end{eqnarray*}
It is clear that the dependence of the components of the \linebreak {"}large"\
propagating wave on 
\udensdash{t$\ \atop\ $\hspace{-0.33cm}ime\hspace{-0.05cm}\ }
must of course have the same character, otherwise it is impossible
to satisfy the boundary conditions. Just on the basis of these boundary 
conditions only, nothing can be said 
about the character of the dependence of these components on the 
coordinates. This way, the most general ansatz for these quantities has
the form

----------------------------------------

1/\ We incidentally note that the precise value of the 
propagation velocity of all of these small waves can obviously perhaps
be obtained in considering the second approximation.

\pagebreak
\refstepcounter{ppage}

\setcounter{equation}{88}  % equationnumber 87-->88

\begin{eqnarray}
\label{teq89}
&\left.\begin{array}{rcl}
E_y&=&F_1(x)\ \cos\omega t\ +\ F_2(x)\ \sin\omega t\\[0.3cm]
B_z&=&\Phi_1(x)\ \cos\omega t\ +\ \Phi_2(x)\ \sin\omega t
\end{array}\right\}&\ . \ 
\end{eqnarray}
We will denote the maximal values of the functions $F(x)$ and $\Phi(x)$ 
by $F$ and $\Phi$. Then, we can obviously represent /\ref{teq89}/
in the form: 
\begin{eqnarray*}
E_y&=&F_1\ \cos f_1(x)\ \cos\omega t
\ +\ F_2\ \sin f_2(x)\ \sin\omega t\\[0.3cm]
B_z&=&\Phi_1\ \cos\varphi_1(x)\ \cos\omega t
\ +\ \Phi_2\ \sin\varphi_2(x)\ \sin\omega t\ \ \ \ \ ,
\end{eqnarray*}
where $f_1(x),\ f_2(x),\ \varphi_1(x)\ and\ \varphi_2(x)$
are certain new functions of $x$ chosen in an appropriate manner.

\hspace{1cm}We now make the following completely natural 
assumption that for the propagation along the $ox$-axis the 
electromagnetic energy of our wave does not concentrate and
scatter, i.e.\ the flow of the Poynting vector through a 
unit surface placed perpendicularly on an arbitrary point 
on the $ox$-axis does not depend on $x$ in the temporal average. 

In other words, operating with the zeroth approximation all the time 
we demand that the temporal average of the product $E_y B_z$
did not depend on $x$, i.e.\ that
\begin{eqnarray*}
\overline{E_y B_z}^{\ t}&=&
\frac{1}{2}\left(F_1\ \Phi_1\ \cos f_1(x) \cos\varphi_1
+ F_2\ \Phi_2\ \sin f_2(x) \sin\varphi_2\right)\ =\ const\ .
\end{eqnarray*}
Obviously, this can apply only if  
\begin{eqnarray*}
F_1\ \Phi_1&=&F_2\ \Phi_2\\[0.3cm]
and \hspace{1.5cm}f_1(x)\ =&f_2(x)&=\varphi_1(x)\ =\ 
\varphi_2(x)\ =\ f(x)\ .
\end{eqnarray*}
This way
\begin{eqnarray}
\label{teq90}
&\left.\begin{array}{rcl}
E_y&=&F_1(x)\ \cos f(x) \cos\omega t\ +\ F_2(x)\ \sin f(x) \sin\omega t\\[0.3cm]
B_z&=&\Phi_1(x)\ \cos f(x) \cos\omega t\ +\ \Phi_2(x)\ \sin f(x) \sin\omega t
\end{array}\right\}&\ . \ 
\end{eqnarray}
Inserting /\ref{teq90}/ into the first equation of Born
$\frac{\displaystyle\partial E_y}{\displaystyle\partial x} 
+ \frac{\displaystyle 1}{\displaystyle c}\;\dot{B}_z = 0$

\pagebreak
\refstepcounter{ppage}

we find that
\begin{eqnarray*}
f^\prime(x)&=&const\ ,
\end{eqnarray*}
i.e.\ that $f(x)$ is a linear function of $x$. This way we
immediately arrive at the solution of the type /\ref{teq49}/.\\

\refstepcounter{subsection}
\label{tglavaIIpara4}
\hspace{1cm}\udensdash{\S\ 4.\ A plane wave of light in a homogeneous
magnetic field.\hspace{-0.05cm}\ }\\[-0.3cm]

\hspace{1cm}We consider here a problem analogous to that studied in 
\S\ {\tt\ref{tglavaIIpara2}} with one difference only that, instead of
an electric field, we will be dealing with a homogeneous
\udensdash{m$\ \atop\ $\hspace{-0.33cm}agnetic\hspace{-0.05cm}\ }
field filling the whole space; as also considered earlier, a plane
wave of light is propagating in this space. As the method of solving
this problem as well as the obtained results will be very similar
to the exposition in the preceding paragraphs we will not 
rest for their detailed discussion and will give the results only.

Thus, we have to find the solutions of the equations /\ref{teq50}/ and 
/\ref{teq51}/ for our problem.

We choose, as done earlier, the direction of the wave vector along
the $ox$-axis and the $xy$-plane as the plane spanned by this vector 
and the external field $h$. The angle between these vectors is denoted
by $\alpha$.

We will search for the solution of the equations of Born /\ref{teq50}/ and 
/\ref{teq51}/ in the form
\begin{eqnarray}
\label{teq91}
\ \hspace{-1.5cm}&\left.\begin{array}{rcl}
E_x&=&E^0_x\ \cos\omega_1\left(t - 
\frac{\displaystyle x}{\displaystyle v_1}\right)\hspace{0.7cm}
B_x\ =\ B^0_x\ \cos\nu_1\left(t - 
\frac{\displaystyle x}{\displaystyle w_1}\right)\ +\ h\ \cos\alpha\\[0.3cm]
E_y&=&E^0_y\ \cos\omega_2\left(t - 
\frac{\displaystyle x}{\displaystyle v_2}\right)\hspace{0.7cm}
B_y\ =\ B^0_y\ \cos\nu_2\left(t - 
\frac{\displaystyle x}{\displaystyle w_2}\right)\ +\ h\ \sin\alpha\\[0.3cm]
E_z&=&E^0_z\ \cos\omega_3\left(t - 
\frac{\displaystyle x}{\displaystyle v_3}\right)\hspace{0.7cm}
B_z\ =\ B^0_z\ \cos\nu_3\left(t - 
\frac{\displaystyle x}{\displaystyle w_3}\right)
\end{array}\right\}&
\end{eqnarray}

\pagebreak
\refstepcounter{ppage}

where, for brevity, the initial phases are omitted.

We insert /\ref{teq91}/ into /\ref{teq50}/.

The equation $div\; B = 0$ gives us
\begin{eqnarray}
\label{teq92}
B^0_x&=&0\ .
\end{eqnarray}
The equation $rot\; E + \frac{\displaystyle 1}{\displaystyle c}\;\dot{B}$ 
gives
\begin{eqnarray}
\label{teq93}
&\left.\begin{array}{rcl}
\nu_2&=&\omega_3\ \ ;\ \ w_2\ =\ v_3\ \ ;\ \ 
B^0_y\ =\ -\ \frac{\displaystyle c}{\displaystyle v_3}\ E^0_z\\[0.3cm]
\nu_3&=&\omega_2\ \ ;\ \ w_3\ =\ v_2\ \ ;\ \ 
B^0_z\ =\ \frac{\displaystyle c}{\displaystyle v_2}\ E^0_y
\end{array}\right\}&\ .\ 
\end{eqnarray}
Taking into account /\ref{teq92}/ and /\ref{teq93}/ and
introducing a notation analogous to that applied in the 
formulas /\ref{teq56}/, we obtain instead of /\ref{teq91}/ 
\begin{eqnarray}
\label{teq94}
\ \hspace{-1.5cm}&\left.\begin{array}{rcl}
E_x&=&p\ \cos\nu\left(t - 
\frac{\displaystyle x}{\displaystyle w}\right)\hspace{1.2cm}
B_x\ =\ h\ \cos\alpha\\[0.3cm]
E_y&=&L\ \cos\omega\left(t - 
\frac{\displaystyle x}{\displaystyle v}\right)\hspace{1.15cm}
B_y\ =\ M^\prime\ \cos\omega^\prime\left(t - 
\frac{\displaystyle x}{\displaystyle v^\prime}\right) + h\ \sin\alpha\\[0.3cm]
E_z&=&L^\prime\ \cos\omega^\prime\left(t - 
\frac{\displaystyle x}{\displaystyle v^\prime}\right)\hspace{0.7cm}
B_z\ =\ M\ \cos\omega\left(t - 
\frac{\displaystyle x}{\displaystyle v}\right)
\end{array}\right\}&\hspace{-0.2cm},\ \
\end{eqnarray}
where
\begin{eqnarray}
\label{teq95}
M&=&\frac{c}{v}\ L\hspace{1cm}{\stexttt and}\hspace{1cm}
M^\prime\ =\ -\ \frac{c}{v^\prime}\ L^\prime\ .
\end{eqnarray}
We insert now /\ref{teq94}/ into /\ref{teq51}/. For this purpose,
we first find $F$:
\begin{eqnarray}
\label{teq96}
F&=&\frac{1}{b^2}\ \left(B^2 - E^2\right)\ =\
\frac{1}{b^2}\ \left\{2h\ M^\prime\ \cos\omega^\prime\left(t - 
\frac{\displaystyle x}{\displaystyle v^\prime}\right)
\cdot\sin\alpha + h^2\right\}\ . \ \ \ 
\end{eqnarray}
Then
\begin{eqnarray}
\label{teq97}
\dot{F}&=&-\ \frac{\displaystyle 2h\ M^\prime 
\omega^\prime \sin\alpha}{\displaystyle b^2}\ \sin\omega^\prime\left(t - 
\frac{\displaystyle x}{\displaystyle v^\prime}\right)
\end{eqnarray}
and
\begin{eqnarray}
\label{teq98}
\frac{\displaystyle\partial F}{\displaystyle\partial x}&=&
-\ \frac{1}{v^\prime}\;\dot{F}
\ =\ -\ \frac{1}{c}\;\dot{F}\ \ \ ,
\end{eqnarray}
i.e.\ on the r.h.s.\ of the equations /\ref{teq51}/ we can put
(in expressions not standing in the $\cos$ or $\sin$ symbols)
$v^\prime = c$.

We calculate the r.h.s.\ of the equations  
/\ref{teq51}/ in the first approximation: 
\begin{eqnarray}
\label{teq99}
\frac{1}{c}\ j_x&=&-\ \frac{1}{2c}\ \dot{F} E_x\ =\ 0
\end{eqnarray}

\pagebreak
\refstepcounter{ppage}

\begin{eqnarray}
\label{teq100}
\frac{1}{c}\ j_y&=&-\ \frac{1}{2}
\left(\frac{\partial F}{\partial x}\ B_z + \frac{1}{c}\;\dot{F}\ E_y\right)
\ =\ 0\\[0.3cm]
\label{teq101}
\frac{1}{c}\ j_z&=&\frac{1}{2}
\left(\frac{\partial F}{\partial x}\ B_y - \frac{1}{c}\;\dot{F}\ E_z\right)
\ =\ -\ \frac{1}{2c}\ \dot{F}\ h \sin\alpha\ =\nonumber\\[0.3cm]
&=&\frac{\omega^\prime h^2}{c\ b^2}\ M^\prime\ \sin^2\alpha\
\sin\omega^\prime\left(t - 
\frac{\displaystyle x}{\displaystyle v^\prime}\right)\\[0.3cm]
\label{teq102}
\rho&=&- \frac{1}{2c}\ \dot{F}\ E_y\ =\ 0\hspace{3cm}.
\end{eqnarray}
We consider the first of the equations /\ref{teq51}/. It yields
\begin{eqnarray*}
-\ \frac{1}{c}\;\dot{E}_x\ = \ 0\ .
\end{eqnarray*}
Obviously, this equation is satisfied if we put:
\begin{eqnarray}
\label{teq103}
p&=&0\ .
\end{eqnarray}
The second equation
\begin{eqnarray*}
-\ \frac{\displaystyle\partial B_z}{\displaystyle\partial x}
\ -\ \frac{1}{c}\;\dot{E}_y\ &=&0
\end{eqnarray*}
gives
\begin{eqnarray*}
\frac{M}{v} \sin\omega \left(t - 
\frac{\displaystyle x}{\displaystyle v}\right)&=&
\frac{1}{c}\;L\ \sin\omega\left(t -
\frac{\displaystyle x}{\displaystyle v}\right)\ ,
\end{eqnarray*}
where from, in view of /\ref{teq95}/, we find
\begin{eqnarray}
\label{teq104}
v&=&c\hspace{1cm};\hspace{1cm}M\ =\ L\ .
\end{eqnarray}
The third equation
\begin{eqnarray*}
\frac{\displaystyle\partial B_y}{\displaystyle\partial x}
\ -\ \frac{1}{c}\;\dot{E}_z&=&\frac{1}{c}\; j_z
\end{eqnarray*}

\pagebreak
\refstepcounter{ppage}

gives
\begin{eqnarray*}
\frac{M^\prime}{v^\prime}\ +\ 
\frac{L^\prime}{c}&=&
\frac{\displaystyle h^2\ M^\prime \sin^2\alpha}{\displaystyle c\ b^2}\ , 
\end{eqnarray*}
or, taking into account /\ref{teq95}/,
\begin{eqnarray}
\label{teq105}
\frac{\displaystyle v^{\prime\ 2}}{\displaystyle c^2}&=&
1\ -\ \frac{h^2}{b^2}\ \sin^2\alpha\ .
\end{eqnarray}
The equation $div\; E = \rho$ gives
\begin{eqnarray*}
\frac{\displaystyle\partial E_\alpha}{\displaystyle\partial x}&=&0\ \ \ ,
\end{eqnarray*}
which is automatically fulfilled because we have put $p = 0$.
The formulas /\ref{teq103}/, /\ref{teq104}/, and /\ref{teq105}/
finally allow to rewrite the solution of the equations 
/\ref{teq50}/ and /\ref{teq51}/ for our problem in the following way:
\begin{eqnarray}
\label{teq106}
\ \hspace{-1.5cm}&\left.\begin{array}{rcl}
E_x&=\ 0\hspace{3.4cm}
&B_x\ =\ h\ \cos\alpha\\[0.3cm]
E_y&=\ L\ \cos\omega\left(t - 
\frac{\displaystyle x}{\displaystyle c}\right)\hspace{0.75cm}
&B_y\ =\ M^\prime\ \cos\omega^\prime\left(t - 
\frac{\displaystyle x}{\displaystyle v^\prime}\right) + h\ \sin\alpha\\[0.3cm]
E_z&=\ L^\prime\ \cos\omega^\prime\left(t - 
\frac{\displaystyle x}{\displaystyle v^\prime}\right)\hspace{0.3cm}
&B_z\ =\ L\ \cos\omega\left(t - 
\frac{\displaystyle x}{\displaystyle c}\right)
\end{array}\right\}&\hspace{-0.2cm},\ \ \ 
\end{eqnarray}
where
\begin{eqnarray}
\label{teq107}
&\left.\begin{array}{rcl}
M^\prime&=&-\ \frac{\displaystyle c}{\displaystyle v^\prime}\ L^\prime\\[0.3cm]
and\hspace{1.5cm}
\frac{\displaystyle v^{\prime\ 2}}{\displaystyle c^2}&=&
1\ -\ \frac{\displaystyle h^2}{\displaystyle b^2}\ \sin^2\alpha
\end{array}\right\}&
\end{eqnarray}
whereby, as earlier,  $L,\ L^\prime,\ \omega\ and\ \omega^\prime$ 
remain arbitrary.
The discussion of the formulas /\ref{teq106}/ and /\ref{teq107}/ can
be performed in complete analogy to the discussion corresponding
to the formulas of the preceding paragraphs.

\pagebreak
\refstepcounter{ppage}
\refstepcounter{subsection}
\label{tglavaIIpara5}
\hspace{1cm}\udensdash{\S\ 5.\ Two plane waves in 
vacuo.\hspace{-0.05cm}\ }\\[-0.5cm]

\hspace{1cm}We now go over to the consideration of a more involved
problem of finding such solutions of the equations of Born which
in the zeroth approximation would correspond to
\udensdash{t$\ \atop\ $\hspace{-0.33cm}wo\hspace{-0.05cm}\ } \linebreak 
\udensdash{plane waves in vacuo\hspace{-0.05cm}\ }, 
i.e.\ we will assume that a plane 
wave does not propagate in a constant electric of magnetic field 
but in the electric and magnetic field, periodic in $r$ and $t$, of another
plane wave. We will solve this problem for that particular case only
that the wave vectors of both of the\linebreak  plane waves do have opposite
directions, their electric vectors, in zeroth approximation, do have 
the same direction, and the magnetic - the opposite one. We choose these
three directions as the coordinate axes. More precisely, we take
the direction of the wave vector of the first wave as $ox$-axis,
the direction of the electric vectors for the $oy$-axis, and the 
direction of the magnetic vector of the first wave - for $oz$.
/Fig.\ref{tris2}/.

\begin{figure}[h]
\caption{\label{tris2}}
\unitlength1.mm
\begin{picture}(150,50)
\put(25,25){\line(1,0){90}}
\put(70,0){\line(0,1){50}}
\put(70,25){\line(1,1){20}}
\put(115,25){\line(-2,-1){3.6}}
\put(115,25){\line(-2,1){3.6}}
\put(101,25){\line(-2,-1){3.6}}
\put(101,25){\line(-2,1){3.6}}
\put(40,25){\line(2,-1){3.6}}
\put(40,25){\line(2,1){3.6}}
\put(70,50){\line(-1,-2){1.8}}
\put(70,50){\line(1,-2){1.8}}
\put(70,42){\line(-1,-2){1.8}}
\put(70,42){\line(1,-2){1.8}}
\put(70,11){\line(-1,2){1.8}}
\put(70,11){\line(1,2){1.8}}
\put(90,45){\line(-3,-1){3.6}}
\put(90,45){\line(-1,-3){1.2}}
\put(84,39){\line(-3,-1){3.6}}
\put(84,39){\line(-1,-3){1.2}}
\put(95,5){\it Fig 2.}
\put(65,26){$0$}
\put(72,49){$z$}
\put(93,44){$y$}
\put(117,24){$x$}
\put(63,40){$B_1$}
\put(86,36){$E_1 E_2$}
\put(40,20){$n_2$}
\put(100,20){$n_1$}
\put(72,10){$B_2$}

\end{picture}
\end{figure}
\vspace{-0.5cm}
We are trying to find the solution of the equations of Born for this
case by means of a method analogous to that considered in the preceding
paragraphs, however, taking into account the possible appearance 
of small additive terms yet playing the role of $x^\prime$ in formula
/\ref{teq47}/. Specifically, we will \linebreak search the solution of the
equations /\ref{teq50}/ and /\ref{teq51}/ in the form
\begin{eqnarray}
\label{teq108}
\ \hspace{-1.2cm}&\left.\begin{array}{rcl}
E_x&=&0\;;\ E_y\ =\ Q\;\cos\Omega\left(t - 
\frac{\displaystyle x}{\displaystyle v}\right) +
q\;\cos\omega\left(t + 
\frac{\displaystyle x}{\displaystyle v}\right) + E^\prime\;;\
E_z\ =\ 0\\[0.3cm]
B_x&=&0\;;\ B_y\ =\ 0\;;\ B_z\ =\ \frac{\displaystyle c}{\displaystyle V}\;
Q\;\cos\Omega\left(t - 
\frac{\displaystyle x}{\displaystyle V}\right) -
\frac{\displaystyle c}{\displaystyle v}\
q\;\cos\omega\left(t + 
\frac{\displaystyle x}{\displaystyle v}\right) + B^\prime\hspace{-0.5cm}\
\end{array}\right\}.&\ \ \ \
\end{eqnarray}

\pagebreak 
\refstepcounter{ppage}

Such a manner of notation$^{\displaystyle 1)}$, 
firstly, introduces the assumption that 
as in the unperturbed problem too, independently of the {"}interaction"\ 
of our waves the quantities\linebreak 
$E_x,\ E_z,\ B_x\ and\ B_y$ can be 
considered to be equal to zero.\linebreak This assumption is completely 
natural for symmetry reasons; a strict proof of its validity will 
rely on the fact that, by means of the expressions /\ref{teq108}/,
the Born equations can indeed be satisfied as we will see further below.
Secondly, we have assumed earlier that the connection between the 
amplitudes $E_y\ and\ B_z$ has the usual form - which, of course,
also has its perfect reason - because this connection derives from
the \linebreak first group of field equations which has the same form for all
the problems. Incidentally, one could instead of /\ref{teq108}/ also
write down more general formulas of the type /\ref{teq49}/,
considering the field components $E\ and\ B$ as arbitrary harmonic
functions of the coordinates and time, but is is not difficult 
to convince oneself that such a more involved formulation would not
give anything essentially new, at least within our setting of the
problem.

Thus, we insert /\ref{teq108}/ into /\ref{teq50}/.
The equation $div\; B = 0$ is satisfied if 
\begin{eqnarray*}
div\; B^\prime&=&0\ ,
\end{eqnarray*}
\hspace{1cm}or
\begin{eqnarray}
\label{teq109}
\frac{\displaystyle\partial B^\prime }{\displaystyle\partial z}&=&0\ .
\end{eqnarray}
The equation
$rot\; E + \frac{\displaystyle 1}{\displaystyle c}\;\dot{B}$
is satisfied if
\begin{eqnarray}
\label{teq110}
\frac{\displaystyle\partial E^\prime }{\displaystyle\partial z}&=&0
\end{eqnarray}
-------------------

1/\ For simplicity, the initial phases of the waves are put 
equal to zero again.

\pagebreak
\refstepcounter{ppage}

and

\vspace{-1.cm}

\begin{eqnarray}
\label{teq111}
\frac{\displaystyle\partial E^\prime }{\displaystyle\partial x}
\ +\ \frac{\displaystyle 1}{\displaystyle c}\;\dot{B}&=&0\ .
\end{eqnarray}
We insert now /\ref{teq108}/ into /\ref{teq51}/.

For doing so, we first calculate the quantity $F$ in first approximation:
\begin{eqnarray}
\label{teq112}
F&=&-\ \frac{4}{b^2}\ Q\; q\;\cos\Gamma\; \cos\gamma\ \ \ ,
\end{eqnarray}
where
\begin{eqnarray}
\label{teq113}
&\left.\begin{array}{rcl}
\Gamma&=&\Omega \left(t - \frac{\displaystyle x}{\displaystyle V}\right)
\\[0.2cm]
\gamma&=&\omega \left(t + \frac{\displaystyle x}{\displaystyle v}\right)
\end{array}\right\}&\ .
\end{eqnarray}
We find $\dot{F}$ and 
$\frac{\displaystyle\partial F}{\displaystyle\partial x}$\ :
\begin{eqnarray}
\label{teq114} 
\dot{F}&=&\frac{4}{b^2}\ Q\; q\;\left(\Omega\;\sin\Gamma\;\cos\gamma
+ \omega\;\cos\Gamma\;\sin\gamma\right)\\[0.1cm]
\label{teq115} 
\frac{\displaystyle\partial F}{\displaystyle\partial x}&=&
\frac{1}{c}\;\frac{4}{b^2}\ Q\; q\;\left(- \Omega\;\sin\Gamma\;\cos\gamma
+ \omega\;\cos\Gamma\;\sin\gamma\right)\ .
\end{eqnarray}
We calculate the r.h.s.\ of the equations /\ref{teq51}/
\begin{eqnarray}
\label{teq116} 
\frac{1}{c}\; j_x&=&\frac{1}{2c}\ \dot{F}\ E_x\ =\ 0\\[0.3cm]
\label{teq117} 
\frac{1}{c}\; j_y&=&-\ \frac{4\; Q\; q}{c\; b^2}\ 
\left(\Omega\; q\;\sin\Gamma\;\cos^2\gamma
+ \omega\; Q\; \sin\gamma\;\cos^2\Gamma\right)\ .
\end{eqnarray}
Noting that
\begin{eqnarray*}
\sin\Gamma\;\cos^2\gamma\ =\ \frac{1}{2}
\ \left[\sin\Gamma + \frac{1}{2}\;\sin\left(\Gamma + 2\gamma\right)
+ \frac{1}{2}\;\sin\left(\Gamma - 2\gamma\right)\right]\\[0.1cm]
and\hspace{0.8cm}
\sin\gamma\;\cos^2\Gamma\ =\ \frac{1}{2}
\ \left[\sin\gamma + \frac{1}{2}\;\sin\left(\gamma + 2\Gamma\right)
+ \frac{1}{2}\;\sin\left(\gamma - 2\Gamma\right)\right]
\end{eqnarray*}
we find
\begin{eqnarray}
\label{teq118}
\frac{1}{c}\; j_y&=&-\ \frac{2}{c\; b^2}\ Q\; q\;\left\{q\;\Omega\
\left[\sin\Gamma + \frac{1}{2}\;\sin\left(\Gamma + 2\gamma\right)
+ \frac{1}{2}\;\sin\left(\Gamma - 2\gamma\right)\right] 
+\right.\nonumber\\[0.1cm]
&&\hspace{2.5cm}\left. + Q\;\omega\ 
\left[\sin\gamma + \frac{1}{2}\;\sin\left(\gamma + 2\Gamma\right)
+ \frac{1}{2}\;\sin\left(\gamma - 2\Gamma\right)\right]\right\}\ .\ \ \ \ \ 
\end{eqnarray}
Furthermore,
\begin{eqnarray}
\label{teq119}
\frac{1}{c}\; j_z&=&0
\end{eqnarray}

\pagebreak
\refstepcounter{ppage}

and finally
\begin{eqnarray}
\label{teq120}
\rho&=&\frac{1}{2}\ 
\frac{\displaystyle\partial F}{\displaystyle\partial x}\ E_x\ =\ 0\ .
\end{eqnarray}
From eq.\ /\ref{teq123}/, we recognize that in difference to the problems
considered earlier in the present problem on the r.h.s.\ of the 
equations appear not only terms containing $sin\;\Gamma\ and\ sin\;\gamma$,
which are analogous to the $cos\;\omega t$ term in equation /\ref{teq46}/,
yielding a {"}resonance{"}, but also terms with different frequencies.
They also occur just at the r.h.s.\ of the equations for determining
$E^\prime\ and\ B^\prime$. With the knowledge of the r.h.s., just 
calculated, of the equations, we can insert /\ref{teq108}/ into  /\ref{teq51}/.
The first of the equations /\ref{teq51}/
\begin{eqnarray*}
\frac{\displaystyle\partial B_z}{\displaystyle\partial y}\ -\ 
\frac{\displaystyle\partial B_y}{\displaystyle\partial z}\ -\
\frac{1}{c}\;\dot{E_x}&=&0
\end{eqnarray*}
is satisfied if it is put
\begin{eqnarray}
\label{teq121}
\frac{\displaystyle\partial B^\prime}{\displaystyle\partial y}&=&0\ .
\end{eqnarray}
The second equation
\begin{eqnarray*}
\hspace{-0.5cm}\frac{\displaystyle\partial B_x}{\displaystyle\partial z} - 
\frac{\displaystyle\partial B_z}{\displaystyle\partial x} -
\frac{\displaystyle 1}{\displaystyle c}\;\dot{E_y}&=&
-\ \frac{\displaystyle 2}{\displaystyle c\; b^2}\ Q\; q\;\left\{q\;\Omega\;
\left[\sin\Gamma + \frac{1}{2}\;\sin\left(\Gamma + 2\gamma\right)
+ \frac{1}{2}\;\sin\left(\Gamma - 2\gamma\right)\right] 
+\right.\\[0.1cm]
&&\hspace{2cm}\left. +\ Q\;\omega\; 
\left[\sin\gamma + \frac{1}{2}\;\sin\left(\gamma + 2\Gamma\right)
+ \frac{1}{2}\;\sin\left(\gamma - 2\Gamma\right)\right]\right\}
\end{eqnarray*}
is satisfied if
\begin{eqnarray}
\label{teq122}
-\ \frac{\displaystyle c\;\Omega}{\displaystyle V^2}\ Q&\sin\Gamma&-\ 
\frac{\displaystyle c\;\omega}{\displaystyle v^2}\; \sin\gamma\ +\
\frac{1}{c}\ Q\;\Omega\;\sin\Gamma\ +\
\frac{1}{c}\ q\;\omega\;\sin\gamma\ =\nonumber\\[0.3cm]
&&=\ -\ \frac{\displaystyle 2}{\displaystyle c\; b^2}\
Q\;q\ \left(q\;\Omega\;\sin\Gamma + Q\;\omega\;\sin\gamma\right)
\end{eqnarray}
and
\begin{eqnarray}
\label{teq123}
\frac{\displaystyle\partial B^\prime}{\displaystyle\partial x}\ +\ 
\frac{1}{c}\;\dot{E^\prime}&=&
\frac{1}{c\; b^2}\ Q\; q\;\left\{q\;\Omega\;
\left[\sin\left(\Gamma + 2\gamma\right)
+ \sin\left(\Gamma - 2\gamma\right)\right] 
+\right.\nonumber\\[0.1cm]
&&\hspace{1.cm}\left. + Q\;\omega\; 
\left[\sin\left(\gamma + 2\Gamma\right)
+ \sin\left(\gamma - 2\Gamma\right)\right]\right\}\ .
\end{eqnarray}

\pagebreak
\refstepcounter{ppage}

The equation  /\ref{teq122}/ gives us
\begin{eqnarray}
\label{teq124}
&\left.\begin{array}{rcl}
1\ -\ \frac{\displaystyle c^2}{\displaystyle V^2}&=&
-\ \frac{\displaystyle 2}{\displaystyle b^2}\ q^2\\[0.2cm]
1\ -\ \frac{\displaystyle c^2}{\displaystyle v^2}&=&
-\ \frac{\displaystyle 2}{\displaystyle b^2}\ Q^2
\end{array}\right\}&\ .
\end{eqnarray}
The third equation
\begin{eqnarray*}
\frac{\displaystyle\partial B_y}{\displaystyle\partial x}\ -\ 
\frac{\displaystyle\partial B_x}{\displaystyle\partial y}\ -\
\frac{1}{c}\;\dot{E_z}&=&0
\end{eqnarray*}
is satisfied automatically.

Finally, the last equation $div\; E = \rho$ gives us
\begin{eqnarray}
\label{teq125}
\frac{\displaystyle\partial E^\prime}{\displaystyle\partial y}&=&0\ .
\end{eqnarray}
For the final solution of the problem, we need to find the 
additive terms $E^\prime\ and\ B^\prime$ yet. For their calculation,
we have obtained the equations /\ref{teq109}/, /\ref{teq110}/, /\ref{teq111}/,
/\ref{teq121}/, /\ref{teq123}/, and /\ref{teq125}/.

We will search the solution of these equations in the form
\begin{eqnarray}
\label{teq126}
\ \hspace{-1.2cm}&\left.\begin{array}{rcl}
E^\prime&=&a_1\cos\left(\Gamma + 2\gamma\right) +
a_2\cos\left(\Gamma - 2\gamma\right) +
a_3\cos\left(\gamma + 2\Gamma\right) +
a_4\cos\left(\gamma - 2\Gamma\right)\\[0.3cm]
B^\prime&=&d_1\cos\left(\Gamma + 2\gamma\right) +
d_2\cos\left(\Gamma - 2\gamma\right) +
d_3\cos\left(\gamma + 2\Gamma\right) +
d_4\cos\left(\gamma - 2\Gamma\right)
\end{array}\hspace{-0.2cm}\right\},&\ \ \  
\end{eqnarray}
whereby $a_1\ ...\ d_4$ are constant coefficients.

Here, the equations /\ref{teq109}/, /\ref{teq110}/,
/\ref{teq121}/, and /\ref{teq125}/ are satisfied automatically,
and the equations /\ref{teq111}/ and /\ref{teq123}/ provide
us with the possibility to determine the coefficients in /\ref{teq126}/.

\hspace{1cm}The first of these equations give us

\pagebreak
\refstepcounter{ppage}

\begin{eqnarray}
\label{teq127}
&\left.\begin{array}{rcl}
a_1&=&\frac{\displaystyle\Omega + 2\omega}{\displaystyle\Omega - 2\omega}\ d_1
\hspace{1cm};\hspace{1cm}a_3\ =\ 
-\ \frac{\displaystyle\omega + 2\Omega}{\displaystyle\omega - 2\Omega}\ d_3
\\[0.3cm]
a_2&=&\frac{\displaystyle\Omega - 2\omega}{\displaystyle\Omega + 2\omega}\ d_2
\hspace{1cm};\hspace{1cm}a_4\ =\ 
-\ \frac{\displaystyle\omega - 2\Omega}{\displaystyle\omega + 2\Omega}\ d_4
\end{array}\right\}&
\end{eqnarray}
and the second
\begin{eqnarray}
\label{teq128}
&\left.\begin{array}{rcl}
a_1&=&\frac{\displaystyle\Omega - 2\omega}{\displaystyle\Omega + 2\omega}\ d_1
\ -\ \frac{\displaystyle 1}{\displaystyle b^2}\ Q\;q^2\
\frac{\displaystyle\Omega}{\displaystyle\Omega + 2\omega}\\[0.3cm]
a_2&=&\frac{\displaystyle\Omega + 2\omega}{\displaystyle\Omega - 2\omega}\ d_2
\ -\ \frac{\displaystyle 1}{\displaystyle b^2}\ Q\;q^2\
\frac{\displaystyle\Omega}{\displaystyle\Omega - 2\omega}\\[0.3cm]
a_3&=&-\ 
\frac{\displaystyle\omega - 2\Omega}{\displaystyle\omega + 2\Omega}\ d_3
\ -\ \frac{\displaystyle 1}{\displaystyle b^2}\ q\;Q^2\
\frac{\displaystyle\omega}{\displaystyle\omega + 2\Omega}\\[0.3cm]
a_4&=&-\ 
\frac{\displaystyle\omega + 2\Omega}{\displaystyle\omega - 2\Omega}\ d_4
\ -\ \frac{\displaystyle 1}{\displaystyle b^2}\ q\;Q^2\
\frac{\displaystyle\omega}{\displaystyle\omega - 2\Omega}
\end{array}\right\}&\ .
\end{eqnarray}
This way, we have obtained a system of eight equations /\ref{teq127}/ and
/\ref{teq128}/ for determining $a_1,\ ...\ a_4,d_1\ ...\ d_4$.

Solving this system of equations, we obtain
\begin{eqnarray}
\label{teq129}
&\left.\begin{array}{rcl}
a_1&=&-\ \frac{\displaystyle Q q^2}{\displaystyle 8 b^2}\ 
\frac{\displaystyle\Omega + 2\omega}{\displaystyle\omega}
\hspace{1cm};\hspace{1cm}
d_1\ =\ -\ \frac{\displaystyle Q q^2}{\displaystyle 8 b^2}\ 
\frac{\displaystyle\Omega - 2\omega}{\displaystyle\omega}\\[0.3cm]
a_2&=&\frac{\displaystyle Q q^2}{\displaystyle 8 b^2}\ 
\frac{\displaystyle\Omega - 2\omega}{\displaystyle\omega}
\hspace{1.6cm};\hspace{1.05cm}
d_2\ = \ \frac{\displaystyle Q q^2}{\displaystyle 8 b^2}\ 
\frac{\displaystyle\Omega + 2\omega}{\displaystyle\omega}\\[0.3cm]
a_3&=&-\ \frac{\displaystyle q Q^2}{\displaystyle 8 b^2}\ 
\frac{\displaystyle\omega + 2\Omega}{\displaystyle\Omega}
\hspace{1cm};\hspace{1cm}
d_3\ =\ \frac{\displaystyle q Q^2}{\displaystyle 8 b^2}\ 
\frac{\displaystyle\omega - 2\Omega}{\displaystyle\Omega}\\[0.3cm]
a_4&=&\frac{\displaystyle q Q^2}{\displaystyle 8 b^2}\ 
\frac{\displaystyle\omega - 2\Omega}{\displaystyle\Omega}
\hspace{1.6cm};\hspace{1.05cm}
d_4\ =\ -\ \frac{\displaystyle q Q^2}{\displaystyle 8 b^2}\ 
\frac{\displaystyle\omega + 2\Omega}{\displaystyle\Omega}
\end{array}\right\}&\ .\ \ 
\end{eqnarray}

\pagebreak
\refstepcounter{ppage}

We now have got a complete solution of our problem which has the form
/\ref{teq108}/, whereby $V\ and\ v$ are determined by means of 
the formula /\ref{teq124}/ and $E^\prime\ and\ B^\prime$ can be found
from formula /\ref{teq126}/ with the coefficients /\ref{teq129}/.

As in the Maxwell theory, the quantities $Q,\ q,\ \Omega\ and\ \omega$ 
remain arbitrary.

The solution obtained by us can be characterized the following way:
If, in empty space, two electromagnetic waves with arbitrarily given
amplitudes and frequencies propagate in opposite directions, whereby
in zeroth approximation their electric and magnetic vectors are 
oriented as indicated in Fig.\ \ref{tris2}, the 
\udensdash{d$\ \atop\ $\hspace{-0.33cm}ifference\hspace{-0.05cm}\ }
from the field which exists in this space according to the theory of
\udensdash{B$\ \atop\ $\hspace{-0.33cm}orn\hspace{-0.05cm}\ } 
to that one obtained from the
\udensdash{M$\ \atop\ $\hspace{-0.33cm}axwell\hspace{-0.05cm}\ }
theory is characterized by two moments:

1/ by the change in propagation velocity of these waves, and also
by the appearance of a difference of the sizes of the electric
and magnetic vectors of each wave.

This effect is completely analogous to the {"}distortion"\ phenomenon
of waves of light we have obtained in the preceding paragraphs.

2/ by the appearance of four small {"}scattered"\ waves with different
frequencies and propagation velocities, strongly differing
from the Maxwellian ones in general.

This effect is specific for the latter problem and, in a well-known sense,
can be characterized as {"}scattering of light by light{"}.

However, it is essential to note that the {"}interaction"\ of two
waves of light calculated according to the theory of Born 

\pagebreak
\refstepcounter{ppage}

can by far not be identified with that {"}scattering of light by light"\ 
repeatedly mentioned in the literature - namely, in view of the existence of
the first type of effects mentioned above; in fact, each wave acquires
peculiar properties because it propagates in a medium {"}polarized"\
under the influence of the other wave.

\hspace{1cm}Finally, I would like to express my deep gratitude to Prof.\ 
S.\ P.\ Shubin for the guidance of this dissertation and also
for a number of highly valuable recommendations and comments.

\vspace{0.5cm}

\hspace{5cm}-------------

\vspace{0.5cm}

\hspace{10cm}February 1936

\pagebreak
\refstepcounter{ppage}
\label{tpage67}

\udensdash{N$\ \atop\ $\hspace{-0.33cm}ote\hspace{-0.05cm}\ }:

\vspace{0.5cm}

\hspace{1cm}After the present dissertation had been finished (i.e.\ after
February 1936), in the literature appeared a number of works 
connected with some problems studied here.

\hspace{1cm}Some of the results of these works will be studied 
in an addendum to the dissertation which presently is under preparation.

}

\newpage
\refstepcounter{ppage}

\renewcommand{\thepage}{\theppage}
\chead{\rm - \theppage\ -\hspace{3cm}\ }

\ \\[5cm]
\begin{center}
{\color{gray}\large blank page\hspace{3cm}\ }
\end{center}

\newpage
\refstepcounter{ppage}

\phantomsection
\addcontentsline{toc}{section}{Supplementary information}
{\Large\bf Supplementary information}\\[0.3cm]

\phantomsection
\addcontentsline{toc}{subsection}{Translator's notes}
{\large\bf Translator's notes}:

\begin{itemize}
\item[1.]
Concerning the Russian original of the thesis:
A copy of the thesis is filed at the 
Institute of Archival Studies 
\begin{otherlanguage}{ukrainian}
(Iнститут архiвознавства
\end{otherlanguage}
%({\cyrrm {\=I}nstitut arkh\=\i voznavstva}
[\={I}nstitut arkh\={\i}vo\-znavst\-va])
of the National Library of Ukraine
\begin{otherlanguage}{ukrainian}
(Національна бібліотека України імені В.І.\ Вернадського
\end{otherlanguage}
%{\cyrrm Nats\=\i onal\cprime na b\=\i bl\=\i oteka Ukra\"\i ni
%\=\i men\=\i\ V.I.\ Vernad\cprime skogo}
[Nats\={\i}onal'na b\={\i}bl\={\i}oteka Ukra\"{\i}ni
\={\i}men\={\i} V.I.\ Vernad'skogo]), 
archival collection
\begin{otherlanguage}{ukrainian}
(архівні фонд
\end{otherlanguage}
%({\cyrrm arkh\=\i vny\=\i\ fond} 
[arkh\={\i}vny\={\i}\ fond]) 
no.\ 167, section
\begin{otherlanguage}{ukrainian}
(росділ
\end{otherlanguage}
%({\cyrrm rosd\=\i l} 
[rosd\={\i}l]) 1, item no.\ 3 (see 
{\tt \url{http://irbis-nbuv.gov.ua/fond/ia/ires/OPYS/167_1.pdf}}).
The available copy
of the original thesis has turned out not to be suited for easy reading
due to its poor quality ((copy of a) carbon copy; 
cf.\ the attached facsimile of 
the original thesis, p.\ \pageref{facsimile}, which we have appended here
for archival reasons). Therefore, the original 
Russian text has been rekeyed (without any correction) as LaTeX file 
whereby, as far as possible, the original graphical layout 
(including the page numbering) has been preserved.
In rekeying the original thesis, 
the interpunctuation has also been preserved and not been
corrected (for example, missing hyphens have not been added in general).
However, in the English translation of the thesis, interpunctuation
has been added in a number of cases (in a mild manner, not always). 
Handwritten equations and other handwritten pieces of text
are displayed in the rekeyed Russian text in blue colour (exceptions are 
minor legibility corrections to typewriter letters done in the 
original which are not indicated in the rekeyed text) while in the 
English translation no colour coding for handwritten items has been
applied. Also, for handwritten
text the font type `italic' has been chosen. 
Hyperlinking for sections and equations has been introduced (coloured mahogany).
If necessary for improving printout quality, the colours can 
easily be set to `black' in the preamble of the LaTeX file.
Full details of the cited literature can be found in a separately 
added list on p.\ \pageref{literature} of the present translation.

\item[2.]
Note: On p.\ \pageref{page50} of the original Russian text, 
first row below from eq.\ (\ref{eq80}) ,
the index {\tt 1} reads in the original thesis {\tt I}. This change has been
made to let the index conform to the index display in the LaTeX
equations.

\item[3.]
The results presented in the thesis in \S\ \ref{glavaIIpara5} 
of chap.\ \ref{glavaII} have
later been published in \cite{1940smirnov1,1941smirnov}.

\item[4.]
The English translation preserves, as far as possible, the 
page structure of the original Russian thesis. For example, 
the page number T-7 of the English translation corresponds to
the page number 7 of the original Russian thesis. 

\end{itemize}

\newpage
\refstepcounter{ppage}

\phantomsection
\addcontentsline{toc}{subsection}{Noticed misprints}
{\large\bf Noticed misprints}\\

The following misprints have been noticed (but not corrected) 
in the original Russian text:

\begin{itemize}
\item[-]
p.\ \pageref{page2}, 6.\ row: 
\begin{otherlanguage}{russian}
{\tt крантовым},
\end{otherlanguage}
%{\cyrtt krantovymi}, 
should read correctly: 
\begin{otherlanguage}{russian}
{\tt квантовым}
\end{otherlanguage}
%{\cyrtt kvantovymi}

\item[-]
p.\ \pageref{page11}, 11.\ row from below: 
\begin{otherlanguage}{russian}
{\tt также},
\end{otherlanguage}
%{\cyrtt takzhe}, 
should read correctly: 
\begin{otherlanguage}{russian}
{\tt такие}
\end{otherlanguage}
%{\cyrtt takie}

\item[-]
p.\ \pageref{page17}, 2.\ row:
\begin{otherlanguage}{russian}
{\tt то\hspace{-0.1cm}\textasciiacute\hspace{0.1cm}это},
\end{otherlanguage}
%{\cyrtt to\hspace{-0.1cm}\textasciiacute\hspace{0.1cm}\protect{\`{e}}to}, 
should read correctly: 
\begin{otherlanguage}{russian}
{\tt то это}
\end{otherlanguage}
%{\cyrtt to \protect{\`{e}}to}

\item[-]
p.\ \pageref{page22}, 2.\ row below from eq.\ (\ref{eq34}):
\begin{otherlanguage}{russian}
{\tt выясниться},
\end{otherlanguage} 
%{\cyrtt vyyasnit\cprime sya}, 
should read correctly:\hfill\ \linebreak
\begin{otherlanguage}{russian}
{\tt выяснится}
\end{otherlanguage}
%{\cyrtt vyyasnit{s}ya}

\item[-]
p.\ \pageref{page25}, 12.\ row below from eq.\ (\ref{eq38}): 
\begin{otherlanguage}{russian}
{\tt содерживающих},
\end{otherlanguage}
%{\cyrtt soderzhivayushchikh}, 
should read correctly: 
\begin{otherlanguage}{russian}
{\tt сдерживающих}
\end{otherlanguage}
%{\cyrtt sderzhivayushchikh}

\item[-]
p.\ \pageref{page26}, 6./7.\ row: 
\begin{otherlanguage}{russian}
{\tt прямолинено},
\end{otherlanguage}
%{\cyrtt pryamolineno}, 
should read correctly: 
\begin{otherlanguage}{russian}
{\tt прямолинейно},
\end{otherlanguage}
%{\cyrtt pryamoline\u ino}

\item[-]
p.\ \pageref{page28}, 1.\ row: 
\begin{otherlanguage}{russian}
{\tt искажение},
\end{otherlanguage}
%{\cyrtt iskazhenie}, 
should read correctly: 
\begin{otherlanguage}{russian}
{\tt искажения}
\end{otherlanguage}
%{\cyrtt iskazheniya}

\item[-]
p.\ \pageref{page29}, 10.\ row from below: 
\begin{otherlanguage}{russian}
{\tt производится},
\end{otherlanguage}
%{\cyrtt proizvodit{s}ya}, 
should read correctly:\hfill\ \linebreak
\begin{otherlanguage}{russian}
{\tt производиться}
\end{otherlanguage} 
%{\cyrtt proiz\-vo\-dit\cprime s\-ya}

\item[-]
p.\ \pageref{page30}, 6.\ row: 
\begin{otherlanguage}{russian}
{\tt расстояних},
\end{otherlanguage}
%{\cyrtt rasstoyanikh}, 
should read correctly:
\begin{otherlanguage}{russian}
{\tt расстояниях},
\end{otherlanguage} 
%{\cyrtt rasstoyaniyakh}

\item[-]
p.\ \pageref{page40}, bottom row: 
\begin{otherlanguage}{russian}
{\tt вне сет},
\end{otherlanguage}
%{\cyrtt vne set}, 
should read correctly:
\begin{otherlanguage}{russian}
{\tt внесет}
\end{otherlanguage} 
%{\cyrtt vneset}

\item[-]
p.\ \pageref{page46}, 3./2.\ row from below: 
\begin{otherlanguage}{russian}
{\tt кристаллоптики},
\end{otherlanguage}
%{\cyrtt kristalloptiki}, 
should read correctly:\hfill\ \linebreak 
\begin{otherlanguage}{russian}
{\tt кристаллооптики}
\end{otherlanguage}
%{\cyrtt kristallo\-optiki}

\item[-]
p.\ \pageref{page48}, 7.\ row below from eq.\ (\ref{eq78}): 
\begin{otherlanguage}{russian}
{\tt тойже},
\end{otherlanguage}
%{\cyrtt to\u izhe}, 
should read correctly:
\begin{otherlanguage}{russian}
{\tt той же},
\end{otherlanguage} 
%{\cyrtt to\u i zhe}

\item[-]
p.\ \pageref{page53}, 8.\ row from below: 
\begin{otherlanguage}{russian}
{\tt носит},
\end{otherlanguage}
%{\cyrtt nosit}, 
should read correctly: 
\begin{otherlanguage}{russian}
{\tt носить}
\end{otherlanguage}
%{\cyrtt nosit\cprime }

\item[-]
p.\ \pageref{page53}, 5.\ row from below: 
\begin{otherlanguage}{russian}
{\tt нельзы},
\end{otherlanguage}
%{\cyrtt nel\cprime zy}, 
should read correctly: 
\begin{otherlanguage}{russian}
{\tt нельзя}
\end{otherlanguage}
%{\cyrtt nel\cprime zya}

\end{itemize}

\vspace{0.5cm}

\phantomsection
\addcontentsline{toc}{subsection}{Acknowledgements}
{\large\bf Acknowledgements}\\

I am grateful to the daughter of A.\ A.\ Smirnov, 
Irina Adrianovna Smirnova (Kiev), for her kind permission to make
the Ph.D.\ thesis of her father publicly available on the arXiv,
and to the grandson of A.\ A.\ Smirnov, Sergei Valentinovich 
Smirnov (Helsinki), for his help in obtaining a copy of the thesis
and a careful reading of the rekeyed version of it.
I am indebted to A.\ P.\ Nosov of the Institute of Metal Physics,
Ekaterinburg, for providing me with a copy of \cite{shub1}.
Kind hospitality at the Theoretical Particle Physics Group of the 
Vrije Universiteit Amsterdam is also gratefully acknowledged.

\pagebreak
\refstepcounter{ppage}
\label{literature}

\phantomsection
\addcontentsline{toc}{subsection}{Literature}
{\large\bf Literature}\\[0.3cm]

{\small 
[For references in Cyrillic letters, 
we apply the (new) {\it Mathematical Reviews} transliteration

(transcription) scheme
(to be found at the end of index issues of {\it Mathematical Reviews}).]
}\\[0.3cm]

\phantomsection
\addcontentsline{toc}{subsubsection}{Full details of the cited literature}
{\bf Full details of the cited literature} (ordered
alphabetically by author names):\\[-0.3cm]

%\bibitem[()]{1934born1}
M.\ Born:
On the quantum theory of the electromagnetic field.
{\it Proceedings of the Royal Society of London. Series A,
Containing Papers of Mathematical and Physical Character}
{\bf 143}:849(1934)410-437
(\href{http://dx.doi.org/10.1098/rspa.1934.0010}{DOI: 
10.1098/rspa.1934.0010}, stable JSTOR\hfill\ \linebreak 
URL: {\tt \url{http://www.jstor.org/stable/96101}}).
Reprinted in: \cite{1963born}, 
item 73, pp.\ 486-513.\\ 
%%CITATION = PRSLA,A143,410;%%

%\bibitem[()]{1934born3}
M.\ Born, L.\ Infeld:
Foundations of the new field theory.
{\it Proceedings of the Royal Society of London. Series A,
Containing Papers of Mathematical and Physical Character}
{\bf 144}:852(1934)425-451
(\href{http://dx.doi.org/10.1098/rspa.1934.0059}{DOI: 
10.1098/rspa.1934.0059}, stable JSTOR URL:\hfill\ \linebreak
{\tt \url{http://www.jstor.org/stable/2935568}}).
Reprinted in: 1.\ \cite{1963born}, item 74, pp.\ 514-540.
2.\ \cite{1978infeld}, pp.\ 55-78. For the annotation of
some misprints see E.\ Schr\"odinger. {\it Proc.\ Roy.\ Soc.\ London
Ser.\ A} {\bf 150}(1935)465, footnote on
p.\ 472.\\
%%CITATION = PRSLA,A144,425;%%

%\bibitem[()]{1934born4}
M.\ Born, L.\ Infeld:
On the quantization of the new field equations. I.
{\it Proceedings of the Royal Society of London. Series A,
Containing Papers of Mathematical and Physical Character}
{\bf 147}:862(1934)522-546
(\href{http://dx.doi.org/10.1098/rspa.1934.0234}{DOI: 
10.1098/rspa.1934.0234}, stable\hfill\ \linebreak 
JSTOR URL: {\tt \url{http://www.jstor.org/stable/96309}}).
Reprinted in:
1.\ \cite{1963born}, item 75, pp.\ 541-565.
2.\ \cite{1978infeld}, pp.\ 79-100.\\
%%CITATION = PRSLA,A147,522;%%

%\bibitem[()]{1935born1}
M.\ Born, L.\ Infeld:
On the quantization of the new field equations. II.
{\it Proceedings of the Royal Society of London. Series A,
Containing Papers of Mathematical and Physical Character}
{\bf 150}:869(1935)141-166
(\href{http://dx.doi.org/10.1098/rspa.1935.0093}{DOI: 
10.1098/rspa.1935.0093}, stable\hfill\ \linebreak 
JSTOR URL: {\tt \url{http://www.jstor.org/stable/96328}}).
Reprinted in: \cite{1978infeld}, pp.\ 101-122.\\
%%CITATION = PRSLA,A150,141;%%

%\bibitem[()]{1935born2}
M.\ Born, E.\ Schr\"odinger:
The absolute field constant in the new field theory.
{\it Nature} {\bf 135}:3409(1935)342
(\href{http://dx.doi.org/10.1038/135342a0}{DOI: 
10.1038/135342a0}).
Reprinted in: \cite{1984schroedinger}, p.\ 195.\\
%%CITATION = NATUA,135,342;%%

%\bibitem[()]{1935euler}
H.\ Euler, B.\ Kockel:
\"Uber die Streuung von Licht an Licht nach der Diracschen Theorie.
%[On the scattering of light by light according to Dirac's theory].
{\it Die Naturwissenschaften} {\bf 23}:15(1935)246-247
(\href{http://dx.doi.org/10.1007/BF01493898}{DOI: 
10.1007/BF01493898}).\hfill\ \linebreak[5]
[in German] English translation (by D.\ H.\ Delphenich): 
{\it The scattering of light by light in Dirac's theory} (The translation is
freely available at the URL:\hfill\ \linebreak
{\tt
  \url{http://neo-classical-physics.info/uploads/3/0/6/5/3065888/euler-koeckel_-
_scattering_of_light_by_light.pdf}}).\\
%%CITATION = NATWA,23,246;%%

%\bibitem[()]{1934heisenberg}
W.\ Heisenberg:
Bemerkungen zur\ $\,$D$\,$i$\,$r$\,$a$\,$c$\,$schen Theorie des Positrons.\hfill
%[Remarks on Dirac's theory of the positron].
{\it Zeitschrift f\"ur Physik} {\bf 90}:3(1934)209-231
(\href{http://dx.doi.org/10.1007/BF01333516}{DOI: 10.1007/BF01333516}), 
erratum {\it ibid.}\ {\bf 92}:9(1934)692\hfill\ \linebreak
(\href{http://dx.doi.org/10.1007/BF01340782}{DOI: 
10.1007/BF01340782}). [in German]
Reprint of the original article in: \cite{1989heisenberg}, 
pp.\ 132-154, erratum p.\ 161.
English transl.\ (by D.\ H.\ Delphenich):
{\it Remarks on the Dirac theory of positron} (The translation is
freely available at the URL:\hfill\ \linebreak
{\tt \url{http://neo-classical-physics.info/uploads/3/0/6/5/3065888/heisenberg-_remarks_on_the_dirac_theory_of_positrons.pdf}}).\\
%%CITATION = ZEPYA,90,209;%%
%%CITATION = ERRAT,92,692;%%

%\bibitem[()]{1937heisenberg}
W.\ Heisenberg: Bemerkungen zur Theorie des Atomkerns 
[Remarks concerning the theory of the atomic nucleus].
In: {\it Pieter Zeeman, 1865 -- 25 Mei -- 1935: Verhandelingen op
25 Mei 1935 aangeboden aan Prof.\ Dr.\ P.\ Zeeman}.
Martinus Nijhoff, The Hague, 1935, pp.\ 108-116. [in German]
Reprinted in: \cite{1984heisenberg}, pp.\ 238-246.
Russian transl.: 
\begin{otherlanguage}{russian}
В.\ Гейзенберг
\end{otherlanguage}
%{\cyrrm V.\ Ge\u izenberg}
[V.\ Ge\u\i zenberg]:
\begin{otherlanguage}{russian}
Замечания к теории атомного ядра
\end{otherlanguage}
%{\cyrrm Zamechaniya k teorii atomnogo yadra}
[Zamechaniya k teorii atomnogo yadra].
{\it 
\begin{otherlanguage}{russian}
Успехи Физических Наук
\end{otherlanguage}
%{\cyrit Uspekhi Fizicheskikh Nauk} 
[Uspekhi Fizicheskikh Nauk]}
{\bf 16}:1(1936)1-7
(The translation is freely available online at the journal website given by the
\href{http://dx.doi.org/10.3367/UFNr.0016.193601a.0001}{DOI:
  10.3367/UFNr.0016.193601a.0001} .).\\
%%CITATION = UFNAA,16,1;%%

%\bibitem[()]{1935schroedinger2}
E.\ Schr\"odinger:
Contributions to Born's new theory of the electromagnetic field.
{\it Proceedings of the Royal Society of London. Series A,
Containing Papers of Mathematical and Physical Character}
{\bf 150}:870(1935)465-477
(\href{http://dx.doi.org/10.1098/rspa.1935.0116}{DOI: 
10.1098/rspa.1935.0116},
stable JSTOR URL: {\tt \url{http://www.jstor.org/stable/96399}}).
Reprinted in: \cite{1984schroedinger}, pp.\ 196-208.\\
%%CITATION = PRSLA,A150,465;%%

%\bibitem[()]{1936shubin}
\begin{otherlanguage}{russian}
С.\ Шубин
\end{otherlanguage}
%{\cyrrm S.\ Shubin} 
[S.\ Shubin],
\begin{otherlanguage}{russian}
А.\ Смирнов
\end{otherlanguage}
%{\cyrrm A.\ Smirnov} 
[A.\ Smirnov]:
\begin{otherlanguage}{russian}
Простой пример из электродинамики Борна
\end{otherlanguage}
%{\cyrrm Prosto\u i primer iz \`{e}lektrodinamiki Borna} 
[Prosto\u\i\ primer iz \`{e}lektrodinamiki Borna]/[A 
simple example from the electrodynamics of Born].
{\it
\begin{otherlanguage}{russian}
Доклады Академии Наук СССР, Новая Серия
\end{otherlanguage}
%{\cyrit Doklady Akademii Nauk SSSR, Novaya Seriya}
[Doklady Akademii Nauk SSSR, Novaya Seriya]}
{\bf 1 (X)}:2(79)(1936)65-68. [in Russian] 
Reprinted in: \cite{shub1}, pp.\ 243-246. German transl.:
S.\ \v{S}ubin (Schubin), A.\ Smirnow: Ein einfaches Beispiel aus der 
Bornschen Elektrodynamik. 
{\it Comptes Rendus (Doklady) de l'Acad\'emie des Sciences de l'URSS, 
Nouvelle S\'erie} {\bf 1 (X)}:2(79)\linebreak[3](1936)69-72.\\
%%CITATION = DANKA,1,65;%%
%%CITATION = DANKA,10,65;%%
%%CITATION = 00199,1,69;%%
%%CITATION = 00199,10,69;%%

\refstepcounter{ppage}

\phantomsection
\addcontentsline{toc}{subsubsection}{Further references}
\def\refname{\normalsize Further references:}

\refstepcounter{ppage}

\newpage
\refstepcounter{ppage}

\refstepcounter{section}
\addcontentsline{toc}{section}{Original Russian text of the thesis}

\newcounter{rsection}
\renewcommand{\thesection}{\Roman{rsection}}
\newcounter{rsubsection}
\renewcommand{\thesubsection}{\arabic{rsubsection}}

\renewcommand{\thesubsubsection}{\asbuk{subsubsection}}
\newcounter{requation}
\renewcommand{\theequation}{\arabic{requation}}
\newcounter{rfigure}
\renewcommand{\caption}{\refstepcounter{rfigure}}

\ \\[3.5cm]

\begin{center}
{\Large\begin{otherlanguage}{russian}
Адриан Анатольевич Смирнов
\end{otherlanguage}
%\cyrrm Adrian Anatol\cprime evich Smirnov
}\\[0.2cm]

[Adrian Anatol'evich Smirnov]\\[0.2cm]

[16.11.(O.S.\ 3.11.)1908-6.12.1992] \cite{1988smirnov,1996smirnov}\\[1cm]

{\Large\begin{otherlanguage}{russian}
Применение электродинамики Борна к теории
\end{otherlanguage}
%\cyrrm Primenenie \protect{\`{e}}lektrodinamiki Borna k teorii
\\[0.2cm]

\begin{otherlanguage}{russian}
распространения света в электромагнитных полях
\end{otherlanguage}
%rasprostraneniya sveta v \protect{\`{e}}lektromagnitnykh polyakh
}\\[0.2cm]

[Primenenie \`{e}lektrodinamiki 
Borna k teorii rasprostraneniya sveta v
\`{e}lektromagnitnykh polyakh]/[The application of the 
electrodynamics of Born to the theory
of the propagation of light in electromagnetic fields].\\[3.5cm]

\end{center}

\begin{otherlanguage}{russian}
Кандидатская диссертация
\end{otherlanguage}
%{\cyrrm Kandidat{s}kaya dissertatsiya}

[Kandidat$\cdot$skaya dissertatsiya]/[Ph.D.\ thesis],

\begin{otherlanguage}{russian}
Московский государственный университет
\end{otherlanguage}
%{\cyrrm Moskovski\u i gosudarstvenny\u i universi\-tet}

[Moskovski\u\i\ gosudarstvenny\u\i\ universitet]/[Moscow
State University], 

Moscow, 1936, 67 pp.. [in Russian]\\[2cm]

Thesis advisor: 
\begin{otherlanguage}{russian}
С.\ П.\ Шубин
\end{otherlanguage}
%{\cyrrm S.\ P.\ Shubin} 
[S.\ P.\ Shubin] 
[31.7.(O.S.\ 18.7.)1908-20.11.1938] \cite{shub1,vons1}.\\[0.1cm]

Thesis defense: 28.6.1936 (approved: 17.2.1937), Moscow State University

\hspace{2.9cm}(source: \cite{zaio1}, p.\ 120, item 656).\\

\newpage
\refstepcounter{ppage}
\addcontentsline{toc}{subsection}{Contents of the thesis}
\ \\[1cm]

{\bf\large Contents of the thesis / \begin{otherlanguage}{russian}
Содержание диссертации
\end{otherlanguage}
%{\cyrrm Soderzhanie dissertatsii}
}\\

\begin{center} % just for vertical spacing and killing indent
\begin{tabular*}{\textwidth}{@{}l@{\extracolsep{\fill}}r@{}}
\begin{otherlanguage}{russian}
Первая страница
\end{otherlanguage}
%{\cyrrm Pervaya stranitsa}
&\pageref{page1}\\[1cm]
{\bf\begin{otherlanguage}{russian}
Глава
\end{otherlanguage}
%{\cyrrm Glava}
 I}&\pageref{glavaI}\\[3mm]
\ \ \S\ 1\ \uloosdot{\ \hspace{12.5cm}\ }&\pageref{glavaIpara1}\\[3mm]
\ \ \S\ 2\ \uloosdot{\ \hspace{12.5cm}\ }&\pageref{glavaIpara2}\\[3mm]
\ \ \S\ 3\ \uloosdot{\ \hspace{12.5cm}\ }&\pageref{glavaIpara3}\\[3mm]
\ \ \S\ 4\ \uloosdot{\ \hspace{12.5cm}\ }&\pageref{glavaIpara4}\\[3mm]
\ \ \S\ 5\ \uloosdot{\ \hspace{12.5cm}\ }&\pageref{glavaIpara5}\\[3mm]
\ \ \S\ 6\ \uloosdot{\ \hspace{12.5cm}\ }&\pageref{glavaIpara6}\\[3mm]
\ \ \S\ 7\ \uloosdot{\ \hspace{12.5cm}\ }&\pageref{glavaIpara7}\\[3mm]
\ \ \S\ 8\ \uloosdot{\ \hspace{12.5cm}\ }&\pageref{glavaIpara8}\\[1cm]
{\bf\begin{otherlanguage}{russian}
Глава
\end{otherlanguage}
%{\cyrrm Glava}
 II}&\pageref{glavaII}\\[3mm]
\ \ \S\ 1. \begin{otherlanguage}{russian}
Введение
\end{otherlanguage}
%{\cyrrm Vvedenie}
&\pageref{glavaIIpara1}\\[3mm]
\ \ \S\ 2. \begin{otherlanguage}{russian}
Плоская световая волна в однородном электростатическом поле
\end{otherlanguage}
%{\cyrrm Ploskaya svetovaya volna v odnorodnom 
%\protect{\`{e}}lektrostaticheskom pole}
&\pageref{glavaIIpara2}\\[3mm]
\ \ \S\ 3. \begin{otherlanguage}{russian}
"Рассеяние"\ света на постоянном поле плоского конденсатора
\end{otherlanguage}
%{\cyrrm "Rasseyanie" sveta na postoyannom pole ploskogo kondensatora}
&\pageref{glavaIIpara3}\\[3mm]
\ \ \S\ 4. \begin{otherlanguage}{russian}
Плоская световая волна в однородном магнитном поле
\end{otherlanguage}
%{\cyrrm Ploskaya svetovaya volna v odnorodnom magnitnom pole} 
&\pageref{glavaIIpara4}\\[3mm]
\ \ \S\ 5. \begin{otherlanguage}{russian}
Две плоские волны в вакууме
\end{otherlanguage}
%{\cyrrm Dve ploskie volny v vakuume}
&\pageref{glavaIIpara5}\\[1cm]
{\bf\begin{otherlanguage}{russian}
Примечание
\end{otherlanguage}
%{\cyrrm Primechanie}
}&\pageref{page67}
\end{tabular*}
\end{center}

\newpage
\refstepcounter{ppage}
\label{page1}
\addcontentsline{toc}{subsection}{Text [pp.\ 1-67 (= P-81 - P-147)]}

% color Russian equation numbering
\makeatletter
\let\mytagform@=\tagform@
\def\tagform@#1{\maketag@@@{$\color{blue} (#1)$}}
\makeatother

\renewcommand{\thepage}{\mbox{\arabic{page}}}
\chead{{\rm\color{gray}- \theppage\ -}\hspace{3.2cm}\ }

\setcounter{page}{1}
\newgeometry{top=5cm,textheight=23cm,textwidth=16.3cm}

{\large\tt\
%\stexttt\ {\cyrtt\

\hspace{0.3cm}
\begin{otherlanguage}{russian}
\udensdash{\ \ ПРИМЕНЕНИЕ ЭЛЕКТРОДИНАМИКИ 
БОРНА$\ \atop\ $\hspace{-0.1cm}К ТЕОРИИ\ \hspace{-0.05cm}\ }
%\udensdash{\ \ PRIMENENIE \protect{\`{E}}LEKTRODINAMIKI 
%BORNA$\ \atop\ $\hspace{-0.1cm}K TEORII\ \hspace{-0.05cm}\ }
\\[-0.4cm]

\udensdash{РАСПРОСТРАНЕНИЯ СВЕТА$\ \atop\ $\hspace{-0.1cm}В
ЭЛЕКТРОМАГНИТНЫХ ПОЛЯХ.\hspace{-0.05cm}\ }
%\udensdash{RASPROSTRANENIYA SVETA$\ \atop\ $\hspace{-0.1cm}V
%\protect{\`{E}}LEKTROMAGNITNYKH POLYAKH.\hspace{-0.05cm}\ }
\\[-0.1cm]

\hspace{7.4cm}
\udensdash{A.A.$\ \atop\ $\hspace{-0.1cm}Смирнов.\hspace{-0.05cm}\ }
%\udensdash{A.A.$\ \atop\ $\hspace{-0.1cm}Smirnov.\hspace{-0.05cm}\ }
\\[1cm]

\hspace{1cm} Предметом настоящей диссертацией является рассмо-
%Predmetom nastoyashche\u i dissertatsie\u i  yavlyaet{sya} rassmo-

трение некоторых проблем теории электромагнитного поля,
%trenie nekotorykh problem teorii \protect{\`{e}}lektromagnitnogo polya,

преложенной недавно \udensdash{Борном\hspace{-0.05cm}\ } и
\udensdash{Инфельдом\hspace{-0.05cm}\ }$^{\mbox{\large\stexttt{I/}}}$.
Эти рассмо-
%prelozhenno\u i nedavno \udensdash{Bornom\hspace{-0.05cm}\ } i 
%\udensdash{Infel\cprime dom\hspace{-0.05cm}\ }}$^{\mbox{\large\stexttt{I/}}}$. 
%{\cyrtt\protect{\`{E}}ti rassmo-

трения мы будем вести исключительно в рамках классичес-
%treniya my budem vesti isklyuchitel\cprime no v ramkakh klassiches-

кого / не квантового / варианта теории; вопрос о том,
%kogo / ne kvantovogo / varianta teorii; vopros o tom,

как видоизменится постановка разобранных здесь проблем
%kak vidoizmenit{s}ya postanovka razobrannykh zdes\cprime\ problem

при введении квантования поля, нами не обсуждается.
%pri vvedenii kvantovaniya polya, nami ne obsuzhdaet{s}ya.

Диссертация делится на две части. В первой части мы да-
%Dissertatsiya delit{s}ya na dve chasti. V pervo\u i chasti my da-  

ем общий обзор современного состояния Борновской теории
%em obshchi\u i obzor sovremennogo sostoyaniya Bornovsko\u i teorii

не останавливаясь, при этом, на отдельных частностях
%ne ostanavlivayas\cprime , pri \protect{\`{e}}tom, na otdel\cprime nykh 
%chastnostyakh 

вычислений, а стремясь выявить только основной ход мыс-
%vychisleni\u i, a stremyas\cprime\ vyyavit\cprime\ tol\cprime ko osnovno\u i 
%khod mys-

ли. В частности, мы пытаемся установить, какие общие
%li. V chastnosti, my pytaemsya ustanovit\cprime , kakie obshchie

вопросы следовало бы еще выяснить для того чтобы полу-
%voprosy sledovalo by eshche vyyasnit\cprime\ dlya togo chtoby polu-

чить лучшее представление о ценности теории и ее даль-
%chit\cprime\ luchshee predstavlenie o tsennosti teorii i ee dal\cprime -

нейших перспективах - разумеется, отнюдь не ставя себе
%ne\u ishikh perspektivakh - razumeet{s}ya, otnyud\cprime\ ne stavya sebe

целью дать ответ на эти вопросы в рамках настоящей
%tsel\cprime yu dat\cprime\ otvet na \protect{\`{e}}ti voprosy v ramkakh
%nastoyashche\u i 

диссертации. Вторая часть представляет собою изложение
%dissertatsii. Vtoraya chast\cprime\ predstavlyaet soboyu izlozhenie

решения трех примеров на применение электродинамики
\end{otherlanguage}
%resheniya trekh primerov na primenenie \protect{\`{e}}lektrodinamiki
%}

-----------------------------

I/
{\ith M.\ Born. Proc.\ Roy.\ Soc.\ }\hspace{2.cm} 
{\handcolor\underline{\color{black}I43}}.\ I934.\ 4I0\hfill\ 

\hspace{0.5cm} 
{\ith M.\ Born and L.\ Infeld. Proc.\ Roy.\ Soc.\ }\hspace{1.15cm}
{\handcolor\underline{\color{black}I44}}.\ I934.\ 425

\hspace{1.5cm} {\large\ith''\ \hspace{2.6cm}''\ \hspace{2cm}''\ }
\hspace{3.2cm}{\handcolor\underline{\color{black}I47}}.\ I934.\ 522\hfill\

\hspace{1.5cm} {\large\ith''\ \hspace{2.6cm}''\ \hspace{2cm}''\ }
\hspace{3.2cm}{\handcolor\underline{\color{black}I50}}.\ I935.\ I4I\hfill\

\hspace{0.5cm} {\large\ith E.\ Schr\"odinger  Proc.\ Roy.\ Soc.\ }
\hspace{2.45cm} {\handcolor\underline{\color{black}I50}}.\ I935.\ 465}\hfill\
{\large\tt

\pagebreak
\refstepcounter{ppage}
\label{page2}
\chead{{\rm\color{gray}- \theppage\ -}\hspace{3.2cm}\ \linebreak
\large\tt - \thepage\ -\hspace{3cm}\ }

\begin{otherlanguage}{russian}
%\stexttt {\cyrtt

Борна к теории распространения света в электромагнит-
%Borna k teorii rasprostraneniya sveta v \protect{\`{e}}lektromagnit-

ных полях. В этих примерах особенно ярко проявляются
%nykh polyakh. V \protect{\`{e}}tikh primerakh osobenno yarko proyavlyayut{s}ya

предсказываемые теорией Борна своеобразные отклонения
%predskazyvaemye teorie\u i Borna svoeobraznye otkloneniya

от законов электродинамики Максвелла, отклонения сов-
%ot zakonov \protect{\`{e}}lektrodinamiki Maksvella, otkloneniya sov-

сем иного характера чем те, которые связаны с хорошо
%sem inogo kharaktera chem te, kotorye svyazany s khorosho

известными крантовыми эффектами.
%izvestnymi krantovymi \protect{\`{e}}ffektami.
\\[0.5cm]

\refstepcounter{rsection}
\label{glavaI}

\hspace{3.5cm}
\udensdash{\ Г$\ \atop\ $\hspace{-0.1cm}л\ а\ в\ а\ \ I.\ 
%\udensdash{\ G$\ \atop\ $\hspace{-0.1cm}l\ a\ v\ a\ \ {\tt I}.\ 
\hspace{-0.1cm}\ }\\[-0.2cm]

\refstepcounter{rsubsection}
\label{glavaIpara1}
\hspace{1cm}\S\ I. Работы \udensdash{Борна\hspace{-0.05cm}\ } и 
\udensdash{Инфельда\hspace{-0.05cm}\ } впервые дали решение
%{\tt I}. Raboty \udensdash{Borna\hspace{-0.05cm}\ } i 
%\udensdash{Infel\cprime da\hspace{-0.05cm}\ } vpervye dali reshenie

задачи, много занимавшей умы теоретиков в первое двад-
%zadachi, mnogo zanimavshe\u i umy teoretikov v pervoe dvad-

цатилетие ХХ века - задачи построения, в рамках класси-
%tsatiletie {\tt XX} veka - zadachi postroeniya, v ramkakh klassi-

ческой электродинамики, так называемой 
"\udensdash{единой / или\ \hspace{-0.05cm}\ }
%chesko\u i \protect{\`{e}}lektrodinamiki, tak nazyvaemo\u i 
%"\udensdash{edino\u i / ili\ \hspace{-0.05cm}\ }

\udensdash{унитарной / теории поля и материи"\hspace{-0.05cm}\ }
%\udensdash{unitarno\u i / teorii polya i materii"\hspace{-0.05cm}\ }}
\ $^{\mbox{\large\stexttt{I/}}}$.

\hspace{1cm} %{\cyrtt 
Основная идея унитарной теории - при наиболее ради-
%Osnovnaya ideya unitarno\u i teorii - pri naibolee radi-

кальном / отчасти даже сознательно утрированном / ее
%kal\cprime nom / otchasti dazhe soznatel\cprime no utrirovannom / ee

толковании - заключается в том, что существование вся-
%tolkovanii - zaklyuchaet{s}ya v tom, chto sushchestvovanie vsya-

кой электрически заряженной
\udensdash{ч$\ \atop\ $\hspace{-0.33cm}астицы\hspace{-0.05cm}\ }
может быть пол-
%ko\u i \protect{\`{e}}lektricheski zaryazhenno\u i 
%\udensdash{ch$\ \atop\ $\hspace{-0.33cm}astitsy\hspace{-0.05cm}\ } 
%mozhet byt\cprime\ pol-

ностью описано, как особое состояние
\udensdash{электромагнитного\hspace{-0.05cm}\ }
%nost\cprime yu opisano, kak osoboe sostoyanie 
%\udensdash{\protect{\`{e}}lektromagnitnogo\hspace{-0.05cm}\ }

\udensdash{по$\ \atop\ $\hspace{-0.33cm}ля\hspace{-0.05cm}\ }.
Стандардная схема, которая более или менее одно-
%\udensdash{po$\ \atop\ $\hspace{-0.33cm}lya\hspace{-0.05cm}\ }. 
%Standardnaya skhema, kotoraya bolee ili menee odno-

значно диктуется для построения такой теории, выглядит,
%znachno diktuet{s}ya dlya postroeniya tako\u i teorii, vyglyadit,

в самых общих чертах, - так.
%v samykh obshchikh chertakh, - tak.

----------------------------------

I/ Мы пользуемся здесь установившейся среди физиков тер-
%{\tt I}/ My pol\cprime zuemsya zdes\cprime\ ustanovivshe\u isya sredi
%fizikov ter- 

минологией, согласно которой "поле"\ и "материя"\ проти-
%minologie\u i, soglasno kotoro\u i "pole" i "materiya" proti-

вопоставляются друг другу. С точки зрения общефилософ-
%vopostavlyayut{s}ya drug drugu. S tochki zreniya obshchefilosof-

ской, такая терминология, разумеется, не выдерживает
%sko\u i, takaya terminologiya, razumeet{s}ya, ne vyderzhivaet

критики, так как само электромагнитное поле тоже-есть
%kritiki, tak kak samo \protect{\`{e}}lektromagnitnoe pole 
%tozhe-est\cprime

\udensdash{материя\hspace{-0.05cm}\ }, как и всякая 
реальность, существующая объектив-
%\udensdash{materiya\hspace{-0.05cm}\ }, kak i vsyakaya 
%real\cprime nost\cprime , sushchestvuyushchaya ob"ektiv-

но вне нашего сознания.
%no vne nashego soznaniya.

\pagebreak 
\refstepcounter{ppage}

В основу кладутся законы электромагнитного поля, вы-
%V osnovu kladut{s}ya zakony \protect{\`{e}}lektromagnitnogo polya, vy-

раженные лучше всего, в виде вариационного принципа.
%razhennye luchshe vsego, v vide variatsionnogo printsipa.

Поле, как обычно, описывается двумя векторами - электри-
%Pole, kak obychno, opisyvaet{s}ya dvumya vektorami - \protect{\`{e}}lektri-

ческим и магнитным; его лагранжиан представляет собою
%cheskim i magnitnym; ego lagranzhian predstavlyaet soboyu

функцию от этих величин. Из вариационного принципа вы-
%funktsiyu ot \protect{\`{e}}tikh velichin. Iz variatsionnogo printsipa vy-

текают, в качестве уравнений
\udensdash{Эйлера\hspace{-0.05cm}\ } - 
\udensdash{Лагранжа\hspace{-0.05cm}\ }, так
%tekayut, v kachestve uravneni\u i 
%\udensdash{\protect{\`{E}}\u ilera\hspace{-0.05cm}\ } - 
%\udensdash{Lagranzha\hspace{-0.05cm}\ }, tak

называемые уравнения поля /причем, при проведении ва-
%nazyvaemye uravneniya polya /prichem, pri provedenii va-

риации, вводится, обычно, добавочное предположение о
%riatsii, vvodit{s}ya, obychno, dobavochnoe predpolozhenie o

существовании четырехмерного вектор - потенциала, ком-
%sushchestvovanii chetyrekhmernogo vektor - potentsiala, kom-

поненты которого и служат варьируемыми функциями/. Эти
%ponenty kotorogo i sluzhat var\cprime iruemymi funktsiyami/. \protect{\`{E}}ti 

уравнения позволяют сразу-же получить
\udensdash{законы сохране-\hspace{-0.05cm}\ }
%uravneniya pozvolyayut srazu-zhe poluchit\cprime\ 
%\udensdash{zakony sokhrane-\hspace{-0.05cm}\ }

\udensdash{ни$\ \atop\ $\hspace{-0.33cm}я\hspace{-0.05cm}\ }
для электромагнитного поля, вводя понятие электро-
%\udensdash{ni$\ \atop\ $\hspace{-0.33cm}ya\hspace{-0.05cm}\ } 
%dlya \protect{\`{e}}lektromagnitnogo polya, vvodya ponyatie 
%\protect{\`{e}}lektro-

магнитной энергии и электромагнитного количеста дви-
%magnitno\u i \protect{\`{e}}nergii i \protect{\`{e}}lektromagnitnogo
%kolichesta dvi-

жения /импульса/.
%zheniya /impul\cprime sa/.

Решающий шаг - включение в теорию материи производится
%Reshayushchi\u i shag - vklyuchenie v teoriyu materii proizvodit{s}ya

следующим образом. Рассматривается определенный класс
%sleduyushchim obrazom. Rassmatrivaet{s}ya opredelenny\u i klass

решений уравнений поля, а именно такие решения, для
%resheni\u i uravneni\u i polya, a imenno takie resheniya, dlya

которых, в какой-то выбранной системе координат, магнит-
%kotorykh, v kako\u i-to vybranno\u i sisteme koordinat, magnit-

ный вектор равен нулю, а электрический - сравнительно сим-
%ny\u i vektor raven nulyu, a \protect{\`{e}}lektricheski\u i - 
%sravnitel\cprime no sim-

метричен. В силу самой структуры уравнений поля, эти
%metrichen. V silu samo\u i struktury uravneni\u i polya, \protect{\`{e}}ti

два требования достаточно для того, чтобы определить
%dva trebovaniya dostatochny dlya togo, chtoby opredelit\cprime

тип решения полностью$^{\displaystyle\handcolor\ast)}$;
произвольным остается лишь
%tip resheniya polnost\cprime yu$^{\displaystyle\handcolor\ast)}$;
%proizvol\cprime nym ostaet{s}ya lish\cprime

один постоянный фактор. Физическое состояние, изобра-
%odin postoyanny\u i faktor. Fizicheskoe sostoyanie, izobra-

жаемое каким нибудь решением этого класса интерпрети-
%zhaemoe kakim nibud\cprime\ resheniem \protect{\`{e}}togo klassa
%interpreti-

руется, как состояние, соответствующее наличию электри-
%ruet{s}ya, kak sostoyanie, sootvet{s}tvuyushchee 
%nalichiyu \protect{\`{e}}lektri-

чески заряженной
\udensdash{материальной частицы,\hspace{-0.05cm}\ }
покоющейся в
%cheski zaryazhenno\u i 
%\udensdash{material\cprime no\u i chastitsy,\hspace{-0.05cm}\ }
%pokoyushche\u isya v

начале координат выбранной системы отсчета. Оставшийся
%nachale koordinat vybranno\u i sistemy ot{s}cheta. Ostavshi\u isya

произвольный фактор определяет, как говорят, полный
%proizvol\cprime ny\u i faktor opredelyaet, kak govoryat, polny\u i 

заряд нашей частицы; эта величина никак не фиксируется
%zaryad nashe\u i chastitsy; \protect{\`{e}}ta velichina
%nikak ne fiksiruet{s}ya

теорией.
%teorie\u i.
\linebreak\nopagebreak[5]
\ \hspace{-1.3cm}
{\handcolor\underline{\ \hspace{12.5cm}\ }}
\linebreak\nopagebreak[5]
\ \hspace{-1.5cm}
$\handcolor *)$ 
{\ith Недавно Инфельдом был предложен новый вариант
%\cyrit Nedavno Infel\cprime dom byl predlozhen novy\u i variant

унитарной теории в которой эта однозначность не
%unitarno\u i teorii v kotoro\u i \protect{\`{e}}ta odnoznachnost\cprime\ ne 

имеет места (см.\ пример в конце).}
%imeet mesta (sm.\ primer v kontse).}.

\pagebreak
\refstepcounter{ppage}

\hspace{1cm}Конкретное содержание фразы: "данное состояние
%Konkretnoe soderzhanie frazy: "dannoe sostoyanie

интерпретируется, как состояние, соответсвующее на-
%interpretiruet{s}ya, kak sostoyanie, sootvet{s}vuyushchee na-

личию частицы"\ заключается, прежде всего, в том, что
%lichiyu chastitsy" zaklyuchaet{s}ya, prezhde vsego, v tom, chto

координаты "особой точки"\ решения толкуются как коорди-
%koordinaty "osobo\u i tochki" resheniya tolkuyut{s}ya kak koordi-

наты частицы /или ее "центра"/ а соответствующие рас-
%naty chastitsy /ili ee "tsentra"/ a sootvet{s}tvuyushchie ras-

матриваемому состоянию поля полная энергия и полный
%matrivaemomu sostoyaniyu polya 
%polnaya \protect{\`{e}}nergiya i polny\u i

импульс толкуются, соответственно, как
\udensdash{механическая энер-\hspace{-0.05cm}\ }
%impul\cprime s tolkuyut{s}ya, sootvet{s}tvenno, kak 
%\udensdash{mekhanicheskaya \protect{\`{e}}ner-\hspace{-0.05cm}\ }

гия и \udensdash{механический импульс\hspace{-0.05cm}\ }
частицы. Значение этих величин
%giya i \udensdash{mekhanicheski\u i impul\cprime s\hspace{-0.05cm}\ }
%chastitsy. Znachenie \protect{\`{e}}tikh velichin

зависит,конечно, от выбора системы отсчета; в частности
%zavisit,konechno, ot vybora sistemy ot{s}cheta; v chastnosti

в исходной системе, т.е., в той, относительно которой
%v iskhodno\u i sisteme, t.e., v to\u i, otnositel\cprime no kotoro\u i

частица покоится, полный импульс равен нулю без чего
%chastitsa pokoit{s}ya, polny\u i impul\cprime s raven nulyu
%/bez chego

механическая интерпретация была бы, разумеется, невоз-
%mekhanicheskaya interpretatsiya byla by, razumeet{s}ya, nevoz-

можной/, полная-же энергия представляет собою релятиви-
%mozhno\u i/, polnaya-zhe \protect{\`{e}}nergiya predstavlyaet soboyu
%relyativi-

стскую энергию покоя рассматриваемой частицы.
%st{s}kuyu \protect{\`{e}}nergiyu pokoya rassmatrivaemo\u i chastitsy.

После того, как таким образом введены основные
\udensdash{пон$\ \atop\ $\hspace{-0.33cm}ятия\hspace{-0.05cm}\ }
%Posle togo, kak takim obrazom vvedeny osnovnye 
%\udensdash{pon$\ \atop\ $\hspace{-0.33cm}yatiya\hspace{-0.05cm}\ }

механики, необходимо, для того, чтобы теория была пол-
%mekhaniki, neobkhodimo, dlya togo, chtoby teoriya byla pol-

ной, установить основные ее /механики/
\udensdash{за$\ \atop\ $\hspace{-0.33cm}коны\hspace{-0.05cm}\ }. В рамках
%no\u i, ustanovit\cprime\ osnovnye ee /mekhaniki/ 
%\udensdash{za$\ \atop\ $\hspace{-0.33cm}kony\hspace{-0.05cm}\ }. V ramkakh

рассматриваемой схемы от этих законов можно, очевидно,
%rassmatrivaemo\u i skhemy ot \protect{\`{e}}tikh zakonov mozhno, ochevidno,

требовать ответа только на один вопрос: как меняются
%trebovat\cprime\ otveta tol\cprime ko na odin vopros: kak menyayut{s}ya

энергия и импульс частицы или, общее, характеризующее
%\protect{\`{e}}nergiya i impul\cprime s chastitsy ili, obshchee,
%kharakterizuyushchee 

наличие частицы состояние поля, под влиянием внешних
%nalichie chastitsy sostoyanie polya, pod vliyaniem vneshnikh

электромагнитных полей.
%\protect{\`{e}}lektromagnitnykh pole\u i.

При такой ограниченной постановке задачи, законы меха-
%Pri tako\u i ogranichenno\u i postanovke zadachi, zakony mekha-

ники, по крайней мере принципиально, могут быть полу-
%niki, po kra\u ine\u i mere printsipial\cprime no, mogut byt\cprime\ polu-

чены, как частный случай выставленного в самом начале,
%cheny, kak chastny\u i slucha\u i vystavlennogo v samom nachale,

как исходный пункт, вариационного принципа электромаг-
%kak iskhodny\u i punkt, variatsionnogo printsipa \protect{\`{e}}lektromag-

нитизма и, тем самым теория может считаться завершенной.
%nitizma i, tem samym teoriya mozhet schitat\cprime sya zavershenno\u i.

Такого рода схема является "идеальной"\ в том смысле,
%Takogo roda skhema yavlyaet{s}ya "ideal\cprime no\u i" v tom smysle,

что при последовательном ее проведении, роль основных
%chto pri posledovatel\cprime nom ee provedenii, rol\cprime\ osnovnykh

величин, вводимых в теорию с самого начала $\;${\ith и}грают
\udensdash{тол$\ \atop\ $\hspace{-0.37cm}ько}
%velichin, vvodimykh v teoriyu s samogo nachala$\;${\ith\cyrit{i}}grayut
%\udensdash{tol$\ \atop\ $\hspace{-0.37cm}\cprime ko}

\pagebreak
\refstepcounter{ppage}

величины, характеризующие поле. Так как осуществление
%velichiny, kharakterizuyushchie pole. Tak kak osushchestvlenie

этой схемы в рамках Максвелловой электродинамики было,
%\protect{\`{e}}to\u i skhemy v ramkakh Maksvellovo\u i 
%\protect{\`{e}}lektrodinamiki bylo,

по хорошо известным причинам, невозможным, то наряду с
%po khorosho izvestnym prichinam, nevozmozhnym, to naryadu s

нею рассматривались и другие схемы построения унитарной
%neyu rassmatrivalis\cprime\ i drugie skhemy postroeniya unitarno\u i

теории, которые можно в известном условном смысле назвать
%teorii, kotorye mozhno v izvestnom uslovnom smysle nazvat\cprime

более "компромиссными". По таким "компромиссным"\ схемам
%bolee "kompromissnymi". Po takim "kompromissnym" skhemam

построены, в частности, излагаемые во многих учебниках
%postroeny, v chastnosti, izlagaemye vo mnogikh uchebnikakh 

теории \udensdash{Абрагама\hspace{-0.05cm}\ } и
\udensdash{Лоренца\hspace{-0.05cm}\ }, отклонявшиеся от указанной
%teorii \udensdash{Abragama\hspace{-0.05cm}\ } i
%\udensdash{Lorentsa\hspace{-0.05cm}\ }, otklonyavshiesya ot 
%ukazanno\u i 

программы в том отношении, что в них с самого начала,
%programmy v tom otnoshenii, chto v nikh s samogo nachala,

наряду с понятиями электрического и магнитного векторов,
%naryadu s ponyatiyami \protect{\`{e}}lektricheskogo i magnitnogo
%vektorov,

вводились понятия, характеризующие
\udensdash{заряды\hspace{-0.05cm}\ } и - правда,
%vvodilis\cprime\ ponyatiya, kharakterizuyushchie 
%\udensdash{zaryady\hspace{-0.05cm}\ } i - pravda,

только со стороны их электрических
%tol\cprime ko so storony ikh \protect{\`{e}}lektricheskikh

свойств, а именно, в первую очередь, понятие плотности
%svo\u istv, a imenno, v pervuyu ochered\cprime , ponyatie plotnosti

заряда. Место несколько абстрактного представления о
%zaryada. Mesto neskol\cprime ko abstraktnogo predstavleniya o

частице, как об особом состоянии поля, занимали у Абрага-
%chastitse, kak ob osobom sostoyanii polya, zanimali u Abraga-

ма и Лоренца преставления, по внешности более нагляд-
%ma i Lorentsa prestavleniya, po vneshnosti bolee naglyad-

ные: частица рассматривалась как "заряженный шарик"\protect{,}
%nye: chastitsa rassmatrivalas\cprime\ kak  "zaryazhenny\u i sharik",

т.е.\ характеризовалась определенным распределением плот-
%t.e.\ kharakterizovalas\cprime\ opredelennym raspredeleniem plot-

ности электричества в относительно небольшом объеме
%nosti \protect{\`{e}}lektrichestva v otnositel\cprime no nebol\cprime shem
%ob"eme

/причем порядок величины "радиуса"\ частицы определялся
%/prichem poryadok velichiny "radiusa" chastitsy opredelyalsya

из ее заряда и массы по хорошо известному соотношению
%iz ee zaryada i massy po khorosho izvestnomu sootnosheniyu

$\handcolor r_0\sim \frac{e^2}{mc^2}\ /$. 
При этом, в отличие от "идеальной"\
%Pri \protect{\`{e}}tom, v otlichie ot "ideal\cprime no\u i"

схемы, в теорию входила известная неоднозначность, т.к.
%skhemy, v teoriyu vkhodila izvestnaya neodnoznachnost\cprime , t.k.

даже в предположении о радиальной симметрии частицы,
%dazhe v predpolozhenii o radial\cprime no\u i simmetrii chastitsy,

распределение заряда "внутри"\ ее могло носить еще самый
%raspredelenie zaryada "vnutri" ee moglo nosit\cprime\ eshche samy\u i

разнообразный характер. Правда, этот добавочный произ-
%raznoobrazny\u i kharakter. Pravda, \protect{\`{e}}tot dobavochny\u i proiz-

вол значительно смягчался тем, что различные гипотезы
%vol znachitel\cprime no smyagchalsya tem, chto razlichnye gipotezy

о внутреннем "строении"\ частицы совсем не влияли на
%o vnutrennem "stroenii" chastitsy sovsem ne vliyali na

принципиальную сторону выводов теории.
%printsipial\cprime nuyu storonu vyvodov teorii.

\pagebreak
\refstepcounter{ppage}

Как хорошо известно, развитие идеи \udensdash{Абрагама\hspace{-0.05cm}\ }
и \udensdash{Лоренца\hspace{-0.05cm}\ }
%Kak khorosho izvestno, razvitie ide\u i \udensdash{Abragama\hspace{-0.05cm}\ }
%i \udensdash{Lorentsa\hspace{-0.05cm}\ }

встретило в свою очередь ряд затруднений /см.\ \S\
{\tt\ref{glavaIpara3}} этой
%vstretilo v svoyu ochered\cprime\ ryad zatrudneni\u i /sm.\ \S\ 
%{\tt\ref{glavaIpara3}} \protect{\`{e}}to\u i

главы/ и единственная удачная из всех доселе существую-
%glavy/ i edinstvennaya udachnaya iz vsekh dosele sushchestvuyu-

щих унитарных теорий - теория
\udensdash{Борна\hspace{-0.05cm}\ } -
\udensdash{Инфельда\hspace{-0.05cm}\ } - постро-
%shchikh unitarnykh teori\u i - teoriya 
%\udensdash{Borna\hspace{-0.05cm}\ } - 
%\udensdash{Infel\cprime da\hspace{-0.05cm}\ } - postro-

ена как раз на основе "идеальной схемы". Мы упомянули
%ena kak raz na osnove  "ideal\cprime no\u i skhemy". My upomyanuli

здесь о "компромиссных"\ теориях главным образом для
%zdes\cprime\ o  "kompromissnykh" teoriyakh glavnym obrazom dlya

того, чтобы подчеркнуть, что введение основных
\udensdash{поня$\ \atop\ $\hspace{-0.33cm}тий\hspace{-0.05cm}\ }
%togo, chtoby podcherknut\cprime , chto vvedenie osnovnykh 
%\udensdash{ponya$\ \atop\ $\hspace{-0.33cm}ti\u i\hspace{-0.05cm}\ }

\udensdash{механ$\ \atop\ $\hspace{-0.33cm}ики\hspace{-0.05cm}\ }
производилось и в этих теориях совершенно по
%\udensdash{mekhan$\ \atop\ $\hspace{-0.33cm}iki\hspace{-0.05cm}\ } 
%proizvodilos\cprime\ i v \protect{\`{e}}tikh teoriyakh sovershenno po

такому-же приему, какой был только что указан, а именно,
%takomu-zhe priemu, kako\u i byl tol\cprime ko chto ukazan, a imenno,

путем отождествления энергии и импульса электромагнит-
%putem otozhdestvleniya \protect{\`{e}}nergii i impul\cprime sa
%\protect{\`{e}}lektromagnit-

ного поля частицы с ее механической энергией и механи-
%nogo polya chastitsy s ee mekhanichesko\u i \protect{\`{e}}nergie\u i
%i mekhani-

ческим импульсом. Таким образом, именно это отождествле-
%cheskim impul\cprime som. Takim obrazom, imenno \protect{\`{e}}to
%otozhdestvle-

ние должно быть рассматриваемо как центральный пункт вся-
%nie dolzhno byt\cprime\ rassmatrivaemo kak tsentral\cprime ny\u i
%punkt vsya-

кой унитарной теории. По этой причине единую теорию ма-
%ko\u i unitarno\u i teorii. Po \protect{\`{e}}to\u i prichine edinuyu
%teoriyu ma-

терии и поля часто называют "теорией электромагнитной
%terii i polya chasto nazyvayut "teorie\u i \protect{\`{e}}lektromagnitno\u i

массы".\\
%massy".\\

\refstepcounter{rsubsection}
\label{glavaIpara2}
\hspace{1cm}\S\ 2. В какой мере руководящая идея унитарной теории
%V kako\u i mere rukovodyashchaya ideya unitarno\u i teorii

может считаться физически оправданной{\ith ?}
%mozhet schitat\cprime sya fizicheski opravdanno\u i{\ith ?}

\hspace{1cm}В рамках классической физики, ответ на этот
%V ramkakh klassichesko\u i fiziki, otvet na \protect{\`{e}}tot

вопрос представляется более или менее ясным.
%vopros predstavlyaet{s}ya bolee ili menee yasnym.

Тот факт, что всякий электрический заряд обладает эле-
%Tot fakt, chto vsyaki\u i \protect{\`{e}}lektricheski\u i zaryad
%obladaet \protect{\`{e}}le-

ктромагнитной энергией и электромагнитным импульсом, ко-
%ktromagnitno\u i \protect{\`{e}}nergie\u i i 
%\protect{\`{e}}lektromagnitnym impul\cprime som, ko-

торые обязательно должны учитываться при написании зако-
%torye obyazatel\cprime no dolzhny uchityvat\cprime sya pri
%napisanii zako-

нов сохранения для любой системы, в состав которой этот
%nov sokhraneniya dlya lyubo\u i sistemy, v sostav kotoro\u i 
%\protect{\`{e}}tot

заряд входит, является для классической теории совершен-
%zaryad vkhodit, yavlyaet{s}ya dlya klassichesko\u i teorii sovershen-

но бесспорным.
%no besspornym.

\pagebreak
\refstepcounter{ppage}

В самом деле, он вытекает из наиболее фундаментальных
%V samom dele, on vytekaet iz naibolee fundamental\cprime nykh

положений электродинамики Максвелла.
%polozheni\u i \protect{\`{e}}lektrodinamiki Maksvella.

Отсюда, однако, еще отнюдь не следует, что, как это утвер-
%Ot{s}yuda, odnako, eshche otnyud\cprime\ ne sleduet, chto, kak 
%\protect{\`{e}}to utver-

ждает унитарная теория,
\udensdash{вс$\ \atop\ $\hspace{-0.33cm}я\hspace{-0.05cm}\ } масса 
каждой заряженной ча-
%zhdaet unitarnaya teoriya, 
%\udensdash{vs$\ \atop\ $\hspace{-0.33cm}ya\hspace{-0.05cm}\ } massa
%kazhdo\u i zaryazhenno\u i cha-

стицы имеет чисто электромагнитное происхождение. В самом
%stitsy imeet chisto \protect{\`{e}}lektromagnitnoe proiskhozhdenie. V samom

деле, опыт показывает, что не существует таких материаль-
%dele, opyt pokazyvaet, chto ne sushchestvuet takikh material\cprime-

ных частиц, все свойства которых исчерпывались бы тем,
%nykh chastits, vse svo\u istva kotorykh ischerpyvalis\cprime\ by tem,

что эти частицы являются электрическими зарядами.
%chto \protect{\`{e}}ti chastitsy yavlyayut{s}ya 
%\protect{\`{e}}lektricheskimi zaryadami.

Если даже оставить в стороне тяготение /вопросы гравитаци-
%Esli dazhe ostavit\cprime\ v storone tyagotenie /voprosy gravitatsi-

онных действий в этой диссертации вообще совершенно не
%onnykh de\u istvi\u i v \protect{\`{e}}to\u i dissertatsii voobshche
%sovershenno ne 

рассматриваются/, то мы знаем наверное, что кроме электро-
%rassmatrivayut{s}ya/, to my znaem navernoe, chto krome \protect{\`{e}}lektro-

магнетизма и гравитации безусловно существует совершенно
%magnetizma i gravitatsii bezuslovno sushchestvuet sovershenno

особый тип сил, сдерживающих атомное ядро. Между тем, клас-
%osoby\u i tip sil, sderzhivayushchikh atomnoe yadro. Mezhdu tem, klas-

сическая теория утверждает, что "источником"\ массы мате-
%sicheskaya teoriya utverzhdaet, chto "istochnikom" massy mate-

риальной частицы является не только электромагнитное, но
%rial\cprime no\u i chastitsy yavlyaet{s}ya ne tol\cprime ko
%\protect{\`{e}}lektromagnitnoe, no
 
и \udensdash{вся$\ \atop\ $\hspace{-0.33cm}кое\hspace{-0.05cm}\ }
создаваемое ею силовое поле {\ith (гравитация, как
%i \udensdash{vsya$\ \atop\ $\hspace{-0.33cm}koe\hspace{-0.05cm}\ } 
%sozdavaemoe eyu silovoe pole{\ith\cyrit (gravitatsiya, kak 

сказано выше, исключается).}
%skazano vyshe, isklyuchaet{s}ya).}

Таким образом, в этой радикальной форме, в какой она была
%Takim obrazom, v \protect{\`{e}}to\u i radikal\cprime no\u i
%forme, v kako\u i ona byla 

нами высказана в \S\ {\tt\ref{glavaIpara1}},\
унитарная теория наверняка не отве-
%nami vyskazana v \S\ {\tt\ref{glavaIpara1}},\ 
%unitarnaya teoriya navernyaka ne otve-

чает физической реальности. Тем не менее, рассмотрение ее
%chaet fizichesko\u i real\cprime nosti. Tem ne menee, rassmotrenie ee

все же является полезным, исходя из следующих соображений.
%vse zhe yavlyaet{s}ya poleznym, iskhodya iz sleduyushchikh soobrazheni\u i.

Во первых, существует такой тип материальных частиц - еле-
%Vo pervykh, sushchestvuet tako\u i tip material\cprime nykh chastits -
%\protect{\`{e}}le-

ктроны / и позитроны/, которые насколько нам сегодня из-
%ktrony / i pozitrony/, kotorye naskol\cprime ko nam segodnya iz-

вестно, являются источниками только двух видов силовых
%vestno, yavlyayut{s}ya istochnikami tol\cprime ko dvukh vidov silovykh

полей: электромагнитного и гравитационного. При этом, есть
%pole\u i: \protect{\`{e}}lektromagnitnogo i gravitatsionnogo. Pri
%\protect{\`{e}}tom, est\cprime

все основания полагать, что на законы, управляющие поведе-
%vse osnovaniya polagat\cprime , chto na zakony, upravlayushchie povede-

нием электрона, создаваемое им самим гравитационное поле
%niem \protect{\`{e}}lektrona, sozdavaemoe im samym gravitatsionnoe pole

оказывает сравнительно небольшое влияние. Поэтому
\udensdash{к$\ \atop\ $элек-\hspace{-0.05cm}\ }
%okazyvaet sravnitel\cprime no nebol\cprime shoe vliyanie. 
%Po\protect{\`{e}}tomu
%\udensdash{k$\ \atop\ $\protect{\`{e}}lek-\hspace{-0.05cm}\ }

\udensdash{тро$\ \atop\ $\hspace{-0.33cm}ну\hspace{-0.05cm}\ }
унитарную теорию нужно считать в основном
\udensdash{применимой\hspace{-0.05cm}\ }{\ith ;}
%\udensdash{tro$\ \atop\ $\hspace{-0.33cm}nu\hspace{-0.05cm}\ }
%unitarnuyu teoriyu nuzhno schitat\cprime\ v osnovnom 
%\udensdash{primenimo\u i\hspace{-0.05cm}\ }{\ith ;}

во всяком случае, совершенно естественно попытаться постро-
%vo vsyakom sluchae, sovershenno estestvenno popytat\cprime sya postro-

ить для электрона такую теорию. Как это ни странно, но
%it\cprime\ dlya \protect{\`{e}}lektrona takuyu teoriyu. Kak
%\protect{\`{e}}to ni stranno, no

тот по существу совершенно очевидный, факт, что всякая
%tot po sushchestvu sovershenno ochevidny\u i, fakt, chto vsyakaya

\pagebreak
\refstepcounter{ppage}

унитарная чисто электромагнитная теория, в том числе и
%unitarnaya chisto \protect{\`{e}}lektromagnitnaya teoriya, v tom chisle i

теория \udensdash{Борна-\hspace{-0.05cm}\ }
\udensdash{Инфельда,\hspace{-0.05cm}\ } является 
"самое большее"\ теорией
%teoriya \udensdash{Borna-\hspace{-0.05cm}\ }
%\udensdash{Infel\cprime da,\hspace{-0.05cm}\ } yavlyaet{s}ya
%"samoe bol\cprime shee" teorie\u i

электрона, до сих пор еще оспаривается в литературе$^{\mbox{\large\stexttt{I/}}}$
%\protect{\`{e}}lektrona, do sikh por eshche osparivaet{s}ya v
%literature}$^{\mbox{\large\stexttt{I/}}}$

Во вторых, унитарная теория, помимо непосредственно-фи-
%{\cyrtt Vo vtorykh, unitarnaya teoriya, pomimo neposredstvenno-fi-

зического, имеет и глубокое\udensdash{\ методологическое\hspace{-0.05cm}\ }
значение. В са-
%zicheskogo, imeet i glubokoe\udensdash{\ metodologicheskoe\hspace{-0.05cm}\ }
%znachenie. V sa-

мом деле, все развитие теоретической физики позволяет
%mom dele, vse razvitie teoretichesko\u i fiziki pozvolyaet

думать, что построение теории различного вида силовых
%dumat\cprime , chto postroenie teorii razlichnogo vida silovykh 

полей может \udensdash{быть проведено методами, в 
основном аналогич-\hspace{-0.05cm}\ }
%pole\u i mozhet \udensdash{byt\cprime\ provedeno metodami, v 
%osnovnom analogich-\hspace{-0.05cm}\ }

\udensdash{н$\ \atop\ $\hspace{-0.33cm}ыми\hspace{-0.05cm}\ } 
друг другу. Поэтому, теория "чистого электричества"\protect{,}
%\udensdash{n$\ \atop\ $\hspace{-0.33cm}ymi\hspace{-0.05cm}\ } 
%drug drugu. Po\protect{\`{e}}tomu,
%teoriya "chistogo \protect{\`{e}}lektrichestva",

т.е.\ воображаемых частиц, являющихся
\udensdash{то$\ \atop\ $\hspace{-0.33cm}лько\hspace{-0.05cm}\ } 
электриче-
%t.e.\ voobrazhaemykh chastits, yavlyayushchikhsya 
%\udensdash{to$\ \atop\ $\hspace{-0.33cm}l\cprime ko\hspace{-0.05cm}\ } 
%\protect{\`{e}}lektriches-

кими зарядами, может рассматриваться, как прообраз тео-
%kimi zaryadami, mozhet rassmatrivat\cprime sya, kak proobraz teo-

рии реальных частиц, служащих источниками самых разнооб-
%rii real\cprime nykh chastits, sluzhashchikh istochnikami samykh raznoob-

разных полей. В настоящее время среди теоретиков господ-
%raznykh pole\u i. V nastoyashchee vremya sredi teoretikov gospod-

ствует убеждение о том, чо это утверждение в общем при-
%stvuet ubezhdenie o tom, cho \protect{\`{e}}to utverzhdenie v obshchem pri-

менимо и к тем недавно открытым силам, которые действуют
%menimo i k tem nedavno otkrytym silam, kotorye de\u istvuyut

между тяжелыми частицами - протонами и нейтронами и
%mezhdu tyazhelymi chastitsami - protonami i ne\u itronami i

являются ответственными за устойчивость атомного ядра.$^{\mbox{\large\stexttt{2/}}}$
%yavlyayut{s}ya otvet{s}tvennymi za usto\u ichivost\cprime\ atomnogo
%yadra.$^{\mbox{\large\stexttt{2/}}}$

По поводу этого последнего обобщения можно, правда, за-
%Po povodu \protect{\`{e}}togo poslednego obobshcheniya mozhno, pravda, za-

метить, что протонно-нейтронное взаимодействие может быть
%metit\cprime , chto protonno-ne\u itronnoe vzaimode\u istvie mozhet byt\cprime

выражено в терминах теории поля, вообще, повидимому,
%vyrazheno v terminakh teorii polya, voobshche, povidimomu,

уже только на квантовом языке. Мы не будем здесь остана-
%uzhe tol\cprime ko na kvantovom yazyke. My ne budem zdes\cprime\ ostana-

вливаться на анализе этого вопроса, поскольку он не имеет
%vlivat\cprime sya na analize \protect{\`{e}}togo voprosa, poskol\cprime ku
%on ne imeet

непосредственного отношения к нашей теме; укажем лишь,
%neposredstvennogo otnosheniya k nashe\u i teme; ukazhem lish\cprime ,

что сделанное замечание есть только лишнее напоминание
%chto sdelannoe zamechanie est\cprime\ tol\cprime ko  lishnee
%napominanie

о том основном факте, который и помимо него играет решаю-
%o tom osnovnom fakte, kotory\u i i pomimo nego igraet reshayu-

щую роль во всем разбираемом здесь круге вопросов, а
%shchuyu rol\cprime\ vo vsem razbiraemom zdes\cprime\ kruge voprosov, a

-----------------------------------

I/ См.\ хотя бы заметку \udensdash{Борна\hspace{-0.05cm}\ } и
\udensdash{Шредингера\hspace{-0.05cm}\ } в
%{\tt I}/ Sm.\ khotya by zametku \udensdash{Borna\hspace{-0.05cm}\ } i
%\udensdash{Shredingera\hspace{-0.05cm}\ } v 
{\ith Nature 1935.}

2/ См.напр.статью В.Гейзенберга в
%Sm.napr.stat\cprime yu V.Ge\u izenberga v}
{\ith Zeeman Festschrift},

\hspace{0.7cm}I935.%{\cyrtt\

\pagebreak
\refstepcounter{ppage}

именно о том, что истинными законами микромира являются
%imenno o tom, chto istinnymi zakonami mikromira yavlyayut{s}ya

все же не классические, а
\udensdash{к$\ \atop\ $\hspace{-0.33cm}вантовые\hspace{-0.05cm}\ }
законы. Поэтому вся-
%vse zhe ne klassicheskie, a 
%\udensdash{k$\ \atop\ $\hspace{-0.33cm}vantovye\hspace{-0.05cm}\ }
%zakony. Po\protect{\`{e}}tomu vsya-

кая теория, говорящая о поведении элементарных частиц ма-
%kaya teoriya, govoryashchaya o povedenii \protect{\`{e}}lementarnykh
%chastits ma-

терии и не учитывающая этих законов, может рассматривать-
%terii i ne uchityvayushchaya \protect{\`{e}}tikh zakonov, mozhet
%rassmatrivat\cprime -

ся, в лучшем случае, как нечто временное, т.е.\ как пере-
%sya, v luchshem sluchae, kak nechto vremennoe, t.e.\ kak pere-

ходный этап к другой, более правильной, теории.
%khodny\u i \protect{\`{e}}tap k drugo\u i, bolee pravil\cprime no\u i,
%teorii.

\hspace{1cm}Релятивистской квантовой электродинамики сегодня
%Relyativist{s}ko\u i kvantovo\u i \protect{\`{e}}lektrodinamiki
%segodnya

еще не существует. Известно, что основная, стоящая на
%eshche ne sushchestvuet. Izvestno, chto osnovnaya, stoyashchaya na

пути ее развития трудность - так называемая трудность
%puti ee razvitiya trudnost\cprime\ - tak nazyvaemaya trudnost\cprime\

"бесконечной собственной энергии"\ - представляет собою
%"beskonechno\u i sobstvenno\u i \protect{\`{e}}nergii" - predstavlyaet soboyu

непосредственную аналогию той трудности, которую до по-
%neposredstvennuyu analogiyu to\u i trudnosti, kotoruyu do po-

явления работ Борна не могла, сколько нибудь удовлетвори-
%yavleniya rabot Borna ne mogla, skol\cprime ko nibud\cprime\
%udovletvori-

тельно, преодолеть классическая физика, стремясь построить
%tel\cprime no, preodolet\cprime\ klassicheskaya fizika, stremyas\cprime\
%postroit\cprime\ 

унитарную теорию поля и материи. А {\ith priori} возможно
%unitarnuyu teoriyu polya i materii. A {\ith priori} vozmozhny

две точки зрения на тот способ, которым будет в дальней-
%dve tochki zreniya na tot sposob, kotorym budet v dal\cprime ne\u i-

схем разрешена эта трудность. Можно стоять на той точке
%shem razreshena \protect{\`{e}}ta trudnost\cprime . Mozhno stoyat\cprime\
%na to\u i tochke

зрения, что сперва должна быть построена правильня тео-
%zreniya, chto sperva dolzhna byt\cprime\ postroena pravil\cprime nya teo-

рия электромагнитной массы в рамках классических идей,
%riya \protect{\`{e}}lektromagnitno\u i massy v ramkakh klassicheskikh ide\u i,

перевод которой на квантовый язык, произведенный по рецеп-
%perevod kotoro\u i na kvantovy\u i yazyk, proizvedenny\u i po retsep-

там, аналогичными известным рецептам "квантования"\ и даст
%tam, analogichnymi izvestnym retseptam "kvantovaniya" i dast

истинную релятивистскую микро- электродинамику. С этой
%istinnuyu relyativist{s}kuyu mikro- \protect{\`{e}}lektrodinamiku. S 
%\protect{\`{e}}to\u i

точки зрения, исследование Борновских построений пред-
%tochki zreniya, issledovanie Bornovskikh postroeni\u i pred-

ставляет исключительный интерес для всего дальнейшего
%stavlyaet isklyuchitel\cprime ny\u i interes dlya vsego dal\cprime ne\u ishego

развития науки. Но можно стать и на ту точку зрения,
%razvitiya nauki. No mozhno stat\cprime\ i na tu tochku zreniya,

что путь аналогии с классикой, столь плодотворный и для все-
%chto put\cprime\ analogii s klassiko\u i, stol\cprime\ plodotvorny\u i
%dlya vse-

го предшествующего этапа квантовой теории, в основном
%go predshestvuyushego \protect{\`{e}}tapa kvantovo\u i teorii, v
%osnovnom

уже себя исчерпал и преоделение трудностей, стоящих на
%uzhe sebya ischerpal i preodelenie trudnoste\u i, stoyashchikh na

\pagebreak
\refstepcounter{ppage}

пути дальнейшего развития этой теории должно быть и бу-
%puti dal\cprime ne\u ishego razvitiya \protect{\`{e}}to\u i teorii
%dolzhno byt\cprime\ i bu-

дет получено на каких то принципиально новых путях. С
%det polucheno na kakikh to printsipial\cprime no novykh putyakh. S

этой точки зрения занятия 
теорией \udensdash{Борна\hspace{-0.05cm}\ } можно, с боль-
%\protect{\`{e}}to\u i tochki zreniya zanyatiya
%teorie\u i  \udensdash{Borna\hspace{-0.05cm}\ } mozhno, s bol\cprime -

шей или меньшей степенью решительности, считать напрас-
%she\u i ili men\cprime she\u i stepen\cprime yu 
%reshitel\cprime nosti, schitat\cprime\ napras-

ной потерей времени.
%no\u i potere\u i vremeni.

В настоящее время у нас нет однозначного критерия, ко-
%V nastoyashchee vremya u nas net odnoznachnogo kriteriya, ko-

торый позволил бы решить, какая из двух названных точек
%tory\u i pozvolil by reshit\cprime , kakaya iz dvukh nazvannykh tochek

зрения более близка к истине. Во всяком случае, есть все
%zreniya bolee blizka k istine. Vo vsyakom sluchae, est\cprime\ vse

основания сначала испробовать уже изведанный путь анало-
%osnovaniya snachala isprobovat\cprime\ uzhe izvedanny\u i
%put\cprime\ analo-

гии с классической физикой - это либо приведет к успеху,
%gii s klassichesko\u i fiziko\u i - \protect{\`{e}}to libo privedet
%k uspekhu,

ибо, на худой конец, поможет выяснить, в каком пункте
%libo, na khudo\u i konets, pomozhet vyyasnit\cprime , v kakom punkte 
 
этот путь перестает быть надежным.
%\protect{\`{e}}tot put\cprime\ perestaet byt\cprime\ nadezhnym.\\

\refstepcounter{rsubsection}
\label{glavaIpara3}
\hspace{1cm}\S\ 3. После этого небольшого оступления, возвращае-
%Posle \protect{\`{e}}togo nebol\cprime shogo
%ostupleniya, vozvrashchae-

мся снова непосредственно к нашей теме.
%msya snova neposredstvenno k nashe\u i teme.

\hspace{1cm}В
%V 
\S\ {\tt\ref{glavaIpara1}} мы набросали "идеальную"\ схему построения уни-
%my nabrosali "ideal\cprime nuyu" skhemu postroeniya uni-

тарной теории материи и поля. Согласно этой схеме, един-
%tarno\u i teorii materii i polya. Soglasno \protect{\`{e}}to\u i skheme, edin-

ственная свобода, которая остается в руках теоретика,
%stvennaya svoboda, kotoraya ostaet{s}ya v rukakh teoretika,

пытающегося построить унитарную теорию, заключается в
%pytayushchegosya postroit\cprime\ unitarnuyu teoriyu, zaklyuchaet{s}ya v

выборе лагранжиана поля. После того, как этот выбор сде-
%vybore lagranzhiana polya. Posle togo, kak \protect{\`{e}}tot
%vybor sde-

лан, все остальное идет уже более или менее автоматичес-
%lan, vse ostal\cprime noe idet uzhe bolee ili menee avtomatiches-

ки /если, для того, чтобы не очень удаляться от Максвел-
%ki /esli, dlya togo, chtoby ne ochen\cprime\ udalyat\cprime sya ot
%Maksvel-

ловой схемы, принять существование четырехмерного вектор
%lovo\u i skhemy, prinyat\cprime\ sushchestvovanie chetyrekhmernogo vektor

потенциала/.
%potentsiala/.

В связи с этим, естественно спросить себя: каким требо-
%V svyazi s \protect{\`{e}}tim, estestvenno sprosit\cprime\ sebya: kakim trebo-

ваниям должен удовлетворять лагранжиан теории для того,
%vaniyam dolzhen udovletvoryat\cprime\ lagranzhian teorii dlya togo,

чтобы набросанная выше программа могла быть фактически
%chtoby nabrosannaya vyshe programma mogla byt\cprime\ fakticheski

осуществлена. По сути дела, этот вопрос является первым,
%osushchestvlena. Po suti dela, \protect{\`{e}}tot vopros yavlyaet{s}ya pervym,

\pagebreak
\refstepcounter{ppage}
\label{page11}

на который нужно ответить при критическом изучении тео-
%na kotory\u i nuzhno otvetit\cprime\ pri kriticheskom izuchenii teo-

рии.
%rii.

Было -бы педантством стремится перечислить
\udensdash{в$\ \atop\ $\hspace{-0.33cm}се\hspace{-0.05cm}\ }
эти тре-
%Bylo -by pedant{s}tvom stremit\cprime sya perechislit\cprime\ 
%\udensdash{v$\ \atop\ $\hspace{-0.33cm}se\hspace{-0.05cm}\ }
%\protect{\`{e}}ti tre-

бования; полезнее остановиться главным образом на тех
%bovaniya; poleznee ostanovit\cprime sya glavnym obrazom na tekh

из них, которые действительно существенно с"уживают
%iz nikh, kotorye de\u istvitel\cprime no sushchestvenno s"uzhivayut

класс допустимых лагранжианов. Таких требований можно
%klass dopustimykh lagranzhianov. Takikh trebovani\u i mozhno

насчитать по меньшей мере
\udensdash{ш$\ \atop\ $\hspace{-0.33cm}есть\hspace{-0.05cm}\ }.
Мы будем нумеровать
%naschitat\cprime\ po men\cprime she\u i mere 
%\udensdash{sh$\ \atop\ $\hspace{-0.33cm}est\cprime\hspace{-0.05cm}\ }.
%My budem numerovat\cprime\ 

их латинскими буквами.
%ikh latinskimi bukvami.

\setcounter{subsubsection}{0}
\refstepcounter{subsubsection}
\label{suba}
\ \udensdash{\ \ /а/. Теория должна быть релятивистски 
инвариантной\hspace{-0.05cm}\ }
%\udensdash{\ \ /a/. Teoriya dolzhna byt\cprime\ relyativist{s}ki
%invariantno\u i\hspace{-0.05cm}\ }

\ \ \ 
\udensdash{и эйхинвариантной.\hspace{-0.05cm}\ }\\[-0.2cm]
%\udensdash{i \protect{\`{e}}\u ikhinvariantno\u i.\hspace{-0.05cm}\ }\\[-0.2cm]

\hspace{1cm}Первая половина этого требования не нуждается в ком-
%Pervaya polovina \protect{\`{e}}togo trebovaniya ne nuzhdaet{s}ya v kom-

ментариях. Вторая половина его означает, что четырехмер-
%mentariyakh. Vtoraya polovina ego oznachaet, chto chetyrekhmer-

ный вектор-потенциал должен играть в теории только чисто
%ny\u i vektor-potentsial dolzhen igrat\cprime\ v teorii tol\cprime ko chisto

вспомогательную роль. Существование этого потенциала
%vspomogatel\cprime nuyu rol\cprime . Sushchestvovanie  \protect{\`{e}}togo
%potentsiala

должно рассматриваться лишь как сокращенная запись некото-
%dolzhno rassmatrivat\cprime sya lish\cprime\ kak sokrashchennaya zapis\cprime\
%nekoto-

рых свойств, присущих электромагнитному
\udensdash{п$\ \atop\ $\hspace{-0.33cm}олю\hspace{-0.05cm}\ }/конкрет-
%rykh svo\u istv, prisushchikh  \protect{\`{e}}lektromagnitnomu 
%\udensdash{p$\ \atop\ $\hspace{-0.33cm}olyu\hspace{-0.05cm}\ }/konkret-

но - первой группы уравнений Максвелла/.\\
%no - pervo\u i gruppy uravneni\u i Maksvella/.\\

\setcounter{subsubsection}{2}  % Russian letter v
\refstepcounter{subsubsection}
\label{subb}
\ \udensdash{\ \ /в/ Уравнения поля должны допускать 
решения, упо-\hspace{-0.05cm}\ }
%\udensdash{\ \ /v/ Uravneniya polya dolzhny dopuskat\cprime\
%resheniya, upo-\hspace{-0.05cm}\ }

\udensdash{мянутые в \S\ {\tt\ref{glavaIpara1}},}
а именно также, когда магнитное поле
%\udensdash{myanutye v \S\ {\tt\ref{glavaIpara1}},} 
%a imenno takzhe, kogda magnitnoe pole 

равно нулю, электрическое радиально симметрично и
%ravno nulyu, \protect{\`{e}}lektricheskoe radial\cprime no 
%simmetrichno i

решение содержит одну произвольную константу, которую
%reshenie soderzhit odnu proizvol\cprime nuyu konstantu, kotoruyu

можно отождествить с зарядом. Без этого весьма труд-
%mozhno otozhdestvit\cprime\ s zaryadom. Bez
%\protect{\`{e}}togo ves\cprime ma trud-

но было бы ввести естественным образом в теорию за-
%no bylo by vvesti estestvennym obrazom v teoriyu za-

ряженные частицы.
%ryazhennye chastitsy.

\setcounter{subsubsection}{16}   % Russian letter s
\refstepcounter{subsubsection}
\label{subc}
\renewcommand{\thesubsubsection}{\alph{subsubsection}}
/с/ В той системе координат, в которой выполняется тре-
%/s/ V to\u i sisteme koordinat, v kotoro\u i vypolnyaet{s}ya tre-

бование /\ref{subb}/, \udensdash{полное количество движения должно 
равнять-\hspace{-0.05cm}\ }
%bovanie /\ref{subb}/, \udensdash{polnoe kolichestvo dvizheniya dolzhno
%ravnyat\cprime -\hspace{-0.05cm}\ }

\udensdash{ся нулю, а полная энергия должна быть 
конечной.\hspace{-0.05cm}\ }
%\udensdash{sya nulyu, a polnaya \protect{\`{e}}nergiya dolzhna byt\cprime\
%konechno\u i.\hspace{-0.05cm}\ }

Как известно, теория Максвелла не удовлетворяла второй
%Kak izvestno, teoriya Maksvella ne udovletvoryala vtoro\u i

части этого требования. Теории Абрагама-Лоренца,
%chasti \protect{\`{e}}togo trebovaniya. Teorii Abragama-Lorentsa,

\pagebreak
\refstepcounter{ppage}

базировавшиеся на искуственно вводимом представле-
%bazirovavshiesya na iskustvenno vvodimom predstavle-

нии о "радиусе электрона"\protect{,} удовлетворяли ему. Но они,
%nii o "radiuse \protect{\`{e}}lektrona", udovletvoryali emu. No oni,

со своей стороны были плохи тем, что не удовлетворяли
%so svoe\u i storony byli plokhi tem, chto ne udovletvoryali

следующему четвертому требованию, без выполнения кото-
%sleduyushchemu chetvertomu trebovaniyu, bez vypolneniya koto-

рого теория опять таки не может быть построена. Это
%rogo teoriya opyat\cprime\ taki ne mozhet byt\cprime\ postroena. 
%\protect{\`{E}}to

четвертое требование гласит:
%chetvertoe trebovanie glasit:

\setcounter{subsubsection}{3}  % Latin letter d
\refstepcounter{subsubsection}
\label{subd}
\udensdash{/{\ith d}/. Приравнивание 
электромагнитных величин меха-\hspace{-0.05cm}\ }\\[-0.2cm]
%\udensdash{/{\ith d}/. Priravnivanie 
%\protect{\`{e}}lektromagnitnykh velichin mekha-\hspace{-0.05cm}\ }\\[-0.2cm]

\udensdash{ническому количеству движения и энергии 
должно носить \hspace{-0.05cm}\ }\\[-0.2cm]
%\udensdash{nicheskomu kolichestvu dvizheniya i \protect{\`{e}}nergii
%dolzhno nosit\cprime\ \hspace{-0.05cm}\ }\\[-0.2cm]

\udensdash{релятивистски инвариантный 
характер.\hspace{-0.05cm}\ }\\[-0.2cm]
%\udensdash{relyativist{s}ki invariantny\u i 
%kharakter.\hspace{-0.05cm}\ }\\[-0.2cm]

Как мы увидим несколько ниже, это условие, необходи-
%Kak my uvidim neskol\cprime ko nizhe, \protect{\`{e}}to uslovie, neobkhodi-

мость которого очевидна, отнюдь не выполняется автома-
%most\cprime\ kotorogo ochevidna, otnyud\cprime\ ne vypolnyaet{s}ya avtoma-

тически. Нарушение его приводит к нарушению релятивист-
%ticheski. Narushenie ego privodit k narusheniyu relyativist-

ской инвариантности теории в ее механической части,
%sko\u i invariantnosti teorii v ee mekhanichesko\u i chasti,

что, как известно и произошло с теорией Лоренца.
%chto, kak izvestno i proizoshlo s teorie\u i Lorentsa.

Если эти четыре требования выполнены, то унитарная те-
%Esli \protect{\`{e}}ti chetyre trebovaniya vypolneny, to unitarnaya te-

ория принципиально, может быть построена. Но для того,
%oriya printsipial\cprime no, mozhet byt\cprime\ postroena. No dlya togo,

чтобы она имела физическое значение нужно, чтобы те
%chtoby ona imela fizicheskoe znachenie nuzhno, chtoby te

основные уравнения электродинамики и механики, к ко-
%osnovnye uravneniya \protect{\`{e}}lektrodinamiki i mekhaniki, k ko-

торым она приводит, во всяком случае, в известном пре-
%torym ona privodit, vo vsyakom sluchae, v izvestnom pre-

деле переходили соответственно в уравнения Максвелла-
%dele perekhodili sootvet{s}tvenno v uravneniya Maksvella-

Лоренца и Эйнштейна. Причем этот предел должен опре-
%Lorentsa i \protect{\`{E}}\u inshte\u ina. Prichem \protect{\`{e}}tot predel 
%dolzhen opre-

делятся конечно не квантовыми / о коих здесь речь
%delyat\cprime sya konechno ne kvantovymi / o koikh zdes\cprime\ rech\cprime

вообще не идет/, а какими то другими соображениями. Мы
%voobshche ne idet/, a kakimi to drugimi soobrazheniyami. My

должны, следовательно потребовать еще следующее:
%dolzhny, sledovatel\cprime no potrebovat\cprime\  eshche sleduyushchee:

\refstepcounter{subsubsection}
\label{sube}
/е/ Если, при построении теории, придется видоизменить
%/e/ Esli, pri postroenii teorii, pridet{s}ya vidoizmenit\cprime

\pagebreak
\refstepcounter{ppage}

уравнения Максвелла, то эти уравнения все-же должны
%uravneniya Maksvella, to \protect{\`{e}}ti uravneniya vse-zhe dolzhny

остаться справедливыми в известном предельном случае,
%ostat\cprime sya spravedlivymi v izvestnom predel\cprime nom sluchae,

т.е.\ \udensdash{классическая электродинамика должна так или 
иначе\hspace{-0.05cm}\ }
%t.e.\ \udensdash{klassicheskaya \protect{\`{e}}lektrodinamika dolzhna tak ili 
%inache\hspace{-0.05cm}\ }

\udensdash{содержатся в новой теории.\ \hspace{-0.05cm}\ }\\[-0.3cm]
%\udensdash{soderzhat\cprime sya v novo\u i 
%teorii.\ \hspace{-0.05cm}\ }\\[-0.3cm]
 
\refstepcounter{subsubsection}
\label{subf}
/{\ith f}/. Аналогичное условие должно иметь место и по
%Analogichnoe uslovie dolzhno imet\cprime\ mesto i po

отношению к новым уравнениям
\udensdash{ме$\ \atop\ $\hspace{-0.33cm}ханики:\hspace{-0.05cm}\ }
в определенном
%otnosheniyu k novym uravneniyam 
%\udensdash{me$\ \atop\ $\hspace{-0.33cm}khaniki:\hspace{-0.05cm}\ }
%v opredelennom 

пределе они должны переходить в уравнения
\udensdash{релятивистс-\hspace{-0.05cm}\ }
%predele oni dolzhny perekhodit\cprime\ v uravneniya 
%\udensdash{relyativist{s}-\hspace{-0.05cm}\ }

\udensdash{кой ме$\ \atop\ $\hspace{-0.33cm}ханики 
Эйнштейна\hspace{-0.05cm}\ }\\[0.3cm]
%\udensdash{ko\u i me$\ \atop\ $\hspace{-0.33cm}khaniki 
%\protect{\`{E}}\u inshte\u ina.\hspace{-0.05cm}\ }\\[0.3cm]

\refstepcounter{rsubsection}
\label{glavaIpara4}
\hspace{1cm}\S\ 4. \udensdash{Борну\hspace{-0.05cm}\ } и
\udensdash{Инфельду\hspace{-0.05cm}\ } удалось построить Лагранже-
%\udensdash{Bornu\hspace{-0.05cm}\ } i
%\udensdash{Infel\cprime du\hspace{-0.05cm}\ } udalos\cprime\
%postroit\cprime\ Lagranzhe-

ву функцию, удовлетворяющую всем перечисленным выше
%vu funktsiyu, udovletvoryayushchuyu vsem perechislennym vyshe

шести требованиям.
%shesti trebovaniyam.

Постараемся кратко воспроизвести ход их мысли. Вариаци-
%Postaraemsya kratko vosproizvesti khod ikh mysli. Variatsi-

онный принцип, служащий, как всегда, исходным пунктом
%onny\u i printsip, sluzhashchi\u i, kak vsegda, iskhodnym punktom

теории, имеет вид
%teorii, imeet vid
{\handcolor
\begin{eqnarray}
\refstepcounter{requation}
\label{eq1}
\ \hspace{-4cm}\delta\int L\ d\tau &=&0\ ,\\[0.3cm]
\ \hspace{-6cm}\textnormal{\ith где
%\cyrit{gde}
}
\hspace{2cm}d\tau &=&dx^1 dx^2 dx^3 dx^4  
\nonumber
\end{eqnarray}
}Для того, чтобы удовлетворить первой половине требо-
%Dlya togo, chtoby udovletvorit\cprime\ pervo\u i polovine trebo-

вания /\ref{suba}/, величина $\handcolor L\ d\tau$ должна быть инвариантом
%vaniya /\ref{suba}/, velichina $\handcolor L\ d\tau$ dolzhna byt\cprime\
%invariantom

в смысле общей теории относительности. Мы должны,
%v smysle obshche\u i teorii otnositel\cprime nosti. My dolzhny,

следовательно, иметь
%sledovatel\cprime no, imet\cprime
{\handcolor
\begin{eqnarray}
\refstepcounter{requation}
\label{eq2}
\ \hspace{-5cm}L\ d\tau &=&Inv  
\end{eqnarray}
}

\hspace{1cm}Как должно преобразовываться $\handcolor L$ для того, чтобы
%Kak dolzhno preobrazovyvat\cprime sya $\handcolor L$ dlya togo, chtoby

это имело место{\ith ?} Хорошо известный ответ на этот вопрос
%\protect{\`{e}}to imelo mesto{\ith ?} {Kh}orosho izvestny\u i
%otvet na \protect{\`{e}}tot vopros

\pagebreak
\refstepcounter{ppage}

гласит: $\handcolor L$ должно преобразовываться как выражение вида
%glasit:  $\handcolor L$ dolzhno  preobrazovyvat\cprime sya kak 
%vyrazhenie vida
{\handcolor
\begin{eqnarray}
\label{eq3}
\refstepcounter{requation}
\sqrt{\vert a_{ik}\vert}&&  
\end{eqnarray}
}где $\handcolor\vert a_{ik}\vert$- детерминант, образованный из ковариантных
%gde $\handcolor\vert a_{ik}\vert$- determinant, obrazovanny\u i iz 
%kovariantnykh 

компонент любого тензора 2-го ранга.
%komponent lyubogo tenzora 2-go ranga.

\hspace{1cm}По основной идее теории /см.\ \S\ {\tt\ref{glavaIpara1}}/, 
Лагранжева функ-
%Po osnovno\u i idee teorii /sm.\ \S\ {\tt\ref{glavaIpara1}}/,
%Lagranzheva funk-

ция должна зависить от компонент тензора поля
%tsiya dolzhna zavisit\cprime\ ot komponent tenzora polya 
$\handcolor f_{ik}$\ \ %}
$^{\mbox{\stexttt{I/}}}$

и, очевидно, поскольку теория строится в рамках общего
%{\cyrtt i, ochevidno, poskol\cprime ku teoriya stroit{s}ya v ramkakh obshchego

принципа относительности от компонент метрического тен-
%printsipa otnositel\cprime nosti ot komponent metricheskogo ten-

зора $\handcolor g_{ik}$. Принимая это во внимание, естественно искать
%zora $\handcolor g_{ik}$. Prinimaya \protect{\`{e}}to vo vnimanie, 
%estestvenno iskat\cprime

$\handcolor L$ среди выражений типа
%$\handcolor L$ sredi vyrazheni\u i tipa
{\handcolor
\begin{eqnarray}
\refstepcounter{requation}
\label{eq4}
\sum_\alpha\ A_\alpha\ 
\sqrt{\sum_\beta B_{\alpha\beta}\ \vert a_{ik}^{\ (\alpha\beta)}\vert}\ \ ,
\end{eqnarray}
}где $\handcolor A_\alpha$ и $\handcolor B_{\alpha\beta}$ - инварианты, 
а $\handcolor a_{ik}^{\ (\alpha\beta)}$ - ком-
%gde $\handcolor A_\alpha$ i $\handcolor B_{\alpha\beta}$ - invarianty, 
%a $\handcolor a_{ik}^{\ (\alpha\beta)}$
%- kom-

поненты тензоров, как-то зависящие от $\handcolor f_{ik}$ и
%ponenty tenzorov, kak-to zavisyashchie ot $\handcolor f_{ik}$ i 
$\handcolor g_{ik}$.

\udensdash{Борн\hspace{-0.05cm}\ } и
\udensdash{Инфельд\ \hspace{-0.05cm}\ }указывают два 
выражения типа /\ref{eq4}/, которые
%\udensdash{Born\hspace{-0.05cm}\ } i
%\udensdash{Infel\cprime d\ \hspace{-0.05cm}\ }ukazyvayut dva
%vyrazheniya tipa /\ref{eq4}/, kotorye

будучи положены в основу теории, удовлетворяют всем,
%buduchi polozheny v osnovu teorii, udovletvoryayut vsem,

выставленным выше шести требованиям, а именно:
%vystavlennym vyshe shesti trebovaniyam, a imenno:

------------------------------

I/ Другие величины, характеризующие электромагнитное по-
%{\tt I}/ Drugie velichiny, kharakterizuyushchie 
%\protect{\`{e}}lektromagnitnoe po-

ле, напр.\ слагающие вектора-потенциала, не могут вхо-
%le, napr.\ slagayushchie vektora-potentsiala, ne mogut vkho-

дить из за требования эйхинвариантности, которое,
%dit\cprime\ iz za trebovaniya \protect{\`{e}}\u ikhinvariantnosti, kotoroe,

таким образом, в этой теории оказывается автоматичес-
%takim obrazom, v \protect{\`{e}}to\u i teorii okazyvaet\cprime sya
%avtomatiches-

ки выполненным.
%ki vypolnennym.

\pagebreak
\refstepcounter{ppage}

{\handcolor
\begin{eqnarray}
\refstepcounter{requation}
\label{eq5}
L_1&=&\sqrt{-\vert g_{ik} + f_{ik} \vert}\ -\ 
\sqrt{-\vert g_{ik} \vert}\\[0.3cm]
\refstepcounter{requation}
\label{eq6}
L_2&=&\sqrt{-\vert g_{ik} + f_{ik} \vert + \vert f_{ik} \vert }\ -\ 
\sqrt{-\vert g_{ik} \vert}
\end{eqnarray}
}или
%ili
{\handcolor
\begin{eqnarray}
\refstepcounter{requation}
\label{eq7}
L_1&=&\sqrt{-\vert g_{ik} \vert}\ \left(\sqrt{1 + F^2 - G^2} - 1\right)\\[0.3cm]
\refstepcounter{requation}
\label{eq8}
L_2&=&\sqrt{-\vert g_{ik} \vert}\ \left(\sqrt{1 + F^2} - 1\right)\ \ ,
\end{eqnarray}
}где
%gde
{\handcolor
\begin{eqnarray}
\refstepcounter{requation}
\label{eq9}
F&=&\frac{1}{2}\ f_{ik} f^{ik}\\[0.3cm]
\refstepcounter{requation}
\label{eq10}
G&=&\left(f_{23}\, f_{14} + f_{31}\, f_{24} + f_{12}\, f_{34}\right)
\frac{1}{\sqrt{-\vert g_{ik} \vert}}
\end{eqnarray}
}и преставляет собою пространственные инварианты.
%i prestavlyaet soboyu prostranstvennye invarianty.

\hspace{1cm}$\handcolor\bigl[$ В этих формулах поля 
выражены, конечно, в условных
%V \protect{\`{e}}tikh formulakh polya 
%vyrazheny, konechno, v uslovnykh 

единицах. При переходе к обычным единицам, мы должны
%edinitsakh. Pri perekhode k obychnym edinitsam, my dolzhny

ввести в формулу фактор "$\handcolor b$"\protect{,} имеющий размер-
%vvesti v formuly faktor "$\handcolor b$", imeyushchi\u i razmer-

ность напряженности поля. Тогда, напр., формула /\ref{eq8}/
%nost\cprime\ napryazhennosti polya. Togda, napr., formula /\ref{eq8}/

примет вид
%primet vid
{\handcolor
\begin{eqnarray}
\refstepcounter{requation}
\label{eq11}
L_2&=&\sqrt{-\vert g_{ik} \vert}\ 
\left(\sqrt{1 + \frac{f_{ik} f^{ik}}{2 b^2}} - 1\right)
\end{eqnarray}
}Аналогичным образом может быть переписана и формула /\ref{eq7}/
%Analogichnym obrazom mozhet byt\cprime\ perepisana i formula /\ref{eq7}/

В дальнейшем оказывается, что константа "$\handcolor b$"\ играет в те-
%V dal\cprime ne\u ishem okazyvaet{s}ya, chto konstanta "$\handcolor b$" 
%igraet v te-

ории роль "критического"\ поля, в известном смысле анало-
%orii rol\cprime\ "kriticheskogo" polya, v izvestnom smysle analo-

гично тому, как скорость света в теории относительности
%gichno tomu, kak skorost\cprime\ sveta v teorii otnositel\cprime nosti

играет роль критической скорости. Величина "$\handcolor b$"\ оказы-
%igraet rol\cprime\ kritichesko\u i skorosti. Velichina "$\handcolor b$" okazy-

\pagebreak
\refstepcounter{ppage}

вается весьма большой - порядка
%vaet{s}ya ves\cprime ma bol\cprime sho\u i - poryadka}
I0$^{\mbox{\large\stexttt{I6}}}$%{\cyrtt
абс.\ {\ith ед} $\handcolor \bigr]$
%abs.\ {\ith\cyrit ed} $\handcolor \bigr]$

Легко видеть, что для полей, малых по сравнению с "кри-
%Legko videt\cprime , chto dlya pole\u i, malykh po sravneniyu s "kri-

тическим{\ith "\ }полем $\handcolor b$, обе Лагранжевы функции теории Борна
%ticheskim{\ith\cyrit "} polem $\handcolor b$, obe Lagranzhevy funktsii 
%teorii Borna

переходят в Лагранжеву функцию Максвелла, т.е.\ требова-
%perekhodyat v Lagranzhevu funktsiyu Maksvella, t.e.\ trebova-

ние /\ref{sube}/ оказывается выполненным. Напомним, что в теории
%nie /\ref{sube}/ okazyvaet{s}ya vypolnennym. Napomnim, chto v teorii

Максвелла, Лагранжиан имеет вид:
%Maksvella, Lagranzhian imeet vid:
{\handcolor
\begin{eqnarray}
\refstepcounter{requation}
\label{eq12}
&&\frac{1}{4}\ \sqrt{-\vert g_{ik} \vert}\ f_{ik} f^{ik}
\end{eqnarray}
}который тоже подходит под общую формулу /\ref{eq4}/.
%kotory\u i tozhe podkhodit pod obshchuyu formulu /\ref{eq4}/.

Принципиальное отличие этой формулы
/\ref{eq12}/ от /\ref{eq5}/ и /\ref{eq6}/
%Printsipial\cprime noe otlichie \protect{\`{e}}to\u i formuly
%/\ref{eq12}/ ot /\ref{eq5}/ i /\ref{eq6}/

заключается в том, что в /\ref{eq12}/ у Максвелла под корнем
%zaklyuchaet{s}ya v tom, chto v /\ref{eq12}/ u Maksvella pod kornem

остается только метрический тензор, тогда как компоненты
%ostaet{s}ya tol\cprime ko metricheski\u i tenzor, togda kak 
%komponenty

тензора поля входят лишь в один из инвариантов
%tenzora polya vkhodyat lish\cprime\ v odin iz invariantov 
$\handcolor A_\alpha$.

В теории же Борна, как это видно из формул /\ref{eq5}/ и /\ref{eq6}/,
%V teorii zhe Borna, kak \protect{\`{e}}to vidno iz formul
%/\ref{eq5}/ i /\ref{eq6}/,

величины, характеризующие и поле и метрику стоят под кор-
%velichiny, kharakterizuyushchie i pole i metriku stoyat pod kor-

нем, а инварианты $\handcolor A_\alpha$ и $\handcolor B_{\alpha\beta}$
оказываются простыми чис-
%nem, a invarianty $\handcolor A_\alpha$ i $\handcolor B_{\alpha\beta}$ 
%okazyvayut{s}ya prostymi chis-

лами. Как выяснится из дальнейшего, этот факт существен-
%lami. Kak vyyasnit{s}ya iz dal\cprime ne\u ishego, \protect{\`{e}}tot
%fakt sushchestven-

но влияет на вид новых уравнений поля, делая их, в отли-
%no vliyaet na vid novykh uravneni\u i polya, delaya ikh, v otli-

чие от уравнений Максвелла, нелинейными в полях.
%chie ot uravneni\u i Maksvella, neline\u inymi v polyakh.

\hspace{1cm}В какой мере выбор функции /\ref{eq5}/ и /\ref{eq6}/
является одно-
%V kako\u i mere vybor funktsii /\ref{eq5}/ i /\ref{eq6}/
%yavlyaet{s}ya odno-

значным и как вообще можно ставить вопрос об однозначно-
%znachnym i kak voobshche mozhno stavit\cprime\ vopros ob 
%odnoznachno-

сти в данном случае{\ith ?}
%sti v dannom sluchae{\ith ?}

Совершенно очевидно, что одного требования релятивистской
%Sovershenno ochevidno, chto odnogo trebovaniya relyativist{s}ko\u i

инвариантности интеграла действия далеко не достаточно
%invariantnosti integrala de\u istviya daleko ne dostatochno

для того, чтобы однозначно выбрать даже тип Лагранжиана$^{\mbox{\large\stexttt{I/}}}$
%dlya togo, chtoby odnoznachno vybrat\cprime\ dazhe tip
%Lagranzhiana}$^{\mbox{\large\stexttt{I/}}}$

-----------------------

\hspace{-0.8cm}
%{\cyrtt 
{\ith 1)}\ Замечание Борна о том, что обычный выбор Лагранжиана в
%Zamechanie Borna o tom, chto obychny\u i vybor Lagranzhiana v

виде
%vide 
$\handcolor Inv \cdot \sqrt{-\vert g_{ik} \vert}$
/где
%/gde 
$\handcolor Inv$ 
зависит только {\ith от}
%zavisit tol\cprime ko {\ith\cyrit ot} 
$\handcolor f_{ik}$/ 
является
%yavlyaet{s}ya

"формальным"\protect{,} вряд ли является существенным, так как сам
%"formal\cprime nym", vryad li yavlyaet{s}ya sushchestvennym, tak kak sam

критерий "формальности"\ весьма неясен. Любопытно, что одно
%kriteri\u i "formal\cprime nosti" ves\cprime ma neyasen. Lyubopytno, chto odno

из рекомендуемых самым Борном выражений - а именно выра-
%iz rekomenduemykh samim Bornom vyrazheni\u i - a imenno vyra-

жение /\ref{eq6}/ - тоже принадлежит, как это особенно ясно видно
%zhenie /\ref{eq6}/ - tozhe prinadlezhit, kak \protect{\`{e}}to
%osobenno yasno vidno

из /\ref{eq8}/, к критикуемому им типу.
%iz /\ref{eq8}/, k kritikuemomu im tipu.

\pagebreak
\refstepcounter{ppage}
\label{page17}

Но если прибавить к нему еще требование предельного
%No esli pribavit\cprime\ k nemu eshche trebovanie predel\cprime nogo

перехода к теории Максвелла, 
то\hspace{-0.1cm}\textasciiacute\hspace{0.1cm} это уже значительно
%perekhoda k teorii Maksvella, 
%to\hspace{-0.1cm}\textasciiacute\hspace{0.1cm}\protect{\`{e}}to 
%uzhe znachitel\cprime no

с"узит класс допустимых Лагранжианов.
%s"uzit klass dopustimykh Lagranzhianov.

Полной однозначности выбора все же и это, конечно, не
%Polno\u i odnoznachnosti vybora vse zhe i \protect{\`{e}}to, konechno, ne

дает: достаточно указать хотя бы на существование трех
%daet: dostatochno ukazat\cprime\ khotya by na sushchestvovanie trekh

функций /\ref{eq7}/,  /\ref{eq8}/ и /\ref{eq12}/.
%funktsi\u i  /\ref{eq7}/,  /\ref{eq8}/ i  /\ref{eq12}/.

Интересно было бы выяснить, достаточно-ли добавить к
%Interesno bylo by vyyasnit\cprime, dostatochno-li dobavit\cprime\ k

этим двум требованиям еще требование конечности энер-
%\protect{\`{e}}tim dvum trebovaniyam eshche trebovanie konechnosti
%\protect{\`{e}}ner-

гии частицы, для того, чтобы определился хотя бы общий
%gii chastitsy, dlya togo, chtoby opredelilsya khotya by obshchi\u i

тип Лагранжевых функций. Более узко этот вопрос можно
%tip Lagranzhevykh funktsi\u i. Bolee uzko \protect{\`{e}}tot 
%vopros mozhno 

поставить так: существуют-ли такие Лагранжианы, кото-
%postavit\cprime\ tak: sushchestvuyut-li takie Lagranzhiany, koto-

рые удовлетворяют всем шести требованиям выставленным
%rye udovletvoryayut vsem shesti trebovaniyam vystavlennym

в \S\ {\tt\ref{glavaIpara3}} и, вместе с тем, существенно отличаются от выра-
%v \S\ {\tt\ref{glavaIpara3}} i, vmeste s tem, sushchestvenno 
%otlichayut{s}ya ot vyra-

жений /\ref{eq5}/ и /\ref{eq6}/.$^{\displaystyle\handcolor\ast)}$
Заметим, что эта последняя неоднознач-
%zheni\u i /\ref{eq5}/ i /\ref{eq6}/.$^{\displaystyle\handcolor\ast)}$
%Zametim, chto \protect{\`{e}}ta poslednaya neodnoznach-

\vspace{-0.825cm}

\hspace{8.cm}\ {\handcolor\underline{\ \hspace{6.cm}\ }}

\vspace{0.1925cm}

ность не является опасной, так как выражения
/\ref{eq5}/ и /\ref{eq6}/
%nost\cprime\ ne yavlyaet{s}ya opasno\u i, tak kak vyrazheniya  
%/\ref{eq5}/ i /\ref{eq6}/

\vspace{-0.825cm}

{\handcolor\underline{\ \hspace{8.4cm}\ }}

\vspace{0.1925cm}

по существу однотипны и в целом ряде случаев приводят,
%po sushchestvu odnotipny i v tselom ryade sluchaev privodyat,

даже к\ \  полностью одинаковым результатам.
%dazhe k\ \  polnost\cprime yu odinakovym rezul\cprime tatam.

Именно, так обстоит дело в основной статической задаче{\ith .}
%Imenno, tak obstoit delo v osnovno\u i statichesko\u i zadache{\ith .}

\ \ {\ith Т}ак-же во всех случаях малых полей, которые только и
%{\ith\cyrit{T}}ak-zhe vo vsekh sluchayakh malykh pole\u i, kotorye 
%tol\cprime ko i

\hspace{-0.06cm} 
{\ith различие между этими результатами, по\underline{видимому несущественно}}
%{\ith\cyrit razlichie mezhdu \protect{\`{e}}timi rezul\cprime tatami,
%po\underline{vidimomu nesushchestvenno}}

\vspace{-0.09cm}

будут интересовать нас во второй 
главе.\hspace{-0.2cm}${{\handcolor /}\atop }$\ Только для силь-
%budut interesovat\cprime\ nas vo vtoro\u i 
%glave.\hspace{-0.2cm}${{\handcolor /}\atop }$\ Tol\cprime ko dlya sil\cprime-

ных полей /вблизи заряженных частиц/ результаты, к
%nykh pole\u i /vblizi zaryazhennykh chastits/ rezul\cprime taty, k

которым приводят нас оба Лагранжиана существенно раз-
%kotorym privodyat nas oba Lagranzhiana sushchestvenno raz-

личны и вопрос о том какую из этих функций нужно в
%lichny i vopros o tom kakuyu iz \protect{\`{e}}tikh funktsi\u i nuzhno v

этой области предпочесть может быть решен лишь пу-
%\protect{\`{e}}to\u i oblasti predpochest\cprime\ mozhet byt\cprime\ 
%reshen lish\cprime\ pu-

тем сравнения выводов теории с опытными фактами.
%tem sravneniya vyvodov teorii s opytnymi faktami.

В дальнейшем, мы будем пользоваться исключительно
%V dal\cprime ne\u ishem, my budem pol\cprime zovat{s}ya 
%isklyuchitel\cprime no

функцией /\ref{eq8}/.
%funktsie\u i /\ref{eq8}/.

\vspace{1.5cm}

{\handcolor\underline{\ \hspace{10cm}\ }}

$\handcolor *)$\ {\ith См.\ прим.\ в конце.}
%{\ith\cyrit Sm.\ prim.\ v kontse.}

\pagebreak
\refstepcounter{ppage}
\refstepcounter{rsubsection}
\label{glavaIpara5}
\hspace{1cm}\S\ 5.\ Итак, Лагранжева функция выбрана.
%Itak, Lagranzheva funktsiya vybrana.

В декартовой системе координат специальной теории отно-
%V dekartovo\u i sisteme koordinat spetsial\cprime no\u i teorii otno-

сительности она имеет вид:
%sitel\cprime nosti ona imeet vid:
{\handcolor
\begin{eqnarray}
\refstepcounter{requation}
\label{eq13}
L&=&\sqrt{1\, +\, F}\ -\ 1
\end{eqnarray}
}Дальнейшее развитие теории происходит по "идеальной"
%Dalne\u ishee razvitie teorii proiskhodit po "ideal\cprime no\u i"

схеме, изложенной в \S\ {\tt\ref{glavaIpara1}}.
%skheme, izlozhenno\u i v \S\ {\tt\ref{glavaIpara1}}.

Постулируется существование четырехмерного вектор- по-
%Postuliruet{s}ya sushchestvovanie chetyrekhmernogo vektor- po-

тенциал{\ith а} $\handcolor\varphi_i$ удовлетворящего условию:
%tentsial{\ith\cyrit a} $\handcolor\varphi_i$ udovletvoryashchego usloviyu:
{\handcolor
\begin{eqnarray}
\refstepcounter{requation}
\label{eq14}
f_{ik}&=&\frac{\partial\varphi_k}{\partial x^i}\ -\ 
 \frac{\partial\varphi_i}{\partial x^k}\ \ \ ,
\end{eqnarray}
}что эквивалентно установлению первой системы уравнений
%chto \protect{\`{e}}kvivalentno ustanovleniyu pervo\u i sistemy uravneni\u i

поля:
%polya:
{\handcolor
\begin{eqnarray}
\refstepcounter{requation}
\label{eq15}
\frac{\partial f_{ik}}{\partial x^l}\ +\
\frac{\partial f_{kl}}{\partial x^i}\ +\
\frac{\partial f_{li}}{\partial x^k}&=&0
\end{eqnarray}
}Уравнения Эйлера для вариационного принципа
%Uravneniya \protect{\`{E}}\u ilera dlya variatsionnogo printsipa
$\handcolor\delta \int L\, d\tau = 0$

дают вторую группу уравнений поля:
%dayut vtoruyu gruppu uravneni\u i polya:
{\handcolor
\begin{eqnarray}
\refstepcounter{requation}
\label{eq16}
\frac{\partial\ }{\partial x^k}\, \frac{\partial L}{\partial f_{ik}}&=&0\ \ \ ,
\end{eqnarray}
}которые, как было уже указано, нелинейны в полях в силу
%kotorye, kak bylo uzhe ukazano, neline\u iny v polyakh v silu

того, что $\handcolor L$ 
\udensdash{\ не$\ \atop\ $\hspace{-0.33cm} является\hspace{-0.05cm}\ }
вадратичной функцией полей,
%togo, chto $\handcolor L$ 
%\udensdash{\ ne$\ \atop\ $\hspace{-0.33cm} yavlyaet{s}ya\hspace{-0.05cm}\ }
%kvadratichno\u i funktsie\u i pole\u i,

как это имеет место в теории Максвелла. Вводя тензор
%kak \protect{\`{e}}to imeet mesto v teorii Maksvella. Vvodya tenzor

индукции
%induktsii
{\handcolor
\begin{eqnarray}
\refstepcounter{requation}
\label{eq17}
P_{ik}&=&\frac{\partial L}{\partial f_{ik}}\ \ \ ,
\end{eqnarray}
}мы можем переписать вторую группу уравнений поля в
%my mozhem perepisat\cprime\ vtoruyu gruppu uravneni\u i polya v

виде
%vide
{\handcolor
\begin{eqnarray}
\refstepcounter{requation}
\label{eq18}
\frac{\partial P^{ik}}{\partial x^k}&=&0\ \ \ ,
\end{eqnarray}
}

\pagebreak
\refstepcounter{ppage}

Совершенно таким же методом, как в классической электро-
%Sovershenno takim zhe metodom, kak v klassichesko\u i \protect{\`{e}}lektro-

динамике Максвелла - Ленарда, можно, исходя из уравнений
%dinamike Maksvella - Lenarda, mozhno, iskhodya iz uravneni\u i

поля /\ref{eq15}/ и /\ref{eq16}/, построить тензор энергии - импульса
%polya /\ref{eq15}/ i /\ref{eq16}/, postroit\cprime\ tenzor
%\protect{\`{e}}nergii - impul\cprime sa

$\handcolor T_{ik}$, 
и получить законы сохранения:
%i poluchit\cprime\ zakony sokhraneniya:
{\handcolor
\begin{eqnarray}
\refstepcounter{requation}
\label{eq19}
\frac{\partial T_i^{\, l}}{\partial x^l}&=&0\ \ \ ,
\end{eqnarray}
}где
%gde
{\handcolor
\begin{eqnarray}
\refstepcounter{requation}
\label{eq20}
T_i^{\, l}&=&\delta_i^{\, l} L\ -\ f_{ik} P^{lk}
\end{eqnarray}
}Получим теперь уравнения поля и законы сохранения в
%Poluchim teper\cprime\ uravneniya polya i zakony sokhraneniya v

векторной форме. Обозначим через В, Е, $\handcolor D$ и 
Н$^{\displaystyle\handcolor\ \ast)}$ простран-
%vektorno\u i forme. Oboznachim cherez V, E, $\handcolor D$ i 
%N$^{\displaystyle\handcolor\ \ast)}$prostran-

ственные векторы, характеризирующие электромагнитное поле
%stvennye vektory, kharakteriziruyushchie \protect{\`{e}}lektromagnitnoe pole

в обычных Хевисайдовых единицах. Именно, положим:
%v obychnykh Khevisa\u idovykh edinitsakh. Imenno, polozhim:
{\handcolor
\begin{eqnarray}
\refstepcounter{requation}
\label{eq21}
&\left.\begin{array}{lllcc}
f_{23}\ ,&f_{31}\ ,&f_{12}&\ \ \longrightarrow\ \ &B\\[0.2cm]
f_{14}\ ,&f_{24}\ ,&f_{34}&\ \ \longrightarrow\ \ &E\\[0.2cm]
P_{23}\ ,&P_{31}\ ,&P_{12}&\ \ \longrightarrow\ \ &H\\[0.2cm]
P_{14}\ ,&P_{24}\ ,&P_{34}&\ \ \longrightarrow\ \ &D
\end{array}\right\}&
\end{eqnarray}
}Тогда
%Togda
{\handcolor
\begin{eqnarray*}
L&=&\sqrt{1 + F}\ -\ 1\hspace{0.5cm},
\end{eqnarray*}
}\hspace{1cm}где
%gde
{\handcolor
\begin{eqnarray}
\refstepcounter{requation}
\label{eq22}
F&=&\frac{1}{b^2}\left(B^2 - E^2\right).
\end{eqnarray}
}Далее
%Dalee
{\handcolor
\begin{eqnarray}
\refstepcounter{requation}
\label{eq23}
&\left.\begin{array}{rcccl}
H&=&b^2\ \frac{\displaystyle\partial L}{\displaystyle\partial B}
&=&\frac{\displaystyle B}{\displaystyle\sqrt{1+F}}\\[0.8cm]
D&=&-\ b^2\ \frac{\displaystyle\partial L}{\displaystyle\partial E}
&=&\frac{\displaystyle E}{\displaystyle\sqrt{1+F}}
\end{array}\right\}&
\end{eqnarray}
}

\hspace{-1cm}{\handcolor\underline{\ \hspace{10cm}\ }}

\hspace{-1cm}$\handcolor *)$\ 
{\ith Для векторных величин здесь и ниже не
%\cyrit Dlya vektornykh velichin zdes\cprime\ i nizhe ne

\hspace{-0.3cm}вводится специальных обозначений.}
%vvodit{s}ya spetsial\cprime nykh oboznacheni\u i.}

\pagebreak
\refstepcounter{ppage}

и уравнения поля принимают вид уравнений Макс-
%i uravneniya polya prinimayut vid uravneni\u i Maks-

велла для среды, но без зарядов и токов:
%vella dlya sredy, no bez zaryadov i tokov:
{\handcolor
\begin{eqnarray}
\refstepcounter{requation}
\label{eq24}
&\left.\begin{array}{rcccccl}
rot\, E + \frac{\displaystyle 1}{\displaystyle c}\;\dot{B}&=&0&
\hspace{3cm}&div\, B&=&0\\[0.3cm]
rot\, H - \frac{\displaystyle 1}{\displaystyle c}\;\dot{D}&=&0&
\hspace{3cm}&div\, D&=&0
\end{array}\right\}&
\end{eqnarray}
}или, если в последние два уравнения подставить вместо
%ili, esli v poslednie dva uravneniya podstavit\cprime\ vmesto

Н и $\handcolor D$ их значение /\ref{eq23}/ через В и Е, то, окончательно,
%N i $\handcolor D$ ikh znachenie /\ref{eq23}/ cherez V i E, to, 
%okonchatel\cprime no,

система уравнений поля будет иметь вид:
%sistema uravneni\u i polya budet imet\cprime\ vid:
{\handcolor
\begin{eqnarray}
\refstepcounter{requation}
\label{eq25}
&\left.\begin{array}{l}
rot\, E + \frac{\displaystyle 1}{\displaystyle c}\;\dot{B}\ =\ 0\\[0.3cm]
div\, B\ =\ 0\\[0.3cm]
rot\, B - \frac{\displaystyle 1}{\displaystyle c}\;\dot{E}\ =\ 
\frac{\displaystyle 1}{\displaystyle 2}\
\frac{\displaystyle 1}{\displaystyle 1 + F}\ \left\{\left[ grad\, F, B\right]\ 
-\, \frac{\displaystyle 1}{\displaystyle c}\;\dot{F} E\right\}\\[0.6cm]
div\, E\ =\ \frac{\displaystyle 1}{\displaystyle 2}\ 
\frac{\displaystyle 1}{\displaystyle 1 + F}\ 
\left( grad\, F, E\right)
\end{array}\right\}&
\end{eqnarray}
}Для тензора энергии и импульса мы получаем таблицу
%Dlya tenzora \protect{\`{e}}nergii i impul\cprime sa my 
%poluchaem tablitsy 

компонент
%komponent
{\handcolor
\begin{eqnarray}
\refstepcounter{requation}
\label{eq26}
T^{il}&=&\left[\begin{array}{cccc}
X_x&X_y&X_z&c\, G_x\\[0.3cm]
Y_x&Y_y&Y_z&c\, G_y\\[0.3cm]
Z_x&Z_y&Z_z&c\, G_z\\[0.3cm]
\frac{\displaystyle 1}{\displaystyle c}\, 
S_x&\frac{\displaystyle 1}{\displaystyle c}\, 
S_y&\frac{\displaystyle 1}{\displaystyle c}\, S_z&U
\end{array}\right]\ \ \ ,
\end{eqnarray}
}где
%gde
{\handcolor
\begin{eqnarray}
\refstepcounter{requation}
\label{eq27}
&\left.\begin{array}{rcl}
X_x&=&H_y\, B_y\ +\ H_z\, B_z\ -\ D_x\, E_x\ -\ b^2\, L\\[0.3cm]
Y_x&=&X_y\ =\ -\  H_y\, B_x\ -\ D_x\, E_y\\[0.3cm]
\frac{\displaystyle 1}{\displaystyle c}\, S_x&=&c\, 
G_z\ = D_y\, B_z\ -\ D_z\, B_y\\[0.3cm]
U&=&D\, E\ +\ b^2\, L
\end{array}\right\}&
\end{eqnarray}
}

\pagebreak
\refstepcounter{ppage}

{\handcolor
{\it При этом законы сохранения принимают вид:}
%\cyrit Pri \protect{\`{e}}tom zakony sokhraneniya
%prinimayut vid:}
%
\begin{eqnarray}
\refstepcounter{requation}
\label{eq28}
&\left.\begin{array}{rcl}
div\, X&=&-\ \frac{\displaystyle 1}{\displaystyle c^2}\ 
\frac{\displaystyle\partial S_x}{\displaystyle\partial t}\\[0.3cm]
.\ \ .\ &.&\ .\ \ .\ \ .\ \ .\\[0.3cm]
.\ \ .\ &.&\ .\ \ .\ \ .\ \ .\\[0.3cm]
div\, S&=&-\ \frac{\displaystyle\partial U}{\displaystyle\partial t}
\end{array}\right\}&
\end{eqnarray}
}Как подчеркивает Борн, за независимые переменные мож-
%Kak podcherkivaet Born, za nezavisimye peremennye mozh-

но выбрать, конечно любую пару из четырех векторов по-
%no vybrat\cprime , konechno lyubuyu paru iz chetyrekh vektorov po-

ля Е,В,Н и $\handcolor D$. В зависимости от этого выбора мы можем
%lya E,V,N i $\handcolor D$. V zavisimosti ot \protect{\`{e}}togo 
%vybora my mozhem

получать внешне различные формулировки теории с раз-
%poluchat\cprime\ vneshne razlichnye formulirovki teorii s raz-

личными Лагранжианами, но, разумеется, приводящие к
%lichnymi Lagranzhianami, no, razumeet{s}ya, privodyashchie k

одним и тем же, по существу, уравнениям поля. Очевидно,
%odnim i tem zhe, po sushchestvu, uravneniyam polya. Ochevidno,

возможны 4 таких формулировки. Наиболее удобной для нас
%vozmozhny 4 takikh formulirovki. Naibolee udobno\u i dlya nas

в дальнейшем формулировкой теории будет та, которую
%v dal\cprime ne\u ishem formulirovko\u i teorii budet ta, kotoruyu

мы до сих пор и рассматривали /т.е., при которой за
%my do sikh por i rassmatrivali /t.e., pri kotoro\u i za

независимые векторы выбраны Е и В/, хотя и при любом
%nezavisimye vektory vybrany E i V/, khotya i pri lyubom

другом выборе вычисления усложнились бы лишь очень
%drugom vybore vychisleniya uslozhnilis\cprime\ by lish\cprime\ ochen\cprime\

незначительно.
%neznachitel\cprime no.

\hspace{1cm}Отметим, что для построения квантовой теории оказы-
%Otmetim, chto dlya postroeniya kvantovo\u i teorii okazy-

вается более удобным выбрать за независимые векторы
%vaet{s}ya bolee udobnym vybrat\cprime\ za nezavisimye vektory

$\handcolor D$ и В. При этом роль Лагранжевой функции играет плоть-
%$\handcolor D$ i V. Pri \protect{\`{e}}tom rol\cprime\ Lagranzhevo\u i 
%funktsii igraet plot-

ность энергии $\handcolor U$, равная по последней формуле /\ref{eq27}/:
%nost\cprime\ \protect{\`{e}}nergii $\handcolor U$, ravnaya po posledne\u i
%formule /\ref{eq27}/:
{\handcolor
\begin{eqnarray}
\refstepcounter{requation}
\label{eq29}
U&=&b^2\ \left(\sqrt{\left(1 + \frac{D^2}{b^2}\right)
\left(1 + \frac{B^2}{b^2}\right)}\ -\ 1\right)
\end{eqnarray}
}

\refstepcounter{rsubsection}
\label{glavaIpara6}
\hspace{1cm}\S\ 6. Найдем, по указанному в \S\ {\tt\ref{glavaIpara1}} рецепту, 
решение
%Na\u idem, po ukazannomu v \S\ {\tt\ref{glavaIpara1}} retseptu, reshenie

уравнений поля, соответствующее наличию покоющейся
%uravneni\u i polya, sootvet{s}tvuyushchee nalichiyu
%pokoyushche\u isya

заряженной частицы в начале координат.
%zaryazhenno\u i chastitsy v nachale koordinat.

Согласно сказанному в \S\ {\tt\ref{glavaIpara1}},\ мы при этом должны, прежде
%Soglasno skazannomu v \S\ {\tt\ref{glavaIpara1}},\ my pri \protect{\`{e}}tom 
%dolzhny, prezhde 

всего, положить
%vsego, polozhit\cprime
%
\begin{eqnarray*}
{\handcolor B}&=\ H\ =&{\handcolor 0}
\end{eqnarray*}

\pagebreak
\refstepcounter{ppage}
\label{page22}

В силу /\ref{eq24}/, это дает
%V silu /\ref{eq24}/, \protect{\`{e}}to daet
{\handcolor
\begin{eqnarray}
\refstepcounter{requation}
\label{eq30}
rot\, E&=&0\hspace{2cm};\ \ \ \ div\, D\ =\ 0 
\end{eqnarray}
}причем $\handcolor D$ /а, следовательно, и Е/ не зависит от времени.
%prichem $\handcolor D$ /a, sledovatel\cprime no, i E/
%ne zavisit ot vremeni.

Ищем радиально симметричное решение этих уравнений,
%Ishchem radial\cprime no simmetrichnoe reshenie \protect{\`{e}}tikh 
%uravneni\u i,

т.е.\ такое решение, для которого оба вектора Е и $\handcolor D$ ра-
%t.e.\ takoe reshenie, dlya kotorogo oba vektora E i $\handcolor D$ ra-

диальны по направлению и по величине зависят только
%dial\cprime ny po napravleniyu i po velichine zavisyat tol\cprime ko

от $\handcolor r$.
%ot $\handcolor r$.

Второе из уравнений /\ref{eq30}/ в сферических координатах дает
%Vtoroe iz uravneni\u i /\ref{eq30}/ v sfericheskikh koordinatakh daet
{\handcolor
\begin{eqnarray}
\refstepcounter{requation}
\label{eq31}
\frac{1}{r^2}\ \frac{d\ }{d r}\left(r^2\, D_r\right)&=&0 
\end{eqnarray}
}Откуда совершенно однозначно получаем
%Otkuda sovershenno odnoznachno poluchaem
{\handcolor
\begin{eqnarray}
\refstepcounter{requation}
\label{eq32}
D_r&=&\frac{e}{r^2}
\end{eqnarray}
}где е - единственная константа интеграции, отождествля-
%gde e - edinstvennaya konstanta integratsii, otozhdestvlya-

емая с зарядом. Это решение имеет особую точку при $\handcolor r=$0.
%emaya s zaryadom. \protect{\`{E}}to reshenie imeet osobuyu tochku 
%pri $\handcolor r=$0.

Описываемое им состояние поля мы и интерпретируем, как
%Opisyvaemoe im sostoyanie polya my i interpretiruem, kak

соответствующее наличию частицы в начале координат.
%sootvet{s}tvuyushchee nalichiyu chastitsy v nachale koordinat.

\hspace{1cm}Весь ход рассуждений показывает нам, что условие
%Ves\cprime\ khod rassuzhdeni\u i pokazyvaet nam, chto uslovie

/\ref{subb}/ в теории Борна, как и у Максвелла, выполнено.
%/\ref{subb}/ v teorii Borna, kak i u Maksvella, vypolneno.

Вычисляя поле Е, находим
%Vychislyaya pole E, nakhodim
{\handcolor
\begin{eqnarray}
\refstepcounter{requation}
\label{eq33}
E_r&=&\frac{e}{\sqrt{r^4 + r_0^4}}
\end{eqnarray}
}где
%gde
{\handcolor
\begin{eqnarray}
\refstepcounter{requation}
\label{eq34}
r_0&=&\sqrt{\frac{e}{b}}
\end{eqnarray}
}и представляет собою константу, имеющую размерность
%i predstavlyaet soboyu konstantu, imeyushchuyu razmernost\cprime

длины, которая, как ниже выясниться, будет в известной
%dliny, kotoraya, kak nizhe vyyasnit\cprime sya, budet v izvestno\u i

мере аналогична
$^{\handcolor /}$\hspace{-0.25cm}"радиусу$^{\handcolor /}$\hspace{-0.25cm}"
электрона в теории Лоренца.
%mere analogichna 
%$^{\handcolor /}$\hspace{-0.25cm}"radiusu$^{\handcolor /}$\hspace{-0.25cm}"
%\protect{\`{e}}lektrona v teorii Lorentsa.

\pagebreak
\refstepcounter{ppage}

{\ith Из формулы (\ref{eq33}) мы видим, что поле $\handcolor E$ не имеет особых}
%\cyrit Iz formuly (\ref{eq33}) my vidim, chto pole 
%$\handcolor E$ ne imeet osobykh}

точек. Однако сам по себе факт конечности Е еще отнюдь
%tochek. Odnako sam po sebe fakt konechnosti E eshche otnyud\cprime

не означает, что в теории нет расходимостей. Для того,
%ne oznachaet, chto v teorii net raskhodimoste\u i. Dlya togo,

чтобы проверить, удовлетворяется ли для нашего решения
%chtoby proverit\cprime, udovletvoryaet{s}ya li dlya nashego resheniya

условие /\ref{subc}/, необходимо обратиться к непосредственным
%uslovie /\ref{subc}/, neobkhodimo obratit\cprime sya k neposredstvennym

вычислениям по формулам /\ \ref{eq27}/
%vychisleniyam po formulam /\ \ref{eq27}/

Легко видеть, и здесь мы имеем крупный принципиальный
%Legko videt\cprime , i zdes\cprime\ my imeem krupny\u i 
%printsipial\cprime ny\u i 

успех теории, что в данном случае это условие
\udensdash{удовлет-\ \hspace{-0.05cm}\ }
%uspekh teorii, chto v dannom sluchae \protect{\`{e}}to uslovie
%\udensdash{udovlet-\ \hspace{-0.05cm}\ }

\udensdash{воряется.\hspace{-0.05cm}\ }\ В самом деле, импульс 
в нашем случае обращает-
%\udensdash{voryaet{s}ya.\hspace{-0.05cm}\ }\ V samom dele, impul\cprime s
%v nashem sluchae obrashchaet-

ся в нуль в силу того, что В $\handcolor =$ 0; конечность же полной
%sya v nul\cprime\ v silu togo, chto V $\handcolor =$ 0; konechnost\cprime\
%zhe polno\u i

энергии $\handcolor W = \int U dV$ подтверждается непосредственными вы-
%\protect{\`{e}}nergii $\handcolor W = \int U dV$ podverzhdaet{s}ya
%neposredstvennymi vy-

числениями, дающими для $\handcolor W$ значение:
%chisleniyami, dayushchimi dlya $\handcolor W$ znachenie:
%
\begin{eqnarray}
\refstepcounter{requation}
\label{eq35}
{\handcolor W}&=&{\tt I,2361}\ {\handcolor\frac{e^2}{r_0}}
\end{eqnarray}
Этот пункт является, как сказано, весьма существенным.
%\protect{\`{E}}tot punkt yavlyaet{s}ya, kak skazano, ves\cprime ma
%sushchestvennym.

Он показывает, что Борну удалось, оставаясь в рамках
%On pokazyvaet, chto Bornu udalos\cprime , ostavayas\cprime\ v ramkakh

классических идей устранить трудность бесконечной соб-
%klassicheskikh ide\u i ustranit\cprime\ trudnost\cprime\ beskonechno\u i
%sob-

ственной энергии электрона, не прибегая при этом, ни к
%stvenno\u i \protect{\`{e}}nergii \protect{\`{e}}lektrona, ne pribegaya
%pri \protect{\`{e}}tom, ni k

каким искусственным представлениям о его "структур{\ith е}"
%kakim iskusstvennym predstavleniyam o ego  "struktur{\ith\cyrit e}".

Зная заряд и массу электрона, мы можем найти численное
%Znaya zaryad i massu \protect{\`{e}}lektrona, my mozhem na\u iti
%chislennoe 

значение константы $\handcolor r_0$:
%znachenie konstanty $\handcolor r_0$: 
{\handcolor
\begin{eqnarray}
\refstepcounter{requation}
\label{eq36}
r_0&=&1,2361\ \frac{e^2}{m_0 c^2}\ =\ 2,28 \cdot 10^{-13}\ \textnormal{\ith см.}
\end{eqnarray}
}и, следовательно, численное значение "критического"
%i, sledovatel\cprime no, chislennoe znachenie "kriticheskogo"

поля "$\handcolor b$":
%polya "$\handcolor b$":
{\handcolor
\begin{eqnarray}
\refstepcounter{requation}
\label{eq37}
b&=&\frac{e}{r_0^2}\ =\ 9,18 \cdot 10^{15}\ \textnormal{\ith абс.\ ед.}
\end{eqnarray}
}Из этих формул видно, что мы можем, в терминологии те-
%Iz \protect{\`{e}}tikh formul vidno, chto my mozhem, v terminologii te-

ории Лоренца, величину $\handcolor r_0$ интерпретировать как "радиус"
%orii Lorentsa, velichinu $\handcolor r_0$ interpretirovat\cprime\ 
%kak "radius"

\pagebreak
\refstepcounter{ppage}

электрона, а "критическое"\ поле "$\handcolor b$"\ - как напряженность
%\protect{\`{e}}lektrona, a "kriticheskoe" pole "$\handcolor b$" - kak 
%napryazhennost\cprime\

электрического поля на его "поверхности". Величина этого
%\protect{\`{e}}lektricheskogo polya na ego "poverkhnosti". Velichina
%\protect{\`{e}}togo

"критического"\ поля, очевидно, является универсальной кон-
%"kriticheskogo" polya, ochevidno, yavlyaet{s}ya universal\cprime no\u i kon-

стантой теории, независящей от заряда е, который, как
%stanto\u i teorii, nezavisyashche\u i ot zaryada e, kotory\u i, kak

уже упоминалось, никак теорией не фиксируется. Следова-
%uzhe upominalos\cprime , nikak teorie\u i ne fiksiruet{s}ya. Sledova-

тельно, при выбранном е, мы получаем
\udensdash{вполне определенное\hspace{-0.05cm}\ }
%tel\cprime no, pri vybrannom e, my poluchaem 
%\udensdash{vpolne opredelennoe\hspace{-0.05cm}\ }

$\handcolor r_0$ и, в силу /\ref{eq36}/,
\udensdash{вполне определенную массу $\handcolor m_0$\hspace{-0.05cm}\ }.
Это
%$\handcolor r_0$ i, v silu /\ref{eq36}/, 
%\udensdash{vpolne opredelennuyu massu $\handcolor m_0$\hspace{-0.05cm}\ }. 
%\protect{\`{E}}to

замечание убеждает нас в том, что при
\udensdash{данном\hspace{-0.05cm}\ } заряде е,
%zamechanie ubezhdaet nas v tom, chto pri 
%\udensdash{dannom\hspace{-0.05cm}\ } zaryade e,

теория Борна приводит только к \udensdash{одному\hspace{-0.05cm}\ }
значению для массы
%teoriya Borna privodit tol\cprime ko k \udensdash{odnomu\hspace{-0.05cm}\ }
%znacheniyu dlya massy

$\handcolor m_0$, т.е.\ является теорией только одного сорта частиц-
%$\handcolor m_0$, t.e.\ yavlyaet{s}ya teorie\u i tol\cprime ko odnogo sorta 
%chastits-

электронов. Получить теорию протона, как видно отсюда, в
%\protect{\`{e}}lektronov. Poluchit\cprime\ teoriyu protona, kak vidno
%ot{s}yuda, v

рамках вышеизложенных соображений, оказывается невозмож-
%ramkakh vysheizlozhennykh soobrazheni\u i, okazyvaet{s}ya nevozmozh-

ним.\\[-0.2cm]
%nym.\\[-0.2cm]

\refstepcounter{rsubsection}
\label{glavaIpara7}
\hspace{1cm}\S\ 7. Рассмотрим теперь несколько подробнее вопрос,
%Rassmotrim teper\cprime\ neskol\cprime ko podrobnee vopros,

затронуты при формулировке требования /{\it\ref{subd}}/, а именно
%zatronuty\u i pri formulirovke trebovaniya /{\it\ref{subd}}/, a imenno

вопрос о релятивистски ивариантном приравнивании электро-
%vopros o relyativist{s}ki ivariantnom priravnivanii \protect{\`{e}}lektro-

магнитных величин механическим.
%magnitnykh velichin mekhanicheskim.

Как мы уже выяснили, для построения механики в унитарной
%Kak my uzhe vyyasnili, dlya postroeniya mekhaniki v unitarno\u i

теории, мы должны отождествить интегралы по об"ему от
%teorii, my dolzhny otozhdestvit\cprime\ integraly po ob"emu ot

электромагнитного количества движения т.е., от
%\protect{\`{e}}lektromagnitnogo kolichestva dvizheniya t.e., ot
$\handcolor c\, G_x,$

$\handcolor c\, G_y,$\ {\ith и}\ $\handcolor c\, G_z,$ для указанного типа 
состояний поля с механи-
%$\handcolor c\, G_y,\ {\it\cyrit{i}}\ c\, G_z,$ dlya ukazannogo tipa 
%sostoyani\u i polya s mekhani-

ческим количеством движения частицы и интеграл от плотно-
%cheskim kolichestvom dvizheniya chastitsy i integral ot plotno-

сти энергии $\handcolor U$ с полной энергией частицы. Для того, чтобы
%sti \protect{\`{e}}nergii $\handcolor U$ s polno\u i 
%\protect{\`{e}}nergie\u i chastitsy. Dlya togo, chtoby

такое отождествление было релятивистски инвариантно, нуж-
%takoe otozhdestvlenie bylo relyativist{s}ki invariantno, nuzh-

но, чтобы эти интегралы, при преобразовании Лоренца, пре-
%no, chtoby \protect{\`{e}}ti integraly, pri preobrazovanii Lorentsa, pre-

образовались как компоненты четырехмерного вектора. Легко
%obrazovalis\cprime\ kak komponenty chetyrekhmernogo vektora. Legko

убедиться в том, что необходимым условием этого является
%ubedit\cprime sya v tom, chto neobkhodimym usloviem \protect{\`{e}}togo 
%yavlyaet{s}ya

равенство нулю, в покоющейся относительно частицы системе
%ravenstvo nulyu, v pokoyushche\u isya otnositel\cprime no chastitsy sisteme

\pagebreak
\refstepcounter{ppage}
\label{page25}

координат, об"емных интегралов от диагональных компо-
%koordinat, ob"emnykh integralov ot diagonal\cprime nykh kompo-

нент тензора Максвелловых напряжений $\handcolor X_x,$ $
\handcolor Y_y$ и $\handcolor Z_z,$
%nent tenzora Maksvellovykh napryazheni\u i $\handcolor X_x,$ $
%\handcolor Y_y$ i $\handcolor Z_z,$

т.е.\ условие:
%t.e.\ uslovie:
{\handcolor
\begin{eqnarray}
\refstepcounter{requation}
\label{eq38}
\int\, X_x\, dV&=&\int\, Y_y\, dV\ =\ \int\, Z_z\, dV\ =\ 0
\end{eqnarray}
}В теории \udensdash{Абрагама\hspace{-0.05cm}\ } -
\udensdash{Лоренца\hspace{-0.05cm}\ } это условие
\udensdash{н$\ \atop\ $\hspace{-0.33cm}е\hspace{-0.05cm}\ } выполнялось
%V teorii \udensdash{Abragama\hspace{-0.05cm}\ } - 
%\udensdash{Lorentsa\hspace{-0.05cm}\ } \protect{\`{e}}to uslovie
%\udensdash{n$\ \atop\ $\hspace{-0.33cm}e\hspace{-0.05cm}\ } 
%vypolnyalos\cprime\ i

в этом, как раз, заключалась вышеупомянутая трудность по-
%v \protect{\`{e}}tom, kak raz, zaklyuchalas\cprime\ 
%vysheupomyanutaya trudnost\cprime\ po-

строения релятивистски инвариантной механики. С этим-же
%stroeniya relyativist{s}ki invariantno\u i mekhaniki. S \protect{\`{e}}tim-zhe 

была связана необходимость введения сил неэлектрического
%byla svyazana neobkhodimost\cprime\ vvedeniya sil 
%ne\protect{\`{e}}lektricheskogo

происхождения, сдерживающих электрон, т.к.\ результирующие
%proiskhozhdeniya, sderzhivayushchikh \protect{\`{e}}lektron, t.k.\
%rezul\cprime tiruyushchie

Максвелловы напряжения были отличны от нуля $^{\mbox{\large\stexttt{I/}}}$.
%Maksvellovy napryazheniya byli otlichny ot 
%nulya }$^{\mbox{\large\stexttt{I/}}}$.

\hspace{1cm}Теория Борна - Инфельда свободна от этой трудности.
%{\cyrtt Teoriya Borna - Infel\cprime da
% svobodna ot \protect{\`{e}}to\u i trudnosti.

Непосредственная подстановка решения /\ref{eq32}/-/\ref{eq38}/ в
/\ref{eq27}/по-
%Neposredstvennaya podstanovka resheniya /\ref{eq32}/-/\ref{eq38}/ v 
%/\ref{eq27}/po-

\vspace{-0.32cm}

\hspace{7.675cm}{\ith этого}
%\cyrit{\protect{\`{e}}togo}}

\vspace{-0.6cm}

\hspace{8.075cm}{\handcolor\underline{\ \hspace{0.6cm}\ }}

\vspace{-0.32cm}

казывает, что условие /\ref{eq38}/ для
\hspace{-0.2cm}{\ith\small /}\hspace{-0.2cm} решения
\udensdash{выпол$\ \atop\ $\hspace{-0.33cm}няется\hspace{-0.05cm}\ } и,
%kazyvaet, chto uslovie /\ref{eq38}/ dlya
%\hspace{-0.2cm}{\ith\small /}\hspace{-0.2cm} resheniya 
%\udensdash{vypol$\ \atop\ $\hspace{-0.33cm}nyaet{s}ya\hspace{-0.05cm}\ } i,

следовательно, мы можем построить релятивистски инвариант-
%sledovatel\cprime no, my mozhem postroit\cprime\ relyativist{s}ki invariant-

ную механику. Иначе говоря, здесь нет необходимости введе-
%nuyu mekhaniku. Inache govorya, zdes\cprime\ net neobkhodimosti vvede-

ния сил, содерживающих электрон, т.к.\ результирующие Мак-
%niya sil, soderzhivayushchikh \protect{\`{e}}lektron, t.k.\
%rezul\cprime tiruyushchie Mak-

свелловы напряжения обращаются в нуль. Таким образом, в
%svellovy napryazheniya obrashchayut{s}ya v nul\cprime . Takim obrazom, v

электродинамике Борна удовлетворяется и последнее требова-
%\protect{\`{e}}lektrodinamike Borna udovletvoryaet{s}ya i 
%poslednee trebova-

ние пред"являемое к унитарной теории, на котором показа-
%nie pred"yavlyaemoe k unitarno\u i teorii, na kotorom pokaza-

ла свою несостоятельность лучшая из прежних теорий -
%la svoyu nesostoyatel\cprime nost\cprime\ i luchshaya iz prezhnykh
%teori\u i -

теория Лоренца.\\[-0.2cm]
%teoriya Lorentsa.\\[-0.2cm]

\refstepcounter{rsubsection}
\label{glavaIpara8}
\hspace{1cm}\S\ 8. Наметить тепер путь, при помощи которого можно
%Nametim teper\cprime\ put\cprime , pri pomoshchi
%kotorogo mozhno

получать в теории Борна - Инфельда
\udensdash{за$\ \atop\ $\hspace{-0.33cm}коны\hspace{-0.05cm}\ }
механики. Для
%poluchat\cprime\ v teorii Borna - Infel\cprime da 
%\udensdash{za$\ \atop\ $\hspace{-0.33cm}kony\hspace{-0.05cm}\ } 
%mekhaniki. Dlya

этого мы очевидно должны рассмотреть вопрос о том,
%\protect{\`{e}}togo my ochevidno dolzhny rassmotret\cprime\ 
%vopros o tom,

{\ith каким образом будут вести себя решения уравнений поля}
%\cyrit kakim obrazom budut vesti sebya resheniya uravneni\u i polya}

{\handcolor
\hspace{4cm}\underline{\hspace{0.7cm}}
\hspace{0.7cm}\underline{\hspace{0.7cm}}
\hspace{0.7cm}\underline{\hspace{0.7cm}}
\hspace{0.7cm}\underline{\hspace{0.7cm}}
\hspace{0.7cm}\underline{\hspace{0.7cm}}
\hspace{0.7cm}\underline{\hspace{0.7cm}}
\hspace{0.7cm}\underline{\hspace{0.7cm}}
}

\vspace{0.5cm}

\underline{\ \hspace{5cm}\ }
{\rm\small written on the backside of the manuscript leaf /

написано на обратной стороне листки рукописи}
%{\cyrrm napisano na obratno\u i storone listki rukopisi}}
\underline{\ \hspace{5cm}\ }

\vspace{0.5cm}

{\ith 1) Интегралы от недиагональных компонент тензора
%\cyrit 1) Integraly ot nediagonal\cprime nykh komponent tenzora

\hspace{0.55cm}Максвелловых напряжений обращаются в нуль
%Maksvellovykh napryazheni\u i obrashchayut{s}ya v nul\cprime\  

\hspace{0.55cm}и в теории Лоренца, и в теории Борна.}
%i v teorii Lorentsa, i v teorii Borna.}

\pagebreak
\refstepcounter{ppage}
\label{page26}

интерпретируемые нами как наличие электрона, под влия-
%interpretiruemye nami kak nalichie \protect{\`{e}}lektrona, pod vliya-

нием внешнего поля.
%niem vneshnego polya.

Мы знаем, что уравнения Борна допускают решения, соответ-
%My znaem, chto uravneniya Borna dopuskayut resheniya, sootvet-

ствующие \udensdash{покоющемуся\hspace{-0.05cm}\ } электрону.
%stvuyushchie \udensdash{pokoyushchemusya\hspace{-0.05cm}\ }
%\protect{\`{e}}lektronu.

Совершая преобразование Лоренца над координатной систе-
%Sovershaya preobrazovanie Lorentsa nad koordinatno\u i siste-

мой, мы можем получить решения, соответствующие прямоли-
%mo\u i, my mozhem poluchit\cprime\ resheniya, sootvet{s}tvuyushchie 
%pryamoli-

нено и равномерно движущемуся электрону. Правда, для ме-
%neno i ravnomerno dvizhushchemusya \protect{\`{e}}lektronu. Pravda,
%dlya me-

ханики желательно было бы знать и такие решения, которые
%khaniki zhelatel\cprime no bylo by znat\cprime\ i takie resheniya,
%kotorye

соответствуют  \udensdash{ускоренно\hspace{-0.05cm}\ }
движущемуся электрону, ибо имен-
%sootvet{s}tvuyut \udensdash{uskorenno\hspace{-0.05cm}\ }
%dvizhushchemusya \protect{\`{e}}lektronu, ibo imen-

но такое движение будет иметь электрон под влиянием внеш-
%no takoe dvizhenie budet imet\cprime\  \protect{\`{e}}lektron pod
%vliyaniem vnesh-

них электромагнитных полей. Принципиально, можно было бы
%nikh \protect{\`{e}}lektromagnitnykh pole\u i. Printsipial\cprime no,
%mozhno bylo by

построить теорию возмущений, беря решение, соответствую-
%postroit\cprime\ teoriyu vozmushcheni\u i, berya reshenie, sootvet{s}tvuyu-

щее равномерному и прямолинейному движению за нулевое при-
%shchee ravnomernomu i pryamoline\u inomu dvizheniyu za nulevoe pri-

ближение и рассматривая влияние ускорения, как малое воз-
%blizhenie i rassmatrivaya vliyanie uskoreniya, kak maloe voz-

мущение. Но хорошо известно, что учет ускорения, вообще,
%mushchenie. No khorosho izvestno, chto uchet uskoreniya, voobshche,

важен только при рассмотрении эффекта реакции собственного
%vazhen tol\cprime ko pri rassmotrenii \protect{\`{e}}ffekta reaktsii
%sobstvennogo

излучения на электрон. Если этим эффектом не интересовать-
%izlucheniya na \protect{\`{e}}lektron. Esli \protect{\`{e}}tim
%\protect{\`{e}}ffektom ne interesovat\cprime -

ся, то вполне достаточно рассматривать решение, соответ-
%sya, to vpolne dostatochno rassmatrivat\cprime\ reshenie, sootvet-

ствующее ускорению равному нулю, что мы и будем делать.
%stvuyushchee uskoreniyu ravnomu nulyu, chto my i budem delat\cprime .

Исходным пунктом теории должны, очевидно, служить такие
%Iskhodnym punktom teorii dolzhny, ochevidno, sluzhit\cprime\ takie

решения уравнений Борна, которые соответствуют одновре-
%resheniya uravneni\u i Borna, kotorye sootvet{s}tvuyut odnovre-

менному существованию электрона и внешнего поля. Правда,
%mennomu sushchestvovaniyu \protect{\`{e}}lektrona i vneshnego polya. Pravda,

\vspace{-0.32cm}

\hspace{12cm}{\ith тракто}
%\cyrit trakto}

\vspace{-0.32cm}

мы выше видели, что в унитарной теории можно практико-
%my vyshe videli, chto v unitarno\u i teorii mozhno praktiko-

\vspace{-0.8cm}

\hspace{11.7cm}{\handcolor\underline{\ \hspace{1.7cm}\ }}

\vspace{0.19cm}

вать как наличие частицы только некоторые, вполне опре-
%vat\cprime\ kak nalichie chastitsy tol\cprime ko nekotorye, vpolne opre-

деленные решения, не предусматривающие присутствия еще
%delennye resheniya, ne predusmatrivayushchie prisut{s}tviya eshche

внешнего поля. Это последнее как-то изменит наше реше-
%vneshnego polya. \protect{\`{E}}to poslednee kak-to izmenit nashe reshe-

не /\ref{eq32}/ и  /\ref{eq33}/   / или полученное из него 
преобразование{\ith м}
%nie /\ref{eq32}/ i  /\ref{eq33}/   / ili poluchennoe iz nego
%preobrazovanie{\ith\cyrit{m}}

Лоренца/ и {\ith a priori} неясно можно ли будет вообще при
%Lorentsa/ i {\ith a priori} neyasno mozhno li budet voobshche pri

этом продолжать говорить о наличии частицы. Иначе гово-
%\protect{\`{e}}tom prodolzhat\cprime\ govorit\cprime\ o nalichii chastitsy.
%Inache govo-

ря, в теории Борна принципиально мыслим такой случай,
%rya, v teorii Borna printsipial\cprime no myslim tako\u i slucha\u i,

когда под влиянием поля электрон теряет как бы свою
%kogda pod vliyaniem polya \protect{\`{e}}lektron teryaet kak by svoyu

\pagebreak
\refstepcounter{ppage}

индивидуальность, совсем грубо говоря, как бы"разрушается".
%individual\cprime nost\cprime, sovsem grubo govorya, kak 
%by"razrushaet{s}ya".

Физически, однако, ясно, что такое положение вещей может
%Fizicheski, odnako, yasno, chto takoe polozhenie veshche\u i mozhet

иметь место только при очень сильных полях/сравнимых с $\handcolor b$/,
%imet\cprime\ mesto tol\cprime ko pri ochen\cprime\ sil\cprime nykh
%polyakh/sravnimykh s $\handcolor b$/,

или - одно,конечно, связано с другим - при больших ускоре-
%ili - odno,konechno, svyazano s drugim - pri bol\cprime shikh uskore-

ниях электрона. При малых же полях и малых ускорениях,вли-
%niyakh \protect{\`{e}}lektrona. Pri malykh zhe polyakh i malykh 
%uskoreniyakh,vli-

яние поля в основном,очевидно, может быть$\;$описано в классы-
%yanie polya v osnovnom,ochevidno, mozhet byt\cprime$\;$opisano v klassi-

ческих терминах, т.е.\ просто как "\ приведение электрона в
%cheskikh terminakh, t.e.\ prosto kak " privedenie 
%\protect{\`{e}}lektrona v

движение". В математической форме, указанное осложняющее
%dvizhenie". V matematichesko\u i forme, ukazannoe oslozhnyayushchee

обстоятельство выражается так. Пусть 
$\handcolor D_0\ \textnormal{\it и}\ B_0$\ $^{\displaystyle\handcolor 1)}$ 
есть реше-
%obstoyatel\cprime stvo vyrazhaet{s}ya tak. Pust\cprime\
%$\handcolor D_0\ {\it\cyrit i}\ B_0$\ $^{\displaystyle\handcolor 1)}$ 
%est\cprime\ reshe-

ние уравнений Борна, соответствующее наличию электрона,
%nie uravneni\u i Borna, sootvet{s}tvuyushchee nalichiyu
%\protect{\`{e}}lektrona,

движущегося равномерно и прямолинейно по какой то {\ith мировой}
%dvizhushchegosya ravnomerno i pryamoline\u ino po kako\u i to
%{\ith\cyrit mirovo\u i}

траектории{\ith /}величины, характеризующие вид этой траектории
%traektorii{\ith /}velichiny, kharakterizuyushchie vid \protect{\`{e}}to\u i
%traektorii

играют в нижеследующих рассуждениях роль параметров./.Пусть
%igrayut v nizhesleduyushchikh rassuzhdeniyakh rol\cprime\ 
%parametrov./.Pust\cprime

далее $\handcolor D_\textnormal{\it вн}$ и
$\handcolor B_\textnormal{\it вн}$
есть решение этих уравнений, соответ-
%dalee $\handcolor D_{\it\cyrit{vn}}$ i 
%$\handcolor B_{\it\cyrit{vn}}$
%est\cprime\ reshenie \protect{\`{e}}tikh uravneni\u i, sootvet-

ствующее внешнему полю. Тогда сумма вида
$\handcolor D_0\hspace{-0.005cm}+\hspace{-0.03cm}D_\textnormal{\it вн}\
\textnormal{\it и}\ B_0\hspace{-0.005cm}+\hspace{-0.03cm}B_\textnormal{\it вн}$
%stvuyushchee vneshnemu polyu. Togda summa vida 
%$\handcolor D_0\hspace{-0.005cm}+\hspace{-0.03cm}D_{\it\cyrit{vn}}\
%{\it\cyrit i}\ B_0\hspace{-0.005cm}+\hspace{-0.03cm}B_{\it\cyrit{vn}}$

как правило, \udensdash{н$\ \atop\ $\hspace{-0.33cm}е\hspace{-0.05cm}\ }
будет представлять собою решения уравнении
%kak pravilo, \udensdash{n$\ \atop\ $\hspace{-0.33cm}e\hspace{-0.05cm}\ }
%budet predstavlyat\cprime\ soboyu resheniya uravnenii

Борна - в силу их нелинейности - ни при каком выборе миро-
%Borna - v silu ikh neline\u inosti - ni pri kakom vybore miro-

вой траектории. В этом и проявляется специфический для
%vo\u i traektorii. V \protect{\`{e}}tom i proyavlyaet{s}ya
%spetsificheski\u i dlya

теории Борна и чуждый классике эффект искажения свойств
%teorii Borna i chuzhdy\u i klassike \protect{\`{e}}ffekt iskazheniya
%svo\u istv 

электрона внешним полем.
%\protect{\`{e}}lektrona vneshnim polem.

Если $\handcolor D_\textnormal{\it вн}\ \textnormal{\it и}\
B_\textnormal{\it вн}$ сравнимы с $\handcolor b$ - напр.если на расстоянии
%Esli $\handcolor D_{\it\cyrit{vn}}\ {\it\cyrit i}\ 
%B_{\it\cyrit{vn}}$ sravnimy s $\handcolor b$ - 
%napr.esli na rasstoyanii

сравнимом с $\handcolor r_0 $ {\ith от нашего} электрона 
находится другой электрон
%sravnimom s $\handcolor r_0 $ {\ith\cyrit ot nashego} \protect{\`{e}}lektrona
%nakhodit{s}ya drugo\u i \protect{\`{e}}lektron

-----------------------

\ \hspace{-1cm}$\handcolor 1)$ 
Оказывается, что для проведения вычислений здесь,пожалуй
%Okazyvaet{s}ya, chto dlya provedeniya vychisleni\u i 
%zdes\cprime,pozhalu\u i

удобнее всего пользоваться той формулировкой теории, где {\ith за}
%udobnee vsego pol\cprime zovat\cprime sya to\u i formulirovko\u i
%teorii, gde {\ith\cyrit za}

независимые переменные выбрали $\handcolor D$ и В и роль {\ith Лагранжиана}
%nezavisimye peremennye vybrali $\handcolor D$ i V i rol\cprime\
%{\ith\cyrit Lagranzhiana}

играет плотность энергии $\handcolor U$.
%igraet plotnost\cprime\ \protect{\`{e}}nergii $\handcolor U$.

\pagebreak
\refstepcounter{ppage}
\label{page28}

то этот эффект искажение играет основную роль во всем явле-
%to \protect{\`{e}}tot \protect{\`{e}}ffekt iskazhenie igraet 
%osnovnuyu rol\cprime\ vo vsem yavle-

нии и {\ith a priori} не ясно как нужно тогда вести вычисления.
%nii i {\ith a priori} ne yasno kak nuzhno togda vesti vychisleniya.

Если же $\handcolor D_\textnormal{\it вн}$ и
$\handcolor B_\textnormal{\it вн}$ малы по сравнению с
$\handcolor b\ ,$ то можно,оче-
%Esli zhe $\handcolor D_{\it\cyrit{vn}}$ i
%$\handcolor B_{\it\cyrit{vn}}$ maly po sravneniyu s 
%$\handcolor b\ ,$ to mozhno,oche-

видно,положить
%vidno,polozhit\cprime
{\handcolor
\begin{eqnarray}
\refstepcounter{requation}
\label{eq39}
D&=&D_0 + D_\textnormal{\it вн}
%D_{\ith\cyrit{vn}} 
+ D^\prime\hspace{0.8cm};
\hspace{0.8cm}B\ =\ B_0 + B_\textnormal{\it вн}
%B_{\ith\cyrit{vn}} 
+ B^\prime\hspace{0.8cm}
\end{eqnarray}
}и искать $\handcolor D^\prime\ \textnormal{\it и}\ B^\prime$ в виде малых 
поправок к исходным по-
%i iskat\cprime\ $\handcolor D^\prime\ {\it\cyrit i}\ B^\prime$ v vide malykh
%popravok k iskhodnym po-

лям. Эффект искажения будет, но он будет мал. Конкретный
%lyam. \protect{\`{E}}ffekt iskazheniya budet, no on budet mal.
%Konkretny\u i

\vspace{-0.15cm}

пример такого вычисления был дан проф.\ И.Е.\ Таммом
%primer takogo vychisleniya byl dan prof.\ I.E.\ Tammom}
\hspace{-0.6cm}$^{\mbox{\large\stexttt{I/}}\atop\ }$ рассмо-
%{\cyrtt rassmo-

тревшим тот простейший случай когда решение
$\handcolor D_0\ \textnormal{\it и}\ B_0$ соответ-
%trevshim tot proste\u ishi\u i slucha\u i kogda reshenie
%$\handcolor D_0\ {\it\cyrit i}\ B_0$ sootvet-

ствует покоющемуся электрону, а решение
$\handcolor D_\textnormal{\it вн}\ \textnormal{\it и}\ B_\textnormal{\it вн}$ 
- посто-
%stvuet pokoyushchemusya \protect{\`{e}}lektronu, a reshenie
%$\handcolor D_{\it\cyrit{vn}}\ {\it\cyrit i}\ B_{\it\cyrit{vn}}$ - posto-

янному однородному электростатическому полю. В этом случае
%yannomu odnorodnomu \protect{\`{e}}lektrostaticheskomu polyu.
%V \protect{\`{e}}tom sluchae

$\handcolor B=0\ ;$для поправки же $\handcolor D^\prime$ 
получаются довольно простые
%$\handcolor B=0\ ;$dlya popravki zhe $\handcolor D^\prime$ 
%poluchayut{s}ya dovol\cprime no prostye

уравнения, которые показывают, что эта поправка в известном
%uravneniya, kotorye pokazyvayut, chto \protect{\`{e}}ta popravka v
%izvestnom

смысле действительно мала: а именно, хотя в центре электро-
%smysle de\u istvitel\cprime no mala: a imenno, khotya v tsentre
%\protect{\`{e}}lektro-

на она идет в бесконечность, но все время остается гораздо
%na ona idet v beskonechnost\cprime, no vse vremya ostaet{s}ya gorazdo

меньше{\ith й} чем $\handcolor D_0\ ;$ на 
больших же расстояниях от этого центра
%men\cprime she{\ith\cyrit{\u i}} chem $\handcolor D_0\ ;$ na 
%bol\cprime shikh zhe rasstoyaniyakh ot \protect{\`{e}}togo tsentra

$\handcolor D^\prime \ll D_\textnormal{\it вн}$. Разумеется при
%$\handcolor D^\prime \ll D_{\it\cyrit{vn}}$. Razumeet{s}ya pri
$\handcolor b\rightarrow\infty\ ,\ D^\prime$ 
уходит в нуль.
%ukhodit v nul\cprime .

Существенно иметь в виду, что решения типа суще-
%Sushchestvenno imet\cprime\ v vidu, chto resheniya tipa /\ref{eq39}/ sushche-

ствуют очевидно при
\udensdash{лю$\ \atop\ $\hspace{-0.33cm}бом\ \hspace{-0.05cm}\ }\ \
в \hspace{-0.52cm}$/$ виде мировой траектории и поя-
%stvuyut ochevidno pri 
%\udensdash{lyu$\ \atop\ $\hspace{-0.33cm}bom\ \hspace{-0.05cm}\ }\ \ 
%v \hspace{-0.52cm}$/$ vide mirovo\u i traektorii i poya-

вление поправок $\handcolor D^\prime\ \textnormal{\it и}\ B^\prime$ никакого 
непосредственного отношения
%vlenie popravok $\handcolor D^\prime\ {\it\cyrit i}\ B^\prime$ nikakogo
%neposredstvennogo otnosheniya

к тому факту, что электрон приходит в движение не имеет.
%k tomu faktu, chto \protect{\`{e}}lektron prikhodit v 
%dvizhenie ne imeet.

Отсюда вытекает, что исходя из одних только уравнений поля{\ith ,}
%Ot{s}yuda vytekaet, chto iskhodya iz odnikh tol\cprime ko uravneni\u i 
%polya{\ith ,}

законов механики электрона вообще получить нельзя. Особенно
%zakonov mekhaniki \protect{\`{e}}lektrona voobshche poluchit\cprime\
%nel\cprime zya. Osobenno

ярко это видно хотя бы из примера разобранного Таммом, кото-
%yarko \protect{\`{e}}to vidno khotya by iz primera razobrannogo
%Tammom, koto-

рый показывает, что уравнения Борна допускают и такие решения
%ry\u i pokazyvaet, chto uravneniya Borna dopuskayut
%i takie resheniya

\ при которых несмотря на наличие внешнего поля электро{\ith н}$\;$ поко-
%\ pri kotorykh nesmotrya na nalichie vneshnego polya 
%\protect{\`{e}}lektro{\ith\cyrit n}$\;$poko-

ится. В этом факте в \hspace{-0.52cm}$\handcolor /$ 
самом по себе нет ничего странного
%it{s}ya. V \protect{\`{e}}tom fakte v \hspace{-0.52cm}$\handcolor /$
%samom po sebe net nichego strannogo

---------------------------------

I/ Как я узнал из его любезного сообщения.
%{\tt I}/ Kak ya uznal iz ego lyubeznogo soobshcheniya.

\pagebreak
\refstepcounter{ppage}
\label{page29}

- в самом деле физически вполне возможно представить себе такой
%- v samom dele fizicheski vpolne vozmozhno prestavit\cprime\ sebe tako\u i

случай когда электрон удерживается на месте какими нибудь силами
%slucha\u i kogda \protect{\`{e}}lektron uderzhivaet\protect{s}ya 
%na meste kakimi nibud\cprime\ silami

не электрического происхождения. Наличие таки{\ith х}$\;$сил,очевидно, ни
%ne \protect{\`{e}}lektricheskogo proiskhozhdeniya. Nalichie 
%taki{\ith\cyrit{kh}}$\;$sil,ochevidno, ni

\vspace{-0.15cm}

как не может
\hspace{-0.8cm}$^{\mbox{\large\textnormal{\tt быть}}\atop\ }$
\hspace{-0.65cm}учтено в самих уравнения{\ith х}$\;$ поля, которые определяют
%kak ne mozhet
%\hspace{-0.8cm}$^{\mbox{\large\stexttt{\cyrtt byt\cprime }}\atop\ }$
%uchteno v samikh uravneniya{\ith\cyrit{kh}}$\;$polya, 
%kotorye opredelyayut

вообще все электродинамически возможные состояния поля и конечно,
%voobshche vse \protect{\`{e}}lektrodinamicheski vozmozhnye sostoyaniya polya 
%i konechno,

ничего не могут сказать о том в каких условиях то или иное из
%nichego ne mogut skazat\cprime\ o tom v kakikh usloviyakh to ili inoe iz

этих состояний осуществимо механически.
%\protect{\`{e}}tikh sostoyani\u i osushchestvimo mekhanicheski.

Это обстоятельство,однако{\ith ,}$\;$отнюдь не означает, что для полу-
%\protect{\`{E}}to obstoyatel\cprime stvo,odnako{\ith ,}$\;$otnyud\cprime\
%ne oznachaet, chto dlya polu-

чения законов механики необходимо ввести в теорию какие то новые
%cheniya zakonov mekhaniki neobkhodimo vvesti v teoriyu kakie to novye

самостоятельные принципы. Дело в том, что{\ith ,}$\;$как неоднократно под-
%samostoyatel\cprime nye printsipy. Delo v tom, chto{\ith ,}$\;$kak
%neodnokratno pod-

черкивалось в ходе всего изложения{\ith ,}$\;$основным утверждением теории
%cherkivalos\cprime\ v khode vsego izlozheniya{\ith ,}$\;$osnovnym
%utverzhdeniem teorii

Борна является не уравнения поля, а вариационный принцип. Содер-
%Borna yavlyaet{s}ya ne uravneniya polya, a
%variatsionny\u i printsip. Soder-

жание же этого принципа отнюдь не исчерпывается уравнениями поля;
%zhanie zhe \protect{\`{e}}togo printsipa otnyud\cprime\ ne
%ischerpyvaet{s}ya uravneniyami polya;

последние определяют собою всю совокупность экстремалей вариаци-
%poslednie opredelyayut soboyu vsyu sovokupnost\cprime\ 
%\protect{\`{e}}kstremale\u i variatsi-

онного интеграла, но ведь отнюдь не для всех экстремалей этот
%onnogo integrala, no ved\cprime\ otnyud\cprime\ ne dlya vsekh 
%\protect{\`{e}}kstremale\u i \protect{\`{e}}tot

интеграл имеет одинаковое значение. Именно благодаря этому обсто-
%integral imeet odinakovoe znachenie. Imenno blagodarya \protect{\`{e}}tomu
%obsto-

ятельству вариационный принцип оказывется достаточным для полу-
%yatel\cprime stvu variatsionny\u i printsip okazyvyet{s}ya 
%dostatochnym dlya polu-

чения не только законов электродинамики, но и законов механики.
%cheniya ne tol\cprime ko zakonov \protect{\`{e}}lektrodinamiki, no
%i zakonov mekhaniki.

\hspace{1cm}Конкретно нужно очевидно поступить так: подставить решение
%Konkretno nuzhno ochevidno postupit\cprime\ tak:
%podstavit\cprime\ reshenie

/\ref{eq39}/ в Лагранжиан /\ref{eq7}/ и /\ref{eq8}/ и искать 
для какого из этих решений
%/\ref{eq39}/ v Lagranzhian /\ref{eq7}/ i  /\ref{eq8}/ i iskat\cprime{\ith ,}
%dlya kakogo iz \protect{\`{e}}tikh resheni\u i

вариационный интеграл имеет экстремальное значение. Роль пара-
%variatsionny\u i integral imeet \protect{\`{e}}kstremal\cprime noe znachenie.
%Rol\cprime\ para-

метров,по которым при этом должна производится вариация{\ith ,}$\;$ играют
%metrov,po kotorym pri \protect{\`{e}}tom dolzhna proizvodit{s}ya 
%variatsiya{\ith ,}$\;$igrayut

величины характеризующие вид мировой линии электрона, от которого
%velichiny kharakterizuyushchie vid mirovo\u i linii \protect{\`{e}}lektrona,
%ot kotorogo

зависят выражения /\ref{eq39}/. Ясно, что получающееся в результате
%zavisyat vyrazheniya /\ref{eq39}/. Yasno, chto poluchayushcheesya v
%rezul\cprime tate

такого метода вар{\ith ьи}рования уравнение действительно по типу ана-
%takogo metoda var{\ith\cyrit \cprime i}rovaniya uravnenie 
%de\u istvitel\cprime no po tipu ana-

логично обычному вариационному принципу механики. Ясно, далее,
%logichno obychnomu variatsionnomu printsipu mekhaniki. Yasno, dalee,

что добавление{\ith м}$\;$соответствующих членов к Лагранжиану, можно, прин-
%chto dobavlenie{\ith\cyrit m}$\;$sootvet{s}tvuyushchikh chlenov k Lagranzhianu,
%mozhno, prin-

ципиально, учесть и действие на заряд не электрических сил.
%tsipial\cprime no, uchest\cprime\ i de\u istvie na zaryad ne
%\protect{\`{e}}lektricheskikh sil.

\hspace{1cm}Если интересоваться только "нулевым приближением"\ т.е.\ те-
%Esli interesovat\cprime sya tol\cprime ko "nulevym
%priblizheniem" t.e.\ te-

ми случаями когда можно ожидать применимости обычных законов ме-
%mi sluchayami kogda mozhno ozhidat\cprime\ primenimosti
%obychnykh zakonov me-

ханики - то можно,очевидно, отбросить поправки
$\handcolor D^\prime\ \textnormal{\it и}\ B^\prime$ и встав-
%khaniki - to mozhno,ochevidno, otbrosit\cprime\ popravki 
%$\handcolor D^\prime\ {\it\cyrit i}\ B^\prime$ i vstav-

\pagebreak
\refstepcounter{ppage}
\label{page30}

лять в вариационный интеграл непосредственно сумму поля
%lyat\cprime\ v variatsionny\u i integral neposredstvenno summu polya

электрона и внешнего поля. Борн и Инфельд показали, что
%\protect{\`{e}}lektrona i vneshnego polya. Born i Infel\cprime d
%pokazali, chto

идя по только что указанному пути можно действительно полу-
%idya po tol\cprime ko chto ukazannomu puti mozhno de\u istvitel\cprime no
%polu-

чить вариационный принцип, аналогичн{\ith ый}$\;$механическому, кото-
%chit\cprime\ variatsionny\u i printsip, 
%analogichn{\ith\cyrit y\u i}$\;$mekhanicheskomu, koto-

рый - в предположении, что внешнее поле мало меняется на
%ry\u i - v predpolozhenii, chto vneshnee pole malo menyaet{s}ya na

расстояних порядка $\handcolor r_0$ - переходит в обычный вариационный
%rasstoyanikh poryadka $\handcolor r_0$ - perekhodit v obychny\u i 
%variatsionny\u i

принцип релятивистской механики. Тем самым задача построе-
%printsip relyativist{s}ko\u i mekhaniki. Tem samym zadacha postroe-

ния механики в рамках унитарной теории, может считаться, по
%niya mekhaniki v ramkakh unitarno\u i teorii, mozhet schitat\cprime sya, po

крайней мере в общих чертах, разрешенной.
%kra\u ine\u i mere v obshchikh chertakh, razreshenno\u i.

\hspace{1cm}Наиболее интересным было бы, конечно, исследование тех
%Naibolee interesnym bylo by, konechno, issledovanie tekh

\udensdash{отклон$\ \atop\ $\hspace{-0.33cm}ений\hspace{-0.05cm}\ }
от обычного вида механических уравнений, кото-
%\udensdash{otklon$\ \atop\ $\hspace{-0.33cm}eni\u i\hspace{-0.05cm}\ }
%ot obychnogo vida mekhanicheskikh uravneni\u i, koto-

рые должны появиться в теории Борна при более точном прове-
%rye dolzhny poyavit\cprime sya v teorii Borna pri bolee tochnom
%prove-

дении вычислений. К сожалению те приближения, от которых
%denii vychisleni\u i. K sozhaleniyu te priblizheniya, ot kotorykh

нужно при этом освободиться повидимому связаны друг с дру-
%nuzhno pri \protect{\`{e}}tom osvobodit\cprime sya povidimomu
%svyazany drug s dru-

гом- если градиент внешнего поля на расстояниях порядка $\handcolor r_0$
%gom- esli gradient vneshnego polya na rasstoyaniyakh poryadka $\handcolor r_0$

не мал, то вряд ли мало и поправки
$\handcolor D^\prime\ \textnormal{\it и}\ B^\prime\ ,$ а также вряд
%ne mal, to vryad li maly i popravki 
%$\handcolor D^\prime\ {\it\cyrit i}\ B^\prime\ ,$ a takzhe vryad 

ли мала и реакция собственного излучения на электрон. Впро-
%li mala i reaktsiya sobstvennogo izlucheniya na \protect{\`{e}}lektron. Vpro-

чем, если учитывать все эти эффекты только в первом прибли-
%chem, esli uchityvat\cprime\ vse \protect{\`{e}}ti \protect{\`{e}}ffekty
%tol\cprime ko v pervom pribli-

жении , то вычисления остаются принципиально выполнимыми
%zhenii , to vychisleniya ostayut{s}ya printsipial\cprime no
%vypolnimymi

хотя и делаются весьма громо{\ith зд}кими.
%khotya i delayut{s}ya ves\cprime ma gromo{\ith\cyrit zd}kimi.

\pagebreak
\refstepcounter{ppage}
\refstepcounter{rsection}
\label{glavaII}
\hspace{3.5cm}
\udensdash{\ Г$\ \atop\ $\hspace{-0.1cm}Л\ А\ В\ А\ \ П.\ 
%\udensdash{\ G$\ \atop\ $\hspace{-0.1cm}L\ A\ V\ A\ \ P.\ 
\hspace{-0.1cm}\ }\\[-0.2cm]

\newcounter{r2subsection}
\renewcommand{\thesubsection}{\Roman{r2subsection}} % shape of Russian number 1
\refstepcounter{r2subsection}
\label{glavaIIpara1}
\renewcommand{\thesubsection}{\arabic{r2subsection}} % shape of Russian number 1
\hspace{0.4cm}
\udensdash{\ \S\ I. Введение\ \ \hspace{-0.05cm}\ }\\[-0.2cm]
%\udensdash{\ \S\ {\tt I}. Vvedenie\ \ \hspace{-0.05cm}\ }\\[-0.2cm]

\hspace{1.cm}Итак, коротко резюмируя содержание главы {\tt\ref{glavaI}}, можно
%Itak, korotko rezyumiruya soderzhanie glavy {\tt\ref{glavaI}}, mozhno

сказать, что Борну удалось построить унитарную теорию
%skazat\cprime , chto Bornu udalos\cprime\ postroit\cprime\ unitarnuyu teoriyu

поля и материи за счет отказа от Максвелловых уравнений
%polya i materii za schet otkaza ot Maksvellovykh uravneni\u i

и введения в$\;${\ith за}ме{\ith н} их новых уравнений 
поля/формулы /\ref{eq24}/
%i vvedeniya v$\;${\ith\cyrit{za}}me{\ith\cyrit{n}} ikh novykh uravneni\u i 
%polya/formuly /\ref{eq24}/

или /\ref{eq25}//{\ith ,}$\;$основной характерной чертой которых является их
%ili /\ref{eq25}//{\ith ,}$\;$osnovno\u i kharakterno\u i cherto\u i kotorykh
%yavlyaet{s}ya ikh

{\handcolor\underline{\color{black}нелинейность}}.
При переходе от уравнений
\udensdash{Макс$\ \atop\ $\hspace{-0.33cm}велла\hspace{-0.05cm}\ } к уравне-
%{\handcolor\underline{\color{black}neline\u inost\cprime}}. 
%Pri perekhode ot uravneni\u i
%\udensdash{Maks$\ \atop\ $\hspace{-0.33cm}vella\hspace{-0.05cm}\ } k uravne-

ниям {\handcolor\underline{\color{black}Борна}}
"\ механические"\ результаты теории, таким обра-
%niyam {\handcolor\underline{\color{black}Borna}} 
%" mekhanicheskie" rezul\cprime taty teorii, takim obra-

зом, существенно меняются: энергия точечного заряда полу-
%zom, sushchestvenno menyayut{s}ya: \protect{\`{e}}nergiya
%tochechnogo zaryada polu-

чается конечной и т.д. Но нетрудо видеть, что одновремен-
%chaet{s}ya konechno\u i i t.d. No netrudo videt\cprime , chto odnovremen-

но с этим перестает быть справедливым и ряд других резуль-
%no s \protect{\`{e}}tim perestaet byt\cprime\ spravedlivym i ryad 
%drugikh rezul\cprime -

татов обычной электродинамики, на первый взгляд никак не
%tatov obychno\u i \protect{\`{e}}lektrodinamiki, na pervy\u i
%vzglyad nikak ne

связаннях с трудностями построения унитарной теории. А
%svyazannyakh s trudnostyami postroeniya unitarno\u i teorii. A

именно поскольку теория становится нелинейной, как мы
%imenno poskol\cprime ku teoriya stanovit{s}ya neline\u ino\u i, kak my

уже вскользь указывали в \S\ {\tt\ref{glavaIpara8}}
{\tt\ref{glavaI}}-ой главы, теряет силу
\udensdash{при$\ \atop\ $\hspace{-0.33cm}н-\hspace{-0.05cm}\ }
%uzhe vskol\cprime z\cprime\ ukazyvali v \S\ {\tt\ref{glavaIpara8}} 
%{\tt\ref{glavaI}}-o\u i glavy,
%teryaet silu \udensdash{pri$\ \atop\ $\hspace{-0.33cm}n-\hspace{-0.05cm}\ }

\udensdash{цип$\ \atop\ $\hspace{-0.1cm}суперпозиции,\hspace{-0.05cm}\ }
играющий столь важную роль во всей
%\udensdash{tsip$\ \atop\ $\hspace{-0.1cm}superpozitsii,\hspace{-0.05cm}\ } 
%igrayushchi\u i stol\cprime\ vazhnuyu rol\cprime\ vo vse\u i 

Максвелловой электродинамике.
%Maksvellovo\u i \protect{\`{e}}lektrodinamike.

Иначе говоря, согласно теории Борна, отклонения от обычных
%Inache govorya, soglasno teorii Borna, otkloneniya ot obychnykh

законов электродинамики должны проявляться не только в
%zakonov \protect{\`{e}}lektrodinamiki dolzhny proyavlyat\cprime sya
%ne tol\cprime ko v

"механических"\ явлениях, но и в эффектах "взаимного искаже-
%"mekhanicheskikh" yavleniyakh, no i v \protect{\`{e}}ffektakh 
%"vzaimnogo iskazhe-

ния"\ двух /или нескольких/ электромагнитных полей.
%niya" dvukh /ili neskol\cprime kikh/ \protect{\`{e}}lektromagnitnykh
%pole\u i.

Наиболее интересным из этих эффектов является, пожалуй
%Naibolee interesnym iz \protect{\`{e}}tikh \protect{\`{e}}ffektov
%yavlyaet{s}ya, pozhalu\u i

эффект искажения поля 
с\udensdash{ветовой$\ \atop\ $\hspace{-0.1cm}волны\hspace{-0.05cm}\ }
другими внешними
%\protect{\`{e}}ffekt iskazheniya polya 
%s\udensdash{vetovo\u i$\ \atop\ $\hspace{-0.1cm}volny\hspace{-0.05cm}\ } 
%drugimi vneshnimi 

полями. Как непосредственно видно из уравнений /\ref{eq24}/, в
%polyami. Kak neposredstvenno vidno iz uravneni\u i /\ref{eq24}/, v

сочетании с /\ref{eq22}/ и /\ref{eq23}/, в теории Борна
\udensdash{плос$\ \atop\ $\hspace{-0.33cm}кая\hspace{-0.05cm}\ } и
\udensdash{шаро$\ \atop\ $\hspace{-0.33cm}вая\hspace{-0.05cm}\ }
%sochetanii s /\ref{eq22}/ i /\ref{eq23}/, v teorii Borna
%\udensdash{plos$\ \atop\ $\hspace{-0.33cm}kaya\hspace{-0.05cm}\ } i 
%\udensdash{sharo$\ \atop\ $\hspace{-0.33cm}vaya\hspace{-0.05cm}\ }

электромагнитная волна представляют собою два
\udensdash{точ$\ \atop\ $\hspace{-0.33cm}ных\hspace{-0.05cm}\ }
%\protect{\`{e}}lektromagnitnaya volna predstavlyayut soboyu dva
%\udensdash{toch$\ \atop\ $\hspace{-0.33cm}nykh\hspace{-0.05cm}\ }

решения уравнений поля / Это имеет место также и для
%resheniya uravneni\u i polya / \protect{\`{E}}to imeet mesto takzhe i dlya

второго варианта теории, имеющего дело с другой Лангранже-
%vtorogo varianta teorii, imeyushchego delo s drugo\u i Langranzhe-

вой функцией. Проще всего в этом можно убедиться, если
%vo\u i funktsie\u i. Proshche vsego v \protect{\`{e}}tom mozhno
%ubedit\cprime sya, esli

\pagebreak
\refstepcounter{ppage}

вспомнить, что для поля волновой зоны оба пространственн-
%vspomnit\cprime , chto dlya polya volnovo\u i zony oba prostranstvenn-

ых инварианта $\handcolor F\ \textnormal{\it и}\ G$ обращаются в 
нуль /. Таким об-
%ykh invarianta $\handcolor F\ {\it\cyrit i}\ G$ obrashchayut{s}ya v 
%nul\cprime /. Takim ob-

разом, пока речь идет о такого типа волнах, электромаг-
%razom, poka rech\cprime\ idet o takogo tipa volnakh, 
%\protect{\`{e}}lektromag-

нитная теория света формулируется по
\udensdash{Борну\hspace{-0.05cm}\ } точно так-же
%nitnaya teoriya sveta formuliruet{s}ya po 
%\udensdash{Bornu\hspace{-0.05cm}\ } tochno tak-zhe

как и по Максвеллу. Но дело начинает обстоять существен-
%kak i po Maksvellu. No delo nachinaet obstoyat\cprime\ sushchestven-

но иначе, как только мы переходим к несколько усложнен-
%no inache, kak tol\cprime ko my perekhodim k neskol\cprime ko
%uslozhnen-

ным случаям, а именно к таким, когда наряду с полем,
%nym sluchayam, a imenno k takim, kogda naryadu s polem,

скажем, плоской световой волны, имеется в наличии еще
%skazhem, plosko\u i svetovo\u i volny, imeet{s}ya v nalichii eshche

какое нибудь другое электромагнитное поле, хотя бы еле-
%kakoe nibud\cprime\ drugoe \protect{\`{e}}lektromagnitnoe pole,
%khotya by \protect{\`{e}}le-

ктростатическое поле плоского конденсатора или поле дру-
%ktrostaticheskoe pole ploskogo kondensatora ili pole dru-

гой световой волны. По
\udensdash{Макс$\ \atop\ $\hspace{-0.33cm}веллу\hspace{-0.05cm}\ },
наличие этого второго
%go\u i svetovo\u i volny. Po 
%\udensdash{Maks$\ \atop\ $\hspace{-0.33cm}vellu\hspace{-0.05cm}\ },
%nalichie \protect{\`{e}}togo vtorogo

поля не оказывает никакого влияния на свойства исходной
%polya ne okazyvaet nikakogo vliyaniya na svo\u istva iskhodno\u i

волны, в то время как по \udensdash{Борну\hspace{-0.05cm}\ }
такое влияние, вообще го-
%volny, v to vremya kak po \udensdash{Bornu\hspace{-0.05cm}\ } 
%takoe vliyanie, voobshche go-

воря, \udensdash{наверно\hspace{-0.05cm}\ } имеет место, т.к.\ напр.\ сумма 
электромаг-
%vorya, \udensdash{naverno\hspace{-0.05cm}\ } imeet mesto, t.k.\ napr.\ summa
%\protect{\`{e}}lektromag-

нитных полей двух плоских волн уже не представляет со-
%nitnykh pole\u i dvukh ploskikh voln uzhe ne predstavlyaet so-

бою решение уравнений Борна. До тех пор, пока все поля,
%boyu reshenie uravneni\u i Borna. Do tekh por, poka vse polya,

с которыми мы здесь имеем дело, слабы по сравнению с
%s kotorymi my zdes\cprime\ imeem delo, slaby po sravneniyu s

"критическим"\ полем "$\handcolor b$"\protect{,} влияние нелинейности должно,
%"kriticheskim" polem "$\handcolor b$", vliyanie neline\u inosti dolzhno,

очевидно, быть слабым и имеет смысл говорить о том,
%ochevidno, byt\cprime\ slabym i imeet smysl govorit\cprime\ o tom,

что и в присутствии посторонного поля мы все же имеем
%chto i v prisut{s}tvii postoronnogo polya my vse zhe imeem

плоскую волну, несколько только искаженную.
%ploskuyu volnu, neskol\cprime ko tol\cprime ko iskazhennuyu.

Разбор трех примеров такого типа искажения и составляет
%Razbor trekh primerov takogo tipa iskazheniya i sostavlyaet

предмет этой главы.
%predmet \protect{\`{e}}to\u i glavy.

Необходимо отметить, что как раз в этом пункте проявляет-
%Neobkhodimo otmetit\cprime , chto kak raz v \protect{\`{e}}tom punkte
%proyavlyaet-

ся замечательная аналогия между 
электродинамикой \udensdash{Борна\hspace{-0.05cm}\ }
%sya zamechatel\cprime naya analogiya mezhdu 
%\protect{\`{e}}lektrodinamiko\u i \udensdash{Borna\hspace{-0.05cm}\ }

и теорией, построенной, казалось бы, на совершенно дру-
%i teorie\u i,  postroenno\u i, kazalos\cprime\ by, na sovershenno dru-

гих физических основах, а именно - теорией позитрона
%gikh fizicheskikh osnovakh, a imenno - teorie\u i pozitrona

\udensdash{Дирака.\hspace{-0.05cm}\ } В самом деле, как указали впервые 
\udensdash{Гальперн\hspace{-0.05cm}\ }
%\udensdash{Diraka.\hspace{-0.05cm}\ } V samom dele, kak ukazali vpervye
%\udensdash{Gal\cprime pern\hspace{-0.05cm}\ } 

\pagebreak
\refstepcounter{ppage}

и Дебай$^{\mbox{\large\stexttt{I/}}}$, согласно этой последней теории тоже должен
%i Deba\u i}$^{\mbox{\large\stexttt{I/}}}$, {\cyrtt soglasno 
%\protect{\`{e}}to\u i posledne\u i teorii tozhe dolzhen

иметь место эффект взаимного искажения двух световых волн
%imet\cprime\ mesto \protect{\`{e}}ffekt vzaimnogo iskazheniya
%dvukh svetovykh voln

и так-же, как нетрудно убедиться, ряд других, аналогичных
%i tak-zhe, kak netrudno ubedit\cprime sya, ryad drugikh, analogichnykh 

Борновским, эффектов искажения. Более того, ученики Гейзен-
%Bornovskim, \protect{\`{e}}ffektov iskazheniya. Bolee togo, ucheniki
%Ge\u izen-

берга \udensdash{Эйлер\hspace{-0.05cm}\ } и
\udensdash{Кок$\ \atop\ $\hspace{-0.33cm}кель\hspace{-0.05cm}\ }
показали$^{\mbox{\large\stexttt{ 2/}}}$, что эта аналогия имеет,
%berga \udensdash{\protect{\`{E}}\u iler\hspace{-0.05cm}\ } i
%\udensdash{Kok$\ \atop\ $\hspace{-0.33cm}kel\cprime\hspace{-0.05cm}\ } 
%pokazali$^{\mbox{\large\stexttt{ 2/}}}$, chto \protect{\`{e}}ta analogiya imeet,

в известной мере и количественный характер: "рассеяние све-
%v izvestno\u i mere i kolichestvenny\u i kharakter: "rasseyanie sve-

та на свете"\ вычисленное по Дираку, оказывается в общих
%ta na svete" vychislennoe po Diraku, okazyvaet{s}ya v obshchikh

чертах таким-же, какого следовало бы ожидать по прокванто-
%chertakh takim-zhe, kakogo sledovalo by ozhidat\cprime\ po prokvanto-

ванной электродинамике Борна. В связи с существованием
%vanno\u i \protect{\`{e}}lektrodinamike Borna. V svyazi s
%sushchestvovaniem

этой аналогии встает ряд интересных вопросов, кото-
%\protect{\`{e}}to\u i analogii vstaet ryad interesnykh voprosov, koto-

рые мы, однако, здесь, не будем обсуждать, поскольку все
%rye my, odnako, zdes\cprime , ne budem obsuzhdat\cprime , poskol\cprime ku vse

они существенно связаны с переходом к квантованию поля,
%oni sushchestvenno svyazany s perekhodom k kvantovaniyu polya,

лежащим вне рамок настоящей диссертации/тем более, что
%lezhashchim vne ramok nastoyashche\u i dissertatsii/tem bolee, chto

детальный анализ результатов Эйлера и Коккеля пока
%detal\cprime ny\u i analiz rezul\cprime tatov \protect{\`{E}}\u ilera i Kokkely%a poka

невозможен, поскольку их вычисления еще неопубликованы/
%nevozmozhen, poskol\cprime ku ikh vychisleniya eshche neopublikovany/

Во всяком случае само существование аналогии между ре-
%Vo vsyakom sluchae samo sushchestvovanie analogii mezhdu re-

зультатами \udensdash{Борна\hspace{-0.05cm}\ } и 
\udensdash{Дирака,\hspace{-0.05cm}\ } как раз в области нелинейных
%zul\cprime tatami \udensdash{Borna\hspace{-0.05cm}\ } i 
%\udensdash{Diraka,\hspace{-0.05cm}\ } kak raz v oblasti neline\u inykh

эффектов, делает исследование именно этой стороны Бор-
%\protect{\`{e}}ffektov, delaet issledovanie imenno \protect{\`{e}}to\u i
%storony Bor-

новской электродинамики особенно интересным и {\ith с} классиче-
%novsko\u i \protect{\`{e}}lektrodinamiki osobenno interesnym  i
%{\ith\cyrit s} klassiche-

ской точки зрения.\\
%sko\u i tochki zreniya.\\

\refstepcounter{r2subsection}
\label{glavaIIpara2}
\hspace{1cm}\S\ 2.$^{\mbox{\large\stexttt{3/}}}$\ 
\udensdash{Плоская световая волна в однородном$\ \atop\ $\hspace{-0.1cm} 
электро-\hspace{-0.05cm}\ }
%$^{\mbox{\large\stexttt{3/}}}$\ 
%\udensdash{Ploskaya svetovaya volna v odnorodnom$\ \atop\ $\hspace{-0.1cm} 
%\protect{\`{e}}lektro-\hspace{-0.05cm}\ }

\hspace{4.5cm}\udensdash{статическом$\ \atop\ $\hspace{-0.1cm} 
поле\hspace{-0.05cm}\ }\\
%\udensdash{staticheskom$\ \atop\ $\hspace{-0.1cm} 
%pole\hspace{-0.05cm}\ }\\

\hspace{1cm}В качестве первого примера, мы рассмотрим задачу о
%V kachestve pervogo primera, my rassmotrim zadachu o

распространении плоской световой волны в однородном элек-
%rasprostranenii plosko\u i svetovo\u i volny v odnorodnom \protect{\`{e}}lek-

тростатическом поле - скажем в поле плоского конденсато-
%trostaticheskom pole - skazhem v pole ploskogo kondensato-

----------------------------

I/\ {\ith см.
%{\tt I/} {\ith{\cyrit sm.} 
Heisenberg, Z.\ f.\ Phys.\ 
\underline{90}. 1934. 209.}

2/\ {\ith Naturwiss.\ 23. 1935. 246.}

3/\ Результаты этого параграфа опубликованы в статье
%Rezul\cprime taty \protect{\`{e}}togo paragrafa opublikovany v
%stat\cprime e

\hspace{0.8cm}С.Шубина и А.Смирнова. ДАН
%S.Shubina i A.Smirnova. DAN} 
I936.
%{\cyrtt\

\pagebreak
\refstepcounter{ppage}

ра. Эту задачу мы разберем особенно подробно, как про-
%ra. \protect{\`{E}}tu zadachu my razberem osobenno podrobno, kak pro-

стейщую и типичную.
%ste\u ishchuyu i tipichnuyu.

Представим себе,сначала, плоскую световую волну, электро-
%Predstavim sebe,snachala, ploskuyu svetovuyu volnu, \protect{\`{e}}lektro-

магнитное поле которой определяется равенствами:
%magnitnoe pole kotoro\u i opredelyaet{s}ya ravenstvami:
{\handcolor
\begin{eqnarray}
\refstepcounter{requation}
\label{eq40}
&\left.\begin{array}{rcccl}
E&=&D&=&E_0\ \cos\omega 
\left(t - \frac{\displaystyle rn}{\displaystyle c}\right)\\[0.3cm]
B&=&H&=&B_0\ \cos\omega 
\left(t - \frac{\displaystyle rn}{\displaystyle c}\right)
\end{array}\right\}&\ \ \ ,
\end{eqnarray}
}где, как обычно
%gde, kak obychno
{\handcolor
\begin{eqnarray}
\refstepcounter{requation}
\label{eq41}
E_0\ =\ \left[n,B_0\right]\hspace{0.8cm}&;&\hspace{0.8cm}B_0
\ =\ \left[E_0,n\right]
\end{eqnarray}
}Выражение /\ref{eq40}/, как было только-что указано, преставляет
%Vyrazhenie /\ref{eq40}/, kak bylo tol\cprime ko-chto ukazano,
%prestavlyaet

собою \udensdash{точ$\ \atop\ $\hspace{-0.33cm}ное\hspace{-0.05cm}\ }
решение уравнений Борна.
%soboyu \udensdash{toch$\ \atop\ $\hspace{-0.33cm}noe\hspace{-0.05cm}\ }
%reshenie uravneni\u i Borna.

Согласно этой теории, изображаемая этим решением волна
%Soglasno \protect{\`{e}}to\u i teorii, izobrazhaemaya \protect{\`{e}}tim
%resheniem volna

может, как правило, распространяться только в
%mozhet, kak pravilo, rasprostranyat\cprime sya tol\cprime ko v

вакууме, т.е.\ в отсутствии всех других полей. В частности
%vakuume, t.e.\ v ot{s}ut{s}tvii vsekh drugikh pole\u i. V chastnosti

если включить в рассматриваемые нами части пространства
%esli vklyuchit\cprime\ v rassmatrivaemye nami chasti prostranstva

еще поле плоского конденсатора $\handcolor N\ ,$ то выражение вида
%eshche pole ploskogo kondensatora $\handcolor N\ ,$ to
%vyrazhenie vida
{\handcolor
\begin{eqnarray}
\refstepcounter{requation}
\label{eq42}
&\left.\begin{array}{rcl}
E&=&E_0\ \cos\omega 
\left(t - \frac{\displaystyle rn}{\displaystyle c}\right)\ +\ N\\[0.3cm]
B&=&B_0\ \cos\omega 
\left(t - \frac{\displaystyle rn}{\displaystyle c}\right)
\end{array}\right\}&
\end{eqnarray}
}\ уже не будет представлять собою возможного состояния
%uzhe ne budet predstavlyat\cprime\ soboyu vozmozhnogo sostoyaniya

поля. Ясно, однако, что до тех пор пока $\handcolor N\ll b$ будут
%polya. Yasno, odnako, chto do tekh por poka $\handcolor N\ll b$ budut

существовать решения, во всяком случае по типу близкие
%sushchestvovat\cprime\ resheniya, vo vsyakom sluchae po tipu blizkie

к /\ref{eq42}/. Эти решения, или, во всяком случае, некоторый
%k /\ref{eq42}/. \protect{\`{E}}ti resheniya, ili, vo vsyakom sluchae,
%nekotory\u i

достаточно широкий класс их мы и ставим\ \ \ \ \ себе зада-
%dostatochno shiroki\u i klass ikh my i stavim\ \ \ \ \  sebe zada-

\pagebreak
\refstepcounter{ppage}

чей найти$^{\mbox{\large\stexttt{ I/}}}$.
%che\u i na\u iti}$^{\mbox{\large\stexttt{ I/}}}$.

При этом, мы сначала не касаемся вопроса о том, в каких
%{\cyrtt Pri \protect{\`{e}}tom, my snachala ne kasaemsya voprosa 
%o tom, v kakikh

условиях то или иное из этих решений осуществляется фа-
%usloviyakh to ili inoe iz \protect{\`{e}}tikh resheni\u i 
%osushchestvlyaet{s}ya fa-

ктически; обсуждение этой более конкретной задачи мы
%kticheski; obsuzhdenie \protect{\`{e}}to\u i bolee konkretno\u i zadachi my 

отодвигаем до
%otodvigaem do 
\S\ {\tt\ref{glavaIIpara3}}.

\hspace{1cm}Опираясь на неравенство $\handcolor N\ll b$ которое является,
%Opirayas\cprime\ na neravenstvo $\handcolor N\ll b$ kotoroe 
%yavlyaet{s}ya,

как уже сказано, существенным, мы будем в ходе вычисле-
%kak uzhe skazano, sushchestvennym, my budem v khode vychisle-

ний отбрасывать все степени величины
%ni\u i otbrasyvat\cprime\ vse stepeni velichiny 
$\handcolor\frac{\displaystyle N}{\displaystyle b}$ кроме наиниз-
%krome nainiz-

шей. Как выяснится в дальнейшем, этой наинизшей степенью
%she\u i. Kak vyyasnit{s}ya v dal\cprime ne\u ishem, \protect{\`{e}}to\u i
%nainizshe\u i stepen\cprime yu

является \udensdash{вто$\ \atop\ $\hspace{-0.33cm}рая\hspace{-0.05cm}\ }.
Таким образом, все выражения, имеющие
%yavlyaet{s}ya \udensdash{vto$\ \atop\ $\hspace{-0.33cm}raya\hspace{-0.05cm}\ }.
%Takim obrazom, vse vyrazheniya, imeyushchie

в знаменателе $\handcolor b^4$, $\handcolor b^6$ и т.д.\ будут нами в 
дальнейшем
%v znamenatele $\handcolor b^4$, $\handcolor b^6$ i t.d.\ budut nami v 
%dal\cprime ne\u ishem

зачеркиваться.$^{\mbox{\large\stexttt{2/}}}$
%zacherkivat\cprime sya.$^{\mbox{\large\stexttt{2/}}}$

-------------------------

I/ Конечно, можно было-бы исходить не из векторов Е и В,
%{\tt I}/ Konechno, mozhno bylo-by iskhodit\cprime\ ne iz vektorov E i V,

а из какой нибудь другой пары векторов, например, $\handcolor D$ и Н
%a iz kako\u i nibud\cprime\ drugo\u i pary vektorov, naprimer, 
%$\handcolor D$ i N

и поставить задачу так: найти решение, близкое к
%i postavit\cprime\ zadachu tak: na\u iti reshenie, blizkoe k
{\handcolor
\begin{eqnarray*}
&\left.\begin{array}{rcl}
D&=&D_0\ \cos\omega 
\left(t - \frac{\displaystyle rn}{\displaystyle c}\right)\ +\ N\\[0.3cm]
H&=&H_0\ \cos\omega \left(t - \frac{\displaystyle rn}{\displaystyle c}\right)
\end{array}\right\}&
\end{eqnarray*}
}Разумеется, окончательные результаты должны быть при
%Razumeet{s}ya, okonchatel\cprime nye rezul\cprime taty dolzhny byt\cprime\ pri

обоих постановках вопроса \udensdash{по существу\hspace{-0.05cm}\ }
одинаковы.\\
%oboikh postanovkakh voprosa \udensdash{po sushchestvu\hspace{-0.05cm}\ }
%odinakovy.\\

В том, что действительно так, мы убедимся в сле-
%V tom, chto de\u istvitel\cprime no tak, my ubedimsya v sle-

дующем параграфе.
%duyushchem paragrafe.

2/ Если мы встретим, в дальнейшем выращения типа
%Esli my vstretim, v dal\cprime ne\u ishem vyrasheniya tipa 
$\handcolor\frac{\displaystyle E_0^4}{\displaystyle b^4}$

или
%ili 
$\handcolor\frac{\displaystyle E_0^2 N^2}{\displaystyle b^4}\ ,$ 
то ими, очевидно, также сможем пренебречь.
%to imi, ochevidno, takzhe smozhem prenebrech\cprime .

\pagebreak
\refstepcounter{ppage}

Ясно, что в таком приближении, различие между первым и вто-
%Yasno, chto v takom priblizhenii, razlichie mezhdu pervym i vto-

рым вариантом теории Борна пропадает, т.к.\ Лагранжева функ-
%rym variantom teorii Borna propadaet, t.k.\ Lagranzheva funk-

ция /\ref{eq7}/ переходит в /\ref{eq8}/.
%tsiya /\ref{eq7}/ perekhodit v /\ref{eq8}/.

Наиболее естественным казалось бы на первый взгляд искать
%Naibolee estestvennym kazalos\cprime\ by na pervy\u i vzglyad iskat\cprime

решение уравнений Борна /\ref{eq25}/ в виде
%reshenie uravneni\u i Borna /\ref{eq25}/ v vide
{\handcolor
\begin{eqnarray}
\refstepcounter{requation}
\label{eq43}
&\left.\begin{array}{rcl}
E&=&E_0\ \cos\omega 
\left(t - \frac{\displaystyle rn}{\displaystyle c}\right)\ 
+\ N\ +\ E^\prime\\[0.3cm]
B&=&B_0\ \cos\omega 
\left(t - \frac{\displaystyle rn}{\displaystyle c}\right)\ +\ B^\prime
\end{array}\right\}&
\end{eqnarray}
}считая $\handcolor E^\prime\ \textnormal{\it и}\ B^\prime$ малыми величинами, 
стремящимися к нулю
%schitaya $\handcolor E^\prime\ {\it\cyrit i}\ B^\prime$ malymi velichinami,
%stremyashchimisya k nulyu

при
%pri 
$\handcolor\frac{\displaystyle N}{\displaystyle b}\ \rightarrow\ 0$.
Однако, если подставать выражения /\ref{eq43}/ в 
%Odnako, esli podstavat\cprime\ vyrazheniya /\ref{eq43}/ v

уравнения /\ref{eq25}/, то оказывается, что для
$\handcolor E^\prime\ \textnormal{\it и}\ B^\prime$ получаются
%uravneniya /\ref{eq25}/, to okazyvaet{s}ya, chto dlya 
%$\handcolor E^\prime\ {\it\cyrit i}\ B^\prime$ poluchayut{s}ya

уравнения, которые не имеют повсюду конечных решений. Имен-
%uravneniya, kotorye ne imeyut povsyudu konechnykh resheni\u i.
%Imen-

но, вычисления показывают, что
$\handcolor E^\prime\ \textnormal{\it и}\ B^\prime$ содержат члены,
%no, vychisleniya pokazyvayut, chto 
%$\handcolor E^\prime\ {\it\cyrit i}\ B^\prime$ soderzhat chleny,

пропорциональное величине
%proportsional\cprime nye velichine 
$\handcolor t + \frac{\displaystyle rn}{\displaystyle c}\ ,$  
т.е.\ обращающиеся
%t.e.\ obrashchayushchiesya

в $\handcolor\infty$ при $\handcolor r = \infty$ и $\handcolor t = \infty$.
%v $\handcolor\infty$ pri $\handcolor r = \infty$ i $\handcolor t = \infty$.

Неудача этого вычисления с необходимостью приводит к тому
%Neudacha \protect{\`{e}}togo vychisleniya s neobkhodimost\cprime yu
%privodit k tomu

выводу, что не существует таких решений уравнений Борна; ко-
%vyvodu, chto ne sushchestvuet takikh resheni\u i uravneni\u i Borna; ko-

торые при всех $\handcolor r\ \textnormal{\it и}\ t$ были бы близки к 
соответствующим
%torye pri vsekh $\handcolor r\ {\it\cyrit i}\ t$ byli by blizki k 
%sootvet{s}tvuyushchim

Максвелловским решениям /\ref{eq42}/. Иначе говоря, для данного класса
%Maksvellovym resheniyam /\ref{eq42}/. Inache govorya, dlya dannogo klassa

задач, применяемый обычно метод возмущений / неизвестная ве-
%zadach, primenyaemy\u i obychno metod vozmushcheni\u i / neizvestnaya ve-

личина ищется как известная $\handcolor +$ маля добавка/ в таком вы-
%lichina ishchet{s}ya kak izvestnaya $\handcolor +$ malya dobavka/ v takom vi-

де не применим.
%de ne primenim.

Любопытно, что аналогичное положение вещей мы имеем и в
%Lyubopytno, chto analogichnoe polozhenie veshche\u i my imeem i v

некоторых нелинейных задачах классической механики. Рассмо-
%nekotorykh neline\u inykh zadachakh klassichesko\u i mekhaniki. Rassmo-

трим, напр.\ уравнение движения ангармонического осциллятора
%trim, napr.\ uravnenie dvizheniya angarmonicheskogo ostsillyatora

с силой, содержащей члены, пропорциональные
$\handcolor x\ \textnormal{\it и}\ x^3 :$ 
%s silo\u i, soderzhashche\u i chleny, proportsional\cprime nye 
%$\handcolor x\ {\it\cyrit i}\ x^3 :$ 
{\handcolor
\begin{eqnarray}
\refstepcounter{requation}
\label{eq44}
\ddot{x}\ +\ \omega^2 x&=&\kappa x^3  
\end{eqnarray}
}и попытаемся искать его решение в виде
%i popytaemsya iskat\cprime\ ego reshenie v vide
{\handcolor
\begin{eqnarray}
\refstepcounter{requation}
\label{eq45}
x&=&A\ \cos\omega t\ +\ x^\prime 
\end{eqnarray}
}

\pagebreak
\refstepcounter{ppage}

Считая $\handcolor\kappa\ \textnormal{\it и}\ x^\prime$ малыми величинами, 
мы для $\handcolor x^\prime$ получим
%Schitaya $\handcolor\kappa\ {\it\cyrit i}\ x^\prime$ malymi velichinami, 
%my dlya $\handcolor x^\prime$ poluchim

уравнение
%uravnenie
{\handcolor
\begin{eqnarray}
\refstepcounter{requation}
\label{eq46}
\ddot{x}^\prime\ +\ \omega^2 x^\prime&=&
\kappa\ A^3 \cos^3\omega t = \kappa\ A^3\ 
\frac{3\cos\omega t + \cos 3 \omega t}{4}                   
\end{eqnarray}
}Это уравнение можно формально рассматривать как уравнение вы-
%\protect{\`{E}}to uravnenie mozhno formal\cprime no rassmatrivat\cprime\
%kak uravnenie vy-

нужденных колебаний с силой, содержащей член с той же частотой
%nuzhdennykh kolebani\u i s silo\u i, soderzhashche\u i chlen s to\u i
%zhe chastoto\u i

какую имеют собственные колебания. В результате получается "ре-
%kakuyu imeyut sobstvennye kolebaniya. V rezul\cprime tate poluchaet{s}ya "re-

зонанс"\protect{,} т.е.\ $\handcolor x^\prime$ оказывается пропоциональным 
$\handcolor t$ и при $\handcolor t \rightarrow \infty$
%zonans", t.e.\ $\handcolor x^\prime$ okazyvaet{s}ya propotsional\cprime nym 
%$\handcolor t$ i pri $\handcolor t \rightarrow \infty$

мы получаем расходящееся решение, не имеющее физического смысла
%my poluchaem raskhodyashcheesya reshenie, ne imeyushchee fizicheskogo
%smysla

и указывающее, что искать $\handcolor x$ в виде /\ref{eq45}/ нельзя.
%i ukazyvayushchee, chto iskat\cprime\ $\handcolor x$ v vide /\ref{eq45}/ 
%nel\cprime zya.

В данном случае хорошо известно, каким образом нужно изменить
%V dannom sluchae khorosho izvestno, kakim obrazom nuzhno izmenit\cprime

метод решения уравнения /\ref{eq44}/. А именно нужно изменить само
%metod resheniya uravneniya /\ref{eq44}/. A imenno nuzhno izmenit\cprime\
%samo

нулевое приближение и искать решение в виде
%nulevoe priblizhenie i iskat\cprime\ reshenie v vide
{\handcolor
\begin{eqnarray}
\refstepcounter{requation}
\label{eq47}
x&=&A\ \cos\nu t\ +\ x^\prime                  
\end{eqnarray}
}где $\handcolor\nu$ отличается /хотя и мало/ от $\handcolor\omega$ 
и должно быть подоб-
%gde $\handcolor\nu$ otlichaet{s}ya /khotya i malo/ ot $\handcolor\omega$ 
%i dolzhno byt\cprime\ podob-

рано так, чтобы в уравнении для $\handcolor x^\prime$ член, 
дающий "резонанс"\protect{,}
%rano tak, chtoby v uravnenii dlya $\handcolor x^\prime$ chlen, 
%dayushchi\u i "rezonans",

пропадал. Как нетрудно видеть, для этого нужно, в I-м прибли-
%propadal. Kak netrudno videt\cprime , dlya \protect{\`{e}}togo nuzhno,
%v {\tt I}-m pribli-

жении положить
%zhenii polozhit\cprime\
{\handcolor
\begin{eqnarray}
\refstepcounter{requation}
\label{eq48}
\nu&=&\omega \left(1 - \frac{3}{8} \frac{\kappa A^2}{\omega^2}\right)       
\end{eqnarray}
}Если $\handcolor\nu$ выбрано так, то уравнение для $\handcolor x^\prime$
будет при этом
%Esli $\handcolor\nu$ vybrano tak, to uravnenie dlya $\handcolor x^\prime$ 
%budet pri \protect{\`{e}}tom

иметь решение, все время фактически остающееся малым.
%imet\cprime\ reshenie, vse vremya fakticheski ostayushcheesya malym.

Этот простой пример дает ключ для решения нашей задачи. Будем
%\protect{\`{E}}tot prosto\u i primer daet klyuch dlya resheniya nashe\u i
%zadachi. Budem

искать решение уравнений Борна не в виде /\ref{eq43}/, а попробуем как
%iskat\cprime\ reshenie uravneni\u i Borna ne v vide  /\ref{eq43}/, a
%poprobuem kak-

то видоизменить само нулевое приближение, т.е.\ предположим, что
%to vidoizmenit\cprime\ samo nulevoe priblizhenie, t.e.\ predpolozhim, chto

присутствие постоянного электрического поля как-то изменяет
%prisut{s}tvie postoyannogo \protect{\`{e}}lektricheskogo polya kak-to
%izmenyaet

свойства самой исходной плоской волны. Естественнее всего счи-
%svo\u istva samo\u i iskhodno\u i plosko\u i volny. Estestvennee vsego schi-

тать, что это поле вносит некоторую анизотропию в пространство
%tat\cprime , chto \protect{\`{e}}to pole vnosit nekotoruyu anizotropiyu v
%prostranstvo

и попытаться удовлетворить уравнениям Борна выражениями, анало-
%i popytat\cprime sya udovletvorit\cprime\ uravneniyam Borna vyrazheniyami,
%analo-

гичным решениям уравнений Максвелла для света, распространяю-
%gichnym resheniyam uravneni\u i Maksvella dlya sveta, rasprostranyayu-

\pagebreak
\refstepcounter{ppage}

щегося в одноосном кристалле.
%shchegosya v odnoosnom kristalle.

Выберем направление волнового вектора за ось
%Vyberem napravlenie volnovogo vektora za os\cprime\ 
$\handcolor ox\ ,$ 
а плос-
%a plos-

кость, образованную этим вектором и вектором внешнего поля
%kost\cprime , obrazovannuyu \protect{\`{e}}tim vektorom i vektorom
%vneshnego polya 
$\handcolor N$

за плоскость
%za ploskost\cprime\ 
$\handcolor xy$.

Обозначим через $\handcolor\alpha$ угол между этими векторами. Относительно
%Oboznachim cherez $\handcolor\alpha$ ugol mezhdu \protect{\`{e}}timi 
%vektorami. Otnositel\cprime no

компонент поля волны предположим пока только, что они являются
%komponent polya volny predpolozhim poka tol\cprime ko, chto oni
%yavlyayut{s}ya

периодическими функциями $\handcolor x\ \textnormal {\it и}\ t\ ,$
т.е.\ будем искать решение
%periodicheskimi funktsiyami $\handcolor x\ {\it\cyrit i}\ t\ ,$ 
%t.e.\ budem iskat\cprime\ reshenie

уравнений /\ref{eq25}/ в виде
%uravneni\u i /\ref{eq25}/ v vide
{\handcolor
\begin{eqnarray}
\refstepcounter{requation}
\label{eq49}
&\left.\begin{array}{rcl}
\hspace{-0.9cm}E_x&=&
E^0_{\ x} \cos\omega_1 
\left(t - \frac{\displaystyle x}{\displaystyle v_1} + \varphi_1\right)
 + N \cos\alpha\ =\ E^0_{\ x} \cos\gamma_1 + N \cos\alpha\\[0.3cm]
\hspace{-0.9cm}E_y&=&
E^0_{\ y} \cos\omega_2 
\left(t - \frac{\displaystyle x}{\displaystyle v_2} + \varphi_2\right)
+ N \sin\alpha\ =\ E^0_{\ y} \cos\gamma_2 + N \sin\alpha\\[0.3cm]
\hspace{-0.9cm}E_z&=&
E^0_{\ z} \cos\omega_3 
\left(t - \frac{\displaystyle x}{\displaystyle v_3} + \varphi_3\right)
\ =\ E^0_{\ z} \cos\gamma_3\\[0.3cm]
\hspace{-0.9cm}B_x&=&
B^0_{\ x} \cos\nu_1 
\left(t - \frac{\displaystyle x}{\displaystyle w_1} + \theta_1\right)
\ =\ B^0_{\ x} \cos\gamma_4\\[0.3cm]
\hspace{-0.9cm}B_y&=&
B^0_{\ y} \cos\nu_2 
\left(t - \frac{\displaystyle x}{\displaystyle w_2} + \theta_2\right)
\ =\ B^0_{\ y} \cos\gamma_5\\[0.3cm]
\hspace{-0.9cm}B_z&=&
B^0_{\ z} \cos\nu_3 
\left(t - \frac{\displaystyle x}{\displaystyle w_3} + \theta_3\right)
\ =\ B^0_{\ z} \cos\gamma_6\hspace{5.5cm},\hspace{-5.5cm}
\end{array}\right\}&\
\end{eqnarray}
}где $\handcolor\varphi\ \textnormal{\it и}\ \theta$
%gde $\handcolor\varphi\ {\it\cyrit i}\ \theta$ 
постоянные фазы.
%postoyannye fazy.

Таким образом, мы предполагаем заранее, что разные компоненты
%Takim obrazom, my predpolagaem zaranee, chto raznye komponenty

поля волны могут иметь различные амплитуды, частоты, скорости
%polya volny mogut imet\cprime\ razlichnye amplitudy, chastoty, skorosti

и начальные фазы.
%i nachal\cprime nye fazy.

Мы увидим, что уравнениям Борна можно удовлетворить решениями
%My uvidim, chto uravneniyam Borna mozhno udovletvorit\cprime\
%resheniyami

типа /\ref{eq49}/ если распорядиться соответствующим образом входя-
%tipa /\ref{eq49}/ esli rasporyadit\cprime sya sootvet{s}tvuyushchim 
%obrazom vkhodya-

щими туда параметрами; при этом, как окажется, мы можем не
%shchimi tuda parametrami; pri \protect{\`{e}}tom, kak okazhet{s}ya, my
%mozhem ne

добавлять к ним еще аддитивных поправок
%dobavlyat\cprime\ k nim eshche additivnykh popravok 
$\handcolor E^\prime_x\ ,\ E^\prime_y$ 
и т.д..
%i t.d..

\pagebreak
\refstepcounter{ppage}

Подставим /\ref{eq49}/ в /\ref{eq25}/ и, производя вычисления с точностью до
%Podstavim /\ref{eq49}/ v /\ref{eq25}/ i, proizvodya vychisleniya s 
%tochnost\cprime yu do

низшей степени малых величин, определим соотношения между ампли-
%nizshe\u i stepeni malykh velichin, opredelim sootnosheniya mezhdu ampli-

тудами, соотношения между частотами и скорости, входящие в
%tudami, sootnosheniya mezhdu chastotami i skorosti, vkhodyashchie v

/\ref{eq49}/. Заметим, при этом, что мы можем считать малыми все величи-
%/\ref{eq49}/. Zametim, pri \protect{\`{e}}tom, chto my mozhem
%schitat\cprime\ malymi vse velichi-

ны типа
%ny tipa 
$\handcolor\frac{\displaystyle v_1 -c}{\displaystyle c}$,
$\handcolor\frac{\displaystyle E^0_{\ x}}{\displaystyle\sqrt{E^{0\, 2}_{\ y} +
E^{0\, 2}_{\ z}}}$ и.т.д.
%i.t.d. 
{\ith ,}

вообще все постоянные величины, характеризующие отклонения
%voobshche vse postoyannye velichiny, kharakterizuyushchie otkloneniya

нашего решения от Максвелловского.
%nashego resheniya ot Maksvellovskogo.

\hspace{1cm}Перепишем, в нашем приближении, уравнения Борна /\ref{eq25}/:
%Perepishem, v nashem priblizhenii, uravneniya Borna /\ref{eq25}/:
{\handcolor
\begin{eqnarray}
\refstepcounter{requation}
\label{eq50}
\hspace{-2cm}&&\
rot\, E\ +\ \frac{\displaystyle 1}{\displaystyle c}\, 
\dot{B}\ =\ 0\hspace{0.5cm};\hspace{0.5cm}
div\, B\ =\ 0\\[0.3cm]
\refstepcounter{requation}
\label{eq51}
\hspace{-2cm}&&\left.\begin{array}{l}
rot\, B\ -\ \frac{\displaystyle 1}{\displaystyle c}\, \dot{E} = 
\frac{\displaystyle 1}{\displaystyle 2}\, \left\{\left[grad\, F,B\right]
\ -\ \frac{\displaystyle 1}{\displaystyle 2}\, \dot{F}E\right\}\\[0.3cm]
div\, E = \frac{\displaystyle 1}{\displaystyle 2}\, \left(grad\, F\cdot E\right)
\hspace{5cm},\hspace{-5cm}
\end{array}\right\}\hspace{3cm}\ \\[0.3cm]
\refstepcounter{requation}
\label{eq52}
\ \hspace{-1.35cm}\textnormal{\tt\color{black}где}
%{\tt\cyrrm{\color{black}gde}}
\hspace{1.35cm}&&\hspace{3cm}
F\ =\ \frac{\displaystyle B^2 - E^2}{\displaystyle b^2}
\end{eqnarray}
}Уравнение
%Uravnenie 
$\handcolor div\, B = 0$ 
дает нам
%daet nam
{\handcolor
\begin{eqnarray}
\refstepcounter{requation}
\label{eq53}
B^0_x&=&0 
\end{eqnarray}
}Подставляем теперь /\ref{eq49}/ в оставшиеся уравнения /\ref{eq50}/
%Podstavlyaem teper\cprime\ /\ref{eq49}/ v ostavshiesya uravneniya /\ref{eq50}/

Первое из них:
%Pervoe iz nikh:
{\handcolor
\begin{eqnarray*}
\frac{\partial E_z}{\partial y}\, -\, \frac{\partial E_y}{\partial z}
\ +\ \frac{1}{c}\, \dot{B_x}&=&0
\end{eqnarray*}
}удовлетворяется автоматически.
%udovletvoryaet{s}ya avtomaticheski.

Второе:
%Vtoroe:
{\handcolor
\begin{eqnarray*}
\frac{\partial E_x}{\partial z}\, -\, \frac{\partial E_z}{\partial x}
\ +\ \frac{1}{c}\, \dot{B_y}&=&0
\end{eqnarray*}
}дает
%daet
{\handcolor
\begin{eqnarray*}
-\ \frac{\omega_3}{v_3}\, E^0_z\, \sin\gamma_3
\ -\ \frac{\nu_2}{c}\, B^0_y\, \sin\gamma_5&=&0
\end{eqnarray*}
}

\pagebreak
\refstepcounter{ppage}
\label{page40}

Откуда
%Otkuda
{\handcolor
\begin{eqnarray}
\refstepcounter{requation}
\label{eq54}
&\left.\begin{array}{rcl}
\nu_2\ =\ \omega_3&;&w_2\ =\ v_3\ ;\ \theta_2\ =\ \varphi_3\\[0.3cm]
B^0_y&=&-\ \frac{\displaystyle c}{\displaystyle v_3}\ E^0_z
\end{array}\right\}&
\end{eqnarray}
}Третье:
%Tret\cprime e:
{\handcolor
\begin{eqnarray*}
\frac{\partial E_y}{\partial x}\, -\, \frac{\partial E_x}{\partial y}
\ +\ \frac{1}{c}\, \dot{B_z}&=&0
\end{eqnarray*}
}дает
%daet
{\handcolor
\begin{eqnarray*}
\frac{\omega_2}{v_2}\, E^0_y\, \sin\gamma_2
\ -\ \frac{\nu_3}{c}\, B^0_z\, \sin\gamma_6&=&0
\end{eqnarray*}
}Откуда
%Otkuda
{\handcolor
\begin{eqnarray}
\refstepcounter{requation}
\label{eq55}
&\left.\begin{array}{rcl}
\nu_3\ =\ \omega_2&;&w_3\ =\ v_2\ ;\ \theta_3\ =\ \varphi_2\\[0.3cm]
B^0_z&=&-\ \frac{\displaystyle c}{\displaystyle v_2}\ E^0_y
\end{array}\right\}&
\end{eqnarray}
}Принимая во внимание /\ref{eq53}/, /\ref{eq54}/ и /\ref{eq55}/, мы 
можем переписать /\ref{eq49}/
%Prinimaya vo vnimanie /\ref{eq53}/, /\ref{eq54}/ i /\ref{eq55}/, my 
%mozhem perepisat\cprime\  /\ref{eq49}/

в виде
%v vide
{\handcolor
\begin{eqnarray}
\refstepcounter{requation}
\label{eq56}
&\left.\begin{array}{rcl}
\hspace{-0.9cm}E_x&=&
p \cos\nu \left(t - \frac{\displaystyle x}{\displaystyle w} + \varphi_1\right)
 + N \cos\alpha\ =\ p \cos\gamma_1 + N \cos\alpha\\[0.3cm]
\hspace{-0.9cm}E_y&=&
L \cos\omega \left(t - \frac{\displaystyle x}{\displaystyle v} 
+ \varphi_2\right) + N \sin\alpha\ =\ L \cos\gamma_2 + N \sin\alpha\\[0.3cm]
\hspace{-0.9cm}E_z&=&
L^\prime \cos\omega^\prime 
\left(t - \frac{\displaystyle x}{\displaystyle v^\prime} + \varphi_3\right)
\ =\ L^\prime \cos\gamma_3\\[0.3cm]
\hspace{-0.9cm}B_x&=&0\\[0.3cm]
\hspace{-0.9cm}B_y&=&M^\prime \cos\gamma_3\\[0.3cm]
\hspace{-0.9cm}B_z&=&M \cos\gamma_2
\end{array}\right\}&
\end{eqnarray}
}где введены обозначения:
%gde vvedeny oboznacheniya:

-------------------------------------

I/ Некоторые буквы здесь употреблялись уже для других величин
%{\tt I}/ Nekotorye bukvy zdes\cprime\ upotreblyalis\cprime\ uzhe
%dlya drugikh velichin

в гл. {\tt\ref{glavaI}}. Однако, это, конечно, не вне сет путаницы.
%v gl.\ {\tt\ref{glavaI}}. Odnako, \protect{\`{e}}to, konechno, ne vne 
%set putanitsy.

\pagebreak
\refstepcounter{ppage}

\hspace{1.2cm}$\handcolor E^0_x = p$\hspace{0.7cm}{\ith ,} 
$\handcolor\omega_1 = \nu$\hspace{1.3cm}
{\ith и т.д.}
%{\ith\cyrit{i t.d.}}

При этом
%Pri \protect{\`{e}}tom
{\handcolor
\begin{eqnarray}
\refstepcounter{requation}
\label{eq57}
M&=&\frac{c}{v}\, L\hspace{0.7cm};\hspace{0.7cm}
M^\prime\ =\ - \frac{c}{v^\prime}\, L^\prime
\end{eqnarray}
}Подставим /\ref{eq56}/ во вторую систему уравнений 
Борна /\ref{eq51}/. Вычислим
%Podstavim /\ref{eq56}/ vo vtoruyu sistemu uravneni\u i
%Borna /\ref{eq51}/. Vychislim 

сначала, в рамках нашего приближения, правые части уравне-
%snachala, v ramkakh nashego priblizheniya, pravye chasti uravne-

ний. Найдем
%ni\u i. Na\u idem
{\handcolor
\begin{eqnarray*}
F&=&\frac{\displaystyle B^2 - E^2}{\displaystyle b^2}
\end{eqnarray*}
}Т.к.\ $\handcolor F$ содержит в знаменателе $\handcolor b^2$ т.е.\ 
большую вели-
%T.k.\ $\handcolor F$ soderzhit v znamenatele $\handcolor b^2$ t.e.\ 
%bol\cprime shuyu veli-

чину, то мы можем, вычисляя $\handcolor F$ положить
%chinu, to my mozhem, vychislyaya $\handcolor F$ polozhit\cprime
{\handcolor
\begin{eqnarray*}
p&=&0\hspace{0.7cm},\ M\ = L\hspace{0.7cm},\hspace{0.7cm}
M^\prime\ =\ -L^\prime
\end{eqnarray*}
}Тогда
%Togda
{\handcolor
\begin{eqnarray}
\refstepcounter{requation}
\label{eq58}
F&=&-\ \frac{1}{b^2}\ \left(2 L N\, \sin\alpha\, \cos\gamma_2 + N^2\right)
\end{eqnarray}
}Найдем составляющие
%Na\u idem sostavlyayushchie 
$\handcolor grad\; F$:
{\handcolor
\begin{eqnarray}
\refstepcounter{requation}
\label{eq59}
&\left.\begin{array}{rcl}
\frac{\displaystyle\partial F}{\displaystyle\partial x}&=&
- 2\ \frac{\displaystyle\omega}{\displaystyle v}\ 
\frac{\displaystyle L N}{\displaystyle b^2}\
\sin\alpha\, \sin\gamma_2 
\\[0.3cm]
\frac{\displaystyle\partial F}{\displaystyle\partial y}&=&
\frac{\displaystyle\partial F}{\displaystyle\partial z}\ =\ 0
\end{array}\right\}&
\end{eqnarray}
}и наконец, определим
%i nakonets, opredelim 
$\handcolor\dot{F}$:
{\handcolor
\begin{eqnarray}
\refstepcounter{requation}
\label{eq60}
\dot{F}&=&2 \omega\ \frac{L N}{b^2}\ \sin\alpha\, \sin\gamma_2
\end{eqnarray}
}Обозначим правые части уравнений /\ref{eq51}/ формально через
$\handcolor\frac{\displaystyle 1}{\displaystyle c}\, j\ 
\textnormal{\it и}\ \rho$.
%Oboznachim pravye chasti uravneni\u i /\ref{eq51}/ formal\cprime no
%cherez $\handcolor\frac{\displaystyle 1}{\displaystyle c}\, j\ 
%{\it\cyrit{i}}\ \rho$.
 
Тогда
%Togda
{\handcolor
\begin{eqnarray}
\refstepcounter{requation}
\label{eq61}
&\left.\begin{array}{rcl}
\frac{\displaystyle 1}{\displaystyle c}\, j_x&=&
- \frac{\displaystyle\omega}{\displaystyle c}\ 
L\ \frac{\displaystyle N^2}{\displaystyle b^2}\ \sin\alpha\,
\cos\alpha\, \sin\gamma_2\\[0.3cm]
\frac{\displaystyle 1}{\displaystyle c}\, j_y&=&
- \frac{\displaystyle\omega}{\displaystyle c}\ 
L\ \frac{\displaystyle N^2}{\displaystyle b^2}\ \sin^2\alpha\, \sin\gamma_2
\\[0.3cm]
\frac{\displaystyle 1}{\displaystyle c}\, j_z&=&0
\end{array}\right\}&
\end{eqnarray}
}

\pagebreak
\refstepcounter{ppage}

{\handcolor
\begin{eqnarray*}
\ \hspace{2cm}\rho&=&- \frac{\displaystyle\omega}{\displaystyle c}\ 
L\ \frac{\displaystyle N^2}{\displaystyle b^2}\ \sin\alpha\,
\cos\alpha\, \sin\gamma_2\hspace{6.3cm}(61a)
\end{eqnarray*}
}Первое из уравнений /\ref{eq51}/
%Pervoe iz uravneni\u i /\ref{eq51}/
{\handcolor
\begin{eqnarray*}
\frac{\displaystyle\partial B_z}{\displaystyle\partial y}
\ -\ \frac{\displaystyle\partial B_y}{\displaystyle\partial z}
\ -\ \frac{1}{c}\, \dot{E}_x&=&\frac{1}{c}\, j_x
\end{eqnarray*}
}дает\hspace{2cm}
%daet\hspace{2cm} 
$\handcolor\frac{\displaystyle \nu}{\displaystyle c}\ p\ \sin\gamma_1 = 
- \frac{\displaystyle \omega}{\displaystyle c}\ L\
\frac{\displaystyle N^2}{\displaystyle b^2}\ \sin\alpha\, 
\cos\alpha\,\sin\gamma_2$

{\handcolor
{\it Откуда}
%{\it\cyrit{Otkuda}}
%
\begin{eqnarray}
\refstepcounter{requation}
\label{eq62}
&\left.\begin{array}{rcl}
\nu&=&\omega\hspace{0.5cm};\ \ w\ =\ v\hspace{0.5cm};\ 
\varphi_1\ =\ \varphi_2\\[0.3cm]
p&=&-\ L\ \frac{\displaystyle N^2}{\displaystyle b^2}\ \sin\alpha\, \cos\alpha
\end{array}\right\}&
\end{eqnarray}
}Таким образом продольная компонента электрического поля волны
%Takim obrazom prodol\cprime naya komponenta \protect{\`{e}}lektricheskogo
%polya volny

получается отличной от нуля.
%poluchaet{s}ya otlichno\u i ot nulya.

Второе уравнение
%Vtoroe uravnenie
{\handcolor
\begin{eqnarray*}
\frac{\displaystyle\partial B_x}{\displaystyle\partial z}
\ -\ \frac{\displaystyle\partial B_z}{\displaystyle\partial x}
\ -\ \frac{1}{c}\, \dot{E}_y\ =\ \frac{1}{c}\, j_y
\end{eqnarray*}
}дает
%daet
{\handcolor
\begin{eqnarray*}
-\ \frac{\omega}{v}\ M\ \sin\gamma_2\ +\ \frac{\omega}{c}\ L\ \sin\gamma_2&=&
-\ \frac{\omega}{c}\ L\ \frac{N^2}{b^2}\ \sin^2\alpha\,\sin\gamma_2
\end{eqnarray*}
}откуда определим
%otkuda opredelim 
$\handcolor v$:
{\handcolor
\begin{eqnarray}
\refstepcounter{requation}
\label{eq63}
\frac{v^2}{c^2}&=&1\ -\ \frac{N^2}{b^2}\ \sin^2\alpha
\end{eqnarray}
}Третье уравнение
%Tret\cprime e uravnenie
{\handcolor
\begin{eqnarray*}
\frac{\displaystyle\partial B_y}{\displaystyle\partial x}
\ -\ \frac{\displaystyle\partial B_x}{\displaystyle\partial y}
\ -\ \frac{1}{c}\, \dot{E}_z&=&0
\end{eqnarray*}
}дает
%daet
{\handcolor
\begin{eqnarray*}
\frac{\omega^\prime}{v^\prime}\ M^\prime\ \sin\gamma_3\ +\ 
\frac{\omega^\prime}{c}\ L^\prime\ \sin\gamma_3&=&0
\end{eqnarray*}
}откуда, принимая во внимание /\ref{eq57}/ получим
%otkuda, prinimaya vo vnimanie /\ref{eq57}/ poluchim
{\handcolor
\begin{eqnarray}
\refstepcounter{requation}
\label{eq64}
v^\prime&=&c
\end{eqnarray}
}

\pagebreak
\refstepcounter{ppage}

Последнее уравнение $\handcolor div\; E = \rho$ дает:
%Poslednee uravnenie $\handcolor div\; E = \rho$  daet:
{\handcolor
\begin{eqnarray*}
\frac{\nu}{w}\ p\ \sin\gamma_1&=& 
-\ \frac{\omega}{c}\ L\ \frac{N^2}{b^2}\ \sin\alpha\,
\cos\alpha\, \sin\gamma_2\ \ \ \ ,
\end{eqnarray*}
}что автоматически удовлетворяется, в силу /\ref{eq62}/.
%chto avtomaticheski udovletvoryaet{s}ya, v silu /\ref{eq62}/.

Итак, принимая во внимание /\ref{eq62}/,  /\ref{eq63}/ и /\ref{eq64}/,
мы окончательно
%Itak, prinimaya vo vnimanie /\ref{eq62}/, /\ref{eq63}/ i /\ref{eq64}/,
%my okonchatel\cprime no

получаем следующее решение уравнений Борна:
%poluchaem sleduyushchee reshenie uravneni\u i Borna:
{\handcolor
\begin{eqnarray}
\refstepcounter{requation}
\label{eq65}
&\left.\begin{array}{rcl}
E_x&=&p\ \cos\omega 
\left(t - \frac{\displaystyle x}{\displaystyle v} + \varphi\right)
\ +\ N\ \cos\alpha\\[0.3cm]
E_y&=&L\ \cos\omega 
\left(t - \frac{\displaystyle x}{\displaystyle v} + \varphi\right)
\ +\ N\ \sin\alpha\\[0.3cm]
E_z&=&L^\prime\ \cos\omega^\prime 
\left(t - \frac{\displaystyle x}{\displaystyle c} + \varphi^\prime\right)
\\[0.3cm]
B_x&=&0\\[0.3cm]
B_y&=&- L^\prime\ \cos\omega^\prime 
\left(t - \frac{\displaystyle x}{\displaystyle c} + \varphi^\prime\right)
\\[0.3cm]
B_z&=&M \ \cos\omega 
\left(t - \frac{\displaystyle x}{\displaystyle v} + \varphi\right)
\hspace{4cm},\hspace{-4cm}
\end{array}\right\}&\
\end{eqnarray}
}где
%gde
{\handcolor
\begin{eqnarray}
\refstepcounter{requation}
\label{eq66}
&\left.\begin{array}{rcl}
M&=&\frac{\displaystyle c}{\displaystyle v}\, L\\[0.3cm]
\frac{\displaystyle v^2}{\displaystyle c^2}&=&
1\, -\, \frac{\displaystyle N^2}{\displaystyle b^2}\ \sin^2\alpha\\[0.3cm]
p&=&- L\ \frac{\displaystyle N^2}{\displaystyle b^2}\ \sin\alpha\, \cos\alpha
\hspace{1.5cm},\hspace{-1.5cm}
\end{array}\right\}&\
\end{eqnarray}
}причем $\handcolor L,\ L^\prime,\ \omega,\ \omega^\prime,\ \varphi\ 
\textnormal{\it и}\ \varphi^\prime$ остаются произвольными
%prichem $\handcolor L,\ L^\prime,\ \omega,\ \omega^\prime,\ \varphi\ 
%{\it\cyrit{i}}\ \varphi^\prime$ ostayut{s}ya proizvol\cprime nymi\ .

Сравнение этих формул с известными формулами электромагнитной
%Sravnenie \protect{\`{e}}tikh formul s izvestnymi formulami 
%\protect{\`{e}}lektromagnitno\u i

кристаллооптики дает возможность выразить их содержание сле-
%kristallooptiki daet vozmozhnost\cprime\ vyrazit\cprime\ ikh soderzhanie sle-

дующим наглядным образом: в присутствии однородного электро-
%duyushchim naglyadnym obrazom: v prisut{s}tvii odnorodnogo 
%\protect{\`{e}}lektro-

статического поля силы $\handcolor N$ векторы 
$\handcolor E\ \textnormal{\it и}\ B$ плоской
%staticheskogo polya sily $\handcolor N$ vektory 
%$\handcolor E\ {\it\cyrit{i}}\ B$ plosko\u i

\hspace{2cm}световой волны ведут себя, в первом приближении, так
%svetovo\u i volny vedut sebya, v pervom priblizhenii, tak

же как они вели бы себя в одноосном кристалле, оптическая ось
%zhe kak oni veli by sebya v odnoosnom kristalle, opticheskaya os\cprime\

которого была бы направлена по $\handcolor N$ а главное диэлектрические по-
%kotorogo byla by napravlena po $\handcolor N$ a glavnye 
%di\protect{\`{e}}lektricheskie po-

\ стоянные равнялись бы соответственно
%stoyannye ravnyalis\cprime\ by sootvet{s}tvenno
{\handcolor
\begin{eqnarray}
\refstepcounter{requation}
\label{eq67}
\epsilon&=&1\ +\ \frac{N^2}{b^2}
\end{eqnarray}
}и единице.
%i edinitse.

\pagebreak
\refstepcounter{ppage}

В самом деле, решение /\ref{eq65}/ можно представить себе состоящим
%V samom dele, reshenie /\ref{eq65}/ mozhno predstavit\cprime\ sebe
%sostoyashchim

из поля"обыкновенного"\ луча:
%iz polya"obyknovennogo" lucha:
{\handcolor
\begin{eqnarray*}
E_x&=\ 0\hspace{4.85cm}&B_x\ =\ 0\\[0.3cm]
E_y&=\ 0\hspace{4.85cm}&B_x\ =\ - L^\prime\ 
\cos\omega^\prime 
\left(t - \frac{\displaystyle x}{\displaystyle c} 
+ \varphi^\prime\right)\\[0.3cm]
E_z&=\ L^\prime\ 
\cos\omega^\prime 
\left(t - \frac{\displaystyle x}{\displaystyle c} 
+ \varphi^\prime\right)\hspace{1cm}&B_z\ =\ 0
\end{eqnarray*}
}и "необыкновенного":
%i "neobyknovennogo":
{\handcolor
\begin{eqnarray*}
E_x&=\ p\ \cos\omega \left(t - \frac{\displaystyle x}{\displaystyle v} 
+ \varphi\right)\hspace{1cm}&B_x\ =\ 0\\[0.3cm]
E_y&=\ L\ \cos\omega \left(t - \frac{\displaystyle x}{\displaystyle v} 
+ \varphi\right)\hspace{01cm}&B_y\ =\ 0\\[0.3cm]
E_z&=\ 0\hspace{4.6cm}&B_z\ =\ M\ \cos\omega 
\left(t - \frac{\displaystyle x}{\displaystyle v} + \varphi\right)\ ,
\end{eqnarray*}
}проходящих в поле плоского конденсатора\ .
%prokhodyashchikh v pole ploskogo kondensatora\ .

Из полученных нами формул /\ref{eq65}/ видно, между прочим, что при
%Iz poluchennykh nami formul /\ref{eq65}/ vidno, mezhdu prochim, chto pri

прохождении светового луча через конденсатор длины $\handcolor l$ между
%prokhozhdenii svetovogo lucha cherez kondensator dliny $\handcolor l$ mezhdu

$\handcolor y$ и $\handcolor z$ компонентами электрического 
поля этого луча нака-
%$\handcolor y$ i $\handcolor z$ komponentami \protect{\`{e}}lektricheskogo 
%polya \protect{\`{e}}togo lucha naka-

пливается известная разность фаз, равная
%plivaet{s}ya izvestnaya raznost\cprime\ faz, ravnaya
{\handcolor
\begin{eqnarray*}
\Delta\varphi&=&\omega\, l\ \left(\frac{1}{v} - \frac{1}{c}\right)
\ =\ \omega\, l\ \frac{c - v}{c v}
\end{eqnarray*}
}/где читаем, для простоты
%/gde chitaem, dlya prostoty 
{\handcolor$\varphi = \varphi^\prime$/.}

Замечая, что
%Zamechaya, chto 
$\handcolor\frac{\displaystyle v}{\displaystyle c} = 
1 - \frac{\displaystyle N^2}{\displaystyle 2 b^2}$\ ,\ \ \ 
/ в случае, если свет
%/ v sluchae, esli svet

распространяется перпендикулярно к полю конденсатора/ мы по-
%rasprostranyaet{s}ya perpendikulyarno k polyu kondensatora/ my po-

лучим:
%luchim:
{\handcolor
\begin{eqnarray*}
c - v&=&c\ \frac{N^2}{2 b^2}
\end{eqnarray*}
}и
%i
\setcounter{requation}{68}  % equationnumber 67-->68
{\handcolor
\begin{eqnarray}
\refstepcounter{requation}
\label{eq69}
\Delta\varphi&=&\frac{\omega}{c}\ l\ \frac{N^2}{2 b^2}\ =\ 
\pi\ \left(\frac{N}{b}\right)^2\ \frac{l}{\lambda}
\end{eqnarray}
}Иначе говоря, если луч, входя в конденсатор, является линей-
%Inache govorya, esli luch, vkhodya v kondensator, yavlyaet{s}ya line\u i-

но поляризованным, то он должен выйти из него поляризованным
%no polyarizovannym, to on dolzhen vy\u iti iz nego polyarizovannym 

эллиптический. О фактической проверке этого утверждения, ко-
%\protect{\`{e}}llipticheski. O faktichesko\u i proverke 
%\protect{\`{e}}togo utverzhdeniya, ko-

нечно, не может быть и речи, т.к.\
%nechno, ne mozhet byt\cprime\ i rechi, t.k.\ }
${\handcolor b \equiv}$ I0$^{\mbox{\large\stexttt{I6}}}$
{\ith CGSE} и эффект
%{\cyrtt {\ith CGSE} i \protect{\`{e}}ffekt 

\pagebreak
\refstepcounter{ppage}

слишком мал во всех практически осуществимых полях. Тем не
%slishkom mal vo vsekh prakticheski osushchestvimykh polyakh. Tem ne

менее, с чисто теоретической точки зрения, эффект, выражаемый
%menee, s chisto teoretichesko\u i tochki zreniya, 
%\protect{\`{e}}ffekt, vyrazhaemy\u i

формулой /\ref{eq69}/, представляется безусловно интересным потому,
%formulo\u i /\ref{eq69}/, predstavlyaet{s}ya bezuslovno interesnym 
%potomu,

что он дает пример совершенно своеобразного отклонения от зако-
%chto on daet primer sovershenno svoeobraznogo otkloneniya ot zako-

нов Максвелловской электродинамики: отклонение проявляется не
%nov Maksvellovsko\u i \protect{\`{e}}lektrodinamiki: otklonenie 
%proyavlyaet{s}ya ne

в атомных маштабах и не при сильных полях, а при сравнении на-
%v atomnykh mashtabakh i ne pri sil\cprime nykh polyakh, a pri
%sravnenii na-

блюдений, произведенных на
\udensdash{боль$\ \atop\ $\hspace{-0.33cm}ших\hspace{-0.05cm}\ }
расстояниях друг от друга.\\[-0.2cm]
%blyudeni\u i, proizvedennykh na 
%\udensdash{bol\cprime$\ \atop\ $\hspace{-0.33cm}shikh\hspace{-0.05cm}\ }
%rasstoyaniyakh drug ot druga.\\[-0.2cm]

В самом деле, при достаточно большом $\handcolor l$ величина
%V samom dele, pri dostatochno bol\cprime shom $\handcolor l$ velichina 
$\handcolor\Delta\varphi$ может
%mozhet

быть сколь угодно велика при сколь угодно малом
%byt\cprime\ skol\cprime\ ugodno velika pri skol\cprime\ ugodno malom
$\handcolor N/b$. Эта
%\protect{\`{E}}ta

черта теории существенно связана с ее нелинейностью: поправки,
%cherta teorii sushchestvenno svyazana s ee neline\u inost\cprime yu:
%popravki,

вносимые малыми нелинейными членами могут носить существенно
%vnosimye malymi neline\u inymi chlenami mogut nosit\cprime\ 
%sushchestvenno 

иной характер, чем малые линейные поправки.
%ino\u i kharakter, chem malye line\u inye popravki.

Исходя из формул /\ref{eq65}/, нетрудно вычислить для нашего реше-
%Iskhodya iz formul /\ref{eq65}/, netrudno vychislit\cprime\ dlya
%nashego reshe-

ния и вторую пару векторов Борна {\ith D} и Н. Для этого нужно
%niya i vtoruyu paru vektorov Borna {\ith D} i N. Dlya \protect{\`{e}}togo
%nuzhno

воспользоваться формулами
%vospol\cprime zovat\cprime sya formulami
{\handcolor
\begin{eqnarray*}
D&=&\frac{\displaystyle E}{\displaystyle\sqrt{1+F}}\hspace{1.2cm};
\hspace{1.2cm}
H\ =\ \frac{\displaystyle B}{\displaystyle\sqrt{1+F}}\ \ \ .\hspace{5cm}(23)
\end{eqnarray*}
}Положим для простоты $\handcolor\omega = \omega^\prime\ \textnormal{\it и}\
\varphi = \varphi^\prime = 0$.
%Polozhim dlya prostoty $\handcolor\omega = \omega^\prime\ {\it\cyrit{i}}\
%\varphi = \varphi^\prime = 0$.

Имеем
%Imeem:
{\handcolor
\begin{eqnarray}
\refstepcounter{requation}
\label{eq70}
\frac{\displaystyle 1}{\displaystyle\sqrt{1+F}}&=&
1 - \frac{1}{2}\, F\ =\ 
1\ +\ \frac{\displaystyle L N}{\displaystyle b^2}\ \sin\alpha\,
\cos\omega \left(t -\frac{\displaystyle x}{\displaystyle v}\right)
\ +\ \frac{N}{2 b^2}\ \ \ \ 
\end{eqnarray}
}Подставляя /\ref{eq65}/ в /\ref{eq23}/, мы получим
%Podstavlyaya /\ref{eq65}/ v /\ref{eq23}/, my poluchim
{\handcolor
\begin{eqnarray}
\refstepcounter{requation}
\label{eq71}
\hspace{-0.5cm}D_x&=&\frac{\displaystyle E_x}{\displaystyle\sqrt{1+F}} =
\left(p + L 
\frac{\displaystyle N}{\displaystyle b^2}\sin\alpha\,\cos\alpha\right) 
\cos\omega \left(t -\frac{\displaystyle x}{\displaystyle v}\right)
+ N \cos\alpha\cdot\left(1 + \frac{N^2}{2 b^2}\right)\ \ \ 
\end{eqnarray}
}Принимая во внимание /\ref{eq62}/, найдем для $\handcolor D_x$
следующее выраже-
%Prinimaya vo vnimanie /\ref{eq62}/, na\u idem dlya $\handcolor D_x$ 
%sleduyushchee vyrazhe-

ние
%nie
{\handcolor
\begin{eqnarray}
\refstepcounter{requation}
\label{eq72}
D_x&=&N\, \left(1 + \frac{N^2}{2 b^2}\right)\, \cos\alpha
\end{eqnarray}
}Аналогично
%Analogichno
{\handcolor
\begin{eqnarray}
\refstepcounter{requation}
\label{eq73}
\hspace{-2cm}D_y&=&\frac{\displaystyle E_y}{\displaystyle\sqrt{1+F}} =
N\,\left(1 + \frac{N^2}{2 b^2}\right)\,\sin\alpha +
L\,\left(1 + \frac{N^2}{2 b^2}\right)\, \cos\omega 
\left(t -\frac{\displaystyle x}{\displaystyle v}\right) 
+\ \ \ \nonumber\\[0.3cm]
&&+ N\,\frac{L^2}{b^2}\,\sin\alpha\
\frac{1 + \cos 2\omega 
\left(t -\frac{\displaystyle x}{\displaystyle v}\right)}{2}
+ L\,\frac{N^2}{b^2}\,\sin^2\alpha\,\cos\omega
\left(t -\frac{\displaystyle x}{\displaystyle v}\right)
\end{eqnarray}
}

\pagebreak
\refstepcounter{ppage}
\label{page46}

Далее
%Dalee
{\handcolor
\begin{eqnarray}
\refstepcounter{requation}
\label{eq74}
D_z&=&\frac{\displaystyle E_z}{\displaystyle\sqrt{1+F}} =
L^\prime\,\left(1 + \frac{N^2}{2 b^2}\right)\,\cos\omega 
\left(t -\frac{\displaystyle x}{\displaystyle c}\right) 
+\ \ \ \nonumber\\[0.3cm]
&&+ N\,\frac{L L^\prime}{b^2}\,\sin\alpha\
\frac{\cos 2\omega 
\left[t -\frac{\displaystyle x}{\displaystyle 2}
\left(\frac{\displaystyle 1}{\displaystyle c} + 
\frac{\displaystyle 1}{\displaystyle v}\right)\right] +
\cos\omega\, x \left(\frac{\displaystyle 1}{\displaystyle c} - 
\frac{\displaystyle 1}{\displaystyle v}\right)}{2}\ \ \ 
\end{eqnarray}
}Точно так-же можно показать, что
%Tochno tak-zhe mozhno pokazat\cprime , chto
{\handcolor
\begin{eqnarray}
\refstepcounter{requation}
\label{eq75}
H_x&=&0\\[0.3cm]
\refstepcounter{requation}
\label{eq76}
H_y&=&- L^\prime\,\left(1 + \frac{N^2}{2 b^2}\right)\,\cos\omega 
\left(t -\frac{\displaystyle x}{\displaystyle c}\right) 
-\ \ \ \nonumber\\[0.3cm]
&&- N\,\frac{L L^\prime}{b^2}\,\sin\alpha\
\frac{\cos 2\omega 
\left[t -\frac{\displaystyle x}{\displaystyle 2}
\left(\frac{\displaystyle 1}{\displaystyle c} + 
\frac{\displaystyle 1}{\displaystyle v}\right)\right] +
\cos\omega\, x \left(\frac{\displaystyle 1}{\displaystyle v} - 
\frac{\displaystyle 1}{\displaystyle c}\right)}{2}\ \ \ \\[0.3cm]
\refstepcounter{requation}
\label{eq77}
H_z&=&\frac{\displaystyle c}{\displaystyle v}\, L
\,\left(1 + \frac{N^2}{2 b^2}\right)\,
\cos\omega\left(t - \frac{\displaystyle x}{\displaystyle v}\right)
+ N\,\frac{L^2}{b^2}\,\sin\alpha\
\frac{1 + 
\cos 2\omega\left(t - \frac{\displaystyle x}{\displaystyle v}\right)}{2}
\ \ \ \ \ 
\end{eqnarray}
}Из полученных формул /\ref{eq72}/-/\ref{eq77}/ мы видим, что
\udensdash{продольные\hspace{-0.05cm}\ }
%Iz poluchennykh formul /\ref{eq72}/-/\ref{eq77}/ my vidim, chto
%\udensdash{prodol\cprime nye\hspace{-0.05cm}\ }

компоненты обоих векторов $\handcolor D_x$ и $\handcolor H_x$ нашей
{\handcolor\underline{\color{black}волны}} оказываются
%komponenty oboikh vektorov $\handcolor D_x$ i $\handcolor H_x$ nashe\u i
%{\handcolor\underline{\color{black}volny}} okazyvayut{s}ya

равными нулю$^{\mbox{\large\stexttt{I/}}}$.
В самом деле, легко видеть, что формула /\ref{eq72}/
%ravnymi nulyu}$^{\mbox{\large\stexttt{I/}}}$. 
%{\cyrtt V samom dele, legko videt\cprime , chto formula /\ref{eq72}/

дает величину $\handcolor D_x$ обусловленную только полем конденсатора
%daet velichinu $\handcolor D_x$ obuslovlennuyu tol\cprime ko polem kondensatora

$\handcolor E_x = N \cos\alpha\ ,$ т.к.\ в этом случае
% t.k.\ v \protect{\`{e}}tom sluchae
{\handcolor
\begin{eqnarray*}
D_x&=&\frac{\displaystyle N \cos\alpha}{\displaystyle\sqrt{1+F}}
\ =\ \frac{\displaystyle N \cos\alpha}{\displaystyle\sqrt{1- N^2/b^2}}
\ =\ N\,\left(1 + \frac{N^2}{2 b^2}\right)\,\cos\alpha\ . 
\end{eqnarray*}
}Этот факт показывает, что направление вектора Пойнтинга волны-
%\protect{\`{E}}tot fakt pokazyvaet, chto napravlenie
%vektora Po\u intinga volny-

напомним, что в теории Борна вектор Пойнтинга определяется
%napomnim, chto v teorii Borna vektor Po\u intinga opredelyaet{s}ya

произведением
$\handcolor\left[D\ \textnormal{\color{black}\tt х В}\right]$
- в данном случае, в отличие от кристал- 
%proizvedeniem 
%$\handcolor\left[D\ {\color{black}\stexttt{\cyrtt{kh\ V}}}\right]$ 
%- v dannom sluchae, v otlichie ot kristal-

лоптики, совпадает с направлением волнового вектора.
%loptiki, sovpadaet s napravleniem volnovogo vektora.

Кроме этого, векторы $\handcolor D$ и Н отличаются от
$\handcolor E\ \textnormal{\it и}\ B$\ .\\
%Krome \protect{\`{e}}togo, vektory $\handcolor D$ i N otlichayut{s}ya ot
%$\handcolor E\ {\it\cyrit{i}}\ B$\ .\\

----------------------------------------

I/ Что, впрочем, можно было сказать и заранее исходя из равен-
%{\tt I}/ Chto, vprochem, mozhno bylo skazat\cprime\ i zaranee
%iskhodya iz raven-

\ \ \ ства
%stva 
$\handcolor div\; D = 0$

\pagebreak
\refstepcounter{ppage}

наличием небольших аддитивных добавок, в которых есть как посто-
%nalichiem nebol\cprime shikh additivnykh dobavok, v kotorykh est\cprime\
%kak posto-

янная, так и периодические в {\ith x и t}
члены{\ith ,}  причем среди этих по-
%yannaya, tak i periodicheskie v {\ith x} {\ith\cyrit{i}} {\ith t} 
%chleny{\ith\cyrit{,}} prichem sredi \protect{\`{e}}tikh po-

следних имеются {\ith и} члены с удвоенной частотой. Отсюда может
%slednikh imeyut{s}ya {\ith\cyrit i} chleny s udvoenno\u i
%chastoto\u i. Ot{s}yuda mozhet

показаться, что в данном случае вообще не существует таких ре-
%pokazat\cprime sya, chto v dannom sluchae voobshche ne sushchestvuet
%takikh re-

шений уравнений Борна, для которых не компоненты
$\handcolor E\ \textnormal{\it и}\ B\ ,$ а ко{\ith м}-
%sheni\u i uravneni\u i Borna, dlya kotorykh ne komponenty 
%$\handcolor E\ {\it\cyrit{i}}\ B\ ,$ a ko{\ith\cyrit{m}}-

поненты $\handcolor D$ и Н представлялись бы в виде плоских волн вида
/\ref{eq65}/.
%ponenty $\handcolor D$ i N predstavlyalis\cprime\ by v vide ploskikh voln vida
%/\ref{eq65}/.

Это обстоятельство представляется на первый взгляд несколько
%\protect{\`{E}}to obstoyatel\cprime stvo predstavlyaet{s}ya na
%pervy\u i vzglyad neskol\cprime ko 

странным, поскольку /как уже было указано в примечании в начале
%strannym, poskol\cprime ku /kak uzhe bylo ukazano v primechanii v 
%nachale 

этого \S / мы могли бы вообще строить всю теорию не на основе
%\protect{\`{e}}togo \S / my mogli by voobshche stroit\cprime\ vsyu
%teoriyu ne na osnove

векторов $\handcolor E\ \textnormal{\it и}\ B\ ,$ а на основе 
векторов $\handcolor D$ и Н, причем, в частности
%vektorov $\handcolor E\ {\it\cyrit{i}}\ B\ ,$ a na osnove 
%vektorov $\handcolor D$ i N, prichem, v chastnosti

\ для нашей задачи, все вычисления при этом весьма мало чем от-
%\ dlya nashe\u i zadachi, vse vychisleniya pri \protect{\`{e}}tom 
%ves\cprime ma malo chem ot-

личались бы от вышеизложенных. Для того, чтобы разобраться в
%lichalis\cprime\ by ot vysheizlozhennykh. Dlya togo, chtoby
%razobrat\cprime sya v

этом вопросе, необходимо, однако, учесть следующее.
%\protect{\`{e}}tom voprose, neobkhodimo, odnako, uchest\cprime\ sleduyushchee.

Представим себе, что мы нашли какое нибудь решение уравнений
%Predstavim sebe, chto my nashli kakoe nibud\cprime\ reshenie uravneni\u i 

Борна, справедливое в известном приближении, например, подобно
%Borna, spravedlivoe v izvestnom priblizhenii, naprimer, podobno

нашему решению/ \ref{eq65}/, справедливое с точностью до членов, имею-
%nashemu resheniyu/ \ref{eq65}/,  spravedlivoe s tochnost\cprime yu do
%chlenov, imeyu-

щих в знаменателе
%shchikh v znamenatele 
${\handcolor b}^{\mbox{\large\stexttt{2}}}$.
Пусть это решение определяется вектора- 
%Pust\cprime\ \protect{\`{e}}to reshenie opredelyaet{s}ya vektora-

ми $\handcolor E^0\ \textnormal{\it и}\ B^0$ или, в представлении векторов 
индукции, вектора-
%mi $\handcolor E^0\ {\it\cyrit{i}}\ B^0$ ili, v predstavlenii vektorov 
%induktsii, vektora-

ми
%mi 
${\handcolor D}^{\mbox{\large\tt о}}$, 
Н$^{\mbox{\large\tt о}}$.
Пусть , далее $\handcolor E_1\ \textnormal{\it и}\ B_1$ есть любое 
электромагнитное
%N$^{\mbox{\large\stexttt{\cyrtt{o}}}}$.
%Pust\cprime, dalee $\handcolor E_1\ {\it\cyrit{i}}\ B_1$ est\cprime\ lyuboe
%\protect{\`{e}}lektromagnitnoe

поле, обладающее следующими свойствами: /I/ оба 
вектора $\handcolor E_1\ \textnormal{\it и}\ B_1$
%pole, obladayushchee sleduyushchimi svo\u istvami: /{\tt I}/ oba
%vektora $\handcolor E_1\ {\it\cyrit{i}}\ B_1$

по величине мало, т.е.\ имеют в знаменателе
%po velichine maly, t.e.\ imeyut v znamenatele 
${\handcolor b}^{\mbox{\large\stexttt{2}}}$ 
и,следовательно,
%i,sledovatel\cprime no,

в нашем приближении $\handcolor D_1 = E_1$, и $\handcolor H_1 = B_1$\ ; 
/2/ это поле в
%v nashem priblizhenii $\handcolor D_1 = E_1$, i $\handcolor H_1 = B_1$\ ; 
%/2/ \protect{\`{e}}to pole v

нашем приближении является решением уравнений Максвелла, т.е.\
%nashem priblizhenii yavlyaet{s}ya resheniem uravneni\u i Maksvella, t.e.\

величины
%velichiny

\hspace{1cm}$\handcolor rot\; E_1\ 
+\ \frac{\displaystyle 1}{\displaystyle c}\,\dot{B_1}$
\hspace{2cm};\hspace{2cm}
$\handcolor rot\; B_1\ -\ \frac{\displaystyle 1}{\displaystyle c}\,\dot{E_1}$
\hspace{1cm}{\ith и т.д.}
%\hspace{1cm}{\ith\cyrit{i t.d.}}

суть величины, высшего порядка малости /имеют в знаменателе
%sut\cprime\ velichiny, vysshego poryadka malosti /imeyut v znamenatele 
${\handcolor b}^{\mbox{\large\stexttt{4}}}$

или
%ili 
${\handcolor b}^{\mbox{\large\stexttt{6}}}$ и т.д./ 
Тогда ясно, что сумма обоих этих полей, т.е.\
%i t.d./ 
%Togda yasno, chto summa oboikh \protect{\`{e}}tikh pole\u i, t.e.\

поле, изображаемое в векторах $\handcolor E\ \textnormal{\it и}\ B$ как
%pole, izobrazhaemoe v vektorakh  $\handcolor E\ {\it\cyrit{i}}\ B$ kak

\hspace{3cm}$\handcolor E_0 + E_1\hspace{1cm}\textnormal{\it и}\hspace{1cm}\ 
B_0 + B_1\hspace{1cm},$
%\hspace{3cm}$\handcolor E_0 + E_1\hspace{1cm}{\it\cyrit{i}}\hspace{1cm}\ 
%B_0 + B_1\hspace{1cm},$

а в векторах $\handcolor D$ и Н / в нашем приближении/ как
%a v vektorakh $\handcolor D$ i N / v nashem priblizhenii/ kak

\pagebreak
\refstepcounter{ppage}
\label{page48}

\hspace{3cm}$\handcolor D^0 + E_1\hspace{1cm}\textnormal{\it и}\hspace{1cm}\
H^0 + B_1$
%\hspace{3cm}$\handcolor D^0 + E_1\hspace{1cm}{\it\cyrit{i}}\hspace{1cm}\
%H^0 + B_1$

тоже будет, с точностью до величин высших порядков, решением
%tozhe budet, s tochnost\cprime yu do velichin vysshikh poryadkov,
%resheniem 

уравнений Борна /несмотря на их нелинейность/. Это утверждение
%uravneni\u i Borna /nesmotrya na ikh neline\u inost\cprime/.
%\protect{\`{E}}to utverzhdenie 

настолько очевидно, что не имеет смысла специально останавли-
%nastol\cprime ko ochevidno, chto ne imeet smysla spetsial\cprime no 
%ostanavli-

ваться на его доказательстве.
%vat\cprime sya na ego dokazatel\cprime stve.

Так вот, нетрудно убедиться, что упомянутые "добавки"\ к нашим
%Tak vot, netrudno ubedit\cprime sya, chto upomyanutye "dobavki" k nashim

полям $\handcolor D$ и Н, определяемые формулами /\ref{eq72}/-/\ref{eq77}/, 
как раз прина-
%polyam $\handcolor D$ i N, opredelyaemye formulami /\ref{eq72}/-/\ref{eq77}/,
%kak raz prina-

длежат к типу полей $\handcolor E_1\ \textnormal{\it и}\ B_1\ ,$  о которых 
только что шла речь.
%dlezhat k tipu pole\u i $\handcolor E_1\ {\it\cyrit{i}}\ B_1\ ,$ o kotorykh
%tol\cprime ko chto shla rech\cprime.

В самом деле, рассмотрим хотя бы поле
%V samom dele, rassmotrim khotya by pole
{\handcolor
\begin{eqnarray}
\refstepcounter{requation}
\label{eq78}
&\left.\begin{array}{rcl}
E_{1z}&=&D_{1z}\ =\ \frac{\displaystyle L L^\prime}{\displaystyle2 b^2}
\ N\ \sin\alpha\,
\cos 2\omega\left[t - \frac{\displaystyle x}{\displaystyle 2}
\left(\frac{\displaystyle 1}{\displaystyle c} 
+ \frac{\displaystyle 1}{\displaystyle v}\right)\right]
\\[0.3cm]
B_{1y}&=&H_{1y}\ =\ - \frac{\displaystyle L L^\prime}{\displaystyle 2 b^2}
\ N\ \sin\alpha\,
\cos 2\omega\left[t - \frac{\displaystyle x}{\displaystyle 2}
\left(\frac{\displaystyle 1}{\displaystyle c} 
+ \frac{\displaystyle 1}{\displaystyle v}\right)\right]
\end{array}\right\}&\
\end{eqnarray}
}Амплитуда этой волны сама по себе есть величина первого порядка
%Amplituda \protect{\`{e}}to\u i volny sama po sebe est\cprime\
%velichina pervogo poryadka

\vspace{-0.32cm}

\hspace{11.1cm}{\ith ур-м}
%{\ith\cyrit{ur-m}}

\vspace{-0.6cm}

\hspace{11.5cm}{\handcolor\underline{\ \hspace{0.2cm}\ }}

\vspace{-0.32cm}

малости; далее, оно, очевидно удовлетворяет
%malosti; dalee, ono, ochevidno udovletvoryaet 
\hspace{-0.2cm}{\ith\small /}\hspace{-0.2cm}
Максвелла с точно-
%Maksvella s tochno-

тью до величины второго порядка малости /т.е.\ величина, имеющих
%st\cprime yu do velichiny vtorogo poryadka malosti /t.e.\
%velichina, imeyushchikh

v znamenatele 
%v znamenatele 
${\handcolor b}^{\mbox{\large\stexttt{4}}}$ /. 
Следовательно, согласно только что сказан-
%Sledovatel\cprime no, soglasno tol\cprime ko chto skazan-

ному, мы можем спокойно вычесть из /\ref{eq65}/ выражение 
/\ref{eq78}/;в резуль-
%nomu, my mozhem spoko\u ino vychest\cprime\ iz /\ref{eq65}/ vyrazhenie
%/\ref{eq78}/;v rezul\cprime-

тате получим выражение, которое тоже будет решением уравнений
%tate poluchim vyrazhenie, kotoroe tozhe budet resheniem uravneni\u i

Борна с тойже степенью точности, что и /\ref{eq65}/. Ясно, что таким
%Borna s to\u izhe stepen\cprime yu tochnosti, chto i /\ref{eq65}/. Yasno, 
%chto takim

путем можно, в частности, построить и такое решение, в котором
%putem mozhno, v chastnosti, postroit\cprime\ i takoe reshenie, v kotorom

переменные компоненты $\handcolor D$ и Н представлялись бы в виде простых
%peremennye komponenty $\handcolor D$ i N predstavlyalis\cprime\ by v vide 
%prostykh

гармонических функций $\handcolor x\ \textnormal{\it и}\ t$.\\
%garmonicheskikh funktsii $\handcolor x\ {\it\cyrit{i}}\ t$.\\

\refstepcounter{r2subsection}
\label{glavaIIpara3}
\hspace{1cm}\S\ 3.\ 
"\udensdash{Рассеяние"$\ \atop\ $\hspace{-0.1cm}света на 
постоянном поле плоского конденсато\hspace{-0.05cm}\ }-
%"\udensdash{Rasseyanie"$\ \atop\ $\hspace{-0.1cm}sveta na
%postoyannom pole ploskogo kondensato\hspace{-0.05cm}\ }-

\hspace{2.2cm}
\udensdash{\ ра$\ \atop\ $\hspace{-0.33cm}.\ \ \hspace{-0.05cm}\ }
%\udensdash{\ ra$\ \atop\ $\hspace{-0.33cm}.\ \ \hspace{-0.05cm}\ }

\hspace{1cm}До сих пор мы стави{\ith ли}$\;$ себе задачей просто 
нахождение извест-
%Do sikh por my stavi{\ith\cyrit{li}}$\;$sebe zadache\u i prosto 
%nakhozhdenie izvest-

ного класса решений уравнений поля, не выясняя, в каких услови-
%nogo klassa resheni\u i uravneni\u i polya, ne vyyasnyaya, v kakikh uslovi-

ях то или иное из этих решений осуществляется фактически. По
%yakh to ili inoe iz \protect{\`{e}}tikh resheni\u i osushchestvlyaet{s}ya 
%fakticheski. Po-

ставим теперь вопрос более конкретно.
%stavim teper\cprime\ vopros bolee konkretno.

Представим себе, что в пространство между обкладками плоского
%Predstavim sebe, chto v prostranstvo mezhdu obkladkami ploskogo

\pagebreak
\refstepcounter{ppage}

конденсатора пускают световой луч. Что при этом произойдет{\ith ?}
%kondensatora puskayut svetovo\u i luch. Chto pri \protect{\`{e}}tom 
%proizo\u idet{\ith ?}

Ответ на этот вопрос должен быть , очевидно, 
дан$\;${\ith на}$\;$основе гра-
%Otvet na \protect{\`{e}}tot vopros dolzhen byt\cprime , ochevidno,
%dan$\;${\ith\cyrit{na}}$\;$osnove gra-

ничных условий электромагнитного поля.
%nichnykh uslovi\u i \protect{\`{e}}lektromagnitnogo polya.

Сформулируем, прежде всего, в чем должны заключаться эти
%Sformuliruem, prezhde vsego, v chem dolzhny 
%zaklyuchat\cprime sya \protect{\`{e}}ti 

граничные условия.
%granichnye usloviya.

Уравнения \udensdash{Борна\hspace{-0.05cm}\ }, как мы видели, имеют вид, 
аналогичный урав-
%Uravneniya \udensdash{Borna\hspace{-0.05cm}\ }, kak my videli, imeyut vid, 
%analogichny\u i urav-

нениям
\udensdash{Ма\hspace{-0.1cm}$^{\displaystyle\textnormal{\ith к}}$\hspace{-0.1cm}свел$\ \atop\ $\hspace{-0.33cm}ла\hspace{-0.05cm}\ }
для среды с диэлектрической постоянной и маг-
%neniyam 
%\udensdash{Ma\hspace{-0.1cm}$^{\displaystyle\ith\cyrit{k}}$\hspace{-0.1cm}svel$\ \atop\ $\hspace{-0.33cm}la\hspace{-0.05cm}\ }
%dlya sredy s di\protect{\`{e}}lektrichesko\u i postoyanno\u i i mag-

нитной проницаемостью отличными от единицы, но в отутствии
%nitno\u i pronitsaemost\cprime yu otlichnymi ot edinitsy, no v otut{s}tvii

зарядов и токов, т.е.
%zaryadov i tokov, t.e.
{\handcolor
\begin{eqnarray*}
rot\, E\ +\ \frac{1}{c}\,\dot{B}\ =\ 0\hspace{2cm}div\, B\ =\ 0
\\[0.3cm]
rot\, H\ -\ \frac{1}{c}\,\dot{D}\ =\ 0\hspace{2cm}div\, D\ =\ 0
\end{eqnarray*}
}Отсюда, по аналогии с обычной электродинамикой, можно сразу
%Ot{s}yuda, po analogii s obychno\u i \protect{\`{e}}lektrodinamiko\u i,
%mozhno srazu

заключить, что граничные условия должны иметь вид:
%zaklyuchit\cprime , chto granichnye usloviya dolzhny imet\cprime\ vid:
{\handcolor
\begin{eqnarray}
\refstepcounter{requation}
\label{eq79}
&\left.\begin{array}{rclcrcl}
E_{t1}&=&E_{t2}&\hspace{2.5cm}&B_{n1}&=&B_{n2}\hspace{1cm}\\[0.3cm]
H_{t1}&=&H_{t2}&\hspace{2.5cm}&D_{n1}&=&D_{n2}\hspace{1cm}
\end{array}\right\}&\
\end{eqnarray}
}где значки {\ith t и n} указывают тангенциальные и нормальные
%gde znachki {\ith t} {\ith\cyrit{i}} {\ith n} ukazyvayut 
%tangentsial\cprime nye i normal\cprime nye

компоненты векторов к поверхности раздела наших двух "сред"\protect{-}
%komponenty vektorov k poverkhnosti razdela nashikh dvukh
%"sred"-

вакуума и поля конденсатора.
%vakuuma i polya kondensatora.

Применим граничные условия /\ref{eq79}/ для ответа на поставленный во-
%Primenim granichnye usloviya /\ref{eq79}/ dlya otveta na postavlenny\u i vo-

прос. Предположим, для простоты, что луч света падает перпен-
%pros. Predpolozhim, dlya prostoty, chto luch sveta padaet perpen-

дикулярно к силовым линиям поля конденсатора. Тогда наш коор-
%dikulyarno k silovym liniyam polya kondensatora. Togda nash koor-

динатный трехгранник представится в виде
%dinatny\u i trekhgrannik predstavit{s}ya v vide

%
\begin{figure}[h]
\refstepcounter{rfigure}
\label{ris1}
\unitlength1.mm
\begin{picture}(150,50)
{\handcolor
\put(30,30){\line(1,0){70}}
\put(30,30){\line(0,1){20}}
\put(30,30){\line(-1,-1){20}}
\put(30,43){\line(1,0){48}}
\put(30,15){\line(1,0){52.5}}
\put(65,22){\line(0,1){17}}
\put(100,30){\line(-2,-1){3.6}}
\put(100,30){\line(-2,1){3.6}}
\put(30,50){\line(-1,-2){1.8}}
\put(30,50){\line(1,-2){1.8}}
\put(65,39){\line(-1,-2){1.8}}
\put(65,39){\line(1,-2){1.8}}
\put(10,10){\line(1,3){1.2}}
\put(10,10){\line(3,1){3.6}}
\put(30,43){\line(1,1){3.6}}
\put(32.5,43){\line(1,1){3.6}}
\put(35,43){\line(1,1){3.6}}
\put(37.5,43){\line(1,1){3.6}}
\put(40,43){\line(1,1){3.6}}
\put(42.5,43){\line(1,1){3.6}}
\put(45,43){\line(1,1){3.6}}
\put(47.5,43){\line(1,1){3.6}}
\put(50,43){\line(1,1){3.6}}
\put(52.5,43){\line(1,1){3.6}}
\put(55,43){\line(1,1){3.6}}
\put(57.5,43){\line(1,1){3.6}}
\put(60,43){\line(1,1){3.6}}
\put(62.5,43){\line(1,1){3.6}}
\put(65,43){\line(1,1){3.6}}
\put(67.5,43){\line(1,1){3.6}}
\put(70,43){\line(1,1){3.6}}
\put(72.5,43){\line(1,1){3.6}}
\put(75,43){\line(1,1){3.6}}
\put(32.5,15){\line(-1,-1){3.6}}
\put(35,15){\line(-1,-1){3.6}}
\put(37.5,15){\line(-1,-1){3.6}}
\put(40,15){\line(-1,-1){3.6}}
\put(42.5,15){\line(-1,-1){3.6}}
\put(45,15){\line(-1,-1){3.6}}
\put(47.5,15){\line(-1,-1){3.6}}
\put(50,15){\line(-1,-1){3.6}}
\put(52.5,15){\line(-1,-1){3.6}}
\put(55,15){\line(-1,-1){3.6}}
\put(57.5,15){\line(-1,-1){3.6}}
\put(60,15){\line(-1,-1){3.6}}
\put(62.5,15){\line(-1,-1){3.6}}
\put(65,15){\line(-1,-1){3.6}}
\put(67.5,15){\line(-1,-1){3.6}}
\put(70,15){\line(-1,-1){3.6}}
\put(72.5,15){\line(-1,-1){3.6}}
\put(75,15){\line(-1,-1){3.6}}
\put(77.5,15){\line(-1,-1){3.6}}
\put(80,15){\line(-1,-1){3.6}}
\put(82.5,15){\line(-1,-1){3.6}}
\put(70,0)\textnormal{\it Рис 1.}
\put(68,37){$N$}
\put(14,8){$z$}
\put(32,49){$y$}
\put(103,29){$x$}
}
\end{picture}
\end{figure}

Так как нормальные слагающие полей В и {\ith D} при этом отсутствуют,
%Tak kak normal\cprime nye slagayushchie pole\u i V i {\ith D}
%pri \protect{\`{e}}tom ot{s}ut{s}tvuyut,

\pagebreak
\refstepcounter{ppage}
\label{page50}

то граничные условия /\ref{eq79}/ сведутся к
%to granichnye usloviya /\ref{eq79}/ svedut{s}ya k
{\handcolor
\begin{eqnarray}
\refstepcounter{requation}
\label{eq80}
&\left.\begin{array}{rclcrcl}
E_{y1}&=&E_{y2}&\hspace{2.5cm}&H_{y1}&=&H_{y2}\hspace{1cm}\\[0.3cm]
E_{z1}&=&E_{z2}&\hspace{2.5cm}&H_{z1}&=&H_{z2}\hspace{1cm}
\end{array}\right\}&\
\end{eqnarray}
}где значек 1 соответствует полю в вакууме, а значек 2 -
%gde znachek {\tt 1} sootvet{s}tvuet polyu v vakuume, a znachek 2 -

в конденсаторе.
%v kondensatore.

Первое из этих условий включает в себя, в частности требова-
%Pervoe iz \protect{\`{e}}tikh uslovi\u i vklyuchaet v sebya, v
%chastnosti trebova-

ние того, чтобы электростатическое поле самого конденсатора не
%nie togo, chtoby \protect{\`{e}}lektrostaticheskoe pole samogo
%kondensatora ne

обрывалось сразу, а "сходило на нуль"\ постепенно, как это
%obryvalos\cprime\ srazu, a "skhodilo na nul\cprime" postepenno, kak
%\protect{\`{e}}to

должно быть и в обычной электростатике. Т.к.\ вычисление это-
%dolzhno byt\cprime\ i v obychno\u i \protect{\`{e}}lektrostatike.
%T.k.\ vychislenie \protect{\`{e}}to-

го спадания поля не представляет для нас никакого интереса,
%go spadaniya polya ne predstavlyaet dlya nas
%nikakogo interesa,

то мы будем выставлять граничные условия
\udensdash{толь$\ \atop\ $\hspace{-0.33cm}ко\hspace{-0.05cm}\ }
для поля све-
%to my budem vystavlyat\cprime\ granichnye usloviya
%\udensdash{tol\cprime$\ \atop\ $\hspace{-0.33cm}ko\hspace{-0.05cm}\ }
%dlya polya sve-

товой волны, не учитывая при этом поля конденсатора. Тогда
%tovo\u i volny, ne uchityvaya pri \protect{\`{e}}tom polya
%kondensatora. Togda

вычисления могут быть проведены без всяких трудностей.
%vychisleniya mogut byt\cprime\ provedeny bez vsyakikh trudnoste\u i.

\hspace{1cm}Выберем какую-нибудь одну из двух волн, образующих наше
%Vyberem kakuyu-nibud\cprime\ odnu iz dvukh voln,
%obrazuyushchikh nashe

решение, например, соответствующую "\ необыкновенному"\ лучу в
%reshenie, naprimer, sootvet{s}tvuyushchuyu " neobyknovennomu" luchu v

кристаллооптике. Граничные условия /\ref{eq80}/ для нее будут иметь
%kristallooptike. Granichnye usloviya /\ref{eq80}/ dlya nee budut imet\cprime\

вид
%vid:
{\handcolor
\begin{eqnarray}
\refstepcounter{requation}
\label{eq81}
E_{y1}&=&E_{y2}\hspace{1.5cm};\hspace{1.5cm}H_{z1}\ =\ H_{z2}
\end{eqnarray}
}Пусть в вакууме
%Pust\cprime\ v vakuume
{\handcolor
\begin{eqnarray}
\refstepcounter{requation}
\label{eq82}
&\left.\begin{array}{rcl}
E_{y1}&=&A\ \cos\omega\left(t -
\frac{\displaystyle x}{\displaystyle c}\right)\hspace{1cm}\\[0.3cm]
H_{z1}&=&A\ \cos\omega\left(t -
\frac{\displaystyle x}{\displaystyle c}\right)\hspace{1cm}\hspace{1cm}
\end{array}\right\}&\
\end{eqnarray}
}Мы знаем, что в конденсаторе поле волны характеризуется вели-
%My znaem, chto v kondensatore pole volny kharakterizuet{s}ya veli-

чинами
%chinami
{\handcolor
\begin{eqnarray}
\refstepcounter{requation}
\label{eq83}
&\left.\begin{array}{rcl}
\hspace{-1.5cm}E_{y2}&=&L\, \cos\omega\left(t -
\frac{\displaystyle x}{\displaystyle v}\right)\\[0.3cm]
\hspace{-1.5cm}H_{z2}&=&\frac{\displaystyle c}{\displaystyle v}\, L\,
\left(1 + \frac{\displaystyle N^2}{\displaystyle 2 b^2}\right)\,
\cos\omega\left(t -
\frac{\displaystyle x}{\displaystyle v}\right)
+ N\, \frac{\displaystyle L^2}{\displaystyle 2 b^2}\,
\cos 2\omega\left(t -
\frac{\displaystyle x}{\displaystyle v}\right)
+ N\, \frac{\displaystyle L^2}{\displaystyle 2 b^2}
\end{array}\right\}&
\end{eqnarray}
}

\pagebreak
\refstepcounter{ppage}

Выше мы видели, что решения уравнений Борна имеет смысл писать
%Vyshe my videli, chto resheniya uravneni\u i Borna imeet smysl pisat\cprime\

только с точностью до малых полей, являющихся, в нашем прибли-
%tol\cprime ko s tochnost\cprime yu do malykh pole\u i, yavlyayushchikhsya,
%v nashem pribli-

жении, решениями уравнений Максвелла для пустоты.
%zhenii, resheniyami uravneni\u i Maksvella dlya pustoty.

В частности, мы можем,поэтому, в /\ref{eq83}/ отбросить последний по-
%V chastnosti, my mozhem,po\protect{\`{e}}tomu, v /\ref{eq83}/ 
%otbrosit\cprime\ posledni\u i po-

стоянный член, т.к.\ он мал и является точным решением уравне-
%stoyanny\u i chlen, t.k.\ on mal i yavlyaet{s}ya tochnym resheniem
%uravne-

ний Максвелла. Просто отбросать и второй член в выражении для
%ni\u i Maksvella. Prosto otbrosat\cprime\ i vtoro\u i chlen v vyrazhenii dlya

$\handcolor H_{2z}$ мы,очевидно, не можем, т.к.\ переменная в
{\ith x и t} "чисто
%my,ochevidno, ne mozhem, t.k.\ peremennaya v {\ith x}
%{\ith\cyrit{i}} {\ith t} "chisto

магнитная"\ волна не является решением Максвелловых уравнений.
%magnitnaya" volna ne yavlyaet{s}ya resheniem Maksvellovykh uravneni\u i.

Это обстоятельство, на первый поверхностный взгляд, весьма
%\protect{\`{E}}to obstoyatel\cprime stvo, na pervy\u i poverkhnostny\u i 
%vzglyad, ves\cprime ma

затрудняет построение решения, удовлетворяющего краевом услови-
%zatrudnyaet postroenie resheniya, udovletvoryayushchego kraevym uslovi-

ям. В действительности, однако, такое решение можно построить
%yam. V de\u istvitel\cprime nosti, odnako, takoe reshenie mozhno
%postroit\cprime

весьма просто посредством следующего приема.
%ves\cprime ma prosto posredstvom sleduyushchego priema.

Прибавим к решению /\ref{eq83}/ следующее добавочное поле, удовлетво-
%Pribavim k resheniyu /\ref{eq83}/ sleduyushchee dobavochnoe pole, udovletvo-

ряющее обоим выставленными в конце
%ryayushchee oboim vystavlennymi v kontse 
\S\ {\tt\ref{glavaIIpara2}} условиям:
%usloviyam:
{\handcolor
\begin{eqnarray}
\refstepcounter{requation}
\label{eq84}
&\left.\begin{array}{rcl}
E_{y2}^\ast&=&-\ N\ \frac{\displaystyle L^2}{\displaystyle 4 b^2}\
\cos 2\omega\left(t -
\frac{\displaystyle x}{\displaystyle v}\right)\\[0.3cm]
H_{z2}^\ast&=&-\ N\ \frac{\displaystyle L^2}{\displaystyle 4 b^2}\
\cos 2\omega\left(t -
\frac{\displaystyle x}{\displaystyle v}\right)\hspace{0.8cm}
\end{array}\right\}&
\end{eqnarray}
}Тогда мы получим следующее решение.
%Togda my poluchim sleduyushchee reshenie.
{\handcolor
\begin{eqnarray}
\refstepcounter{requation}
\label{eq85}
&\left.\begin{array}{rcl}
E_{y2}&=&L\ \cos\omega\left(t -
\frac{\displaystyle x}{\displaystyle v}\right)
- N\ \frac{\displaystyle L^2}{\displaystyle 4 b^2}\
\cos 2\omega\left(t -
\frac{\displaystyle x}{\displaystyle v}\right)\\[0.3cm]
H_{z2}&=&\frac{\displaystyle c}{\displaystyle v}\, L\,
\left(1 + \frac{\displaystyle N^2}{\displaystyle 2 b^2}\right)\,
\cos\omega\left(t -
\frac{\displaystyle x}{\displaystyle v}\right)
+ N\ \frac{\displaystyle L^2}{\displaystyle 4 b^2}\  
\cos 2\omega\left(t -
\frac{\displaystyle x}{\displaystyle v}\right)
\end{array}\right\}&\ \ 
\end{eqnarray}
}Потребуем, теперь чтобы при x = o решение в вакууме совпада-
%Potrebuem, teper\cprime\ chtoby pri {\tt x = o} reshenie v vakuume sovpada-

ло с решением /\ref{eq85}/ в конденсаторе. Для этого достаточно ввести
%lo s resheniem /\ref{eq85}/ v kondensatore. Dlya \protect{\`{e}}togo
%dostatochno vvesti

две малых отражённых волны: одну с частот{\ith ой} 
$\handcolor\omega$ и вторую с
%dve malykh otrazhennykh volny: odnu s chastot{\ith\cyrit{o\u i}}
%$\handcolor\omega$ i vtoruyu s

удвоенной частотой $\handcolor 2\omega$. Решение в вакууме тогда, вместо
%udvoenno\u i chastoto\u i $\handcolor 2\omega$. Reshenie v vakuume togda, 
%vmesto

/\ref{eq82}/ будет
%budet

\pagebreak
\refstepcounter{ppage}

{\handcolor
\begin{eqnarray}
\refstepcounter{requation}
\label{eq86}
&\left.\begin{array}{rcl}
E_{y1}&=&A\ \cos\omega\left(t - \frac{\displaystyle x}{\displaystyle c}\right)
- m\ \cos\omega\left(t + \frac{\displaystyle x}{\displaystyle c}\right)
- N\ \frac{\displaystyle L^2}{\displaystyle 4\; b^2}\ 
\cos 2\omega\left(t + \frac{\displaystyle x}{\displaystyle c}\right)\\[0.3cm]
H_{z1}&=&A\ \cos\omega\left(t - \frac{\displaystyle x}{\displaystyle c}\right)
+ m\ \cos\omega\left(t + \frac{\displaystyle x}{\displaystyle c}\right)
+ N\ \frac{\displaystyle L^2}{\displaystyle 4\; b^2}\ 
\cos 2\omega\left(t + \frac{\displaystyle x}{\displaystyle c}\right)
\end{array}\right\}&\ \ \ \ 
\end{eqnarray}
}Приравнивая выражения /\ref{eq85}/ и /\ref{eq86}/, мы получим, далее
%Priravnivaya vyrazheniya /\ref{eq85}/ i /\ref{eq86}/, my poluchim, dalee
{\handcolor
\begin{eqnarray*}
A - m&=&L\\[0.3cm]
A + m&=&\frac{\displaystyle c}{\displaystyle v}\ L\ 
\left(1 + \frac{\displaystyle N^2}{\displaystyle 2 b^2}\right)
\end{eqnarray*}
}и, замечая что
%i, zamechaya chto
{\handcolor
\begin{eqnarray*}
\frac{\displaystyle c}{\displaystyle v}&=&1
+ \frac{\displaystyle N^2}{\displaystyle 2 b^2}
\end{eqnarray*}
}окончательно найдем
%okonchatel\cprime no na\u idem
{\handcolor
\begin{eqnarray}
\refstepcounter{requation}
\label{eq87}
&\left.\begin{array}{rcl}
L&=&A\ \left(1 - \frac{\displaystyle N^2}{\displaystyle 2 b^2}\right)\\[0.5cm]
m&=&A\ \frac{\displaystyle N^2}{\displaystyle 2 b^2}
\end{array}\right\}&\ \ 
\end{eqnarray}
}Таким образом, зная амплитуды волны в вакууме и напряжение поля
%Takim obrazom, znaya amplitudy volny v vakuume i napryazhenie polya

конденсатора, мы можем определить амплитуды "проходящих"\ и
%kondensatora, my mozhem opredelit\cprime\ amplitudy "prokhodyashchikh" i 

"отраженных"\ от этого поля волн.Задача решена.
%"otrazhennykh" ot \protect{\`{e}}togo polya voln.Zadacha reshena.

Полученные выражения показывают, что согласно теории Борна по-
%Poluchennye vyrazheniya pokazyvayut, chto soglasno teorii Borna po-

мимо указанного в
%mimo ukazannogo v 
\S\ {\tt\ref{glavaIIpara2}} эффекта искажения поля волны, проходящей
%\protect{\`{e}}ffekta iskazheniya 
%polya volny, prokhodyashche\u i

через конденсатор, должно иметь место также совершенно своеоб-
%cherez kondensator, dolzhno imet\cprime\ mesto takzhe sovershenno svoeob-

разное физическое явление, которое может быть охарактеризовано,
%raznoe fizicheskoe yavlenie, kotoroe mozhet byt\cprime\ 
%okharakterizovano,

как рассеяние света на постоянном поле конденсатора. В самом
%kak rasseyanie sveta na postoyannom pole kondensatora. V samom

деле, мы пришли к тому, что падающая на это поле световая волна
%dele, my prishli k tomu, chto padayushchaya na \protect{\`{e}}to pole
%svetovaya volna

с частотой $\handcolor\omega$ и амплитудой А дает начало двум проходящим
%s chastoto\u i $\handcolor\omega$ i amplitudo\u i A daet nachalo dvum 
%prokhodyashchim

волнам: одно{\ith й} с частотой 
$\handcolor \omega$ и амплитудой, почти равной\ \ \ А
%volnam: odno{\ith\cyrit{\u i}} s chastoto\u i 
%$\handcolor \omega$ i amplitudo\u i, pochti ravno\u i\ \ \  A

и второй - с частотой 
$\handcolor 2\omega$ и малой 
амплитудой /имеющей $\handcolor b^2$ в
%i vtoro\u i - s chastoto\u i
%$\handcolor 2\omega$ i malo\u i 
%amplitudo\u i /imeyushche\u i $\handcolor b^2$ v

знаменателе/ и кроме того двум малым \udensdash{отраженным\hspace{-0.05cm}\ }
волнам с часто-
%znamenatele/ i krome togo dvum malym \udensdash{otrazhennym\hspace{-0.05cm}\ }
%volnam s chasto-

тами $\handcolor\omega\ \textnormal{\it и}\ 2 \omega$. 
Принципиально, эти волны тоже могли бы быть
%tami $\handcolor\omega\ {\it\cyrit{i}}\ 2 \omega$. 
%Printsipial\cprime no, \protect{\`{e}}ti volny tozhe mogli by byt\cprime\ 

наблюдены.
%nablyudeny.

\pagebreak
\refstepcounter{ppage}
\label{page53}

Сам ход приведенных вычислений показывает, что та 
нео\hspace{-0.3cm}$^{\displaystyle\textnormal{\ith дно-}}$\hspace{-0.6cm}знач-
%Sam khod provedennykh vychisleni\u i pokazyvaet, chto ta 
%neo\hspace{-0.3cm}$^{\displaystyle\ith\cyrit{dno-}}$\hspace{-0.6cm}znach-

\hspace{-0.3cm}ность решения уравнений поля, с которой мы встретились в преды-
%nost\cprime\ resheniya uravneni\u i polya, s kotoro\u i my
%vstretilis\cprime\ v predy-

дущем параграфе, и которая заключается в том, что решения урав-
%dushchem paragrafe, i kotoraya zaklyuchaet{s}ya v tom, chto resheniya
%urav-

нений Борна имело смысл писать лишь с точностью до малых выраже-
%neni\u i Borna imelo smysl pisat\cprime\ lish\cprime\ s tochnost\cprime yu
%do malykh vyrazhe-

ний особого типа, совершенно исчезает при той конкретной поста-
%ni\u i osobogo tipa, sovershenno ischezaet pri to\u i konkretno\u i posta-

новке задачи, с которой мы имеем здесь дело. В каждом частном
%novke zadachi, s kotoro\u i my imeem zdes\cprime\ delo. V kazhdom chastnom 

случае мы можем, исходя из граничных условий совершенно однознач-
%sluchae my mozhem, iskhodya iz granichnykh uslovi\u i sovershenno
%odnoznach-

\hspace{-1.1cm}но\ \ сказать, какое из выражений указанного 
типа нужно прибавить к
%no\ \ skazat\cprime, kakoe iz vyrazheni\u i ukazannogo
%tipa nuzhno pribavit\cprime\ k

исходному решению$^{\mbox{\large\stexttt{I/}}}$.
%iskhodnomu resheniyu}$^{\mbox{\large\stexttt{I/}}}$. 

Вопрос об однозначности полученных здесь решений можно, однако,
%{\cyrtt Vopros ob odnoznachnosti poluchennykh zdes\cprime\ resheni\u i mozhno, 
%odnako,

поставить более широко - а именно, спросить насколько однозначен
%postavit\cprime\ bolee shiroko - a imenno, sprosit\cprime\ naskol\cprime ko
%odnoznachen

сам выбор нулевого приближения в виде /\ref{eq49}/, т.е.\ обязательно
%sam vybor nulevogo priblizheniya v vide /\ref{eq49}/, t.e.\ obyazatel\cprime no

ли считать компоненты поля "большой"\ проходящей волны гармо-
%li schitat\cprime\ komponenty polya "bol\cprime sho\u i"
%prokhodyashche\u i volny garmo-

ническими функциями координат и времени. Мы не беремся здесь
%nicheskimi funktsiyami koordinat i vremeni. My ne beremsya zdes\cprime\

дать математически строго решения этого вопроса и приведем лишь
%dat\cprime\ matematicheski strogo resheniya \protect{\`{e}}togo
%voprosa i privedem lish\cprime\ 

несколько соображений чисто физического характера, которые,
%neskol\cprime ko soobrazheni\u i chisto fizicheskogo kharaktera, kotorye,

впрочем, как нам кажется, \udensdash{по существу\hspace{-0.05cm}\ }
полностью решают вопрос.
%vprochem, kak nam kazhet{s}ya, \udensdash{po sushchestvu\hspace{-0.05cm}\ }
%polnost\cprime yu reshayut vopros.

\hspace{1cm}Пусть компоненты поля $\handcolor E_y\ \textnormal{\it и}\ B_z$
падающей волны в ваку-
%Pust\cprime\ komponenty polya $\handcolor E_y\ {\it\cyrit{i}}\ B_z$
%padayushche\u i volny v vaku-

уме имеют вид
%ume imeyut vid
{\handcolor
\begin{eqnarray*}
A\ \cos\omega\left(t - \frac{\displaystyle x}{\displaystyle c}\right)&=&
A\ \cos\omega t\ \cos\frac{\displaystyle \omega x}{\displaystyle c}
+ A\ \sin\omega t\ \sin\frac{\displaystyle \omega x}{\displaystyle c}
\end{eqnarray*}
}Ясно, что зависимость компонент"большой"\ проходящей волны от
%Yasno, chto zavisimost\cprime\ komponent"bol\cprime sho\u i"
%prokhodyashche\u i volny ot

\udensdash{времени\ \hspace{-0.05cm}\ } наверное должна носит
такой же самый характер, иначе
%\udensdash{vremeni\ \hspace{-0.05cm}\ } navernoe dolzhna nosit 
%tako\u i zhe samy\u i kharakter, inache

никак нельзя будет удовлетворить краевым условиям. О характере
%nikak nel\cprime zya budet udovletvorit\cprime\ kraevym usloviyam. O
%kharaktere

же зависимости этих компонент от координат на основе одних
%zhe zavisimosti \protect{\`{e}}tikh komponent ot koordinat na osnove odnikh

только краевых условий ничего сказать нельзы. Наиболее общий
%tol\cprime ko kraevykh uslovi\u i nichego skazat\cprime\ 
%nel\cprime zy. Naibolee obshchi\u i

{\ith Ansatz} для этих величин имеет, таким образом, вид
%dlya \protect{\`{e}}tikh velichin imeet, takim obrazom, vid\\

----------------------------------------

I/ Заметим, между прочим, что точное значение скорости распро-
%{\tt I/}\ Zametim, mezhdu prochim, chto tochnoe znachenie skorosti raspro-

страни\hspace{-0.25cm}{\ith /}ения 
каждой такой и малой волны, может быть, очевидно, полу-
%strani\hspace{-0.25cm}{\ith /}eniya
%kazhdo\u i tako\u i malo\u i volny, mozhet byt\cprime, ochevidno, polu-

чено только из рассмотрения второго приближения.
%cheno tol\cprime ko iz rassmotreniya vtorogo priblizheniya.

\pagebreak
\refstepcounter{ppage}

\setcounter{requation}{88}  % equationnumber 87-->88

{\handcolor
\begin{eqnarray}
\refstepcounter{requation}
\label{eq89}
&\left.\begin{array}{rcl}
E_y&=&F_1(x)\ \cos\omega t\ +\ F_2(x)\ \sin\omega t\\[0.3cm]
B_z&=&\Phi_1(x)\ \cos\omega t\ +\ \Phi_2(x)\ \sin\omega t
\end{array}\right\}&\ \ 
\end{eqnarray}
}Обозначим через $\handcolor F$ и $\handcolor\Phi$ максимальные значения фун-
%Oboznachim cherez $\handcolor F$ i $\handcolor\Phi$ maksimal\cprime nye 
%znacheniya fun-

кций $\handcolor F(x)$ и $\handcolor\Phi(x)$. Тогда, очевидно, мы можем предста-
%ktsi\u i $\handcolor F(x)$ i $\handcolor\Phi(x)$. Togda, ochevidno, my 
%mozhem predsta-

вить /\ref{eq89}/ в виде:
%vit\cprime\ /\ref{eq89}/ v vide:
{\handcolor
\begin{eqnarray*}
E_y&=&F_1\ \cos f_1(x)\ \cos\omega t
\ +\ F_2\ \sin f_2(x)\ \sin\omega t\\[0.3cm]
B_z&=&\Phi_1\ \cos\varphi_1(x)\ \cos\omega t
\ +\ \Phi_2\ \sin\varphi_2(x)\ \sin\omega t\ \ \ \ \ ,
\end{eqnarray*}
}где 
%gde
$\handcolor f_1(x)$, $\handcolor f_2(x)$, 
$\handcolor\varphi_1(x)\ \textnormal{\it и}\ \varphi_2(x)$ -
%$\handcolor\varphi_1(x)\ {\it\cyrit{i}}\ \varphi_2(x)$ -
некие, подобран
%nekie, podobran

ные соответствующим образом, новые функции
%nye sootvet{s}tvuyushchim obrazom, novye funktsii 
$\handcolor x$.

\hspace{1cm}Сделаем теперь следующее весьма естественное предполо-
%Sdelaem teper\cprime\ sleduyushchee ves\cprime ma
%estestvennoe predpolo-

жение{\ith ;} что при распространении вдоль оси ох, электромаг-
%zhenie{\ith\cyrit{;}} chto pri rasprostranenii vdol\cprime\ osi okh,
%\protect{\`{e}}lektromag-

нитная энергия нашей волны нигде не концентрируется и не
%nitnaya \protect{\`{e}}nergiya nashe\u i volny nigde ne kontsentriruet{s}ya
%i ne

рассеивается, т.е.\ что поток вектораПойн{\ith т}инга через единич-
%rasseivaet{s}ya, t.e.\ chto potok vektoraPo\u in{\ith\cyrit{t}}inga cherez
%edinich-

ную поверхность, поставленную в любом месте оси ох, перпен-
%nuyu poverkhnost\cprime , postavlennuyu v lyubom meste osi okh,
%perpen-

дикулярно к ней, в среднем по времени не зависит от х\ .
%dikulyarno k ne\u i, v srednem po vremeni ne zavisit ot kh\ .

Иными словами, оперируя все время с нулевым приближением,
%Inymi slovami, operiruya vse vremya s nulevym priblizheniem,

потребуем, чтобы среднее по времени от произведения
%potrebuem, chtoby srednee po vremeni ot proizvedeniya 
$\handcolor E_y B_z$

не зависело бы и от х\ , т.е.\ чтобы
%ne zaviselo by i ot kh\ , t.e.\ chtoby
{\handcolor
\begin{eqnarray*}
\overline{E_y B_z}^{\ t}&=&
\frac{1}{2}\left(F_1\ \Phi_1\ \cos f_1(x) \cos\varphi_1
+ F_2\ \Phi_2\ \sin f_2(x) \sin\varphi_2\right)\ =\ const
\end{eqnarray*}
}Очевидно это может иметь место только при
%Ochevidno \protect{\`{e}}to mozhet imet\cprime\ mesto tol\cprime ko pri 
{\handcolor
\begin{eqnarray*}
F_1\ \Phi_1&=&F_2\ \Phi_2\\[0.3cm]
\textnormal{\ith и}
%{\ith\cyrit{i}}
\hspace{1.5cm}f_1(x)\ =&f_2(x)&=\varphi_1(x)\ =\ 
\varphi_2(x)\ =\ f(x)
\end{eqnarray*}
}Таким образом
%Takim obrazom
{\handcolor
\begin{eqnarray}
\refstepcounter{requation}
\label{eq90}
&\left.\begin{array}{rcl}
E_y&=&F_1(x)\ \cos f(x) \cos\omega t\ +\ F_2(x)\ \sin f(x) \sin\omega t\\[0.3cm]
B_z&=&\Phi_1(x)\ \cos f(x) \cos\omega t\ +\ \Phi_2(x)\ \sin f(x) \sin\omega t
\end{array}\right\}&\ \ 
\end{eqnarray}
}Подставляя /\ref{eq90}/ в первое уравнение Борна
%Podstavlaya /\ref{eq90}/ v pervoe uravnenie Borna 
$\handcolor\frac{\displaystyle\partial E_y}{\displaystyle\partial x} 
+ \frac{\displaystyle 1}{\displaystyle c}\;\dot{B}_z = 0$

\pagebreak
\refstepcounter{ppage}

мы найдем, что
%my na\u idem, chto
{\handcolor
\begin{eqnarray*}
f^\prime(x)&=&const
\end{eqnarray*}
}т.е.\ что $\handcolor f(x)$ есть линейная функция $\handcolor x$\ . Тем
%t.e.\ chto $\handcolor f(x)$ est\cprime\ line\u inaya funktsiya 
%$\handcolor x$\ . Tem

самым сразу же приходит к решению типа
%samym srazu zhe prikhodim k resheniyu tipa 
/\ref{eq49}/\ .\\

\refstepcounter{r2subsection}
\label{glavaIIpara4}
\hspace{1cm}\udensdash{\S\ 4.\ Плоская световая волна 
в однородном магнитном поле.\hspace{-0.05cm}\ }\\[-0.2cm]
%Ploskaya svetovaya volna 
%v odnorodnom magnitnom pole.\hspace{-0.05cm}\ }\\[-0.2cm]

\hspace{1cm}Рассмотрим здесь задачу, аналогичную той, которая была
%Rassmotrim zdes\cprime\ zadachu, analogichnuyu to\u i,
%kotoraya byla

разобрана в
%razobrana v 
\S\ {\tt\ref{glavaIIpara2}},
с той лишь разницей, что вместо электри-
%s to\u i lish\cprime\ raznitse\u i, 
%chto vmesto \protect{\`{e}}lektri-

ческого, будем иметь дело с однородным 
\udensdash{\ магнит$\ \atop\ $\hspace{-0.33cm}ным\ \hspace{-0.05cm}\ }
полем
%cheskogo, budem imet\cprime\ delo s odnorodnym 
%\udensdash{\ magnit$\ \atop\ $\hspace{-0.33cm}nym\ \hspace{-0.05cm}\ }
%polem

заполняющим все пространство; в этом пространстве, так-же
%zapolnyayushchim vse prostranstvo; v \protect{\`{e}}tom prostranstve, tak-zhe

как и раньше, пусть распространяется плоская световая волна.
%kak i ran\cprime she, pust\cprime\ rasprostranyaet{s}ya ploskaya
%svetovaya volna. 

\ Т.к.\ метод решения этой задачи, а так-же полученные резуль-
%\ T.k.\ metod resheniya \protect{\`{e}}to\u i zadachi, a tak-zhe 
%poluchennye rezul\cprime-

\ таты будут очень сходны с изложениями в предыдущих пара-
%\ taty budut ochen\cprime\ skhodny s izlozheniyami v predydushchikh para-

графах, то мы не будем останавливаться на их подробной диску
%grafakh, to my ne budem ostanavlivat\cprime sya na ikh podrobno\u i disku

ссии и приведем лишь вычисления.
%ssii i privedem lish\cprime\ vychisleniya.

Итак, нам нужно найти решения уравнений /\ref{eq50}/ и /\ref{eq51}/ для
%Itak, nam nuzhno na\u iti resheniya uravneni\u i /\ref{eq50}/ i 
%/\ref{eq51}/ dlya

нашей задачи.
%nashe\u i zadachi.

Выбираем, как и раньше, направление волнового вектора за
%Vybiraem, kak i ran\cprime she, napravlenie volnovogo vektora za

ос ох\ \ и плоскость, образованную этим вектором и внеш-
%os\cprime\ okh\ \  i ploskost\cprime, obrazovannuyu \protect{\`{e}}tim
%vektorom i vnesh-

ним полем $\handcolor h$ за плоскость $\handcolor xy$. Угол между 
этими вектора-
%nim polem $\handcolor h$ za ploskost\cprime\ $\handcolor xy$. Ugol mezhdu 
%\protect{\`{e}}timi vektora-

ми обозначаем через
%mi oboznachaem cherez 
$\handcolor\alpha$.

Будем искать решение уравнений Борна /\ref{eq50}/ и
%Budem iskat\cprime\ reshenie uravneni\u i Borna /\ref{eq50}/ i 
/\ref{eq51}/

в виде:
%v vide:
{\handcolor
\begin{eqnarray}
\refstepcounter{requation}
\label{eq91}
\ \hspace{-1.5cm}&\left.\begin{array}{rcl}
E_x&=&E^0_x\ \cos\omega_1\left(t - 
\frac{\displaystyle x}{\displaystyle v_1}\right)\hspace{0.7cm}
B_x\ =\ B^0_x\ \cos\nu_1\left(t - 
\frac{\displaystyle x}{\displaystyle w_1}\right)\ +\ h\ \cos\alpha\\[0.3cm]
E_y&=&E^0_y\ \cos\omega_2\left(t - 
\frac{\displaystyle x}{\displaystyle v_2}\right)\hspace{0.7cm}
B_y\ =\ B^0_y\ \cos\nu_2\left(t - 
\frac{\displaystyle x}{\displaystyle w_2}\right)\ +\ h\ \sin\alpha\\[0.3cm]
E_z&=&E^0_z\ \cos\omega_3\left(t - 
\frac{\displaystyle x}{\displaystyle v_3}\right)\hspace{0.7cm}
B_z\ =\ B^0_z\ \cos\nu_3\left(t - 
\frac{\displaystyle x}{\displaystyle w_3}\right)
\end{array}\right\}&
\end{eqnarray}
}

\pagebreak
\refstepcounter{ppage}

где, для краткости, отброшены начальные фазы.
%gde, dlya kratkosti, otbrosheny nachal\cprime nye fazy.

Подставим /\ref{eq91}/ в
%Podstavim /\ref{eq91}/ v 
/\ref{eq50}/.

Уравнение $\handcolor div\; B = 0$ дает нам
%Uravnenie $\handcolor div\; B = 0$ daet nam
{\handcolor
\begin{eqnarray}
\refstepcounter{requation}
\label{eq92}
B^0_x&=&0 
\end{eqnarray}
}Уравнение
%Uravnenie 
$\handcolor rot\; E + 
\frac{\displaystyle 1}{\displaystyle c}\;\dot{B}$ дает
%daet
{\handcolor
\begin{eqnarray}
\refstepcounter{requation}
\label{eq93}
&\left.\begin{array}{rcl}
\nu_2&=&\omega_3\ \ ;\ \ w_2\ =\ v_3\ \ ;\ \ 
B^0_y\ =\ -\ \frac{\displaystyle c}{\displaystyle v_3}\ E^0_z\\[0.3cm]
\nu_3&=&\omega_2\ \ ;\ \ w_3\ =\ v_2\ \ ;\ \ 
B^0_z\ =\ \frac{\displaystyle c}{\displaystyle v_2}\ E^0_y
\end{array}\right\}&\ \ 
\end{eqnarray}
}Принимая во внимание /\ref{eq92}/ и /\ref{eq93}/ и вводя 
обозначения, аналогичные
%Prinimaya vo vnimanie /\ref{eq92}/ i /\ref{eq93}/ i vvodya
%oboznacheniya, analogichnye

принятым в формулях /\ref{eq56}/, мы, вместо /\ref{eq91}/ получим
%prinyatym v formulyakh /\ref{eq56}/, my, vmesto /\ref{eq91}/ poluchim
{\handcolor
\begin{eqnarray}
\refstepcounter{requation}
\label{eq94}
\ \hspace{-1.5cm}&\left.\begin{array}{rcl}
E_x&=&p\ \cos\nu\left(t - 
\frac{\displaystyle x}{\displaystyle w}\right)\hspace{1.2cm}
B_x\ =\ h\ \cos\alpha\\[0.3cm]
E_y&=&L\ \cos\omega\left(t - 
\frac{\displaystyle x}{\displaystyle v}\right)\hspace{1.15cm}
B_y\ =\ M^\prime\ \cos\omega^\prime\left(t - 
\frac{\displaystyle x}{\displaystyle v^\prime}\right) + h\ \sin\alpha\\[0.3cm]
E_z&=&L^\prime\ \cos\omega^\prime\left(t - 
\frac{\displaystyle x}{\displaystyle v^\prime}\right)\hspace{0.7cm}
B_z\ =\ M\ \cos\omega\left(t - 
\frac{\displaystyle x}{\displaystyle v}\right)
\end{array}\right\}&\hspace{-0.2cm},\ \
\end{eqnarray}
}где
%gde
{\handcolor
\begin{eqnarray}
\refstepcounter{requation}
\label{eq95}
M&=&\frac{c}{v}\ L\hspace{1cm}\textnormal{\ith и}\hspace{1cm}
M^\prime\ =\ -\ \frac{c}{v^\prime}\ L^\prime\
\end{eqnarray}
}Подставим теперь /\ref{eq94}/ в /\ref{eq51}/. Для этого найдем сначала
%Podstavim teper\cprime\  /\ref{eq94}/ v /\ref{eq51}/. Dlya 
%\protect{\`{e}}togo na\u idem snachala 
$\handcolor F$:
{\handcolor
\begin{eqnarray}
\refstepcounter{requation}
\label{eq96}
F&=&\frac{1}{b^2}\ \left(B^2 - E^2\right)\ =\
\frac{1}{b^2}\ \left\{2h\ M^\prime\ \cos\omega^\prime\left(t - 
\frac{\displaystyle x}{\displaystyle v^\prime}\right)
\cdot\sin\alpha + h^2\right\}\ \ \ \ 
\end{eqnarray}
}Тогда
%Togda
{\handcolor
\begin{eqnarray}
\refstepcounter{requation}
\label{eq97}
\dot{F}&=&-\ \frac{\displaystyle 2h\ M^\prime 
\omega^\prime \sin\alpha}{\displaystyle b^2}\ \sin\omega^\prime\left(t - 
\frac{\displaystyle x}{\displaystyle v^\prime}\right)
\end{eqnarray}
}и
%i
{\handcolor
\begin{eqnarray}
\refstepcounter{requation}
\label{eq98}
\frac{\displaystyle\partial F}{\displaystyle\partial x}&=&
-\ \frac{1}{v^\prime}\;\dot{F}
\ =\ -\ \frac{1}{c}\;\dot{F}\ \ \ ,
\end{eqnarray}
}т.к.\ в правых частях уравнений /\ref{eq51}/ можно положить / в выражени-
%t.k.\ v pravykh chastyakh uravneni\u i /\ref{eq51}/ mozhno
%polozhit\cprime\ / v vyrazheni-

ях не стоящих под знаком
$\handcolor cos\ \textnormal{\it или}\ sin\ /\ v^\prime = c$.
%yakh ne stoyashchikh pod znakom 
%$\handcolor cos\ {\it\cyrit{ili}}\ sin\ /\ v^\prime = c$.

Вычисляем правые части уравнений /\ref{eq51}/ в первом приближении:
%Vychislyaem pravye chasti uravneni\u i /\ref{eq51}/ v pervom
%priblizhenii:
{\handcolor
\begin{eqnarray}
\refstepcounter{requation}
\label{eq99}
\frac{1}{c}\ j_x&=&-\ \frac{1}{2c}\ \dot{F} E_x\ =\ 0
\end{eqnarray}
}

\pagebreak
\refstepcounter{ppage}

{\handcolor
\begin{eqnarray}
\refstepcounter{requation}
\label{eq100}
\frac{1}{c}\ j_y&=&-\ \frac{1}{2}
\left(\frac{\partial F}{\partial x}\ B_z + \frac{1}{c}\;\dot{F}\ E_y\right)
\ =\ 0\\[0.3cm]
\refstepcounter{requation}
\label{eq101}
\frac{1}{c}\ j_z&=&\frac{1}{2}
\left(\frac{\partial F}{\partial x}\ B_y - \frac{1}{c}\;\dot{F}\ E_z\right)
\ =\ -\ \frac{1}{2c}\ \dot{F}\ h \sin\alpha\ =\nonumber\\[0.3cm]
&=&\frac{\omega^\prime h^2}{c\ b^2}\ M^\prime\ \sin^2\alpha\
\sin\omega^\prime\left(t - 
\frac{\displaystyle x}{\displaystyle v^\prime}\right)\\[0.3cm]
\refstepcounter{requation}
\label{eq102}
\rho&=&- \frac{1}{2c}\ \dot{F}\ E_y\ =\ 0\hspace{3cm}.
\end{eqnarray}
}Рассмотрим первое из уравнений /\ref{eq51}/. Оно дает
%Rassmotrim pervoe iz uravneni\u i /\ref{eq51}/. Ono daet
{\handcolor
\begin{eqnarray*}
-\ \frac{1}{c}\;\dot{E}_x\ = \ 0
\end{eqnarray*}
}Очевидно, это уравнение удовлетворяется, если положить
%Ochevidno, \protect{\`{e}}to uravnenie udovletvoryaet{s}ya, esli
%polozhit\cprime
{\handcolor
\begin{eqnarray}
\refstepcounter{requation}
\label{eq103}
p&=&0
\end{eqnarray}
}Второе уравнение
%Vtoroe uravnenie
{\handcolor
\begin{eqnarray*}
-\ \frac{\displaystyle\partial B_z}{\displaystyle\partial x}
\ -\ \frac{1}{c}\;\dot{E}_y\ &=&0
\end{eqnarray*}
}дает
%daet
{\handcolor
\begin{eqnarray*}
\frac{M}{v} \sin\omega \left(t - 
\frac{\displaystyle x}{\displaystyle v}\right)&=&
\frac{1}{c}\;L\ \sin\omega\left(t -
\frac{\displaystyle x}{\displaystyle v}\right)
\end{eqnarray*}
}откуда, в силу /\ref{eq95}/, находим
%otkuda, v silu /\ref{eq95}/, nakhodim
{\handcolor
\begin{eqnarray}
\refstepcounter{requation}
\label{eq104}
v&=&c\hspace{1cm};\hspace{1cm}M\ =\ L
\end{eqnarray}
}Третье уравнение
%Tret\cprime e uravnenie
{\handcolor
\begin{eqnarray*}
\frac{\displaystyle\partial B_y}{\displaystyle\partial x}
\ -\ \frac{1}{c}\;\dot{E}_z&=&\frac{1}{c}\; j_z
\end{eqnarray*}
}

\pagebreak
\refstepcounter{ppage}

дает
%daet
{\handcolor
\begin{eqnarray*}
\frac{M^\prime}{v^\prime}\ +\ 
\frac{L^\prime}{c}&=&
\frac{\displaystyle h^2\ M^\prime \sin^2\alpha}{\displaystyle c\ b^2} 
\end{eqnarray*}
}или, принимая во внимание
%ili, prinimaya vo vnimanie 
/\ref{eq95}/
{\handcolor
\begin{eqnarray}
\refstepcounter{requation}
\label{eq105}
\frac{\displaystyle v^{\prime\ 2}}{\displaystyle c^2}&=&
1\ -\ \frac{h^2}{b^2}\ \sin^2\alpha
\end{eqnarray}
}Уравнение
%Uravnenie 
$\handcolor div\; E = \rho$ 
дает
%daet
{\handcolor
\begin{eqnarray*}
\frac{\displaystyle\partial E_\alpha}{\displaystyle\partial x}&=&0\ \ \ ,
\end{eqnarray*}
}что автоматически удовлетворяется, т.к.\ мы при{\ith няли}
%chto avtomaticheski udovletvoryaet{s}ya, t.k.\ my 
%pri{\ith\cyrit{nyali}}
$\handcolor p = 0$\ .

Формулы /\ref{eq103}/, /\ref{eq104}/ и /\ref{eq105}/
позволяют окончательно пере-
%Formuly /\ref{eq103}/, /\ref{eq104}/ i  /\ref{eq105}/
%pozvolyayut okonchatel\cprime no pere-

писать решение уравнений /\ref{eq50}/ и /\ref{eq51}/ для нашей задачи в
%pisat\cprime\ reshenie uravneni\u i /\ref{eq50}/ i /\ref{eq51}/ dlya
%nashe\u i zadachi v

следующем виде:
%sleduyushchem vide:
{\handcolor
\begin{eqnarray}
\refstepcounter{requation}
\label{eq106}
\ \hspace{-1.5cm}&\left.\begin{array}{rcl}
E_x&=\ 0\hspace{3.4cm}
&B_x\ =\ h\ \cos\alpha\\[0.3cm]
E_y&=\ L\ \cos\omega\left(t - 
\frac{\displaystyle x}{\displaystyle c}\right)\hspace{0.75cm}
&B_y\ =\ M^\prime\ \cos\omega^\prime\left(t - 
\frac{\displaystyle x}{\displaystyle v^\prime}\right) + h\ \sin\alpha\\[0.3cm]
E_z&=\ L^\prime\ \cos\omega^\prime\left(t - 
\frac{\displaystyle x}{\displaystyle v^\prime}\right)\hspace{0.3cm}
&B_z\ =\ L\ \cos\omega\left(t - 
\frac{\displaystyle x}{\displaystyle c}\right)
\end{array}\right\}&\hspace{-0.2cm},\ \ \ 
\end{eqnarray}
}где
%gde
{\handcolor
\begin{eqnarray}
\refstepcounter{requation}
\label{eq107}
&\left.\begin{array}{rcl}
M^\prime&=&-\ \frac{\displaystyle c}{\displaystyle v^\prime}\ L^\prime\\[0.3cm]
\textnormal{\ith и}\hspace{1.5cm}
\frac{\displaystyle v^{\prime\ 2}}{\displaystyle c^2}&=&
1\ -\ \frac{\displaystyle h^2}{\displaystyle b^2}\ \sin^2\alpha
\end{array}\right\}&
\end{eqnarray}
}причем $\handcolor L,\ L^\prime,\ \omega\ \textnormal{\it и}\ \omega^\prime$
остаются, как и раньше про-
%prichem $\handcolor L,\ L^\prime,\ \omega\ {\it\cyrit{i}}\ \omega^\prime$
%ostayut{s}ya, kak i ran\cprime she pro-

извольными.
%izvol\cprime nymi.

Дискуссия формул /\ref{eq106}/ и /\ref{eq107}/ может быть проведена совер-
%Diskussiya formul /\ref{eq106}/ i /\ref{eq107}/ mozhet byt\cprime\
%provedena sover-

шенно аналогично дискуссии соответствующих формул предыду-
%shenno analogichno diskussii sootvet{s}tvuyushchikh formul predydu-

щих параграфов.
%shchikh paragrafov.

\pagebreak
\refstepcounter{ppage}

\refstepcounter{r2subsection}
\label{glavaIIpara5}
\hspace{1cm}\udensdash{\S\ 5.\ Две плоские волны 
в вакууме.\hspace{-0.05cm}\ }\\
%Dve ploskie volny v vakuume.\hspace{-0.05cm}\ }\\

\hspace{1cm}Перейдем теперь к рассмотрению более сложной задачи нахож-
%Pere\u idem teper\cprime\ k rassmotreniyu bolee 
%slozhno\u i zadachi nakhozh-

дения таких решений уравнений Борна, которые соответствовали
%deniya takikh resheni\u i uravneni\u i Borna, kotorye sootvet{s}tvovali

бы в нулевом приближении
\udensdash{двум плоским волнам в вакууме\hspace{-0.05cm}\ }, т.е.\
%by v nulevom priblizhenii 
%\udensdash{dvum ploskim volnam v vakuume\hspace{-0.05cm}\ }, t.e.\

будем считать, что одна плоская волна распространяется не в
%budem schitat\cprime , chto odna ploskaya volna rasprostranyaet{s}ya ne v

постоянном электрическом или магнитном поле, а в периодическом
%postoyannom \protect{\`{e}}lektricheskom ili magnitnom pole, a v
%periodicheskom

с $\handcolor r$ и $\handcolor t$ электрическом и магнитном 
поле другой плоской
%s $\handcolor r$ i $\handcolor t$ \protect{\`{e}}lektricheskom i magnitnom 
%pole drugo\u i plosko\u i

волны.
%volny.

Мы решим эту задачу только для того частного случая, когда вол-
%My reshim \protect{\`{e}}tu zadachu tol\cprime ko dlya togo 
%chastnogo sluchaya, kogda vol-

новые векторы обоих плоских волн имеют противоположные напра-
%novye vektory oboikh ploskikh voln imeyut protivopolozhnye napra-

вления, электрические векторы их в нулевом приближении имеют
%vleniya, \protect{\`{e}}lektricheskie vektory ikh v nulevom priblizhenii imeyut

одинаковое направление, а магнитные- противоположное. Эти
%odinakovoe napravlenie, a magnitnye- protivopolozhnoe. \protect{\`{E}}ti

три направления выберем за оси координат. Именно, возьмем на-
%tri napravleniya vyberem za osi koordinat. Imenno, voz\cprime mem na-

правление волнового вектора первой волны за ось ох, направле-
%pravlenie volnovogo vektora pervo\u i volny za os\cprime\ okh, napravle-

ние электрических векторов за ось {\ith oy} и направление магнитно-
%nie \protect{\`{e}}lektricheskikh vektorov za os\cprime\ {\ith oy} 
%i napravlenie magnitno-

го вектора первой волны-за {\ith oz}\ . /Рис.\ref{ris2}/.
%go vektora pervo\u i volny-za {\ith oz}\ . /Ris.\ref{ris2}/.

%
\begin{figure}[h]
\refstepcounter{rfigure}
\label{ris2}
\unitlength1.mm
\begin{picture}(150,50)
{\handcolor
\put(25,25){\line(1,0){90}}
\put(70,0){\line(0,1){50}}
\put(70,25){\line(1,1){20}}
\put(115,25){\line(-2,-1){3.6}}
\put(115,25){\line(-2,1){3.6}}
\put(101,25){\line(-2,-1){3.6}}
\put(101,25){\line(-2,1){3.6}}
\put(40,25){\line(2,-1){3.6}}
\put(40,25){\line(2,1){3.6}}
\put(70,50){\line(-1,-2){1.8}}
\put(70,50){\line(1,-2){1.8}}
\put(70,42){\line(-1,-2){1.8}}
\put(70,42){\line(1,-2){1.8}}
\put(70,11){\line(-1,2){1.8}}
\put(70,11){\line(1,2){1.8}}
\put(90,45){\line(-3,-1){3.6}}
\put(90,45){\line(-1,-3){1.2}}
\put(84,39){\line(-3,-1){3.6}}
\put(84,39){\line(-1,-3){1.2}}
\put(95,5)\textnormal {\it Рис 2.}
\put(65,26){$0$}
\put(72,49){$z$}
\put(93,44){$y$}
\put(117,24){$x$}
\put(63,40){$B_1$}
\put(86,36){$E_1 E_2$}
\put(40,20){$n_2$}
\put(100,20){$n_1$}
\put(72,10){$B_2$}
}
\end{picture}
\end{figure}

Попытаемся найти решение уравнений Борна для такого случая ме-
%Popytaemsya na\u iti reshenie uravneni\u i Borna dlya takogo sluchaya me-

тодом, аналогичным рассмотренному в предыдущих параграфах, учи-
%todom, analogichnym rassmotrennomu v predydushchikh paragrafakh, uchi-

тывая, однако еще возможность появления малых аддитивных до-
%tyvaya, odnako eshche vozmozhnost\cprime\ poyavleniya malykh additivnykh do-

бавок, играющих роль $\handcolor x^\prime$ в формуле 
/\ref{eq47}/.А именно, будем
%bavok, igrayushchikh rol\cprime\ $\handcolor x^\prime$ v formule 
%/\ref{eq47}/.A imenno, budem

искать решение уравнений /\ref{eq50}/ и /\ref{eq51}/ в виде
%iskat\cprime\ reshenie uravneni\u i /\ref{eq50}/ i /\ref{eq51}/ v vide 

{\handcolor
\begin{eqnarray}
\refstepcounter{requation}
\label{eq108}
\ \hspace{-1.2cm}&\left.\begin{array}{rcl}
E_x&=&0\;;\ E_y\ =\ Q\;\cos\Omega\left(t - 
\frac{\displaystyle x}{\displaystyle v}\right) +
q\;\cos\omega\left(t + 
\frac{\displaystyle x}{\displaystyle v}\right) + E^\prime\;;\
E_z\ =\ 0\\[0.3cm]
B_x&=&0\;;\ B_y\ =\ 0\;;\ B_z\ =\ \frac{\displaystyle c}{\displaystyle V}\;
Q\;\cos\Omega\left(t - 
\frac{\displaystyle x}{\displaystyle V}\right) -
\frac{\displaystyle c}{\displaystyle v}\
q\;\cos\omega\left(t + 
\frac{\displaystyle x}{\displaystyle v}\right) + B^\prime\hspace{-0.5cm}\
\end{array}\right\}.&\ \ \ \
\end{eqnarray}
}

\pagebreak
\refstepcounter{ppage}

При таком расписании$^{\displaystyle\ith 1)}$введено, во первых, то 
предложение, что
%Pri takom napisanii$^{\displaystyle\ith 1)}$vvedeno, vo pervykh, to
%predlozhenie, chto

несмотря{\ith на}"взаимодействие"\ наших волн, 
слагающие $\handcolor E_x,\ E_z,\ B_x\ \textnormal{\it и}\ B_y$
%nesmotrya{\ith\cyrit{na}}"vzaimode\u istvie" nashikh voln,
%slagayushchie $\handcolor E_x,\ E_z,\ B_x\ {\it\cyrit{i}}\ B_y$

можно считать равными нулю, как и в невозмущенной задаче. Это
%mozhno schitat\cprime\ ravnymi nulyu, kak i v nevozmushchenno\u i zadache. 
%\protect{\`{E}}to

предположение представляется вполне естественным из соображе-
%predpolozhenie predstavlyaet{s}ya vpolne estestvennym iz soobrazhe-

ний симметрии; строгое доказательство его законности будет
%ni\u i simmetrii; strogoe dokazatel\cprime stvo ego zakonnosti budet

заключаться в том, что {\ith выражениями} /\ref{eq108}/
можно-как то уви-
%zaklyuchat{s}ya v tom, chto {\ith\cyrit{vyrazheniyami}} /\ref{eq108}/
%mozhno-kak my uvi-

дим ниже- фактически удовлетворить уравнениям Борна. Во вторых,
%dim nizhe- fakticheski udovletvorit\cprime\ uravneniyam
%Borna. Vo vtorykh,

мы заранее предположили, что связь между амплитудами
$\handcolor E_y\ \textnormal{\it и}\ B_z$
%my zaranee predpolozhili, chto svyaz\cprime\ mezhdu amplitudami
%$\handcolor E_y\ {\it\cyrit{i}}\ B_z$

имеет обычный вид - что тоже, конечно, совершенно законно, -
%imeet obychn\u i vid - chto tozhe, konechno, sovershenno zakonno, -

поскольку эта связь определяется из первой группы уравнений
%poskol\cprime ku \protect{\`{e}}ta svyaz\cprime\ opredelyaet{s}ya iz
%pervo\u i gruppy uravneni\u i

поля, которая во всех задачах имеет одинаковый вид. Можно бы-
%polya, kotoraya vo vsekh zadachakh imeet odinakovy\u i vid. Mozhno by-

ло бы,впрочем, {\ith писать вместо} /\ref{eq108}/ и более общие
%lo by,vprochem, {\ith\cyrit{pisat\cprime\ vmesto}} /\ref{eq108}/ i bolee
%obshchie 

формулы типа /\ref{eq49}/, считая компоненты полей 
$\handcolor E\ \textnormal{\it и}\ B$ любыми
%formuly tipa /\ref{eq49}/, schitaya komponenty pole\u i 
%$\handcolor E\ {\it\cyrit{i}}\ B$ lyubymi

гармоническими функциями координат и времени, но не{\ith -}трудно
%garmonicheskimi funktsiyami koordinat i vremeni, no ne{\ith\cyrit{-}}trudno

убедиться, что такое усложнение написания не дало бы - по
%ubedit\cprime sya, chto takoe uslozhnenie napisaniya ne dalo by - po

крайней мере в нашей постановке вопроса - ничего по существу
%kra\u ine\u i mere v nashe\u i postanovke voprosa - nichego po sushchestvu

нового.
%novogo.

Итак, подставим /\ref{eq108}/ в /\ref{eq50}/. Уравнение
%Itak, podstavim /\ref{eq108}/ v /\ref{eq50}/. Uravnenie 
$\handcolor div\; B = 0$

удовлетворяется, если
%udovletvoryaet{s}ya, esli
{\handcolor
\begin{eqnarray*}
div\; B^\prime&=&0
\end{eqnarray*}
}\hspace{1cm}или
%ili
{\handcolor
\begin{eqnarray}
\refstepcounter{requation}
\label{eq109}
\frac{\displaystyle\partial B^\prime }{\displaystyle\partial z}&=&0
\end{eqnarray}
}Уравнение
%Uravnenie 
$\handcolor rot\; E + \frac{\displaystyle 1}{\displaystyle c}\;\dot{B}$ 
удовлетворится, если
%udovletvorit{s}ya, esli
{\handcolor
\begin{eqnarray}
\refstepcounter{requation}
\label{eq110}
\frac{\displaystyle\partial E^\prime }{\displaystyle\partial z}&=&0
\end{eqnarray}
}-------------------

I/\ Начальные фазы волн опять для простоты положены равными
%{\tt I}/\ Nachal\cprime nye fazy voln opyat\cprime\ dlya prostoty
%polozheny ravnymi

\hspace{0.8cm}нулю.
%nulyu.

\pagebreak
\refstepcounter{ppage}

и
%i

\vspace{-1.cm}

{\handcolor
\begin{eqnarray}
\refstepcounter{requation}
\label{eq111}
\frac{\displaystyle\partial E^\prime }{\displaystyle\partial x}
\ +\ \frac{\displaystyle 1}{\displaystyle c}\;\dot{B}&=&0
\end{eqnarray}
}Подставим теперь /\ref{eq108}/ в /\ref{eq51}/\ .
%Podstavim teper\cprime\ /\ref{eq108}/ v /\ref{eq51}/\ .

Для этого вычислим сначала величину $\handcolor F$ в первом прибли-
%Dlya \protect{\`{e}}togo vychislim snachala velichinu $\handcolor F$ v pervom 
%pribli- 

жении:
%zhenii:
{\handcolor
\begin{eqnarray}
\refstepcounter{requation}
\label{eq112}
F&=&-\ \frac{4}{b^2}\ Q\; q\;\cos\Gamma\; \cos\gamma\ \ \ ,
\end{eqnarray}
}где
%gde
{\handcolor
\begin{eqnarray}
\refstepcounter{requation}
\label{eq113}
&\left.\begin{array}{rcl}
\Gamma&=&\Omega \left(t - \frac{\displaystyle x}{\displaystyle V}\right)
\\[0.2cm]
\gamma&=&\omega \left(t + \frac{\displaystyle x}{\displaystyle v}\right)
\end{array}\right\}&
\end{eqnarray}
}Найдем {\handcolor$\dot{F}\ \textnormal{\it и}\ 
\frac{\displaystyle\partial F}{\displaystyle\partial x}\ :$
%Na\u idem {\handcolor$\dot{F}\ {\it\cyrit{i}}\ 
%\frac{\displaystyle\partial F}{\displaystyle\partial x}\ :$
%
\begin{eqnarray}
\refstepcounter{requation}
\label{eq114} 
\dot{F}&=&\frac{4}{b^2}\ Q\; q\;\left(\Omega\;\sin\Gamma\;\cos\gamma
+ \omega\;\cos\Gamma\;\sin\gamma\right)\\[0.1cm]
\refstepcounter{requation}
\label{eq115} 
\frac{\displaystyle\partial F}{\displaystyle\partial x}&=&
\frac{1}{c}\;\frac{4}{b^2}\ Q\; q\;\left(- \Omega\;\sin\Gamma\;\cos\gamma
+ \omega\;\cos\Gamma\;\sin\gamma\right)
\end{eqnarray}
}Вычислим правые части уравнений
%Vychislim pravye chasti uravneni\u i
 /\ref{eq51}/
{\handcolor
\begin{eqnarray}
\refstepcounter{requation}
\label{eq116} 
\frac{1}{c}\; j_x&=&\frac{1}{2c}\ \dot{F}\ E_x\ =\ 0\\[0.3cm]
\refstepcounter{requation}
\label{eq117} 
\frac{1}{c}\; j_y&=&-\ \frac{4\; Q\; q}{c\; b^2}\ 
\left(\Omega\; q\;\sin\Gamma\;\cos^2\gamma
+ \omega\; Q\; \sin\gamma\;\cos^2\Gamma\right)
\end{eqnarray}
}Замечая, что
%Zamechaya, chto
{\handcolor
\begin{eqnarray*}
\sin\Gamma\;\cos^2\gamma\ =\ \frac{1}{2}
\ \left[\sin\Gamma + \frac{1}{2}\;\sin\left(\Gamma + 2\gamma\right)
+ \frac{1}{2}\;\sin\left(\Gamma - 2\gamma\right)\right]\\[0.1cm]
\textnormal{\ith и}\hspace{0.8cm}
\sin\gamma\;\cos^2\Gamma\ =\ \frac{1}{2}
\ \left[\sin\gamma + \frac{1}{2}\;\sin\left(\gamma + 2\Gamma\right)
+ \frac{1}{2}\;\sin\left(\gamma - 2\Gamma\right)\right]
\end{eqnarray*}
}мы найдем
%my na\u idem
{\handcolor
\begin{eqnarray}
\refstepcounter{requation}
\label{eq118}
\frac{1}{c}\; j_y&=&-\ \frac{2}{c\; b^2}\ Q\; q\;\left\{q\;\Omega\
\left[\sin\Gamma + \frac{1}{2}\;\sin\left(\Gamma + 2\gamma\right)
+ \frac{1}{2}\;\sin\left(\Gamma - 2\gamma\right)\right] 
+\right.\nonumber\\[0.1cm]
&&\hspace{2.5cm}\left. + Q\;\omega\ 
\left[\sin\gamma + \frac{1}{2}\;\sin\left(\gamma + 2\Gamma\right)
+ \frac{1}{2}\;\sin\left(\gamma - 2\Gamma\right)\right]\right\}\ \ \ \ \ \ 
\end{eqnarray}
}Далее
%Dalee
{\handcolor
\begin{eqnarray}
\refstepcounter{requation}
\label{eq119}
\frac{1}{c}\; j_z&=&0
\end{eqnarray}
}

\pagebreak
\refstepcounter{ppage}

и наконец,
%i nakonets,
{\handcolor
\begin{eqnarray}
\refstepcounter{requation}
\label{eq120}
\rho&=&\frac{1}{2}\ 
\frac{\displaystyle\partial F}{\displaystyle\partial x}\ E_x\ =\ 0
\end{eqnarray}
}Из формулы / \ref{eq123}/ мы видим, что в этой задаче, в отличие
%Iz formuly / \ref{eq123}/ my vidim, chto v \protect{\`{e}}to\u i zadache, 
%v otlichie

от задач, рассмотренных ранее, в правых частях уравнений по-
%ot zadach, rassmotrennykh ranee, v pravykh chastyakh uravneni\u i po-

являются не только члены, содержащие
$\handcolor sin\;\Gamma\ \textnormal{\it и}\ sin\;\gamma\ ,$ кото-
%yavlyayut{s}ya ne tol\cprime ko chleny, soderzhashchie 
%$\handcolor sin\;\Gamma\ {\it\cyrit{i}}\ sin\;\gamma\ ,$ koto-

рые аналогичны члену с $\handcolor cos\;\omega t$ в формуле /\ref{eq46}/, 
дающему
%rye analogichny chlenu s $\handcolor cos\;\omega t$ v formule /\ref{eq46}/, 
%dayushchemu

"резонанс"\protect{,} но и члены с другими частотами. Они как-раз , и
%"rezonans", no i chleny s drugimi chastotami. Oni kak-raz , i

образуют правые части уравнений для определения
$\handcolor E^\prime\ \textnormal{\it и}\ B^\prime$.
%obrazuyut pravye chasti uravneni\u i dlya opredeleniya 
%$\handcolor E^\prime\ {\it\cyrit{i}}\ B^\prime$.

Зная вычисленные здесь правые части уравнений, мы можем те-
%Znaya vychislennye zdes\cprime\ pravye chasti uravneni\u i, my mozhem te-

перь подставить /\ref{eq108}/ в /\ref{eq51}/. 
Первое из уравнений /\ref{eq51}/
%per\cprime\ podstavit\cprime\  /\ref{eq108}/ v  /\ref{eq51}/.
%Pervoe iz uravneni\u i /\ref{eq51}/
{\handcolor
\begin{eqnarray*}
\frac{\displaystyle\partial B_z}{\displaystyle\partial y}\ -\ 
\frac{\displaystyle\partial B_y}{\displaystyle\partial z}\ -\
\frac{1}{c}\;\dot{E_x}&=&0
\end{eqnarray*}
}удовлетворяется, если положить
%udovletvoryaet{s}ya, esli polozhit\cprime
{\handcolor
\begin{eqnarray}
\refstepcounter{requation}
\label{eq121}
\frac{\displaystyle\partial B^\prime}{\displaystyle\partial y}&=&0
\end{eqnarray}
}Второе уравнение
%Vtoroe uravnenie
{\handcolor
\begin{eqnarray*}
\hspace{-0.5cm}\frac{\displaystyle\partial B_x}{\displaystyle\partial z} - 
\frac{\displaystyle\partial B_z}{\displaystyle\partial x} -
\frac{\displaystyle 1}{\displaystyle c}\;\dot{E_y}&=&
-\ \frac{\displaystyle 2}{\displaystyle c\; b^2}\ Q\; q\;\left\{q\;\Omega\;
\left[\sin\Gamma + \frac{1}{2}\;\sin\left(\Gamma + 2\gamma\right)
+ \frac{1}{2}\;\sin\left(\Gamma - 2\gamma\right)\right] 
+\right.\\[0.1cm]
&&\hspace{2cm}\left. +\ Q\;\omega\; 
\left[\sin\gamma + \frac{1}{2}\;\sin\left(\gamma + 2\Gamma\right)
+ \frac{1}{2}\;\sin\left(\gamma - 2\Gamma\right)\right]\right\}
\end{eqnarray*}
}удовлетворяется если
%udovletvoryaet{s}ya esli
{\handcolor
\begin{eqnarray}
\refstepcounter{requation}
\label{eq122}
-\ \frac{\displaystyle c\;\Omega}{\displaystyle V^2}\ Q&\sin\Gamma&-\ 
\frac{\displaystyle c\;\omega}{\displaystyle v^2}\; \sin\gamma\ +\
\frac{1}{c}\ Q\;\Omega\;\sin\Gamma\ +\
\frac{1}{c}\ q\;\omega\;\sin\gamma\ =\nonumber\\[0.3cm]
&&=\ -\ \frac{\displaystyle 2}{\displaystyle c\; b^2}\
Q\;q\ \left(q\;\Omega\;\sin\Gamma + Q\;\omega\;\sin\gamma\right)
\end{eqnarray}
}и
%i
{\handcolor
\begin{eqnarray}
\refstepcounter{requation}
\label{eq123}
\frac{\displaystyle\partial B^\prime}{\displaystyle\partial x}\ +\ 
\frac{1}{c}\;\dot{E^\prime}&=&
\frac{1}{c\; b^2}\ Q\; q\;\left\{q\;\Omega\;
\left[\sin\left(\Gamma + 2\gamma\right)
+ \sin\left(\Gamma - 2\gamma\right)\right] 
+\right.\nonumber\\[0.1cm]
&&\hspace{1.cm}\left. + Q\;\omega\; 
\left[\sin\left(\gamma + 2\Gamma\right)
+ \sin\left(\gamma - 2\Gamma\right)\right]\right\}
\end{eqnarray}
}

\pagebreak
\refstepcounter{ppage}

Уравнение /\ref{eq122}/ дает нам
%Uravnenie /\ref{eq122}/ daet nam
{\handcolor
\begin{eqnarray}
\refstepcounter{requation}
\label{eq124}
&\left.\begin{array}{rcl}
1\ -\ \frac{\displaystyle c^2}{\displaystyle V^2}&=&
-\ \frac{\displaystyle 2}{\displaystyle b^2}\ q^2\\[0.2cm]
1\ -\ \frac{\displaystyle c^2}{\displaystyle v^2}&=&
-\ \frac{\displaystyle 2}{\displaystyle b^2}\ Q^2
\end{array}\right\}&
\end{eqnarray}
}Третье уравнение
%Tret\cprime e uravnenie
{\handcolor
\begin{eqnarray*}
\frac{\displaystyle\partial B_y}{\displaystyle\partial x}\ -\ 
\frac{\displaystyle\partial B_x}{\displaystyle\partial y}\ -\
\frac{1}{c}\;\dot{E_z}&=&0
\end{eqnarray*}
}удовлетворяется автоматически .
%udovletvoryaet{s}ya avtomaticheski .

Наконец, последнее уравнение $\handcolor div\; E = \rho$ дает нам
%Nakonets, poslednee uravnenie $\handcolor div\; E = \rho$ daet nam
{\handcolor
\begin{eqnarray}
\refstepcounter{requation}
\label{eq125}
\frac{\displaystyle\partial E^\prime}{\displaystyle\partial y}&=&0
\end{eqnarray}
}Для окончательного решения задачи, нам осталось еще найти
%Dlya okonchatel\cprime nogo resheniya zadachi, nam ostalos\cprime\ eshche 
%na\u iti

аддитивные добавки $\handcolor E^\prime\ \textnormal{\it и}\ B^\prime$\ . Для их 
определения мы
%additivnye dobavki $\handcolor E^\prime\ {\it\cyrit{i}}\ B^\prime$\ . Dlya ikh
%opredeleniya my

получили систему уравнений
%poluchili sistemu uravneni\u i 
/\ref{eq109}/, /\ref{eq110}/, /\ref{eq111}/,
/\ref{eq121}/, /\ref{eq123}/

и
%i 
/\ref{eq125}/.

Будем искать решение этих уравнений в виде
%Budem iskat\cprime\ reshenie \protect{\`{e}}tikh uravneni\u i v vide
{\handcolor
\begin{eqnarray}
\refstepcounter{requation}
\label{eq126}
\ \hspace{-1.2cm}&\left.\begin{array}{rcl}
E^\prime&=&a_1\cos\left(\Gamma + 2\gamma\right) +
a_2\cos\left(\Gamma - 2\gamma\right) +
a_3\cos\left(\gamma + 2\Gamma\right) +
a_4\cos\left(\gamma - 2\Gamma\right)\\[0.3cm]
B^\prime&=&d_1\cos\left(\Gamma + 2\gamma\right) +
d_2\cos\left(\Gamma - 2\gamma\right) +
d_3\cos\left(\gamma + 2\Gamma\right) +
d_4\cos\left(\gamma - 2\Gamma\right)
\end{array}\hspace{-0.2cm}\right\},&\ \ \  
\end{eqnarray}
}где
%gde 
$\handcolor a_1\ ...\ d_4$ {\ith --} постоянные коэффициенты.
%postoyannye ko\protect{\`{e}}ffitsienty.

При этом, уравнения /\ref{eq109}/, /\ref{eq110}/,
/\ref{eq121}/ и /\ref{eq125}/ удовлетворяют-
%Pri \protect{\`{e}}tom, uravneniya /\ref{eq109}/, /\ref{eq110}/,
%/\ref{eq121}/ i /\ref{eq125}/ udovletvoryayut-

ся автоматически, а уравнения /\ref{eq111}/ и /\ref{eq123}/ дадут нам воз-
%sya avtomaticheski, a uravneniya /\ref{eq111}/ i /\ref{eq123}/ dadut nam voz-

можность определить коэффициенты в
%mozhnost\cprime\ opredelit\cprime\ ko\protect{\`{e}}ffitsienty v 
/\ref{eq126}/.

\hspace{1cm}Первое из этих уравнений дает нам
%Pervoe iz \protect{\`{e}}tikh uravneni\u i daet nam

\pagebreak
\refstepcounter{ppage}

{\handcolor
\begin{eqnarray}
\refstepcounter{requation}
\label{eq127}
&\left.\begin{array}{rcl}
a_1&=&\frac{\displaystyle\Omega + 2\omega}{\displaystyle\Omega - 2\omega}\ d_1
\hspace{1cm};\hspace{1cm}a_3\ =\ 
-\ \frac{\displaystyle\omega + 2\Omega}{\displaystyle\omega - 2\Omega}\ d_3
\\[0.3cm]
a_2&=&\frac{\displaystyle\Omega - 2\omega}{\displaystyle\Omega + 2\omega}\ d_2
\hspace{1cm};\hspace{1cm}a_4\ =\ 
-\ \frac{\displaystyle\omega - 2\Omega}{\displaystyle\omega + 2\Omega}\ d_4
\end{array}\right\}&
\end{eqnarray}
}и второе
%i vtoroe
{\handcolor
\begin{eqnarray}
\refstepcounter{requation}
\label{eq128}
&\left.\begin{array}{rcl}
a_1&=&\frac{\displaystyle\Omega - 2\omega}{\displaystyle\Omega + 2\omega}\ d_1
\ -\ \frac{\displaystyle 1}{\displaystyle b^2}\ Q\;q^2\
\frac{\displaystyle\Omega}{\displaystyle\Omega + 2\omega}\\[0.3cm]
a_2&=&\frac{\displaystyle\Omega + 2\omega}{\displaystyle\Omega - 2\omega}\ d_2
\ -\ \frac{\displaystyle 1}{\displaystyle b^2}\ Q\;q^2\
\frac{\displaystyle\Omega}{\displaystyle\Omega - 2\omega}\\[0.3cm]
a_3&=&-\ 
\frac{\displaystyle\omega - 2\Omega}{\displaystyle\omega + 2\Omega}\ d_3
\ -\ \frac{\displaystyle 1}{\displaystyle b^2}\ q\;Q^2\
\frac{\displaystyle\omega}{\displaystyle\omega + 2\Omega}\\[0.3cm]
a_4&=&-\ 
\frac{\displaystyle\omega + 2\Omega}{\displaystyle\omega - 2\Omega}\ d_4
\ -\ \frac{\displaystyle 1}{\displaystyle b^2}\ q\;Q^2\
\frac{\displaystyle\omega}{\displaystyle\omega - 2\Omega}
\end{array}\right\}&
\end{eqnarray}
}My получил{\ith и,}таким образом,систему восьми уравнений /\ref{eq127}/ и
%My poluchil{\ith\cyrit{i,}}takim obrazom,sistemu vos\cprime mi 
%uravneni\u i /\ref{eq127}/ i

/\ref{eq128}/ для определния
%dlya opredelniya 
$\handcolor a_1,\ ...\ a_4,d_1\ ...\ d_4$\ .

Решая эту систему уравнений, мы получим
%Reshaya \protect{\`{e}}tu sistemu uravneni\u i, my poluchim
{\handcolor
\begin{eqnarray}
\refstepcounter{requation}
\label{eq129}
&\left.\begin{array}{rcl}
a_1&=&-\ \frac{\displaystyle Q q^2}{\displaystyle 8 b^2}\ 
\frac{\displaystyle\Omega + 2\omega}{\displaystyle\omega}
\hspace{1cm};\hspace{1cm}
d_1\ =\ -\ \frac{\displaystyle Q q^2}{\displaystyle 8 b^2}\ 
\frac{\displaystyle\Omega - 2\omega}{\displaystyle\omega}\\[0.3cm]
a_2&=&\frac{\displaystyle Q q^2}{\displaystyle 8 b^2}\ 
\frac{\displaystyle\Omega - 2\omega}{\displaystyle\omega}
\hspace{1.6cm};\hspace{1.05cm}
d_2\ = \ \frac{\displaystyle Q q^2}{\displaystyle 8 b^2}\ 
\frac{\displaystyle\Omega + 2\omega}{\displaystyle\omega}\\[0.3cm]
a_3&=&-\ \frac{\displaystyle q Q^2}{\displaystyle 8 b^2}\ 
\frac{\displaystyle\omega + 2\Omega}{\displaystyle\Omega}
\hspace{1cm};\hspace{1cm}
d_3\ =\ \frac{\displaystyle q Q^2}{\displaystyle 8 b^2}\ 
\frac{\displaystyle\omega - 2\Omega}{\displaystyle\Omega}\\[0.3cm]
a_4&=&\frac{\displaystyle q Q^2}{\displaystyle 8 b^2}\ 
\frac{\displaystyle\omega - 2\Omega}{\displaystyle\Omega}
\hspace{1.6cm};\hspace{1.05cm}
d_4\ =\ -\ \frac{\displaystyle q Q^2}{\displaystyle 8 b^2}\ 
\frac{\displaystyle\omega + 2\Omega}{\displaystyle\Omega}
\end{array}\right\}&\ \ \ 
\end{eqnarray}
}

\pagebreak
\refstepcounter{ppage}

Мы имеем теперь полное решение нашей задачи, которое имеет вид
%My imeem teper\cprime\ polnoe reshenie nashe\u i zadachi, kotoroe imeet vid

/\ref{eq108}/, причем $\handcolor  V\ \textnormal{\it и}\ v$ определяются из 
формул /\ref{eq124}/,
%/\ref{eq108}/, prichem $\handcolor  V\ {\it\cyrit{i}}\ v$ opredelyayut{s}ya iz
%formul /\ref{eq124}/,

а $\handcolor E^\prime\ \textnormal{\it и}\ B^\prime$ могут быть найдены 
из формул /\ref{eq126}/ с коэффициен-
%a $\handcolor E^\prime\ {\it\cyrit{i}}\ B^\prime$ mogut byt\cprime\ na\u ideny 
%iz formul /\ref{eq126}/ s ko\protect{\`{e}}ffitsien-

тами
%tami 
/\ref{eq129}/.

Величины $\handcolor Q,\ q,\ \Omega\ \textnormal{\it и}\ \omega\ ,$ как и в 
теории Максвелла,
%Velichiny $\handcolor Q,\ q,\ \Omega\ {\it\cyrit{i}}\ \omega\ ,$ kak i v 
%teorii Maksvella, 

остаются произвольными.
%ostayut{s}ya proizvol\cprime nymi.

Полученное нами решение может быть охарактеризовано следующим
%Poluchennoe nami reshenie mozhet byt\cprime\ okharakterizovano sleduyushchim 

образом: если в пустом пространстве распространяются навстре-
%obrazom: esli v pustom prostranstve rasprostranyayut{s}ya navstre-

чу друг другу две плоские электромагнитные волны с произволь-
%chu drug drugu dve ploskie \protect{\`{e}}lektromagnitnye volny s
%proizvol\cprime-

но заданными амплитудами и частотами, причем их электрические
%no zadannymi amplitudami i chastotami, prichem ikh
%\protect{\`{e}}lektricheskie

и магнитые вектора в нулевом приближении ориентированы так,
%i magnitye vektora v nulevom priblizhenii orientirovany tak,

как указано на рис.\ \ref{ris2}., то
\udensdash{от$\ \atop\ $\hspace{-0.33cm}личие\hspace{-0.05cm}\ }
того поля, которое при
%kak ukazano na ris.\ \ref{ris2}., to 
%\udensdash{ot$\ \atop\ $\hspace{-0.33cm}lichie\hspace{-0.05cm}\ }
%togo polya, kotoroe pri

этом будет в пространстве по теории
\udensdash{Борна\hspace{-0.05cm}\ } от того, которое
%\protect{\`{e}}tom budet v prostranstve po teorii 
%\udensdash{Borna\hspace{-0.05cm}\ } ot togo, kotoroe

получается из теории
\udensdash{Макс$\ \atop\ $\hspace{-0.33cm}велла,\hspace{-0.05cm}\ }
сводится к двум моментам:\\
%poluchaet{s}ya iz teorii ``
%\udensdash{Maks$\ \atop\ $\hspace{-0.33cm}vella,\hspace{-0.05cm}\ }
%svodit{s}ya k dvum momentam:}\\

I/ к изменению скорости распространения этих волн, а так-же
%{\cyrtt\ k izmeneniyu skorosti rasprostraneniya \protect{\`{e}}tikh voln,
%a tak-zhe

к появлению разности амплитуд электрического и магнитного век-
%k poyavleniyu raznosti amplitud \protect{\`{e}}lektricheskogo i magnitnogo vek-

тора каждой волны.
%tora kazhdo\u i volny.

Этот эффект вполне аналогичен тому явлению "искажения"\ свето-
%\protect{\`{E}}tot \protect{\`{e}}ffekt vpolne analogichen tomu
%yavleniyu "iskazheniya" sveto-

вой волны, которое было получено нами в предыдущих параграфах.\\
%vo\u i volny, kotoroe bylo polucheno nami v predydushchikh paragrafakh.\\

2/ К появлению четырех малых "рассеянных"\ волн с другими часто-
%K poyavleniyu chetyrekh malykh "rasseyannykh" voln s drugimi chasto-

тами и скоростями распространения, вообще говоря сильно отлича-
%tami i skorostyami rasprostraneniya, voobshche govorya sil\cprime no otlicha-

ющимися от Максвелловских.
%yushchimisya ot Maksvellovskikh.

Этот эффект является специфическим для последней задачи и в
%\protect{\`{E}}tot \protect{\`{e}}ffekt yavlyaet{s}ya spetsificheskim dlya
%posledne\u i zadachi i v

известном смысле может быть охарактеризован, как"рассеяние
%izvestnom smysle mozhet byt\cprime\ okharakterizovan, kak"rasseyanie

света на свете".
%sveta na svete".

Однако, существенно отметить, что "\ взаимодействие"\ двух
%Odnako, sushchestvenno otmetit\cprime, chto " vzaimode\u istvie" dvukh

световых волн, высчитанное по теории Борна, отнюдь не может
%svetovykh voln, vyschitannoe po teorii Borna, otnyud\cprime\ ne mozhet

\pagebreak
\refstepcounter{ppage}

быть сведено к тому "рассеянию света на свете"\protect{,} о котором
%byt\cprime\ svedeno k tomu "rasseyaniyu sveta na svete", o kotorom

неоднократно упоминалось в литературе - именно, в силу суще-
%neodnokratno upominalos\cprime\ v literature - imenno, v silu sushche-

ствования эффектов первого из указанных выше типов; в самом
%stvovaniya \protect{\`{e}}ffektov pervogo iz ukazannykh vyshe tipov; v samom

деле каждая волна приобретает своеобразные свойства потому,
%dele kazhdaya volna priobretaet svoeobraznye svo\u istva potomu,

что распространяется в среде "поляризованной"\ под влиянием
%chto rasprostranyaet{s}ya v srede "polyarizovanno\u i" pod vliyaniem

другой волны.
%drugo\u i volny.

\hspace{1cm}В заключение, я хочу выразить глубокую благодарность проф.
%V zaklyuchenie, ya khochu vyrazit\cprime\ glubokuyu
%blagodarnost\cprime\ prof.

С.П.Шубину за руководство настоящей диссертацией, а так же за
%S.P.Shubinu za rukovodstvo nastoyashche\u i dissertatsie\u i, a tak zhe za

ряд весьма ценных советов и замечаний.
%ryad ves\cprime ma tsennykh sovetov i zamechani\u i.

\vspace{0.5cm}

\hspace{5cm}-------------

\vspace{0.5cm}

\hspace{10cm}Февраль
%Fevral\cprime\ }
I936 г.
%{\cyrtt\ g.

\pagebreak
\refstepcounter{ppage}
\label{page67}

\udensdash{ПРИ$\ \atop\ $\hspace{-0.33cm}МЕЧАНИЕ:\hspace{-0.05cm}\ }
%\udensdash{PRI$\ \atop\ $\hspace{-0.33cm}MECHANIE:\hspace{-0.05cm}\ }

\vspace{0.5cm}

\hspace{1cm}После того как настоящая диссертация была уже закончена
%Posle togo kak nastoyashchaya dissertatsiya byla uzhe zakonchena

/т.е.\ после февраля 1936 года/ в печати появился ряд работ,
%/t.e.\ posle fevralya 1936 goda/ v pechati poyavilsya ryad rabot,

связанных с некоторыми разбираемыми здесь вопросами.
%svyazannykh s nekotorymi razbiraemymi zdes\cprime\ voprosami.

\hspace{1cm}Некоторые результаты этих работ будут мною разобраны в до-
%Nekotorye rezul\cprime taty \protect{\`{e}}tikh rabot budut mnoyu
%razobrany v do-

полнении$^{\displaystyle\textnormal{\ith к}}$диссертации, которое в настоящее 
время составляется.
%polnenii$^{\displaystyle\ith\cyrit{k}}$dissertatsii, kotoroe v nastoyashchee
%vremya sostavlyaet{s}ya.

\end{otherlanguage}
%}
}

\newpage
\refstepcounter{ppage}

\newgeometry{top=5cm,textheight=22.5cm,textwidth=15cm}

\renewcommand{\thepage}{\theppage}
\chead{\rm - \theppage\ -\hspace{3cm}\ }

\ \\[5cm]
\begin{center}
{\color{gray}\large blank page\hspace{3cm}\ }
\end{center}

\newpage
\label{facsimile}
\refstepcounter{ppage}

\refstepcounter{section}
\addcontentsline{toc}{section}{Facsimile of the original thesis}

\ \\[5cm]

\begin{center}
{\bf\Large Facsimile of the original thesis}:
\end{center}

\vspace{2cm}
\begin{otherlanguage}{russian}
А.\ А.\ Смирнов
\end{otherlanguage}
%{\cyrrm A.\ A.\ Smirnov} 
[A.\ A.\ Smirnov]:

\begin{otherlanguage}{russian}
Применение электродинамики 
Борна к теории распространения света в 
электромагнитных полях
\end{otherlanguage}
%{\cyrrm
%Primenenie \protect{\`{e}}lektrodinamiki 
%Borna k teorii rasprostraneniya sveta v
%\protect{\`{e}}lektromagnitnykh polyakh} 
[Primenenie \`{e}lektrodinamiki 
Borna k teorii rasprostraneniya sveta v
\`{e}lektromagnitnykh polyakh]/[The application of the 
electrodynamics of Born to the theory
of the propagation of light in electromagnetic fields].
\begin{otherlanguage}{russian}
Кандидатская диссертация
\end{otherlanguage}
%{\cyrrm Kan\-di\-dat{s}\-kaya dis\-ser\-ta\-tsiya}
[Kandidat$\cdot$skaya dissertatsiya]/[Ph.D.\ thesis],
\begin{otherlanguage}{russian}
Московский государственный университет
\end{otherlanguage}
%{\cyrrm Moskovski\u i gosudarstvenny\u i universi\-tet}
[Moskovski\u\i\ gosudarstvenny\u\i\ universitet]/[Moscow
State University], Moscow, 1936, 67 pp.. [in Russian]

\newpage
\refstepcounter{ppage}

\ \\[5cm]
\begin{center}
{\color{gray}\large blank page\hspace{3cm}\ }
\end{center}

\pagebreak
\renewcommand{\thepage}{\arabic{page}}

% pp. 66, 67 on one leaf
\includepdf[pages=1-66,lastpage=66,scale=1,offset=0.3cm -1cm,
pagecommand={\thispagestyle{empty}}]{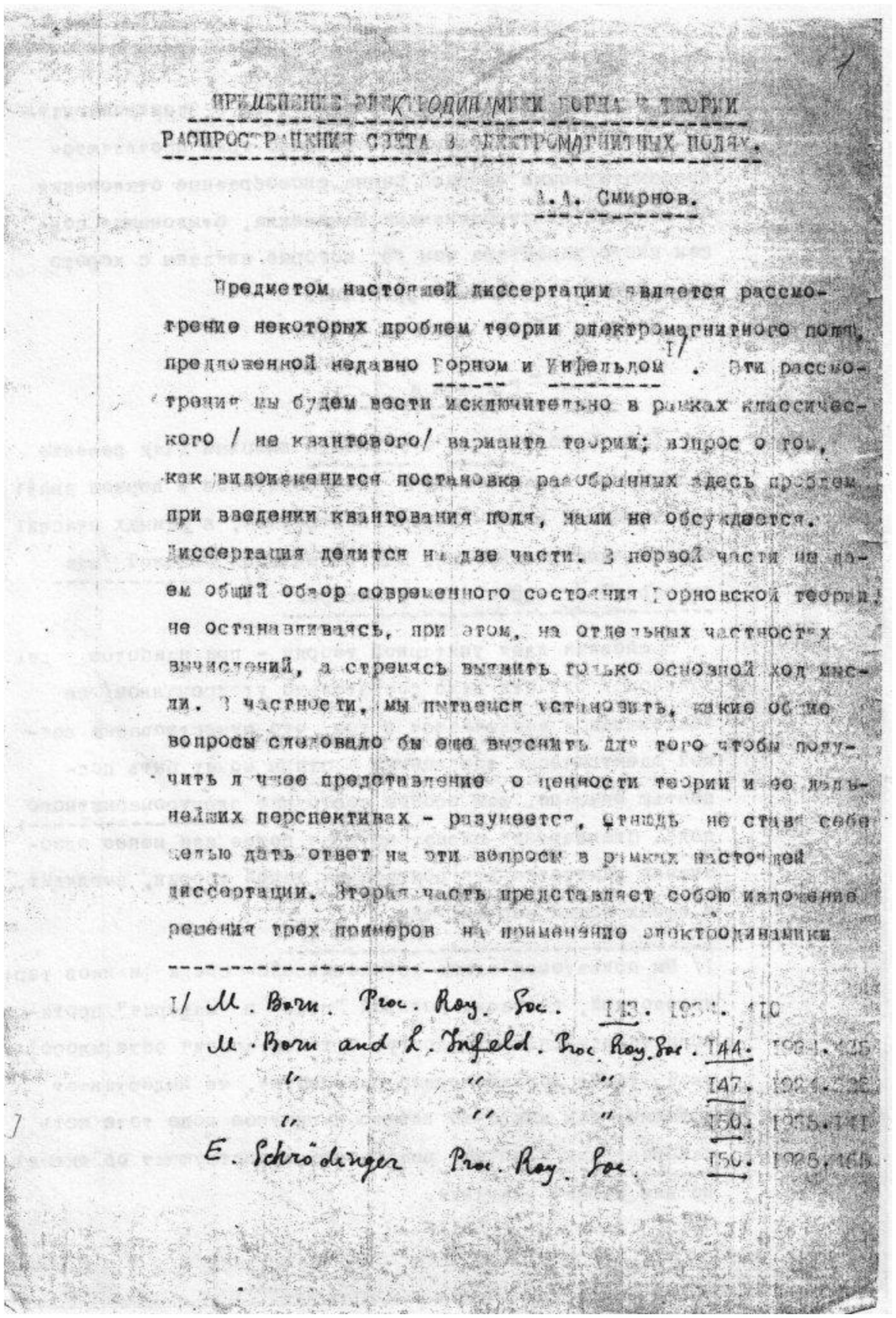}


\begin{thebibliography}{11}

\bibitem{1988smirnov}
{\it 
\begin{otherlanguage}{russian}
Адриан Анатольевич Смирнов
\end{otherlanguage}
%{\cyrit Adrian Anatol\cprime evich Smirnov} 
[Adrian Anatol'evich Smirnov]}.
\begin{otherlanguage}{russian}
Биобиблиография ученых Украинской ССР
\end{otherlanguage}
%{\cyrrm Biobibliografiya uchenykh Ukrainsko\u i SSR}
[Biobibliografiya uchenykh Ukrainsko\u\i\ SSR].
\begin{otherlanguage}{russian}
Сост.
\end{otherlanguage} 
%{\cyrrm Sost.} 
[Sost.]/[Compil.] 
\begin{otherlanguage}{russian}
М.\ Н.\ Верещак
\end{otherlanguage} 
%{\cyrrm M.\ N.\ Vereshchak} 
[M.\ N.\ Vereshchak],
\begin{otherlanguage}{russian}
авт.\ вступ.\ ст.
\end{otherlanguage}
%{\cyrrm avt.\ vstup.\ st.} 
[avt.\ vstup.\ st.]/[auth.\ of the introd.\ art.] 
\begin{otherlanguage}{russian}
В.\ Г.\ Барьяхтар
\end{otherlanguage}
%{\cyrrm V.\ G.\ Bar\cprime yakhtar} 
[V.\ G.\ Bar'yakhtar],
\begin{otherlanguage}{russian}
отв.\ ред.
\end{otherlanguage}
%{\cyrrm otv.\ red.} 
[otv.\ red.]/[main red.]
\begin{otherlanguage}{russian}
В.\ Б.\ Молодкин
\end{otherlanguage}
%{\cyrrm V.\ B.\ Molodkin} 
[V.\ B.\ Molodkin].
\begin{otherlanguage}{russian}
Наукова думка
\end{otherlanguage}
%{\cyrrm Naukova dumka} 
[Naukova dumka], Kiev, 1988.
%%CITATION = NONE;%%

\bibitem{1996smirnov}
\begin{otherlanguage}{russian}
Е.\ С.\ Юшкова-Смирнова
\end{otherlanguage}
%{\cyrrm E.\ S.\ Yushkova-Smirnova} 
[E.\ S.\ Yushkova-Smirnova]:
{\it 
\begin{otherlanguage}{russian}
Адриан Анатольевич Смирнов. Биографический очерк.
\end{otherlanguage}
%{\cyrit Adrian Anatol\cprime evich Smirnov. Biograficheski\u i ocherk.} 
[Adrian Anatol'evich Smirnov. Biograficheski\u\i\ ocherk]}.
Preprint 
\begin{otherlanguage}{russian}
ИМФ НАНУ
\end{otherlanguage}
%{\cyrrm IMF NANU} 
[IMF NANU], \textnumero\ 1.1996.
\begin{otherlanguage}{ukrainian}
Інститут металофізики
\end{otherlanguage}
%I{\cyrrm nstitut metalof\={\i}ziki}
[\={I}nstitut metalof\={\i}ziki]/[Institute of Metal Physics],
\begin{otherlanguage}{ukrainian}
Національна Академія наук України
\end{otherlanguage}  
%{\cyrrm Nats\={\i}onal\cprime na Akadem\={\i}ya nauk Ukra\={\i}ni}
[Nats\={\i}onal'na Akadem\={\i}ya nauk Ukra\={\i}ni]/[National Academy of
Sciences of the Ukraine], Kiev, 1996. 
%%CITATION = NONE;%%

\bibitem{shub1}
\begin{otherlanguage}{russian}
С.\ П.\ Шубин
\end{otherlanguage}
%{\cyrrm S.\ P.\ Shubin} 
[S.\ P.\ Shubin]:
{\it 
\begin{otherlanguage}{russian}
Избранные труды по теоретической физике. 
Очерки жизни. Воспоминания. Статьи
\end{otherlanguage}
%{\cyrit Izbrannye trudy po teoretichesko\u i fizike. 
%Ocherki zhizni. Vospominaniya. Stat\cprime i}
[Izbrannye trudy po teoreti\-chesko\u\i\ fizike. 
Ocherki zhizni. Vospominaniya. Stat'i]}.
Edited by 
\begin{otherlanguage}{russian}
С.\ В.\ Вонсовский
\end{otherlanguage}
%{\cyrrm S.\ V.\ Vonsovski\u i} 
[S.\ V.\ Vonsovski\u\i ], 
\begin{otherlanguage}{russian}
М.\ И.\ Кацнельсон
\end{otherlanguage}
%{\cyrrm M.\ I.\ Katsnel\cprime son} 
[M.\ I.\ Katsnel'son].
\begin{otherlanguage}{russian}
Инстиут физики металлов
\end{otherlanguage}
%{\cyrrm Insti\-tut fiziki metallov} 
[Institut fiziki metallov]/[Institute of Metal Physics],
\begin{otherlanguage}{russian}
Уральское отделение Академии наук СССР
\end{otherlanguage}
%{\cyrrm Ural\cprime skoe otdelenie Akademii nauk SSSR} 
[Ural'skoe otdelenie Akademii nauk SSSR]/[Ural Branch of the Academy of 
Sciences of the USSR], Sverd\-lovsk, 1991.
%%CITATION = NONE;%%

\bibitem{vons1}
\begin{otherlanguage}{russian}
С.\ В.\ Вонсовский
\end{otherlanguage}
%{\cyrrm S.\ V.\ Vonsovski\u i} 
[S.\ V.\ Vonsovski\u\i ],
\begin{otherlanguage}{russian}
М.\ А.\ Леонтович
\end{otherlanguage}
%{\cyrrm M.\ A.\ Leontovich} 
[M.\ A.\ Leontovich],
\begin{otherlanguage}{russian}
И.\ Е.\ Тамм
\end{otherlanguage}
%{\cyrrm I.\ E.\ Tamm} 
[I.\ E.\ Tamm]:
\begin{otherlanguage}{russian}
Семен Петрович Шубин (К пятидесятилетию 
со дня рождения и двадцатилетию со дня смерти
\end{otherlanguage}
%{\cyrrm Semen Petrovich Shubin (K pyatidesyatiletiyu
%so dnya rozhdeniya i dvadtsatiletiyu so dnya smerti)}
[Semen Petrovich Shubin (K pyatidesyatiletiyu
so dnya rozhdeniya i dvadtsatiletiyu so dnya smerti)]/[On the occasion of
the 50th birthday and the 20th day of death].
{\it
\begin{otherlanguage}{russian}
Успехи Физических Наук
\end{otherlanguage}
%{\cyrit Uspekhi Fizicheskikh Nauk} 
[Uspekhi Fizicheskikh Nauk]}
{\bf 65}:4(1958)733-737
(The article is freely available online at the journal website given by the
\href{http://dx.doi.org/10.3367/UFNr.0065.195808h.0733}{DOI: 
10.3367/UFNr.0065.195808h.0733} .). [in Russian]
%%CITATION = UFNAA,65,733;%%

\bibitem{zaio1}
\begin{otherlanguage}{russian}
П.\ А.\ Зайончковского
\end{otherlanguage}
%{\cyrrm P.\ A.\ Za\u ionchkovskogo} 
[P.\ A.\ Za\u\i onchkovskogo],
\begin{otherlanguage}{russian}
Э.\ А.\ Нерсесовой
\end{otherlanguage}
%{\cyrrm \protect{\`{E}}.\ A.\ Nersesovo\u i} 
[\`{E}.\ A.\ Nersesovo\u\i],
\begin{otherlanguage}{russian}
К.\ Р.\ Симона
\end{otherlanguage}
%{\cyrrm K.\ R.\ Simona} 
[K.\ R.\ Simona] 
\begin{otherlanguage}{russian}
(Ред.
\end{otherlanguage}
%({\cyrrm Red.} 
[Red.]/[Eds.]):
{\it
\begin{otherlanguage}{russian}
Докторские и кандидатские диссертации защищенные 
в Московском государственном университете с 1934 по 1954 г.. 
Библиографический указатель.
\end{otherlanguage}
%{\cyrit Doktorskie i kandidat${\it\cyrit{s}}$kie dissertatsii 
%zashchishchennye
%v Moskovskom gosudarstvennom universitete s 1934 po 1954 g..
%Bibliograficheski\u i ukazatel\cprime .} 
[Doktorskie i kandidat$\cdot$skie dissertatsii zashchishchennye
v Moskovskom gosudarstvennom universitete s 1934 po 1954 g..
Bibliograficheski\u\i\ ukazatel'.]}. 3 Vols..
Vol.\ 1: {\it
\begin{otherlanguage}{russian}
Выпуск первый: Факультеты: 
Механико-математический, физический, химический
\end{otherlanguage}
%{\cyrit Vypusk pervy\u i: Fakul\cprime tety: 
%Mekhaniko-matematicheski\u i, fizicheski\u i, khimicheski\u i}
[Vypusk pervy\u\i : Fakul'tety: Mekhaniko-matematicheski\u\i ,
fizicheski\u\i , khimicheski\u\i ]}.
\begin{otherlanguage}{russian}
Издательство Московского университета
\end{otherlanguage}
%{\cyrrm Izdatel\cprime stvo Moskovskogo universiteta}
[Izdatel'stvo Moskovskogo universiteta], Moscow, 1956.
%%CITATION = NONE;%%

\refstepcounter{ppage}

\bibitem{1940smirnov1}
A.\ A.\ Smirnov:
The problem of two plane waves in classical non-linear
electrodynamics.
{\it Journal of Physics (Academy of Sciences of the
\begin{otherlanguage}{russian}
USSR)/Физический Журнал (Академия Наук Союза ССР)
\end{otherlanguage}
%{\cyrit Fiziches\-ki\u i \protect{Zh}urnal (Akademiya Nauk Soyuza SSR)} 
[Fizicheski\u\i\ Zhurnal (Akademiya Nauk Soyuza SSR)]} 
{\bf 3}:6(1940)447-453 (The article
is a slightly revised English version of \cite{1941smirnov}.).
%%CITATION = JOPYA,3,447;%%

\bibitem{1941smirnov}
\begin{otherlanguage}{russian}
А.\ А.\ Смирнов
\end{otherlanguage}
%{\cyrrm A.\ A.\ Smirnov} 
[A.\ A.\ Smirnov]:
\begin{otherlanguage}{russian}
Задача двух плоских волн в классической 
нелинейной электродинамике
\end{otherlanguage}
%{\cyrrm Zadacha dvukh ploskikh voln v klassichesko\u i 
%neline\u ino\u i \protect{\`{e}}lektrodinamike}
[Zadacha dvukh ploskikh voln v klassichesko\u\i\ 
neline\u\i no\u\i\ \`{e}lektrodinamike]/[The problem of two plane 
waves in classical nonlinear electrodynamics].
{\it 
\begin{otherlanguage}{russian}
Ученые Записки Свердловского Государственного 
Университета им.\ А.\ М.\ Горького
\end{otherlanguage}
%{\cyrit Uchenye Zapiski Sverdlovskogo Gosudarstvennogo 
%Universiteta im.\ A.\ M.\ Gor\cprime kogo}
[Uchenye Zapiski Sverdlovskogo Gosudarstvennogo Universiteta 
im.\ A.\ M.\ Gor'kogo]} 
No.\ 3 (1941)44-56 (A slightly revised English translation of the
article is printed in \cite{1940smirnov1}). [in Russian]
%%CITATION = UUUGA,1941N3,44;%%
%%CITATION = UUUGA,3,44;%%

\bibitem{1963born}
M.\ Born: {\it Ausgew\"ahlte Abhandlungen. Mit einem
Verzeichnis der wissenschaft\-lichen Schriften. Zweiter Band}.
Vandenhoeck \& Ruprecht, G\"ottingen, 1963.
%%CITATION = NONE;%%

\bibitem{1978infeld}
E.\ Infeld, I.\ Bia\l ynicki-Birula, A.\ Trautman (Eds.): 
{\it Leopold Infeld, his Life
and Scientific Work}. Polish Men of Science. Polish Scientific Publishers,
Warszaw, 1978.
%%CITATION = NONE;%%

\bibitem{1984schroedinger}
E.\ Schr\"odinger: {\it Collected Papers/Gesammelte Abhandlungen.
Volume 2: Contributions to Field Theory/Band 2: Beitr\"age zur Feldtheorie}.
Verlag der \"Osterreichischen Akade\-mie der Wissenschaften, Friedrich Vieweg
\& Sohn Braunschweig/Wiesbaden, Wien, 1984.
%%CITATION = NONE;%%

\bibitem{1989heisenberg}
W.\ Blum, H.-P.\ D\"urr, H.\ Rechenberg (Eds.): 
{\it Werner Heisenberg --
Gesammelte Werke/Collected Works, Series A/Part II: Original Scientific
Papers -- Wissenschaftliche Originalarbeiten}. Springer-Verlag, Berlin, 1989
(\href{http://dx.doi.org/10.1007/978-3-642-70078-1}{DOI: 
10.1007/978-3-642-70078-1}).
%%CITATION = NONE;%%

\bibitem{1984heisenberg}
W.\ Blum, H.-P.\ D\"urr, H.\ Rechenberg (Eds.): 
{\it Werner Heisenberg --
Gesammelte Werke/Collected Works, Series B: Scientific Review Papers, 
Talks, and Books -- Wissenschaftliche \"Ubersichtsartikel, Vortr\"age 
und B\"ucher}. Springer-Verlag, Berlin, 1984
(\href{http://dx.doi.org/10.1007/978-3-642-61742-3}{DOI: 
10.1007/978-3-642-61742-3}).
%%CITATION = NONE;%%

\end{thebibliography}
\end{document}